\documentclass{article}
\usepackage{authblk}
\usepackage[T1]{fontenc}
\usepackage[utf8]{inputenc}
\usepackage{amsmath,amssymb,amsfonts,amsthm,mathtools,bm,mathrsfs}
\usepackage{graphicx,xcolor}
\usepackage[numbers,sort&compress,square]{natbib}
\usepackage[margin=1in]{geometry}
\usepackage[expansion=false]{microtype}
\usepackage{enumitem,booktabs,array,tabularx,environ}
\usepackage{tikz}
\usepackage{comment}
\usetikzlibrary{arrows.meta,calc,positioning,decorations.pathmorphing,backgrounds}
\usepackage{caption,subcaption,placeins,needspace}
\usepackage[linktocpage]{hyperref}
\hypersetup{colorlinks=true,citecolor=blue,linkcolor=blue,urlcolor=blue,
  breaklinks=true,
  pdftitle={Invertible subalgebras for chiral semions},
  pdfauthor={Yu-An Chen, Meng Sun, Nathanan Tantivasadakarn, Zongyuan Wang, and Bowen Yang}}
\NewEnviron{eqs}{
\begin{equation}\begin{split}
\BODY
\end{split}\end{equation}}
\makeatletter
\def\fps@figure{t}
\makeatother
\newtheorem{theorem}{Theorem}
\newtheorem{proposition}{Proposition}
\newtheorem{lemma}{Lemma}
\newtheorem{corollary}{Corollary}
\newcommand{\rvec}{\mathbf r}
\newcommand{\xhat}{\hat x}
\newcommand{\yhat}{\hat y}
\newcommand{\id}{\mathbf 1}
\newcommand{\Alg}{\operatorname{Alg}}
\newcommand{\Mat}{\operatorname{Mat}}
\newcommand{\supp}{\operatorname{supp}}
\newcommand{\CZ}{\mathrm{CZ}}
\newcommand{\CNOT}{\mathrm{CNOT}}
\newcommand{\SWAP}{\mathrm{SWAP}}
\newcommand{\Asalg}{\mathfrak A_s}
\newcommand{\Abalg}{\mathfrak A_{\bar s}}
\newcommand{\ww}{\omega_8}

\tikzset{
  every picture/.style={line width=.5pt,line cap=round,line join=round},
  gridline/.style={draw=black!40,line width=.5pt},
  edgefocus/.style={draw=black,line width=.58pt},
  op/.style={fill=white,inner sep=.35pt,text=blue,font=\small},
  oplabel/.style={op},
  dualpath/.style={draw=black,line width=.65pt}
}

\colorlet{Zcol}{blue}
\colorlet{Xcol}{blue}
\colorlet{loopcol}{black!70}
\colorlet{CNOTcol}{black}
\colorlet{origincol}{black}
\colorlet{gridgray}{black!35}
\colorlet{faceblue}{blue!60!cyan}

\title{Invertible subalgebras for chiral semion topological order}

\author[1,*]{Yu-An~Chen}
\author[1]{Meng~Sun}
\author[2]{Nathanan~Tantivasadakarn}
\author[1]{Zongyuan~Wang}
\author[3]{Bowen~Yang}

\affil[1]{International Center for Quantum Materials, School of Physics, Peking University, Beijing 100871, China}
\affil[2]{C. N. Yang Institute for Theoretical Physics, Stony Brook University, Stony Brook, NY 11794, USA}
\affil[3]{Center of Mathematical Sciences and Applications, Harvard University, Cambridge, Massachusetts 02138, USA}

\date{\today}

\begin{document}

\maketitle
\renewcommand*{\thefootnote}{*}
\footnotetext[1]{Corresponding author: \href{mailto:yuanchen@pku.edu.cn}{yuanchen@pku.edu.cn}.}
\renewcommand{\thefootnote}{\arabic{footnote}}

\begin{abstract}
Chiral topological order is incompatible with strictly two-dimensional local commuting-projector Hamiltonians on finite-dimensional tensor-product Hilbert spaces. Invertible subalgebras (ISAs) provide an alternative framework. We construct an ISA supporting chiral semions, together with its anti-semion commutant, on a square lattice with one qubit per edge. We give explicit non-Pauli generators for both algebras and prove that together they generate all local operators on the Hilbert space. Starting from the $\mathbb{Z}_4$ toric code, we condense the boson $e^2m^2$ and identify the semion and anti-semion algebras. We then decompose each $\mathbb{Z}_4$ qudit into two qubits and use a finite-depth circuit to disentangle one qubit per edge fixed by the condensation constraints. Pumping the semion ISA along the $z$ direction defines an explicit three-dimensional non-Clifford quantum cellular automaton (QCA). We show that pumping eight times is equivalent to a translation followed by a finite-depth quantum circuit with ancillary qubits. We also give an explicit finite-depth circuit relating this construction to the semion QCA previously obtained from topological quantum field theory. A similar condensation construction yields a $U(1)_4$ ISA on a square lattice. Together with the semion and earlier odd-prime ISAs, finite stacking realizes every pointed Witt class of bosonic Abelian anyon theories. Finally, we extend both constructions to all spatial dimensions $d=4k{-}2$, with one qubit or $\mathbb{Z}_4$ qudit on each $(2k{-}1)$-cell for the semion or $U(1)_4$ ISA, respectively.
\end{abstract}

\tableofcontents

%%%%%%%%%%%%%%%%%%%%%%%%
\section{Introduction}

Local commuting-projector Hamiltonians realize broad families of nonchiral topological order~\cite{Levin2005Stringnet}, while a nonzero chiral central charge obstructs such a realization on a strictly two-dimensional tensor-product Hilbert space with finite-dimensional local factors~\cite{kitaev2006anyons, LinLevin2014, KapustinSpodyneiko2020}. Walker-Wang models realize chiral anyons at the boundary of a three-dimensional commuting Hamiltonian~\cite{Walker2012TQFT, Keyserlingk2013Threedimensional}. These boundary constructions motivate a direct description of chiral observables by local operators on a two-dimensional lattice.

Invertible subalgebras (ISAs) provide this operator-algebraic framework~\cite{Haah2023}. An ISA and its commutant reconstruct every local physical operator as a finite sum of products supported in a uniformly bounded neighborhood. The semion and anti-semion subalgebras commute and together generate the full matrix algebra of local operators. Intuitively speaking, these subalgebras factor the full matrix algebra. Within either subalgebra, one can write down a Hamiltonian with mutually commuting local terms describing the corresponding chiral anyon sector. This factorization also defines a quantum cellular automaton (QCA) in one higher dimension by transporting one ISA factor while fixing its commutant~\cite{Haah2023}, similar to how the ``half translation'' of Majorana fermions implements a QCA that prepares the Kitaev chain. A QCA is an invertible automorphism for which both the map and its inverse spread local operators by at most a fixed distance~\cite{schumacher2004reversibleQCA, Arrighi2011Localizability,haah_QCA_23}.

Non-trivial QCAs capture information flow that cannot be generated by a finite-depth circuit. Index and flow invariants organize their classification in one and two dimensions~\cite{Gross2012GNVWindex, Freedman2020ClassificationQCA}. In three dimensions, an earlier Clifford QCA disentangles the three-fermion Walker-Wang model~\cite{haah_QCA_23}. QCAs also describe the anomalous chiral boundary dynamics of Floquet phases~\cite{PoChiral}.

The construction in Ref.~\cite{Haah2023} gives Clifford QCAs from Pauli ISAs, including a chiral $\mathbb{Z}_3$ ISA. Algebraic Witt groups govern the stable classification of translation-invariant Clifford QCAs on prime-dimensional qudits modulo Clifford circuits and shifts~\cite{Haah2021CliffordQCA,haah2025topological}. An algebraic $L$-theory formulation also treats composite qudit dimensions and lattices without translation symmetry~\cite{Yang2026CategorifyingCliffordQCA}. Beyond the Clifford category, stable QCA classes have recently been organized into a generalized cohomology theory~\cite{JiYang2026Spectrum}. Topological quantum field theories (TQFTs) offer another route through discrete gauge fields and higher cup products~\cite{Kapustin2014theta, ChenTata2023,chen2023exactly}, which we previously used to construct Clifford QCAs and ISAs~\cite{Sun2026, SunNonClifford2026}. For a construction with non-Pauli generators, one must explicitly show how the two commuting algebras reconstruct every local microscopic operator with a uniform support bound.

The semion is a minimal chiral example beyond this Pauli framework. Its nontrivial anyon satisfies $s\times s=1$ and has topological spin $i$. Its orientation reverse has spin $-i$, and together they form double-semion order. A three-dimensional non-Clifford semion QCA was constructed by condensation in a Walker-Wang model~\cite{Shirley2022}, and our subsequent TQFT construction supplied explicit operators for the complete QCA~\cite{SunNonClifford2026}. A complementary perspective relates the three-fermion QCA to chirality pumping in fermionic systems~\cite{FidkowskiHastings2024PumpingChirality}. Related work connects these models to higher-form gauging, dualities, modular relations, and non-invertible symmetries~\cite{zhang2026quantumcellularautomatakramerswannier,inamura2026noninvertiblesymmetriesmixingwitt}. The boundary-algebra correspondence predicts a semion ISA~\cite{Haah2023}. The remaining task is a direct qubit realization with explicit local hopping generators and a constructive decomposition of arbitrary physical operators.

In this work, we construct the semion ISA and its anti-semion commutant on a square lattice with one qubit per edge. Our starting point is boson condensation~\cite{Kong2014Anyon} in the $\mathbb{Z}_4$ toric code. Condensing $e^2m^2$ identifies the dyons $em$ and $em^3$ as semions and anti-semions, respectively, where $e$ and $m$ denote the elementary electric and magnetic anyons, and yields a Pauli stabilizer code for the double semion model~\cite{Ellison2022}. Turning this anyon-level description into complementary operator algebras requires additional microscopic structure. We construct two commuting dyon-hopping families for $em$ and $em^3$ and identify their common condensation algebra. Although the condensed boson has order two, the initial constraint generators have order four and therefore cannot individually be transformed into single-qubit Paulis. We overcome this issue by writing down equivalent local Hermitian involutions that generate the same constraint algebra and select the same condensed Hilbert space. Representing each $\mathbb Z_4$ qudit as two qubits, we then construct a finite-depth circuit that disentangles one ancillary qubit per edge fixed by these constraints. Removing the ancillas leaves unrestricted physical qubits carrying the two chiral algebras.

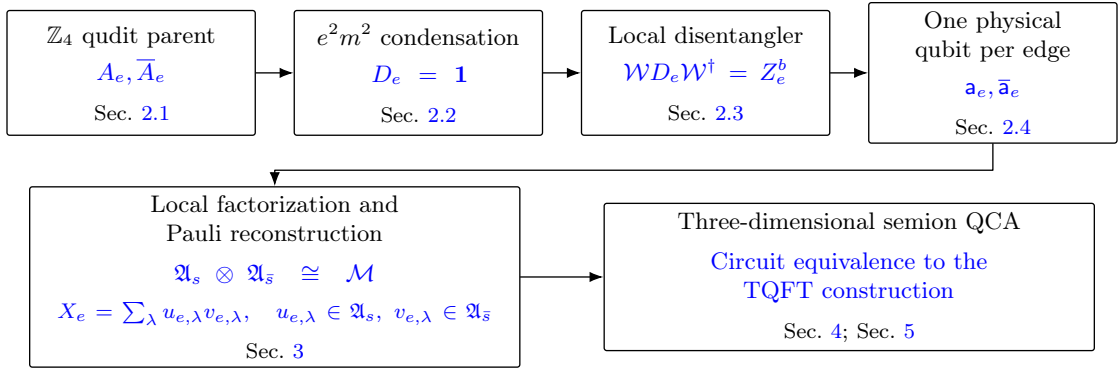
\begin{figure}[t]
\centering
\begin{tikzpicture}[>=Latex,
  box/.style={draw=black,line width=.5pt,rounded corners=1pt,
    text width=3.05cm,minimum height=1.65cm,align=center,font=\small},
  wide/.style={box,text width=6.25cm,minimum height=1.85cm},
  flowarrow/.style={->,line width=.5pt}]
 \node[box] (parent) at (0,0) {$\mathbb Z_4$ qudit parent\\[3pt]
   $\color{blue}A_e,\overline A_e$\\[4pt]
   {\footnotesize Sec.~\ref{sec:transport}}};
 \node[box] (constraints) at (3.8,0) {$e^2m^2$ condensation\\[3pt]
   $\color{blue}D_e=\id$\\[4pt]
   {\footnotesize Sec.~\ref{sec:condensation}}};
 \node[box] (reduce) at (7.6,0) {Local disentangler\\[3pt]
   $\color{blue}\mathcal W D_e\mathcal W^\dagger=Z_e^b$\\[4pt]
   {\footnotesize Sec.~\ref{sec:disentangler}}};
 \node[box] (qubits) at (11.4,0) {One physical qubit per edge\\[3pt]
   $\color{blue}\mathsf a_e,\overline{\mathsf a}_e$\\[4pt]
   {\footnotesize Sec.~\ref{sec:primary}}};
 \draw[flowarrow] (parent)--(constraints);
 \draw[flowarrow] (constraints)--(reduce);
 \draw[flowarrow] (reduce)--(qubits);
 \node[wide] (pump) at (1.9,-2.7) {Local factorization and Pauli reconstruction\\[4pt]
   $\color{blue}\Asalg\otimes\Abalg\cong\mathcal M$\\[3pt]
   {\footnotesize $\color{blue}X_e=\sum_\lambda u_{e,\lambda}v_{e,\lambda},\quad
   u_{e,\lambda}\in\Asalg,\ v_{e,\lambda}\in\Abalg$}\\[4pt]
   {\footnotesize Sec.~\ref{sec:completeness}}};
 \node[wide] (qca) at (9.5,-2.7) {Three-dimensional semion QCA\\[4pt]
   {\color{blue}Circuit equivalence to the\\
   TQFT construction}\\[4pt]
   {\footnotesize Sec.~\ref{sec:microscopicQCA}; Sec.~\ref{sec:TQFTcomparison}}};
 \draw[flowarrow] (qubits.south)--++(0,-.35)-| (pump.north);
 \draw[flowarrow] (pump)--(qca);
\end{tikzpicture}
\caption{Construction of the semion ISA and its three-dimensional QCA. In the parent model, each edge $e$ of the square lattice carries one $\mathbb Z_4$ qudit. The two parent hopping families $A_e$ and $\overline A_e$ commute with one another, and the Hermitian constraints $D_e$ condense the boson $e^2m^2$ of the $\mathbb Z_4$ toric code. Decomposing each qudit into $a$ and $b$ qubits, the finite-depth circuit $\mathcal W$ maps $D_e$ to the ancillary Pauli $Z_e^b$. Fixing $Z_e^b=+1$ and removing the $b$ qubits leaves one unrestricted physical $a$ qubit per edge. The resulting Hermitian semion and anti-semion hopping operators $\mathsf a_e$ and $\overline{\mathsf a}_e$ generate $\Asalg$ and $\Abalg$, respectively. Together they reconstruct the full physical local operator algebra $\mathcal M$ through the multiplication isomorphism $\Asalg\otimes\Abalg\cong\mathcal M$, with uniformly bounded support enlargement. The displayed decomposition of the physical Pauli $X_e$ illustrates this factorization as a finite sum of products with local factors $u_{e,\lambda}\in\Asalg$ and $v_{e,\lambda}\in\Abalg$. Translating $\Asalg$ upward by one layer in a stack while fixing $\Abalg$ defines the three-dimensional QCA. A finite-depth circuit relates its inverse, the downward pump, to the TQFT QCA constructed in Ref.~\cite{SunNonClifford2026}, up to lattice translations and the addition of ancillary qubits.}
\label{fig:overview}
\end{figure}

Each algebra is generated by one local Hermitian hopping operator per edge, which squares to the identity. We prove that the two algebras are mutual commutants and reconstruct every physical operator locally. In particular, we give explicit decompositions of each Pauli $X$ and $Z$ into the semion subalgebra and its anti-chiral copy. All these statements hold on the full physical-qubit Hilbert space. We also give ordinary qubit-gate implementations of the hopping operators and derive their semion and anti-semion exchange statistics. Fig.~\ref{fig:overview} summarizes the construction.

Pumping the semion subalgebra through stacked layers while fixing the other anti-semion factor defines a three-dimensional non-Clifford QCA. The Pauli decompositions make its action on every local operator explicit. We further show that pumping eight times is equivalent to lattice translations and a finite-depth circuit, allowing additional idle copies. More precisely, the eighth power differs from a four-layer translation by a circuit whose depth and gate range are independent of system size. The construction begins with a local exchange of four semion and anti-semion factors and uses the stable group structure of QCAs~\cite{FreedmanHaahHastings2022} to obtain the cancellation circuit.

We construct a finite-depth circuit relating the inverse pump to the semion QCA obtained from TQFT in Ref.~\cite{SunNonClifford2026}, after adding ancillary qubits and applying some lattice shifts. The equivalence holds on the full local operator algebra. It therefore transfers the eightfold cancellation to that previously constructed QCA modulo circuits and translations. The inverse pump is fixed by the relative orientation and preparation conventions of the two constructions. These local algebraic results are independent of the separate chiral-boundary obstruction used to infer stable nontriviality as in Ref.~\cite{Shirley2022}.

Finally, we extend the construction to a $U(1)_4$ ISA on a square lattice with one $\mathbb{Z}_4$ qudit per edge. Together with the semion and earlier odd-prime ISAs, finite stacking realizes every bosonic Abelian anyon theory up to anyon condensation. These equivalence classes form the pointed Witt group~\cite{Davydov2013Witt,davydov2013structure}. Our cup-product formulation extends both constructions directly to all spatial dimensions $d=4k{-}2$. The resulting complementary ISAs retain uniformly local reconstruction, with one physical qubit or $\mathbb{Z}_4$ qudit on each $(2k{-}1)$-cell for the semion or $U(1)_4$ construction, respectively.

The paper is organized as follows. Sec.~\ref{sec:construction} constructs the semion ISA by condensation and local disentangling. Sec.~\ref{sec:completeness} proves local reconstruction and the mutual commutant relations, gives the lattice-level Pauli decompositions, and identifies the two chiralities. Sec.~\ref{sec:microscopicQCA} constructs the pump, determines its Pauli images, and proves eightfold cancellation. Sec.~\ref{sec:TQFTcomparison} establishes equivalence with the TQFT QCA, and Sec.~\ref{sec:discussion} discusses the scope and possible extensions. Appendices~\ref{app:pauliDerivations}--\ref{sec:cochains} provide the detailed local proofs, gate implementations, pump-cancellation circuits, TQFT comparison identities, and cup-product formalism. Appendix~\ref{app:hdsem} gives the higher-dimensional semion ISA construction, while Appendix~\ref{app:u14} develops the $U(1)_4$ ISA and its higher-dimensional extension.

\section{A square-lattice semion invertible subalgebra}
\label{sec:construction}

The construction begins with two families of local hopping operators on the qudit lattice. Operators from different families commute. The two generated algebras share local constraints that condense the boson $e^2m^2$ of the $\mathbb Z_4$ toric code. We replace these constraints by operators that square to the identity, without changing the local constraint algebra or its simultaneous $+1$ subspace. A local circuit then isolates one constrained qubit per edge. Removing these qubits in the condensation sector leaves one physical qubit per edge, carrying the two chiral algebras. This section constructs their generators. Sec.~\ref{sec:completeness} proves that together they generate the full local algebra and identifies their excitation statistics.
Appendix~\ref{app:hdsem} extends this condensation and disentangling construction to spatial dimensions $d=4k{-}2$.
%while Appendix~\ref{app:u14} adapts it to a $U(1)_4$ ISA and its higher-dimensional generalization.

We write $\mathcal M$ for the algebra of operators with finite support. For a subalgebra $\mathfrak B\subset\mathcal M$, its commutant $\mathfrak B^\perp$ consists of all operators in $\mathcal M$ that commute with every element of $\mathfrak B$. A subalgebra $\mathfrak A$ is invertible if every local operator is a finite sum of products from $\mathfrak A$ and $\mathfrak A^\perp$, with both factors supported within a fixed neighborhood of the original operator~\cite[Definition~1.1]{Haah2023}. For our two chiral algebras, this means
\begin{equation}
    O=\sum_\lambda u_\lambda v_\lambda,\qquad 
    u_\lambda\in\Asalg,\quad v_\lambda\in\Abalg,\qquad 
    \supp(u_\lambda),\supp(v_\lambda)\subseteq N_R(\supp O),
    \label{eq:ISAdefinition}
\end{equation}
The sum is finite. The set $N_R(\supp O)$ is the radius-$R$ neighborhood of the support of $O$. The radius $R$ does not depend on $O$. The two algebras commute, and each is the commutant of the other. Both contain the identity. Throughout, $\Alg$ denotes the algebra obtained from the listed operators by taking finite complex linear combinations, products, and adjoints, together with the identity. Appendix~\ref{app:pauliDerivations} supplies the parent commutant proof and local Pauli derivations used to establish this factorization.

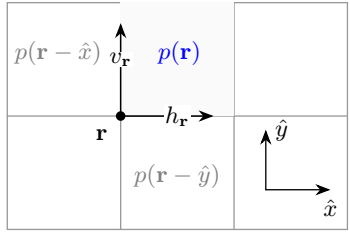
\begin{figure}[t]
\centering
\begin{tikzpicture}[x=1.5cm,y=1.5cm,line cap=round]
 \foreach \x in {-1,0,1,2} \draw[gridline] (\x,-1)--(\x,1);
 \foreach \y in {-1,0,1} \draw[gridline] (-1,\y)--(2,\y);
 \fill[black!2] (0,0) rectangle (1,1);
 \draw[-{Stealth[length=2mm]},edgefocus] (0,0)--(.82,0);
 \draw[-{Stealth[length=2mm]},edgefocus] (0,0)--(0,.82);
 \node[oplabel,text=black] at (.5,0) {$h_{\mathbf r}$};
 \node[oplabel,text=black] at (0,.5) {$v_{\mathbf r}$};
 \node[font=\small,text=blue] at (.52,.56) {$p(\mathbf r)$};
 \node[font=\small,text=gray] at (.5,-.53) {$p(\mathbf r-\hat y)$};
 \node[font=\small,text=gray] at (-.55,.56) {$p(\mathbf r-\hat x)$};
 \fill[black] (0,0) circle (1.8pt);
 \node[font=\small,text=black,below left=2pt] at (0,0) {$\mathbf r$};
 \draw[-{Stealth[length=2mm]}] (1.28,-.65)--(1.85,-.65) node[below,font=\small] {$\hat x$};
 \draw[-{Stealth[length=2mm]}] (1.28,-.65)--(1.28,-.12) node[right,font=\small] {$\hat y$};
\end{tikzpicture}
\caption{Square-lattice conventions. The marked vertex is $\rvec$. $h_{\rvec}$ points right and $v_{\rvec}$ points up. The plaquette $p(\rvec)$ has lower-left vertex $\rvec$, and the neighboring plaquettes show the two faces incident to each marked edge.}
\label{fig:conventions}
\end{figure}

Vertices are $\rvec=(r_x,r_y)\in\mathbb Z^2$, with positively oriented edges (Fig.~\ref{fig:conventions}):
\begin{equation}
    h_{\rvec}:\rvec\longrightarrow\rvec+\xhat,\qquad 
    v_{\rvec}:\rvec\longrightarrow\rvec+\yhat.
\end{equation}

The plaquette $p(\rvec)$ has lower-left vertex $\rvec$. We also use this vertex to index the plaquette and its paired star. We write $p\supset e$ when the edge $e$ lies on the boundary $\partial p$. For an operator indexed by a plaquette, the face subscript and its lower-left vertex subscript denote the same operator. For example, $G_{p(\rvec)}=G_{\rvec}$. This convention also applies to the barred loops, their normalized and reduced versions, and the loop-controlled gates. Initially, there is one four-level $\mathbb Z_4$ qudit per edge, with generalized Pauli operators
\begin{equation}
    X|j\rangle=|j+1\bmod4\rangle, \qquad
    Z|j\rangle=i^j|j\rangle,\qquad
    ZX=iXZ, \qquad 
    \ww=e^{i\pi/4}.
\label{eq:quditPaulis}
\end{equation}
In all lattice diagrams, edges point right and up, the marked vertex is $\rvec$, and complete operator labels are blue. An unlabeled edge carries the identity. Products on a common edge are written in $X$-then-$Z$ order. Operators at different edges commute. After the reduction, $X_e,Z_e$ denote the physical qubit Paulis, with $ZX=-XZ$. The Hilbert space in use will always be specified at the transition.

Our statements concern the infinite plane and local identities on periodic lattices larger than the displayed supports. Infinite products of local gates denote their action by conjugation on finite-support operators. At fixed depth, only finitely many gates affect each such operator. All operator identities act on the full microscopic Hilbert space unless a restriction is written explicitly. The ancilla reduction fixes only the condensation sector.

\subsection{Commuting dyon hopping operators and loops}
\label{sec:transport}

In the parent $\mathbb Z_4$ toric code, a dyon binds an electric charge to a magnetic flux. Condensing the boson $e^2m^2$ turns the dyons $em$ and $em^3$ into the semion and anti-semion, respectively~\cite{Ellison2022}. We first describe their local motion by two families of Pauli strings, chosen so that every hopping operator in one family commutes with every hopping operator in the other. Products of hopping operators give transport paths; closing a path gives a loop that measures the enclosed excitation content. These commuting hopping algebras are the starting point for the two chiral algebras constructed below.

\Needspace{12\baselineskip}
For each oriented edge $e$, let $A_e$ and $\overline A_e$ denote the $em$ and $em^3$ hopping operators. Their horizontal and vertical representatives are

\begin{eqs}\label{eq:parentHops}
 A_{h_{\rvec}}&=\ww\,\vcenter{\hbox{
\begin{tikzpicture}[x=1.15cm,y=1.15cm,line cap=round,line join=round]
\draw[gridline] (0,-1) -- (0,1);
\draw[gridline] (1,-1) -- (1,1);
\draw[gridline] (0,-1) -- (1,-1);
\draw[gridline] (0,0) -- (1,0);
\draw[gridline] (0,1) -- (1,1);
\fill[black!35] (0,-1) circle (0.75pt);
\fill[black!35] (0,0) circle (0.75pt);
\fill[black!35] (0,1) circle (0.75pt);
\fill[black!35] (1,-1) circle (0.75pt);
\fill[black!35] (1,0) circle (0.75pt);
\fill[black!35] (1,1) circle (0.75pt);
\draw[edgefocus] (0,0) -- (1,0);
\node[fill=white,inner sep=.35pt,text=blue,oplabel] at (0.5,0) {${X}{Z^{\dagger}}$};
\node[fill=white,inner sep=.35pt,text=blue,oplabel] at (0.5,1) {${Z}$};
\node[fill=white,inner sep=.35pt,text=blue,oplabel] at (0,-0.5) {${Z^{\dagger}}$};
\node[fill=white,inner sep=.35pt,text=blue,oplabel] at (0,0.5) {${Z}$};
\node[fill=white,inner sep=.35pt,text=blue,oplabel] at (1,0.5) {${Z^{\dagger}}$};
\node[font=\small,inner sep=0pt] at (0.5,-0.35) {$h_{\rvec}$};
\node[font=\small,inner sep=0pt,anchor=east] at (-0.15,0.5) {$v_{\rvec}$};
\node[font=\small,inner sep=0pt,anchor=north east] at (-0.07,-0.07) {$\rvec$};
\fill[black] (0,0) circle (1.6pt);
\path[use as bounding box] (-0.33,-1.25) rectangle (1.33,1.25);
\end{tikzpicture}
}},\qquad
 A_{v_{\rvec}}=\ww\,\vcenter{\hbox{
\begin{tikzpicture}[x=1.15cm,y=1.15cm,line cap=round,line join=round]
\draw[gridline] (-1,0) -- (-1,1);
\draw[gridline] (0,0) -- (0,1);
\draw[gridline] (1,0) -- (1,1);
\draw[gridline] (-1,0) -- (1,0);
\draw[gridline] (-1,1) -- (1,1);
\fill[black!35] (-1,0) circle (0.75pt);
\fill[black!35] (-1,1) circle (0.75pt);
\fill[black!35] (0,0) circle (0.75pt);
\fill[black!35] (0,1) circle (0.75pt);
\fill[black!35] (1,0) circle (0.75pt);
\fill[black!35] (1,1) circle (0.75pt);
\draw[edgefocus] (0,0) -- (0,1);
\node[fill=white,inner sep=.35pt,text=blue,oplabel] at (-0.5,1) {${Z}$};
\node[fill=white,inner sep=.35pt,text=blue,oplabel] at (-1,0.5) {${Z}$};
\node[fill=white,inner sep=.35pt,text=blue,oplabel] at (0,0.5) {${X}{Z^{\dagger}}$};
\fill[black] (0,0) circle (1.6pt);
\path[use as bounding box] (-1.33,-0.25) rectangle (1.33,1.25);
\end{tikzpicture}
}},\\[2mm]
 \overline A_{h_{\rvec}}&=\ww\,\vcenter{\hbox{
\begin{tikzpicture}[x=1.15cm,y=1.15cm,line cap=round,line join=round]
\draw[gridline] (0,-1) -- (0,1);
\draw[gridline] (1,-1) -- (1,1);
\draw[gridline] (0,-1) -- (1,-1);
\draw[gridline] (0,0) -- (1,0);
\draw[gridline] (0,1) -- (1,1);
\fill[black!35] (0,-1) circle (0.75pt);
\fill[black!35] (0,0) circle (0.75pt);
\fill[black!35] (0,1) circle (0.75pt);
\fill[black!35] (1,-1) circle (0.75pt);
\fill[black!35] (1,0) circle (0.75pt);
\fill[black!35] (1,1) circle (0.75pt);
\draw[edgefocus] (0,0) -- (1,0);
\node[fill=white,inner sep=.35pt,text=blue,oplabel] at (0.5,-1) {${Z}$};
\node[fill=white,inner sep=.35pt,text=blue,oplabel] at (0.5,0) {${X}{Z^{\dagger}}$};
\node[fill=white,inner sep=.35pt,text=blue,oplabel] at (1,-0.5) {${Z}$};
\fill[black] (0,0) circle (1.6pt);
\path[use as bounding box] (-0.33,-1.25) rectangle (1.33,1.25);
\end{tikzpicture}
}},\qquad
 \overline A_{v_{\rvec}}=\ww\,\vcenter{\hbox{
\begin{tikzpicture}[x=1.15cm,y=1.15cm,line cap=round,line join=round]
\draw[gridline] (-1,0) -- (-1,1);
\draw[gridline] (0,0) -- (0,1);
\draw[gridline] (1,0) -- (1,1);
\draw[gridline] (-1,0) -- (1,0);
\draw[gridline] (-1,1) -- (1,1);
\fill[black!35] (-1,0) circle (0.75pt);
\fill[black!35] (-1,1) circle (0.75pt);
\fill[black!35] (0,0) circle (0.75pt);
\fill[black!35] (0,1) circle (0.75pt);
\fill[black!35] (1,0) circle (0.75pt);
\fill[black!35] (1,1) circle (0.75pt);
\draw[edgefocus] (0,0) -- (0,1);
\node[fill=white,inner sep=.35pt,text=blue,oplabel] at (-0.5,0) {${Z^{\dagger}}$};
\node[fill=white,inner sep=.35pt,text=blue,oplabel] at (0.5,0) {${Z}$};
\node[fill=white,inner sep=.35pt,text=blue,oplabel] at (0.5,1) {${Z^{\dagger}}$};
\node[fill=white,inner sep=.35pt,text=blue,oplabel] at (0,0.5) {${X}{Z^{\dagger}}$};
\node[fill=white,inner sep=.35pt,text=blue,oplabel] at (1,0.5) {${Z}$};
\fill[black] (0,0) circle (1.6pt);
\path[use as bounding box] (-1.33,-0.25) rectangle (1.33,1.25);
\end{tikzpicture}}}.
\end{eqs}
The first diagram labels $\rvec$ and its outgoing edges $h_{\rvec},v_{\rvec}$ in black. The blue labels specify all nonidentity Pauli factors. The common prefactor $\ww$ fixes the phase convention so that $A_e^4=\overline A_e^4=\id$. Without it, the fourth powers would be $-\id$. Multiplying the generators by this scalar leaves their generated algebras unchanged.

The same operators have a compact expression in terms of integer cochains introduced in Refs.~\cite{Chen2018Bosonization, Chen2019Bosonization, Chen2020Bosonization, Chen2021Disentangling}. Bold symbols denote the elementary cochains supported on a single cell. $\mathbf v$ has value $+1$ at vertex $v$, $\mathbf e$ has value $+1$ on the positively oriented edge $e$, and $\mathbf f$ has value $+1$ on the oriented face $f$, with zero values elsewhere. Their degrees are zero, one, and two, respectively. We write $\delta$ for the coboundary and $\int$ for the sum of a degree-two cochain over the oriented square lattice. We use the integer cubical conventions of Ref.~\cite[Appendix B]{Sun2026}, evaluated over integers before reducing the Pauli exponents modulo four.

With these conventions, Eq.~\eqref{eq:parentHops} becomes
\begin{equation}
    \begin{aligned}
        A_e&=\ww X_e\prod_{e'}Z_{e'}^{\int (\mathbf e'\cup\mathbf e-\delta\mathbf e'\cup_1\mathbf e)},\\
     \overline A_e&=\ww X_e\prod_{e'}Z_{e'}^{\int (\mathbf e\cup\mathbf e'-\delta\mathbf e\cup_1\mathbf e')}.
    \end{aligned}
 \label{eq:parentCupHops}
\end{equation}
The products run over all edges $e'$, including $e$, but only finitely many exponents are nonzero. Exchanging the two input cochains exchanges the two hopping families. The Pauli commutation relations imply
\begin{equation}
    [A_e,\overline A_{e'}]=0 \qquad \forall e, e'.
\label{eq:cross}
\end{equation}
Thus, the two families generate commuting algebras. Appendix~\ref{app:parentPatterns} gives the equivalent matrix notation used in the local factorization proof.

The squares of the hopping operators are Hermitian involutions that enter both the condensation constraints and the local disentangler:
\begin{equation}\label{eq:parentHopSquares}
    \begin{aligned}
        A_e^2&=X_e^2\prod_{e'}Z_{e'}^{2\int (\mathbf e'\cup\mathbf e-\delta\mathbf e'\cup_1\mathbf e)},\\
        \overline A_e^{\,2}&=X_e^2\prod_{e'}Z_{e'}^{2\int (\mathbf e\cup\mathbf e'-\delta\mathbf e\cup_1\mathbf e')},
        \qquad A_e^4=\overline A_e^{\,4}=\id.
    \end{aligned}
\end{equation}

\Needspace{10\baselineskip}
We next close the transport paths to obtain the elementary dyon loops. Define the toric-code star $T_{\rvec}$ at $\rvec$ and the plaquette $W_{\rvec}$ on $p(\rvec)$ by
\begin{equation}\label{eq:toricStarPlaquette}
 T_{\rvec}=\vcenter{\hbox{
\begin{tikzpicture}[x=1.15cm,y=1.15cm,line cap=round,line join=round]
\draw[gridline] (-1,-1) -- (-1,1);
\draw[gridline] (0,-1) -- (0,1);
\draw[gridline] (1,-1) -- (1,1);
\draw[gridline] (-1,-1) -- (1,-1);
\draw[gridline] (-1,0) -- (1,0);
\draw[gridline] (-1,1) -- (1,1);
\fill[black!35] (-1,-1) circle (0.75pt);
\fill[black!35] (-1,0) circle (0.75pt);
\fill[black!35] (-1,1) circle (0.75pt);
\fill[black!35] (0,-1) circle (0.75pt);
\fill[black!35] (0,0) circle (0.75pt);
\fill[black!35] (0,1) circle (0.75pt);
\fill[black!35] (1,-1) circle (0.75pt);
\fill[black!35] (1,0) circle (0.75pt);
\fill[black!35] (1,1) circle (0.75pt);
\node[fill=white,inner sep=.35pt,text=blue,oplabel] at (-0.5,0) {${X}$};
\node[fill=white,inner sep=.35pt,text=blue,oplabel] at (0.5,0) {${X^{\dagger}}$};
\node[fill=white,inner sep=.35pt,text=blue,oplabel] at (0,-0.5) {${X}$};
\node[fill=white,inner sep=.35pt,text=blue,oplabel] at (0,0.5) {${X^{\dagger}}$};
\fill[black] (0,0) circle (1.6pt);
\path[use as bounding box] (-1.33,-1.25) rectangle (1.33,1.25);
\end{tikzpicture}
}},\qquad
 W_{\rvec}=\vcenter{\hbox{
\begin{tikzpicture}[x=1.15cm,y=1.15cm,line cap=round,line join=round]
\fill[black!2] (0,0) rectangle (1,1);
\draw[gridline] (-1,-1) -- (-1,1);
\draw[gridline] (0,-1) -- (0,1);
\draw[gridline] (1,-1) -- (1,1);
\draw[gridline] (-1,-1) -- (1,-1);
\draw[gridline] (-1,0) -- (1,0);
\draw[gridline] (-1,1) -- (1,1);
\fill[black!35] (-1,-1) circle (0.75pt);
\fill[black!35] (-1,0) circle (0.75pt);
\fill[black!35] (-1,1) circle (0.75pt);
\fill[black!35] (0,-1) circle (0.75pt);
\fill[black!35] (0,0) circle (0.75pt);
\fill[black!35] (0,1) circle (0.75pt);
\fill[black!35] (1,-1) circle (0.75pt);
\fill[black!35] (1,0) circle (0.75pt);
\fill[black!35] (1,1) circle (0.75pt);
\node[fill=white,inner sep=.35pt,text=blue,oplabel] at (0.5,0) {${Z}$};
\node[fill=white,inner sep=.35pt,text=blue,oplabel] at (0.5,1) {${Z^{\dagger}}$};
\node[fill=white,inner sep=.35pt,text=blue,oplabel] at (0,0.5) {${Z^{\dagger}}$};
\node[fill=white,inner sep=.35pt,text=blue,oplabel] at (1,0.5) {${Z}$};
\fill[black] (0,0) circle (1.6pt);
\path[use as bounding box] (-1.33,-1.25) rectangle (1.33,1.25);
\end{tikzpicture}
}}.
\end{equation}
Our star convention places $X$ on the incoming edges and $X^\dagger$ on the outgoing edges. The plaquette follows the counterclockwise boundary orientation. Every star commutes with every plaquette. The two dyon loops are therefore the mutually commuting products
\begin{equation}\label{eq:B}
 G_{\rvec}=T_{\rvec}W_{\rvec}=\vcenter{\hbox{
\begin{tikzpicture}[x=1.15cm,y=1.15cm,line cap=round,line join=round]
\fill[black!2] (0,0) rectangle (1,1);
\draw[gridline] (-1,-1) -- (-1,1);
\draw[gridline] (0,-1) -- (0,1);
\draw[gridline] (1,-1) -- (1,1);
\draw[gridline] (-1,-1) -- (1,-1);
\draw[gridline] (-1,0) -- (1,0);
\draw[gridline] (-1,1) -- (1,1);
\fill[black!35] (-1,-1) circle (0.75pt);
\fill[black!35] (-1,0) circle (0.75pt);
\fill[black!35] (-1,1) circle (0.75pt);
\fill[black!35] (0,-1) circle (0.75pt);
\fill[black!35] (0,0) circle (0.75pt);
\fill[black!35] (0,1) circle (0.75pt);
\fill[black!35] (1,-1) circle (0.75pt);
\fill[black!35] (1,0) circle (0.75pt);
\fill[black!35] (1,1) circle (0.75pt);
\node[fill=white,inner sep=.35pt,text=blue,oplabel] at (-0.5,0) {${X}$};
\node[fill=white,inner sep=.35pt,text=blue,oplabel] at (0.5,0) {${X^{\dagger}}{Z}$};
\node[fill=white,inner sep=.35pt,text=blue,oplabel] at (0.5,1) {${Z^{\dagger}}$};
\node[fill=white,inner sep=.35pt,text=blue,oplabel] at (0,-0.5) {${X}$};
\node[fill=white,inner sep=.35pt,text=blue,oplabel] at (0,0.5) {${X^{\dagger}}{Z^{\dagger}}$};
\node[fill=white,inner sep=.35pt,text=blue,oplabel] at (1,0.5) {${Z}$};
\fill[black] (0,0) circle (1.6pt);
\path[use as bounding box] (-1.33,-1.25) rectangle (1.33,1.25);
\end{tikzpicture}
}},\qquad \overline G_{\rvec}=T_{\rvec}W_{\rvec}^{-1}=\vcenter{\hbox{
\begin{tikzpicture}[x=1.15cm,y=1.15cm,line cap=round,line join=round]
\fill[black!2] (0,0) rectangle (1,1);
\draw[gridline] (-1,-1) -- (-1,1);
\draw[gridline] (0,-1) -- (0,1);
\draw[gridline] (1,-1) -- (1,1);
\draw[gridline] (-1,-1) -- (1,-1);
\draw[gridline] (-1,0) -- (1,0);
\draw[gridline] (-1,1) -- (1,1);
\fill[black!35] (-1,-1) circle (0.75pt);
\fill[black!35] (-1,0) circle (0.75pt);
\fill[black!35] (-1,1) circle (0.75pt);
\fill[black!35] (0,-1) circle (0.75pt);
\fill[black!35] (0,0) circle (0.75pt);
\fill[black!35] (0,1) circle (0.75pt);
\fill[black!35] (1,-1) circle (0.75pt);
\fill[black!35] (1,0) circle (0.75pt);
\fill[black!35] (1,1) circle (0.75pt);
\node[fill=white,inner sep=.35pt,text=blue,oplabel] at (-0.5,0) {${X}$};
\node[fill=white,inner sep=.35pt,text=blue,oplabel] at (0.5,0) {${X^{\dagger}}{Z^{\dagger}}$};
\node[fill=white,inner sep=.35pt,text=blue,oplabel] at (0.5,1) {${Z}$};
\node[fill=white,inner sep=.35pt,text=blue,oplabel] at (0,-0.5) {${X}$};
\node[fill=white,inner sep=.35pt,text=blue,oplabel] at (0,0.5) {${X^{\dagger}}{Z}$};
\node[fill=white,inner sep=.35pt,text=blue,oplabel] at (1,0.5) {${Z^{\dagger}}$};
\fill[black] (0,0) circle (1.6pt);
\path[use as bounding box] (-1.33,-1.25) rectangle (1.33,1.25);
\end{tikzpicture}}} .
\end{equation}
The marked vertex is $\rvec$, and the shaded face is $p(\rvec)$. Each loop combines the star at this vertex with the adjacent plaquette and hence has six-edge support. All scalar phases in Eqs.~\eqref{eq:toricStarPlaquette} and~\eqref{eq:B} are $+1$.

For a vertex $v$ at coordinate $\rvec$, write $G_v=G_{\rvec}$ and similarly for the barred loop. The cup-product form makes the vertex--plaquette pairing explicit:
\begin{equation}\label{eq:Gcup}
G_v=X_{\delta\mathbf v}
\prod_{e'}Z_{e'}^{\int\mathbf v\cup\delta\mathbf e'},
\qquad
\overline G_v=X_{\delta\mathbf v}
\prod_{e'}Z_{e'}^{-\int\mathbf v\cup\delta\mathbf e'}.
\end{equation}
Here, $X_{\delta\mathbf v}=\prod_{e'}X_{e'}^{(\delta\mathbf v)(e')}$ is the star at $v$. The cup pairing selects the plaquette with lower-left vertex $v$, and the displayed products place all $X$ factors before all $Z$ factors, as in the diagrams.

An ordered product of four neighboring hopping operators recovers these loops exactly, including their scalar phases (see Appendix \ref{sec:cochains} for the expressions using cup products):
\begin{align}
G_{\rvec}
&=-i A_{h_{\rvec-\xhat}}A_{h_{\rvec}}^\dagger
      A_{v_{\rvec-\yhat}}A_{v_{\rvec}}^\dagger,
\label{eq:Bloop}\\
\overline G_{\rvec}
&=i\overline A_{h_{\rvec-\xhat}}
     \overline A_{h_{\rvec}}^\dagger
     \overline A_{v_{\rvec-\yhat}}
     \overline A_{v_{\rvec}}^\dagger.
\label{eq:barBloop}
\end{align}
The order of the factors is part of these identities. In particular, each loop belongs to the algebra generated by its own hopping family; no additional loop generators are needed. When indexed by a face $f$, $G_f$ and $\overline G_f$ denote the loops at its lower-left vertex.

The relations needed for condensation are
\begin{equation}
    G_{\rvec}^4=\overline G_{\rvec}^4=\id,\qquad G_{\rvec}^2=\overline G_{\rvec}^2,\qquad [A_e^2,G_{\rvec}]=0.
\label{eq:Bprops}
\end{equation}
The common loop squares $G_{\rvec}^2=\overline G_{\rvec}^2$ commute with both hopping families. In the next subsection, these shared central operators provide the local phase corrections that turn the condensation generators into Hermitian involutions.

\subsection{An involutive generating set for the condensate algebra}
\label{sec:condensation}

The condensation constraints lie in the common part of the two commuting hopping algebras. Our aim is to replace their order-four generators by local Hermitian involutions that generate the same constraint algebra and select the same simultaneous $+1$ sector. This will allow a local circuit to isolate the constrained degrees of freedom on ancillary qubits.

For an edge $e$, the local condensation operator multiplies the squared hopping operator $A_e^2$ by the dyon loops associated with its two incident faces. Using either hopping family, we obtain
\begin{equation}\label{eq:Cold}
\begin{gathered}
\begin{aligned}
C_e
&=A_e^2\prod_f G_f^{\int(\mathbf f\cup_1\mathbf e-\mathbf e\cup_1\mathbf f)}
 =A_e^2\prod_{f\supset e}G_f
\\
&=\overline A_e^{\,2}\prod_f\overline G_f^{\int(\mathbf f\cup_1\mathbf e-\mathbf e\cup_1\mathbf f)}
 =\overline A_e^{\,2}\prod_{f\supset e}\overline G_f,
\end{aligned}
\\[2mm]
C_{h_{\rvec}}=\vcenter{\hbox{
\begin{tikzpicture}[x=1.05cm,y=1.05cm,line cap=round,line join=round]
\draw[gridline] (-1,-2) -- (-1,1);
\draw[gridline] (0,-2) -- (0,1);
\draw[gridline] (1,-2) -- (1,1);
\draw[gridline] (-1,-2) -- (1,-2);
\draw[gridline] (-1,-1) -- (1,-1);
\draw[gridline] (-1,0) -- (1,0);
\draw[gridline] (-1,1) -- (1,1);
\fill[black!35] (-1,-2) circle (0.75pt);
\fill[black!35] (-1,-1) circle (0.75pt);
\fill[black!35] (-1,0) circle (0.75pt);
\fill[black!35] (-1,1) circle (0.75pt);
\fill[black!35] (0,-2) circle (0.75pt);
\fill[black!35] (0,-1) circle (0.75pt);
\fill[black!35] (0,0) circle (0.75pt);
\fill[black!35] (0,1) circle (0.75pt);
\fill[black!35] (1,-2) circle (0.75pt);
\fill[black!35] (1,-1) circle (0.75pt);
\fill[black!35] (1,0) circle (0.75pt);
\fill[black!35] (1,1) circle (0.75pt);
\draw[edgefocus] (0,0) -- (1,0);
\node[fill=white,inner sep=.35pt,text=blue,oplabel] at (-0.5,-1) {${X}$};
\node[fill=white,inner sep=.35pt,text=blue,oplabel] at (-0.5,0) {${X}$};
\node[fill=white,inner sep=.35pt,text=blue,oplabel] at (0.5,-1) {${X^{\dagger}}{Z}$};
\node[fill=white,inner sep=.35pt,text=blue,oplabel] at (0.5,0) {${X}{Z^{2}}$};
\node[fill=white,inner sep=.35pt,text=blue,oplabel] at (0.5,1) {${Z}$};
\node[fill=white,inner sep=.35pt,text=blue,oplabel] at (0,-1.5) {${X}$};
\node[fill=white,inner sep=.35pt,text=blue,oplabel] at (0,-0.5) {${Z}$};
\node[fill=white,inner sep=.35pt,text=blue,oplabel] at (0,0.5) {${X^{\dagger}}{Z}$};
\node[fill=white,inner sep=.35pt,text=blue,oplabel] at (1,-0.5) {${Z}$};
\node[fill=white,inner sep=.35pt,text=blue,oplabel] at (1,0.5) {${Z^{\dagger}}$};
\fill[black] (0,0) circle (1.6pt);
\path[use as bounding box] (-1.33,-2.25) rectangle (1.33,1.25);
\end{tikzpicture}
}}, \qquad C_{v_{\rvec}}=\vcenter{\hbox{
\begin{tikzpicture}[x=1.05cm,y=1.05cm,line cap=round,line join=round]
\draw[gridline] (-2,-1) -- (-2,1);
\draw[gridline] (-1,-1) -- (-1,1);
\draw[gridline] (0,-1) -- (0,1);
\draw[gridline] (1,-1) -- (1,1);
\draw[gridline] (-2,-1) -- (1,-1);
\draw[gridline] (-2,0) -- (1,0);
\draw[gridline] (-2,1) -- (1,1);
\fill[black!35] (-2,-1) circle (0.75pt);
\fill[black!35] (-2,0) circle (0.75pt);
\fill[black!35] (-2,1) circle (0.75pt);
\fill[black!35] (-1,-1) circle (0.75pt);
\fill[black!35] (-1,0) circle (0.75pt);
\fill[black!35] (-1,1) circle (0.75pt);
\fill[black!35] (0,-1) circle (0.75pt);
\fill[black!35] (0,0) circle (0.75pt);
\fill[black!35] (0,1) circle (0.75pt);
\fill[black!35] (1,-1) circle (0.75pt);
\fill[black!35] (1,0) circle (0.75pt);
\fill[black!35] (1,1) circle (0.75pt);
\draw[edgefocus] (0,0) -- (0,1);
\node[fill=white,inner sep=.35pt,text=blue,oplabel] at (-1.5,0) {${X}$};
\node[fill=white,inner sep=.35pt,text=blue,oplabel] at (-0.5,0) {${Z}$};
\node[fill=white,inner sep=.35pt,text=blue,oplabel] at (-0.5,1) {${Z}$};
\node[fill=white,inner sep=.35pt,text=blue,oplabel] at (0.5,0) {${X^{\dagger}}{Z}$};
\node[fill=white,inner sep=.35pt,text=blue,oplabel] at (0.5,1) {${Z^{\dagger}}$};
\node[fill=white,inner sep=.35pt,text=blue,oplabel] at (-1,-0.5) {${X}$};
\node[fill=white,inner sep=.35pt,text=blue,oplabel] at (-1,0.5) {${X^{\dagger}}{Z}$};
\node[fill=white,inner sep=.35pt,text=blue,oplabel] at (0,-0.5) {${X}$};
\node[fill=white,inner sep=.35pt,text=blue,oplabel] at (0,0.5) {${X}{Z^{2}}$};
\node[fill=white,inner sep=.35pt,text=blue,oplabel] at (1,0.5) {${Z}$};
\fill[black] (0,0) circle (1.6pt);
\path[use as bounding box] (-2.33,-1.25) rectangle (1.33,1.25);
\end{tikzpicture}}},
\end{gathered}
\end{equation}
where $f$ runs over faces, and $G_f$ denotes the loop at the lower-left vertex of $f$. With our cubical conventions, the exponent is $1$ when $e\subset\partial f$ and $0$ otherwise, so precisely the two adjacent loops enter the product. The diagrams display the complete Pauli strings for both edge orientations. The emphasized edge is $e$, the marked vertex is $\rvec$, and both scalar phases are $+1$.

Eq.~\eqref{eq:Cold} places each $C_e$ in both hopping algebras. Since the two families commute, $C_e$ commutes with every hopping operator in either family. Thus the hopping operators preserve each simultaneous constraint sector. More strongly, the constraints form the entire common central algebra. Define
\begin{equation}\label{eq:parentAlgebras}
 \mathfrak H:=\Alg\{A_e,\overline A_e\},\qquad
 \mathfrak C:=\Alg\{C_e\}.
\end{equation}
Then
\begin{equation}\label{eq:parentCommutants}
 \mathfrak H=\mathfrak C^\perp,\qquad
 \mathfrak H^\perp=\mathfrak C,\qquad
 \operatorname{Center}(\mathfrak H)=\mathfrak C.
\end{equation}
These commutants are taken in the full parent-qudit local operator algebra, before selecting a constraint sector. The equalities hold on the infinite plane and sufficiently large periodic lattices. Appendix~\ref{app:localPauliFactorization} proves them using the Laurent-polynomial representation, with a uniform bound on support enlargement. In particular, every parent Pauli commuting with all $C_e$ is generated by the two hopping families within a fixed neighborhood of its support.

The condensate algebra is also exactly the intersection of the two hopping algebras, i.e.
\begin{equation}\label{eq:parentIntersection}
 \mathfrak C=\Alg\{A_e\}\cap\Alg\{\overline A_e\}.
\end{equation}
One inclusion follows from Eq.~\eqref{eq:Cold}. Conversely, an operator belonging to both hopping algebras commutes with each family, because the families commute. It therefore belongs to $\mathfrak H^\perp=\mathfrak C$.

The boson $e^2m^2$ has order-two fusion, whereas its microscopic condensation generators have order four. On the full parent Hilbert space,
\begin{equation}\label{eq:Csquare}
    C_e^2=\prod_{f\supset e}G_f^2\ne\id,\qquad C_e^4=\id.
\end{equation}
This distinction matters for the local reduction. Unitary conjugation preserves operator order, so an individual $C_e$ cannot be mapped to an ancillary qubit Pauli. Imposing only $C_e^2=\id$ leaves both the $C_e=+1$ and $C_e=-1$ eigenspaces and therefore does not impose the desired condensation constraint. We therefore change the local generating set of $\mathfrak C$, constructing Hermitian involutions with the same generated algebra and simultaneous $+1$ sector. The disentangler below will establish that the new constraints are independent.

The spectral map $r$ sends the eigenvalues $(1,i,-1,-i)$ to $(1,1,-1,-1)$, retracting them onto $\{1,-1\}$. It can be viewed through the group extension
\begin{align}
    0 \rightarrow \mathbb Z_2 \xrightarrow{\iota} \mathbb Z_4 \xrightarrow{\pi} \mathbb Z_2 \rightarrow 0
\end{align}
Writing $j=j_0+2j_1$ with $j_0,j_1\in\{0,1\}$, the map removes the quotient parity $j_0$ and sends $i^j$ to $(-1)^{j_1}$. This is a set-theoretic retraction, not a group homomorphism: $r(i)^2=1$, whereas $r(i^2)=-1$.

For a Hermitian involution $P$, i.e., $P^2=\id$ and $P^\dagger=P$, define
\begin{equation}
    S(P):=i^{(1-P)/2}=\frac{1+i}{2}\id+\frac{1-i}{2}P, \qquad S(P)^2=P.
\label{eq:S}
\end{equation}
For the single-qubit Pauli operator $P=Z=\operatorname{diag}(1,-1)$, this is the standard qubit $S$ gate: $S(Z)=\operatorname{diag}(1,i)$. More generally, $S(P)$ multiplies the $P=-1$ eigenspace by $i$ and leaves the $P=+1$ eigenspace unchanged.

For an order-four unitary $Q$, we denote its spectral retraction in sans serif by
\begin{equation}
    \mathsf{Q}:= r(Q) =QS(Q^2)^\dagger =\frac{1-i}{2}Q+\frac{1+i}{2}Q^{-1}.
\label{eq:hatQ}
\end{equation}
To obtain equivalent local generators of the full condensate algebra, we retract each neighboring loop separately and define
\begin{equation}
\begin{aligned}
    D_e&:=A_e^2\prod_{f\supset e}\mathsf{G}_f
    =A_e^2\prod_f\mathsf{G}_f^{\int\mathbf f\cup_2\delta\mathbf e}
    \\
    &=C_e\prod_{f\supset e}S(G_f^2)^\dagger
    =C_e\prod_f\bigl[S(G_f^2)^\dagger\bigr]^{
        \int(\mathbf f\cup_2\delta\mathbf e-2\mathbf e\cup_1\mathbf f)}.
\end{aligned}
\label{eq:D}
\end{equation}
For the two orientations,
\begin{equation}
    D_{h_{\rvec}}=A_{h_{\rvec}}^2\mathsf{G}_{\rvec}\mathsf{G}_{\rvec-\yhat}, \qquad D_{v_{\rvec}}=A_{v_{\rvec}}^2\mathsf{G}_{\rvec}\mathsf{G}_{\rvec-\xhat}.
\label{eq:Dorient}
\end{equation}
The same corrected operator is also obtained in the barred algebra as
\begin{equation}
    D_e=\overline A_e^{\,2}\prod_{f\supset e}\overline{\mathsf{G}}_f.
\label{eq:Dbar}
\end{equation}
The common loop squares are central, and therefore
\begin{equation}
    D_e^\dagger=D_e,\qquad D_e^2=\id,\qquad [D_e,D_{e'}]=[D_e,A_{e'}]=[D_e,\overline A_{e'}]=0.
\label{eq:Dproperties}
\end{equation}
These hold on the full qudit Hilbert space. We will now show that enforcing $D_e=1$ and $C_e=1$ on all edges are actually equivalent.

%%%%%%%%%%%%%%%%%%%%%%%%
\begin{lemma}[Equivalent condensation constraints] \label{prop:samespace}
    The families $\{C_e\}$ and $\{D_e\}$ generate the same local algebra $\mathfrak C$. That is, each family can be recovered from the other within a fixed neighborhood of each edge. Moreover, imposing $C_e=1$ on every edge is equivalent to imposing $D_e=1$ on every edge. 
\end{lemma}

\begin{proof}
    The defining expressions give the local star identities
    \begin{align}
        G_{\rvec}^2
        &=
        D_{h_{\rvec-\xhat}}D_{h_{\rvec}} D_{v_{\rvec-\yhat}}D_{v_{\rvec}},
        \label{eq:Dstar}\\
        G_{\rvec-\xhat-\yhat}^2
        &=
        C_{h_{\rvec-\xhat}}C_{h_{\rvec}}^{-1} C_{v_{\rvec-\yhat}}C_{v_{\rvec}}^{-1}.
        \label{eq:Cstar}
    \end{align}
    In Eq.~\eqref{eq:Dstar}, each normalized loop occurs twice and cancels because $\mathsf{G}_p^{\,2}=1$, leaving the four $A_e^2$ factors, whose product is $G_{\rvec}^2$. In Eq.~\eqref{eq:Cstar}, the same $A_e^2$ factors remain, since $A_e^{-2}=A_e^2$, while the commuting loop factors contribute $G_{\rvec-\xhat-\yhat}^2G_{\rvec}^{-2}$. The displacement follows from the lower-left-vertex plaquette convention.
    
    Thus, every $G_p^2$ is generated locally by either family. Since $S$ is linear in its involutive argument by Eq.~\eqref{eq:S}, the same holds for $S(G_p^2)$ and its adjoint. Therefore, Eq.~\eqref{eq:D} and its inverse,
    \begin{equation}
        C_e=D_e\prod_{f\supset e}S(G_f^2),
        \label{eq:CfromD}
    \end{equation}
    give uniformly local changes of generators, proving that the two local algebras coincide.
    
    Lastly, imposing $D_e=1$ on every edge forces $G_p^2=1$ on every plaquette by Eq.~\eqref{eq:Dstar}. Since $S(1)=1$, Eq.~\eqref{eq:CfromD} then gives $C_e=1$ on every edge. Conversely, imposing all $C_e=1$ forces all $G_p^2=1$ by Eq.~\eqref{eq:Cstar}, and hence all $D_e=1$ by Eq.~\eqref{eq:D}. The two families select the same condensed Hilbert space.
\end{proof}

By Lemma~\ref{prop:samespace}, the involutions $D_e$ generate the same condensate algebra, $\mathfrak C=\Alg\{D_e\}$. The disentangler below maps them to independent ancillary $Z$ operators. On a finite periodic lattice, their common $+1$ sector carries the full physical-qubit algebra. Sec.~\ref{sec:physicalFactorization} describes this reduction explicitly, and Sec.~\ref{sec:completeness} proves that the reduced chiral algebras are complementary mutual commutants.

Each $G_p$ retains both eigenvalues $+1$ and $-1$ in the condensed sector. The operator $D_e$ is obtained by retracting the incident loops separately; it differs from $r(C_e)$ by the controlled sign in Eq.~\eqref{eq:rootproductwarning}. We use $D_e$ because Lemma~\ref{prop:samespace} ensures that this replacement preserves the entire condensate algebra and hence its commutant.

\subsection{Removing one ancillary qubit per edge}
\label{sec:disentangler}

Condensing $e^2m^2$ gives the double-semion theory, a twisted $\mathbb Z_2$ gauge theory~\cite{Ellison2022}. A qubit description stores its binary degrees of freedom in one qubit per edge, while local hopping phases encode the twist. We obtain this description by isolating the degree of freedom fixed by the condensation constraint so that it can be disentangled from the remaining system. The circuit first removes the loop factors from $D_e$, then its Pauli decorations. In the resulting basis, $D_e=Z_e^b$: the $b$ qubit is fixed to $|0\rangle$ and can be discarded, while the $a$ qubit remains unrestricted. Using Eq.~\eqref{eq:hatQ}, we retract the plaquette loop to an order-two operator as
\begin{equation}
    \mathsf{W}_{\rvec}=W_{\rvec}S(W_{\rvec}^2)^\dagger
\label{eq:Rhat}
\end{equation}
For commuting Hermitian involutions $P,Q$, define the generalized controlled phase gate
\begin{equation}\label{eq:CZ}
    \CZ(P,Q):=\frac{\id+P+Q-PQ}{2} = (-1)^{\frac{1-P}{2}\frac{1-Q}{2}}.
\end{equation}
It multiplies the joint $P=Q=-1$ eigenspace by $-1$ and is the identity on the other three sectors. The disentangling gate is defined as
\begin{equation}\label{eq:ugate}
    u_{\rvec}:=\CZ(\mathsf{G}_{\rvec},\mathsf{W}_{\rvec}).
\end{equation}
% \begin{center}
% \renewcommand{\arraystretch}{1.15}
% \begin{tabular}{c|cccc}
% $G_{\rvec}\backslash W_{\rvec}$ & $1$ & $i$ & $-1$ & $-i$\\\hline
% $1$  & $1$ & $1$ & $1$ & $1$\\
% $i$  & $1$ & $1$ & $1$ & $1$\\
% $-1$ & $1$ & $1$ & $-1$ & $-1$\\
% $-i$ & $1$ & $1$ & $-1$ & $-1$
% \end{tabular}
% \end{center}
Since these operators commute,
\begin{equation}
    u_{\rvec}^\dagger=u_{\rvec},\qquad u_{\rvec}^2=\id, \qquad [u_{\rvec},u_{\rvec'}]=0,
\end{equation}
the unitary
\begin{equation}\label{eq:U}
    \mathcal U=\prod_{\rvec}u_{\rvec}, \qquad \mathcal U^\dagger=\mathcal U
\end{equation}
is a finite-depth quantum circuit.

The operator $A_e^2$ commutes with every normalized dyon loop $\mathsf{G}_{\rvec}$. It anticommutes with $\mathsf{W}_{\rvec}$ at the two plaquettes incident to $e$ and commutes with it elsewhere. A controlled phase attaches its first observable to an operator that flips its second. Thus, 
\begin{equation}
    u_{\rvec}A_e^2u_{\rvec}=
    \begin{cases}
        A_e^2\mathsf{G}_{\rvec},&e\in\partial p(\rvec),\\
        A_e^2,&\text{otherwise},
    \end{cases}
\end{equation}
and the two incident plaquettes give
\begin{equation}
    \mathcal UA_e^2\mathcal U^\dagger=D_e,\qquad \mathcal UD_e\mathcal U^\dagger=A_e^2.
\label{eq:Uconjugation}
\end{equation}
Thus, $\mathcal U$ turns the original condition $D_e=1$ into $A_e^2=1$ by removing the two incident normalized-loop factors from the constraint.

\begin{figure}[t]
\centering
\begin{subfigure}[b]{.47\linewidth}\centering
\begin{tikzpicture}[x=1.75cm,y=1.75cm,line cap=round,idlelabel/.style={font=\scriptsize,text=black!55,fill=white,inner sep=.2pt}]
\draw[gridline] (0,-1)--(0,1);
\draw[gridline] (1,-1)--(1,1);
\draw[gridline] (0,-1)--(1,-1);
\draw[gridline] (0,0)--(1,0);
\draw[gridline] (0,1)--(1,1);
\coordinate (ah0) at (0.5,0.14);   % a on h_r
\coordinate (ahp) at (0.5,1.14);   % a on h_{r+y}
\coordinate (avm) at (-0.14,-0.5); % a on v_{r-y}
\coordinate (av0) at (-0.14,0.5);  % a on v_r
\coordinate (avx) at (0.86,0.5);   % a on v_{r+x}
\coordinate (target) at (0.5,-0.14); % b on h_r
\draw[-{Stealth[length=1.7mm]},black,thin] (ah0)--(target);
\draw[-{Stealth[length=1.7mm]},black,thin] (ahp)
 .. controls (0.27,1.03) and (0.30,0.16) .. (target);
\draw[-{Stealth[length=1.7mm]},black,thin] (avm)--(target);
\draw[-{Stealth[length=1.7mm]},black,thin] (av0)--(target);
\draw[-{Stealth[length=1.7mm]},black,thin] (avx)--(target);
\foreach \p in {(0.5,-0.86),(0.86,-0.5)}
\fill[black!45] \p circle (1.5pt);
\foreach \p in {(0.5,-1.14),(0.5,0.86),(0.14,-0.5),(0.14,0.5),(1.14,-0.5),(1.14,0.5)}
 \draw[draw=black!45,fill=white,line width=.45pt] \p circle (1.7pt);
\node[idlelabel,right=3pt] at (0.5,-0.86) {$a$};
\node[idlelabel,below left=3pt] at (0.86,-0.5) {$a$};
\node[idlelabel,right=3pt] at (0.5,-1.14) {$b$};
\node[idlelabel,right=3pt] at (0.5,0.86) {$b$};
\node[idlelabel,right=3pt] at (0.14,-0.5) {$b$};
\node[idlelabel,right=3pt] at (0.14,0.5) {$b$};
\node[idlelabel,below right=3pt] at (1.14,-0.5) {$b$};
\node[idlelabel,right=3pt] at (1.14,0.5) {$b$};
% Draw active markers and their full indices after the connections.
\foreach \p in {ah0,ahp,avm,av0,avx} \fill[blue] (\p) circle (2pt);
\node[oplabel] at (.56,.40) {$a_{h_{\rvec}}$};
\node[oplabel,above=3pt] at (ahp) {$a_{h_{\rvec+\yhat}}$};
\node[oplabel,left=3pt] at (avm) {$a_{v_{\rvec-\yhat}}$};
\node[oplabel,left=3pt] at (av0) {$a_{v_{\rvec}}$};
\draw[blue!55,line width=.4pt] (avx)--(1.28,0.75);
\node[oplabel,anchor=west] at (1.30,0.75) {$a_{v_{\rvec+\xhat}}$};
\node[fill=white,inner sep=0pt,text=blue,font=\Large] at (target) {$\oplus$};
\node[oplabel,below right=3pt] at (target) {$b_{h_{\rvec}}$};
\fill[black] (0,0) circle (1.8pt);
\node[font=\small,fill=white,inner sep=.35pt,below left=3pt] at (0,0) {$\rvec$};
\path[use as bounding box] (-1.15,-1.34) rectangle (2.35,1.42);
\end{tikzpicture}
\caption{Target $b_{h_{\rvec}}$.}
\end{subfigure}
\hfill
\begin{subfigure}[b]{.47\linewidth}\centering
\begin{tikzpicture}[x=1.75cm,y=1.75cm,line cap=round,
 idlelabel/.style={font=\scriptsize,text=black!55,fill=white,inner sep=.2pt}]
\draw[gridline] (-1,0)--(-1,1);
\draw[gridline] (0,0)--(0,1);
\draw[gridline] (1,0)--(1,1);
\draw[gridline] (-1,0)--(1,0);
\draw[gridline] (-1,1)--(1,1);
\coordinate (ahp) at (-0.5,1.14);
\coordinate (avm) at (-1.14,0.5);
\coordinate (av0) at (-0.14,0.5);
\coordinate (target) at (0.14,0.5); % b on v_r
\draw[-{Stealth[length=1.7mm]},black,thin] (ahp)--(target);
\draw[-{Stealth[length=1.7mm]},black,thin] (avm)
 .. controls (-0.96,0.16) and (-0.06,0.18) .. (target);
\draw[-{Stealth[length=1.7mm]},black,thin] (av0)--(target);
\foreach \p in {(-0.5,0.14),(0.5,0.14),(0.5,1.14),(0.86,0.5)}
 \fill[black!45] \p circle (1.5pt);
\foreach \p in {(-0.5,-0.14),(0.5,-0.14),(-0.5,0.86),(0.5,0.86),(-0.86,0.5),(1.14,0.5)}
 \draw[draw=black!45,fill=white,line width=.45pt] \p circle (1.7pt);
\node[idlelabel,right=3pt] at (-0.5,0.14) {$a$};
\node[idlelabel,right=3pt] at (0.5,0.14) {$a$};
\node[idlelabel,right=3pt] at (0.5,1.14) {$a$};
\node[idlelabel,above left=3pt] at (0.86,0.5) {$a$};
\node[idlelabel,right=3pt] at (-0.5,-0.14) {$b$};
\node[idlelabel,right=3pt] at (0.5,-0.14) {$b$};
\node[idlelabel,right=3pt] at (-0.5,0.86) {$b$};
\node[idlelabel,right=3pt] at (0.5,0.86) {$b$};
\node[idlelabel,above right=3pt] at (-0.86,0.5) {$b$};
\node[idlelabel,right=3pt] at (1.14,0.5) {$b$};
\foreach \p in {ahp,avm,av0} \fill[blue] (\p) circle (2pt);
\node[oplabel,above=3pt] at (ahp) {$a_{h_{\rvec-\xhat+\yhat}}$};
\node[oplabel,left=3pt] at (avm) {$a_{v_{\rvec-\xhat}}$};
\node[oplabel,above left=3pt] at (av0) {$a_{v_{\rvec}}$};
\node[fill=white,inner sep=0pt,text=blue,font=\Large] at (target) {$\oplus$};
\node[oplabel,below right=3pt] at (target) {$b_{v_{\rvec}}$};
\fill[black] (0,0) circle (1.8pt);
\node[font=\small,fill=white,inner sep=.35pt,below left=3pt] at (0,0) {$\rvec$};
\path[use as bounding box] (-2.20,-0.38) rectangle (1.45,1.42);
\end{tikzpicture}
\caption{Target $b_{v_{\rvec}}$.}
\end{subfigure}
\caption{The local CNOT patterns in $V$, with both encoded qubits shown on every displayed edge. Filled dots denote $a$ qubits and open circles denote $b$ qubits. Blue markers identify the controls and selected $\oplus$ target. Gray qubits are untouched by these CNOTs. Horizontal-edge pairs place $a$ above and $b$ below the midpoint. Vertical-edge pairs place $a$ left and $b$ right. Both markers belong to that same physical edge. The black vertex is $\rvec$. The horizontal and vertical targets have five and three controls, respectively, including their own-edge $a$ qubit. Translating these patterns gives Eq.~\eqref{eq:VM}.} \label{fig:cnot}
\end{figure}
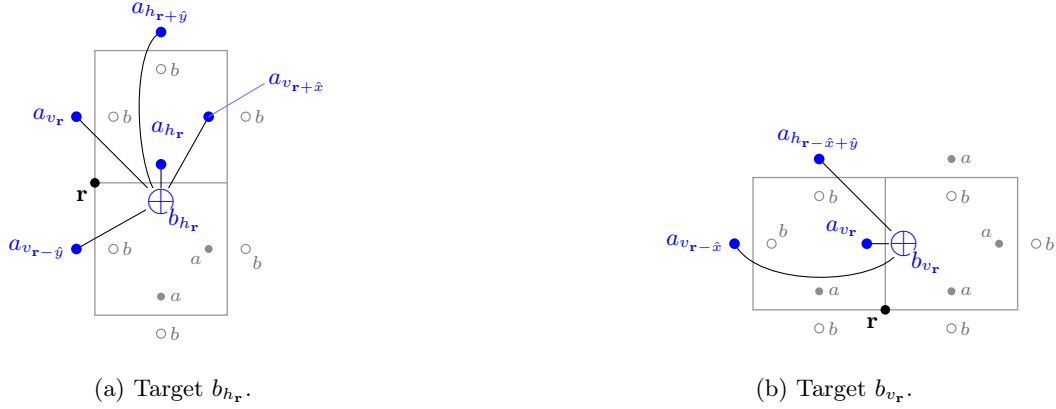

Encode a qudit as two qubits by the on-site identification
\begin{equation}
    E:\ |j=a+2b\rangle\longmapsto |a\rangle_a\otimes H|b\rangle_b, \qquad a,b\in\{0,1\}.
\label{eq:encoding}
\end{equation}
Writing $S_{a_e}=S(Z_e^a)$ as in Eq.~\eqref{eq:S}, the encoded qudit Pauli operators are
\begin{equation}\label{eq:encodedXZ}
    \begin{aligned}
    EX_eE^\dagger&=X_e^a\CZ_{a_eb_e},& EZ_eE^\dagger&=X_e^bS_{a_e},\\
    EX_e^2E^\dagger&=Z_e^b,&EZ_e^2E^\dagger&=Z_e^a.
    \end{aligned}
\end{equation}
In particular,
\begin{equation}\label{eq:encodedHopSquare}
    EA_e^2E^\dagger=Z_e^b\prod_{e'}(Z_{e'}^a)^{\int(\mathbf{e}'\cup\mathbf{e}-\delta\mathbf e'\cup_1\mathbf{e})}.
\end{equation}
In this basis, $A_e^2=1$ ties the encoded $b_e$ bit to the indicated parity of the neighboring $a$ bits. The following CNOTs add that parity to $b_e$, leaving the simple condition $b_e=0$ as
\begin{equation}
    V=\prod_{e,e'}\CNOT_{a_{e'}\to b_e}^{\int(\mathbf e'\cup\mathbf e-\delta\mathbf e'\cup_1\mathbf e)}.
\label{eq:VM}
\end{equation}
All controls lie on $a$ qubits and all targets on $b$ qubits, so the gates commute. The product is locally finite: the five horizontal and three vertical control offsets give at most eight disjoint CNOT layers. Fig.~\ref{fig:cnot} draws exactly these cup-product pairings, including the encoded constraint's $Z^a$ decorations. The identity $Z_b\mapsto Z_aZ_b$ gives
\begin{equation}
    V(EA_e^2E^\dagger)V^\dagger=Z_e^b.
\end{equation}
Let $\overline V$ be the alternate CNOT disentangling circuit obtained by replacing every gate $a_{e'}\to b_e$ in Fig.~\ref{fig:cnot} with $a_e\to b_{e'}$. It transposes the edge pairing while keeping $a$ qubits as controls and $b$ qubits as targets.

The complete disentangler is therefore
\begin{equation}\label{eq:W}
 \mathcal W:=VE\mathcal U,\qquad
 \mathcal WD_e\mathcal W^\dagger=Z_e^b.
\end{equation}

On a periodic system with $N_e$ edges, Eq.~\eqref{eq:W} proves that the $D_e$ are independent constraints. Their common $+1$ space has dimension $4^{N_e}/2^{N_e}=2^{N_e}$ and obeys
\begin{equation}
    \mathcal W\mathcal H_{D=1}=\left(\bigotimes_e\mathbb C^2_{a_e}\right)\otimes|0\rangle_b^{\otimes N_e}.
\label{eq:reducedH}
\end{equation}
We remove the disentangled $b$ qubits, fixed in $|0\rangle_b$, from the system. One unrestricted physical $a$ qubit remains per edge. The reduced loops $g_{\rvec},\overline g_{\rvec}$ remain physical observables with both eigenvalues.

\subsection{Hermitian hopping generators}
\label{sec:primary}

We now express the parent dyon hopping operators on the remaining physical qubits. We also remove their loop-dependent phases so that applying the same hopping operator twice gives the identity. These Hermitian hopping operators will generate the two chiral algebras.

For an operator commuting with all $D_e$, define its qubit reduction as
\begin{equation}
    \mathcal R(O):={}_b\langle0|\mathcal WO\mathcal W^\dagger|0\rangle_b.
\label{eq:reduction}
\end{equation}
This reduction is a unital $*$-homomorphism on the constraint commutant: after conjugation, its inputs preserve every joint ancillary sector. We denote the corresponding reduced order-four hopping operators and loops by the corresponding lower case symbol:
\begin{equation}
    a_e=\mathcal R(A_e),\qquad \overline a_e=\mathcal R(\overline A_e),\qquad g_{\rvec}=\mathcal R(G_{\rvec}),\qquad\overline g_{\rvec}=\mathcal R(\overline G_{\rvec}).
\label{eq:reducedhops}
\end{equation}
They are local unitaries on the unrestricted qubit Hilbert space. The constraints imply
\begin{equation}
    g_{\rvec}^2=\overline g_{\rvec}^2=\id,\qquad a_e^2=\prod_{f \supset e} g_f,\qquad \overline a_e^2=\prod_{f \supset e}\overline g_f.
\label{eq:physicalB}
\end{equation}
The reduced hopping operators have order four because their squares are products of neighboring loops. Multiplying each hopping operator by a phase determined by those loops makes it square to the identity while preserving its action on the excitation endpoints. We retract the order-four hopping operators in Eq.~\eqref{eq:hatQ} to order-two ones as
\begin{equation}
\mathsf{A}_e=A_eS(A_e^2)^\dagger,\qquad\overline{\mathsf{A}}_e=\overline A_eS(\overline A_e^{\,2})^\dagger.
\label{eq:precondInvolutions}
\end{equation}
Since $S(A_e^2)^2=A_e^2$, these are Hermitian involutions on the full qudit Hilbert space and commute with all $D_{e'}$. Their qubit reductions are
\begin{equation}\label{eq:involutivehop}
\mathsf{a}_e:=\mathcal{R}(\mathsf{A}_e)=a_e S(a_e^2)^\dagger, \qquad\overline{\mathsf a}_e:=\mathcal R(\overline{\mathsf{A}}_e)=\overline a_eS(\overline{a}_e^2)^\dagger.
\end{equation}
They satisfy
\begin{equation}\label{eq:involutiveProperties}
    \mathsf{a}_e^\dagger=\mathsf{a}_e,\quad \mathsf{a}_e^2=\id, \qquad \overline{\mathsf{a}}_e^\dagger=\overline{\mathsf{a}}_e,\quad\overline{\mathsf{a}}_e^{2}=\id,\quad[\mathsf{a}_e,\overline{\mathsf{a}}_{e'}]=0.
\end{equation}
Equivalently,
\begin{equation}\label{eq:involutivePolynomial}
    \mathsf{a}_e=\frac{1-i}{2}a_e+\frac{1+i}{2}a_e^\dagger, \qquad \overline{\mathsf{a}}_e=\frac{1-i}{2}\overline{a}_e +\frac{1+i}{2}\overline{a}_e^\dagger.
\end{equation}
Appendix~\ref{app:gates} discusses deterministic, ancilla-free implementations of the order-four and Hermitian hopping operators using ordinary $X,H,S,T$, CNOT, $\CZ$, and $\mathrm{CCZ}$ gates and their inverses, including their local supports and exact scalar phases. The Hermitian implementations square to the identity on the unrestricted physical-qubit Hilbert space. To establish that the normalization retains the full chiral algebras, we next reconstruct the local loops and invert Eq.~\eqref{eq:involutivePolynomial}.

\section{Local generation and complementary operator algebras}
\label{sec:completeness}

In this section, we show that the operators $\mathsf{a}_e$ and $\overline{\mathsf{a}}_e$ commute and describe the a short string of semions and anti-semions. We show that the algebras generate an arbitrary logcal operator in the tensor product qubit Hilbert space and therefore an ISA.

We proceed in three steps.
\begin{enumerate}
    \item  In Sec.~\ref{sec:hopOnly}, we show that the loop operators $g_\rvec$ can be reconstructed from the hopping operators.
    \item In Sec.~\ref{sec:physicalFactorization}, and \ref{sec:pauliDictionary}, we establish the mutual-commutant relations and the reconstruction of single Pauli operators from the combined algebra.
    \item In Sec.~\ref{sec:statistics}, we derive order-two fusion and exchange statistics of the semion and antisemion hopping operators.
\end{enumerate}

The higher-dimensional semion construction and the $U(1)_4$ generalization are presented in Appendices~\ref{app:hdsem} and~\ref{app:u14}, respectively.

\subsection{Local loop reconstruction}
\label{sec:hopOnly}

For the first step, we recover each physical loop from four neighboring Hermitian hopping operators. The ordered loop products were already derived in the parent model, in Eqs.~\eqref{eq:Bloop} and~\eqref{eq:barBloop}. Applying the reduction and the normalization in Eq.~\eqref{eq:involutivehop} gives the following formulas directly.

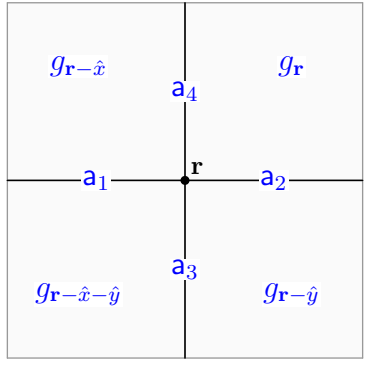
\begin{figure}[t]
\centering
\begin{tikzpicture}[x=2.35cm,y=2.35cm]
 \fill[black!2] (-1,-1) rectangle (1,1);
 \foreach \x in {-1,0,1}{\draw[gridline] (\x,-1)--(\x,1);}
 \foreach \y in {-1,0,1}{\draw[gridline] (-1,\y)--(1,\y);}
 \draw[edgefocus] (-1,0)--(1,0);
 \draw[edgefocus] (0,-1)--(0,1);
 \node[fill=white,inner sep=.35pt,text=blue,oplabel,font=\large] at (-.5,0) {$\mathsf a_1$};
 \node[fill=white,inner sep=.35pt,text=blue,oplabel,font=\large] at (.5,0) {$\mathsf a_2$};
 \node[fill=white,inner sep=.35pt,text=blue,oplabel,font=\large] at (0,-.5) {$\mathsf a_3$};
 \node[fill=white,inner sep=.35pt,text=blue,oplabel,font=\large] at (0,.5) {$\mathsf a_4$};
 \node[fill=white,inner sep=.35pt,font=\large,text=blue] at (-.60,-.64) {$g_{\rvec-\xhat-\yhat}$};
 \node[fill=white,inner sep=.35pt,font=\large,text=blue] at (.60,-.64) {$g_{\rvec-\yhat}$};
 \node[fill=white,inner sep=.35pt,font=\large,text=blue] at (-.60,.64) {$g_{\rvec-\xhat}$};
 \node[fill=white,inner sep=.35pt,font=\large,text=blue] at (.60,.64) {$g_{\rvec}$};
 \fill[black] (0,0) circle(1.7pt);
 \node[font=\normalsize,text=black,above right,inner sep=2pt] at (0,0) {$\rvec$};
\end{tikzpicture}
\caption{Reconstructing $g_{\rvec}$ from four neighboring hopping operators. Each edge label denotes the full local unitary $\mathsf a_j$, including its finite decorations. The face labels identify the loops appearing in the proof. Translating this neighborhood reconstructs every local loop.} \label{fig:hopOnly}
\end{figure}

Around a vertex $\rvec$, label the four edges as in Fig.~\ref{fig:hopOnly}
\begin{equation}\label{eq:fourLocalHops}
\begin{gathered}
    e_1=h_{\rvec-\xhat},\quad e_2=h_{\rvec},\quad e_3=v_{\rvec-\yhat},\quad e_4=v_{\rvec}, \\
    \mathsf a_j:=\mathsf a_{e_j},\qquad \overline{\mathsf a}_j:=\overline{\mathsf a}_{e_j}.
\end{gathered}
\end{equation}
The loop is a finite sum of products of these four hopping generators. The relations \eqref{eq:Bloop} and \eqref{eq:barBloop}  hold through the reduction to a qubit. But we must substitute the retracted version $\mathsf{a}$ and $\overline{\mathsf{a}}$ into the expressions. Explicitly,
\begin{equation}\label{eq:loopPolynomialExpanded}
    g_\rvec =-\frac{i}{2}\mathsf{a}_1 \mathsf{a}_2 \mathsf{a}_3 \mathsf{a}_4 \Big[\id+i(\mathsf{a}_1 \mathsf{a}_4)^2 -i(\mathsf{a}_1 \mathsf{a}_3)^2 -(\mathsf{a}_1 \mathsf{a}_4)^2 (\mathsf{a}_1 \mathsf{a}_3)^2\Big].
\end{equation}
\begin{equation}\label{eq:barredLoopPolynomialExpanded}
    \overline g_\rvec =+\frac{i}{2}\overline{\mathsf a}_1\overline{\mathsf a}_2 \overline{\mathsf a}_3\overline{\mathsf a}_4 \Big[\id-i(\overline{\mathsf a}_1\overline{\mathsf a}_4)^2 +i(\overline{\mathsf a}_1\overline{\mathsf a}_3)^2 -(\overline{\mathsf a}_1\overline{\mathsf a}_4)^2 (\overline{\mathsf a}_1\overline{\mathsf a}_3)^2\Big].
\end{equation}

\begin{lemma}[Local equivalence of hopping generators]\label{prop:hopOnly}
    The Hermitian hopping operators and the order-four hopping operators generate the same local algebras:
    \begin{equation}\label{eq:hopOnlyEquality}
     \Alg\{\mathsf a_e\}=\Alg\{a_e\}, \qquad \Alg\{\overline{\mathsf a}_e\}=\Alg\{\overline a_e\}.
    \end{equation}
    Both changes of generators increase support by at most a fixed radius.
\end{lemma}

\begin{proof}
    One inclusion follows from Eq.~\eqref{eq:involutivePolynomial}. For the other, 
    Combining \eqref{eq:physicalB} and \eqref{eq:involutivehop} gives
    \begin{equation}\label{eq:recoverAuxiliary}
        a_e=\mathsf{a}_eS \left (\prod_{f \supset e} g_f\right ),\qquad \overline{a}_e=\overline{\mathsf{a}}_e S\left (\prod_{f \supset e} \bar g_f \right ).
    \end{equation}
    where $g$ and $\overline{g}$ are given in terms of $\mathsf{a}$ and $\overline{\mathsf{a}}$ respectively in
    Eqs.~\eqref{eq:loopPolynomialExpanded} and~\eqref{eq:barredLoopPolynomialExpanded}. Explicitly, 
\begin{align}
    S \left (\prod_{f \supset e} g_f\right) &=\frac{1+i}{2}\id+\frac{1-i}{2}\prod_{f \supset e} g_f,\label{eq:SfromHops}\\
    S \left (\prod_{f \supset e} \overline{g}_f\right) &=\frac{1+i}{2}\id+\frac{1-i}{2}\prod_{f \supset e} \overline{g}_f
    \label{eq:barredSfromHops}
\end{align}
\end{proof}

\subsection{Uniformly local factorization and mutual commutants}
\label{sec:physicalFactorization}

In this subsection, we first prove uniformly local reconstruction by the two algebras, then identify them as mutual commutants. Together, these statements establish invertibility on the unrestricted physical-qubit algebra.

We can now use either generating set and define
\begin{equation}\label{eq:algebras}
    \Asalg:=\Alg\{\mathsf a_e\}=\Alg\{a_e\},\qquad
    \Abalg:=\Alg\{\overline{\mathsf a}_e\}=\Alg\{\overline a_e\}.
\end{equation}
The two algebras commute by Eq.~\eqref{eq:involutiveProperties}. Let us show that they form an invertible subalgebra.

\begin{proposition}[Uniformly local physical factorization]\label{prop:physicalFactorization}
There is a fixed radius $R$ such that every finitely supported physical-qubit operator $o$ has a finite decomposition
\begin{equation}\label{eq:localfactorization}
 o=\sum_\lambda u_\lambda v_\lambda,
 \qquad u_\lambda\in\Asalg,\quad
 v_\lambda\in\Abalg,\qquad
 \supp(u_\lambda),\supp(v_\lambda)
 \subseteq N_R(\supp o).
\end{equation}
The same factorization holds on sufficiently large periodic lattices, with a radius independent of the system size.
\end{proposition}
\begin{proof}
Appendix~\ref{app:localPauliFactorization} shows that each parent qudit Pauli operator commuting with all $C_e$ is, up to a scalar phase, a product of $A_e$ and $\overline A_e$ operators, with uniformly bounded support enlargement. These Pauli operators span $\mathfrak C^\perp$.
%Hence a general operator in the parent commutant is a finite sum of such products.

The qubit reduction $\mathcal R$ maps this commutant onto the full physical-qubit algebra. Indeed, any physical operator $o$ has the local preimage
\begin{equation}\label{eq:lift}
 O=\mathcal W^\dagger(o\otimes\id_b)\mathcal W
 \in\mathfrak C^\perp,\qquad \mathcal R(O)=o.
\end{equation}
Even when $o$ is a physical-qubit Pauli, its parent lift $O$ can be a sum of qudit Paulis. Each parent hopping product separately preserves the constraints, so the homomorphism $\mathcal R$ preserves its products. Reducing the parent factorization therefore gives Eq.~\eqref{eq:localfactorization}, since $\mathcal R(A_e)=a_e$ and $\mathcal R(\overline A_e)=\overline a_e$. Lemma~\ref{prop:hopOnly} expresses these order-four hopping products in terms of the Hermitian hopping operators.

For the support bound, let $r_{\mathcal W}$ bound the enlargement under either $\mathcal W$ or $\mathcal W^\dagger$, let $r_{\rm p}$ bound the parent factorization, and let $r_{\rm g}$ bound the replacement of order-four hopping generators by Hermitian ones. We may take $R=2r_{\mathcal W}+r_{\rm p}+r_{\rm g}$. Each radius is fixed independently of $O$ and the system size. The same construction and bounds apply on sufficiently large periodic lattices.
\end{proof}

\paragraph{Mutual commutants.}

The factorization now lets us identify the local observables that commute with either chiral algebra. On a sufficiently large finite periodic lattice, the two algebras commute and jointly generate every physical operator. Each has scalar center, and the finite-dimensional tensor-factor decomposition identifies either algebra with the commutant of the other.

This conclusion extends to local operators on the infinite plane with the same support bound. For an operator $O$ commuting with $\Asalg$, use Eq.~\eqref{eq:localfactorization} and embed its bounded reconstruction neighborhood in a sufficiently large periodic lattice. Choose a buffer that also contains the support of every hopping generator meeting this neighborhood, so their local actions agree with those on the plane. Then $O$ commutes with the finite semion factor. Averaging over that factor fixes $O$ and replaces each $u_\lambda$ by a scalar, leaving a sum of anti-semion operators in the same neighborhood. Appendix~\ref{app:localCommutants} supplies the detailed finite- and infinite-lattice argument.

Consequently,
\begin{equation}\label{eq:ISA}
 [\Asalg,\Abalg]=0,\qquad
 \Asalg\vee\Abalg=\mathcal M_{\mathrm{qubits}},\qquad
 \Asalg^\perp=\Abalg,\quad\Abalg^\perp=\Asalg.
\end{equation}
Here the commutants are taken within the physical local operator algebra. Together with the uniform bound in Eq.~\eqref{eq:localfactorization}, this is the invertibility condition for an ISA~\cite{Haah2023}.

\subsection{Explicit decomposition of each microscopic Pauli}
\label{sec:pauliDictionary}

We now give finite local hopping polynomials for the physical single-qubit Paulis $X_e$ and $Z_e$. These coordinates are defined after the complete disentangler $\mathcal W=VE\mathcal U$ and removal of the $b$ ancillas, so the formulas include the loop-controlled circuit as well as the on-site encoding. The decomposition reconstructs arbitrary local operators and determines the QCA images in Sec.~\ref{sec:explicitPauliQCA}.

Write $\mathsf a_e,\overline{\mathsf a}_e$ for the Hermitian hopping operators. The order-four hopping operators $a_e,\overline a_e$ are the inverse-normalization polynomials in Eq.~\eqref{eq:recoverAuxiliary}. The spectral functions below are evaluated as local hopping polynomials, using $S$ and $\CZ$ as defined in Eqs.~\eqref{eq:S} and~\eqref{eq:CZ}.

The decomposition uses both the two plaquettes whose boundaries contain $e$ and the two endpoint vertices of $e$, denoted $\mathcal V(e)$. Let $p_+(e)$ be the plaquette whose counterclockwise-oriented boundary contains $e$ with coefficient $+1$. It lies above a horizontal edge and to the left of a vertical edge. The geometry is
\begin{equation}\label{eq:pauliGeometry}
\begin{array}{c|c|c|c|c}
 e&j(e)&\varepsilon_e&\mathcal V(e)&p_+(e)\\\hline
 h_{\rvec}&v_{\rvec+\xhat}&+1&\{\rvec,\rvec+\xhat\}&p(\rvec)\\
 v_{\rvec}&h_{\rvec+\yhat}&-1&\{\rvec,\rvec+\yhat\}&p(\rvec-\xhat)
\end{array}
\end{equation}
Here, $\varepsilon_e=+1$ for a horizontal edge and $-1$ for a vertical edge. The two plaquettes whose boundaries contain $j(e)$ are $p(v)$ for $v\in\mathcal V(e)$.

\paragraph{The $Z$ decomposition.}
The form using the order-four hopping operators is
\begin{equation}\label{eq:pauliZaux}
Z_e=a_{j(e)}^{\varepsilon_e}\overline a_{j(e)}^{-\varepsilon_e}
 \prod_{v\in\mathcal V(e)}S(g_v\overline g_v)^{\varepsilon_e}.
\end{equation}
To verify it, pull the physical Pauli back to the parent qudit space. The CNOT disentangling circuit $V$ fixes $Z_e^a$, and Eq.~\eqref{eq:encodedXZ} gives $E^\dagger Z_e^aE=Z_e^2$. The parent clock square $Z_e^2$ flips $\mathsf{G}_v$ precisely at the endpoint vertices $v\in\mathcal V(e)$ and commutes with every $\mathsf{W}_v$. Each corresponding controlled-loop gate therefore attaches $\mathsf{W}_v$, giving
\begin{align}
 \mathcal W^\dagger(Z_e^a\otimes\id_b)\mathcal W
 &=Z_e^2\prod_{v\in\mathcal V(e)}\mathsf{W}_v,
 \nonumber\\
 Z_e^2&=A_{j(e)}^{\varepsilon_e}
 \overline A_{j(e)}^{-\varepsilon_e}
 \prod_{v\in\mathcal V(e)}W_v^{\varepsilon_e}.
 \label{eq:pauliZparentDerivation}
\end{align}
Here $Z_e$ on the right is the parent qudit clock. The second identity follows directly from the hopping diagrams: the common shift factor and scalar phase cancel, and the endpoint plaquettes cancel the remaining clocks except $Z_e^2$. For either sign $\varepsilon$, $W^\varepsilon\mathsf{W}=S(W^2)^\varepsilon$. Applying the reduction homomorphism gives Eq.~\eqref{eq:pauliZaux}, since $W_v^2=G_v\overline G_v^{-1}$ reduces to $g_v\overline g_v$. This is also the Pauli-$Z$ decomposition used in the TQFT comparison below.

There is a particularly useful version involving only the \emph{order-two} hopping operators. Write $\mathcal V(e)=\{u,v\}$ and $j=j(e)$. Then
\begin{equation}\label{eq:pauliZprimary}
 Z_e=
 \mathsf a_j\overline{\mathsf a}_j\,(g_ug_v)
 \CZ(g_u,g_v)\CZ(\overline g_u,\overline g_v)
 \CZ(g_u,\overline g_u)\CZ(g_v,\overline g_v).
\end{equation}
The same formula applies to both edge orientations: $\varepsilon_e$ has canceled. The hopping factors are written to the left of the loop functions. Arbitrary same-chirality reorderings must not be made.

For an explicit sum of separated products, define
\begin{equation}
 \Pi_v^\mu:=\frac{\id+(-1)^\mu g_v}{2},\qquad \mu\in\{0,1\}.
\end{equation}
Expanding the two mixed gates, but keeping the within-sector factors intact, gives the four-term decomposition
\begin{equation}\label{eq:pauliZfour}
 Z_e=\sum_{\mu,\nu=0}^1
 \underbrace{\left[
 \mathsf a_j g_ug_v\CZ(g_u,g_v)\Pi_u^\mu\Pi_v^\nu
 \right]}_{\in\Asalg}
 \underbrace{\left[
 \overline{\mathsf a}_j\CZ(\overline g_u,\overline g_v)
 \overline g_u^{\mu}\overline g_v^{\nu}
 \right]}_{\in\Abalg}.
\end{equation}
A sum of products is exactly the factorization required in the definition of an ISA~\cite{Haah2023}.
%The projectors in this expansion are summands of an operator identity. No measurement or extra constraint is imposed.

\paragraph{The $X$ decomposition.}

\begin{figure}[t]
 \centering
 \begin{subfigure}[c]{.47\linewidth}
  \centering
  \begin{tikzpicture}[x=2.25cm,y=2.25cm,line cap=round,line join=round,
   phase/.style={font=\normalsize,text=blue,fill=white,inner sep=.7pt},
   czlink/.style={draw=black!80,line width=.65pt}]
   \foreach \x in {0,1} \draw[gridline] (\x,-1)--(\x,1);
   \foreach \y in {-1,0,1} \draw[gridline] (0,\y)--(1,\y);
   \draw[draw=black,line width=.9pt] (0,0)--(1,0);

   % CZ(\mathcal O_e,\mathcal O_{e'}): all four curves share only their central endpoint.
   \draw[czlink] (.5,0)
    .. controls (.58,.30) and (.58,.72) .. (.5,1);
   \draw[czlink] (.5,0)
    .. controls (.27,.08) and (.09,.29) .. (0,.5);
   \draw[czlink] (.5,0)
    .. controls (.73,.08) and (.91,.29) .. (1,.5);
   \draw[czlink] (.5,0)
    .. controls (.27,-.08) and (.09,-.29) .. (0,-.5);

   % Single-observable phases. The central label is offset from its CZ dot.
   \node[phase,anchor=north west] at (.40,-.16) {$S(\mathcal O_{h_{\rvec}})^\dagger$};
   \node[phase] at (.5,-1) {$S(\mathcal O_{h_{\rvec-\yhat}})$};
   \node[phase,anchor=west] at (1.05,-.5) {$S(\mathcal O_{v_{\rvec+\xhat-\yhat}})$};
   \foreach \p in {(.5,0),(.5,1),(0,.5),(1,.5),(0,-.5)}
    \fill[blue] \p circle (1.8pt);

   \fill[black] (0,0) circle (1.7pt);
   \node[font=\normalsize,fill=white,inner sep=.4pt,anchor=north east]
    at (-.055,-.045) {$\rvec$};
   \path[use as bounding box] (-.40,-1.18) rectangle (2.25,1.18);
  \end{tikzpicture}
  \caption{$\mathscr D_{h_{\rvec}}[\mathcal O]$.}
 \end{subfigure}
 \hfill
 \begin{subfigure}[c]{.47\linewidth}
  \centering
  \begin{tikzpicture}[x=2.25cm,y=2.25cm,line cap=round,line join=round,
   phase/.style={font=\normalsize,text=blue,fill=white,inner sep=.7pt},
   czlink/.style={draw=black!80,line width=.65pt}]
   \foreach \x in {-1,0,1} \draw[gridline] (\x,0)--(\x,1);
   \foreach \y in {0,1} \draw[gridline] (-1,\y)--(1,\y);
   \draw[draw=black,line width=.9pt] (0,0)--(0,1);

   % CZ(\mathcal O_e,\mathcal O_{e'}): these two curves stay left of the central S marker.
   \draw[czlink] (0,.5)
    .. controls (-.04,.73) and (-.27,.90) .. (-.5,1);
   \draw[czlink] (0,.5)
    .. controls (-.32,.36) and (-.69,.35) .. (-1,.5);

   \node[phase,anchor=west] at (.055,.5) {$S(\mathcal O_{v_{\rvec}})^\dagger$};
   \node[phase] at (1,.5) {$S(\mathcal O_{v_{\rvec+\xhat}})$};
   \node[phase] at (.5,0) {$S(\mathcal O_{h_{\rvec}})$};
   \node[phase] at (-.5,0) {$S(\mathcal O_{h_{\rvec-\xhat}})^\dagger$};
   \node[phase] at (.5,1) {$S(\mathcal O_{h_{\rvec+\yhat}})^\dagger$};
   \foreach \p in {(0,.5),(-.5,1),(-1,.5)}
    \fill[blue] \p circle (1.8pt);

   \fill[black] (0,0) circle (1.7pt);
   \node[font=\normalsize,fill=white,inner sep=.4pt,anchor=north west]
    at (.055,-.045) {$\rvec$};
   \path[use as bounding box] (-1.25,-.68) rectangle (1.60,1.68);
  \end{tikzpicture}
  \caption{$\mathscr D_{v_{\rvec}}[\mathcal O]$.}
 \end{subfigure}
 \caption{The local diagonal functionals in Eq.~\eqref{eq:diagFunctional},
 expanded in Eqs.~\eqref{eq:diagH} and~\eqref{eq:diagV}. Edges point right
 and up. The black vertex is $\rvec$, and the emphasized edge is $e$.
 Each blue edge-midpoint dot denotes the full observable $\mathcal O_{e'}$ on that
 edge label, and each dark connection supplies $\CZ(\mathcal O_e,\mathcal O_{e'})$.
 Each phase marker displays its full spectral function $S(\mathcal O_{e'})$ or its adjoint;
 the marker on the emphasized edge is displaced slightly to reveal its
 blue dot. These are spectral functions of the specified observables,
 with no assumption that the $\mathcal O_{e'}$ are independent on-site Paulis.}
 \label{fig:diagFunctional}
\end{figure}
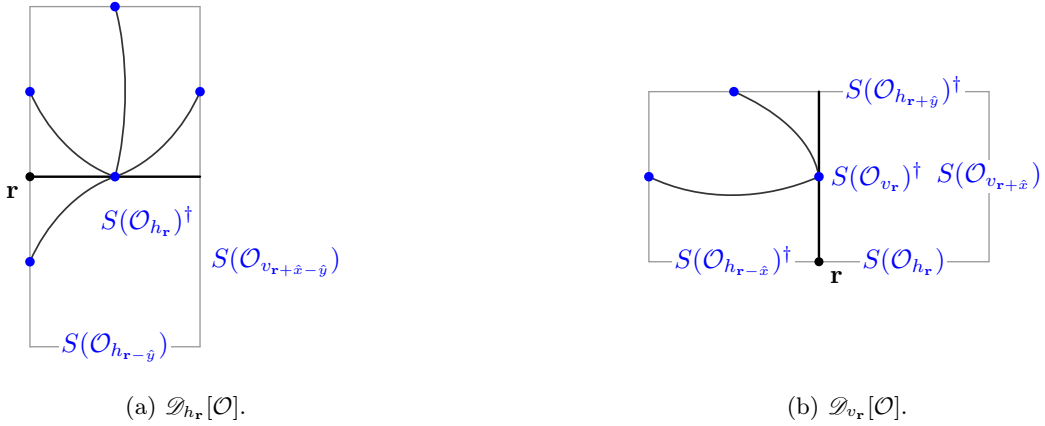

Having reconstructed $Z_e$, we now recover the basis flip $X_e$ by canceling the local phase decorations of a hopping operator. Together, the two Paulis generate the single-edge operator algebra and determine the pump's action on arbitrary local observables. For any mutually commuting Hermitian involutions $\{\mathcal O_{e'}\}$, use the signed cup-product pairings of Eq.~\eqref{eq:parentCupHops} and define
\begin{equation}\label{eq:diagFunctional}
 \mathscr D_e[\mathcal O]:=\prod_{e'}S(\mathcal O_{e'})^{\int
 (\mathbf e\cup\mathbf e'-\delta\mathbf e\cup_1\mathbf e')}
 \prod_{e'\ne e}\CZ(\mathcal O_e,\mathcal O_{e'})^{\int
 (\mathbf e'\cup\mathbf e-\delta\mathbf e'\cup_1\mathbf e)}.
\end{equation}
All products are finite and all factors within $\mathscr D_e$ commute. The $S$ factors use the exchanged cup pairing, while the controlled signs use the original pairing. Since $\CZ^2=\id$, its integer exponent already depends only on parity. Fig.~\ref{fig:diagFunctional} draws both neighborhoods. Explicitly,
\begin{align}
 \mathscr D_{h_{\rvec}}[\mathcal O]
 ={}&S(\mathcal O_{h_{\rvec}})^\dagger S(\mathcal O_{h_{\rvec-\yhat}})
 S(\mathcal O_{v_{\rvec+\xhat-\yhat}})
 \prod_{e'\in\{h_{\rvec+\yhat},v_{\rvec},v_{\rvec+\xhat},v_{\rvec-\yhat}\}}
 \CZ(\mathcal O_{h_{\rvec}},\mathcal O_{e'}),
 \label{eq:diagH}\\
 \mathscr D_{v_{\rvec}}[\mathcal O]
 ={}&S(\mathcal O_{v_{\rvec}})^\dagger S(\mathcal O_{v_{\rvec+\xhat}})
 S(\mathcal O_{h_{\rvec}})S(\mathcal O_{h_{\rvec-\xhat}})^\dagger
 S(\mathcal O_{h_{\rvec+\yhat}})^\dagger
 \prod_{e'\in\{h_{\rvec-\xhat+\yhat},v_{\rvec-\xhat}\}}
 \CZ(\mathcal O_{v_{\rvec}},\mathcal O_{e'}).
 \label{eq:diagV}
\end{align}

\noindent For $\mathcal O_{e'}=Z_{e'}$, these are ordinary local $S$ and $\CZ$ gates.

\noindent For the two plaquettes $p,q$ with $e\in\partial p\cap\partial q$, recall that $p_+(h_{\rvec})=p(\rvec)$ and $p_+(v_{\rvec})=p(\rvec-\xhat)$, as in Eq.~\eqref{eq:pauliGeometry}. The physical Pauli is
\begin{equation}\label{eq:pauliXaux}
    X_e=\ww^{-1}\overline a_e\,g_{p_+(e)} \, \CZ(g_p,\overline g_p)\CZ(g_q,\overline g_q)\,\mathscr D_e[\mathcal Z]^\dagger.
\end{equation}
In particular, its expression using the Hermitian hopping operators is
\begin{equation}\label{eq:pauliXprimary}
 X_e=\ww^{-1}\overline{\mathsf a}_e
 S(\overline g_p\overline g_q)\,g_{p_+(e)}
 \CZ(g_p,\overline g_p)\CZ(g_q,\overline g_q)\,\mathscr D_e[\mathcal Z]^\dagger.
\end{equation}
Here, $\mathscr D_e[\mathcal Z]$ means substituting the explicit polynomial Eq.~\eqref{eq:pauliZprimary} at every occurrence of $\mathcal O_{e'}$. The reconstruction is sequential: first recover $\mathcal Z_e$ from the hopping generators, then use those polynomials to recover $\mathcal X_e$.
Thus, Eq.~\eqref{eq:pauliXprimary} uses only hopping and reconstructed loop operators of the two subalgebras.
%Preserve the displayed order: $\mathscr D_e[\mathcal Z]$ need not commute with the preceding factors.

To turn Eq.~\eqref{eq:pauliXprimary} into a literal sum $\sum_\lambda u_\lambda v_\lambda$, expand $\CZ(g_v,\overline g_v)=\Pi_v^0+\Pi_v^1\overline g_v$, expand every $S$ and $\CZ$ by Eqs.~\eqref{eq:S} and~\eqref{eq:CZ}, and insert Eq.~\eqref{eq:pauliZfour}. Distribute the resulting finite products and group unbarred and barred factors without changing their relative order within either chirality. Exact cross-commutation permits this grouping. This gives a completely specified finite local decomposition without solving another linear system. The factored expression is much shorter than the fully distributed polynomial.

\subsection{Fusion and exchange statistics}\label{sec:statistics}

For the final step, we verify order-two fusion and opposite exchange phases using the local hopping operators. This identifies the two operator factors with semion and anti-semion order. The commuting loop observables detect their endpoints, and the Hermitian hopping operators have the same syndrome as the order-four hopping operators:
\begin{align}
\mathsf a_e g_{\rvec}
 &=\begin{cases}
 -g_{\rvec}\mathsf a_e,&e\in\partial p(\rvec),\\
 \phantom{-}g_{\rvec}\mathsf a_e,&e\notin\partial p(\rvec),
 \end{cases}\label{eq:syndrome}\\
 \overline{\mathsf a}_e\overline g_{\rvec}
 &=\begin{cases}
 -\overline g_{\rvec}\overline{\mathsf a}_e,&e\in\partial p(\rvec),\\
 \phantom{-}\overline g_{\rvec}\overline{\mathsf a}_e,&e\notin\partial p(\rvec).
 \end{cases}\notag
\end{align}
The opposite chiralities commute:
\begin{equation}
 [\mathsf a_e,\overline g_{\rvec}]
 =[\overline{\mathsf a}_e,g_{\rvec}]=0.
\end{equation}
The normalization does not change the syndrome because its argument is a product of commuting loops. Thus, $\mathsf a_e$ flips the two incident $g$ eigenvalues and transports the first excitation along a dual-lattice edge. The barred hopping operator transports only the second excitation.

\paragraph{Order-two fusion and endpoint labels.}
Squaring a string of order-four hopping operators gives, up to a scalar phase, a product of local loop operators by Eq.~\eqref{eq:physicalB}. This establishes order-two fusion for the string excitations. To show that finite hopping products carry no labels beyond their endpoint syndrome, consider a finite product of hopping operators with trivial endpoint syndrome, and let $q$ denote its edge-exponent pattern. Reducing $q$ modulo two gives a closed dual-lattice cycle with finite support. On the plane, every such cycle is a sum of plaquette boundaries. Writing $d\lambda$ for the signed boundary of a finite collection of dual plaquettes, we obtain
\begin{equation}\label{eq:cycleexact}
    q=d\lambda+2w\pmod4
\end{equation}
for finitely supported patterns $\lambda,w$. Both contributions reduce to products of local loop operators: the $d\lambda$ part by Eq.~\eqref{eq:Bloop}, and the $2w$ part by Eq.~\eqref{eq:physicalB}. Thus, hopping products with the same endpoint syndrome differ by a loop product and a scalar phase. On the common $+1$ loop vacuum, they create the same state up to phase. The complete anyon count follows from the parent condensation described below.
%Each chiral factor consequently has only two anyon types: the vacuum and the semion or anti-semion, respectively.

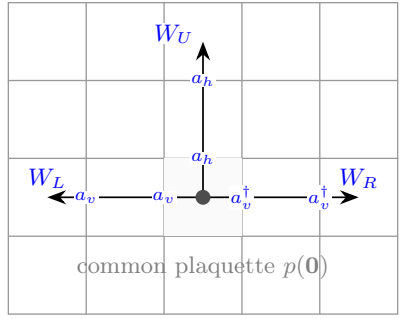
\begin{figure}[t]
\centering
\begin{tikzpicture}[x=1.03cm,y=1.03cm,line cap=round]
 \foreach \x in {-2,-1,0,1,2,3} \draw[gridline] (\x,-1)--(\x,3);
 \foreach \y in {-1,0,1,2,3} \draw[gridline] (-2,\y)--(3,\y);
 \fill[black!2] (0,0) rectangle (1,1);
 \draw[dualpath,-{Stealth[length=2.5mm]}] (.5,.5)--(2.5,.5);
 \draw[dualpath,-{Stealth[length=2.5mm]}] (.5,.5)--(.5,2.5);
 \draw[dualpath,-{Stealth[length=2.5mm]}] (.5,.5)--(-1.5,.5);
 \fill[black!70] (.5,.5) circle (2.8pt);
 \foreach \x in {1,2} \node[fill=white,inner sep=.35pt,font=\scriptsize,text=blue] at (\x,.5) {$a_v^\dagger$};
 \foreach \x in {-1,0} \node[fill=white,inner sep=.35pt,font=\scriptsize,text=blue] at (\x,.5) {$a_v$};
 \foreach \y in {1,2} \node[fill=white,inner sep=.35pt,font=\scriptsize,text=blue] at (.5,\y) {$a_h$};
 \node[fill=white,inner sep=.35pt,text=blue,font=\small,above] at (2.5,.60) {$W_R$};
 \node[fill=white,inner sep=.35pt,text=blue,font=\small,left] at (.40,2.6) {$W_U$};
 \node[fill=white,inner sep=.35pt,text=blue,font=\small,above] at (-1.5,.60) {$W_L$};
 \node[font=\small,text=gray] at (.5,-.4) {common plaquette $p(\mathbf0)$};
\end{tikzpicture}
\caption{The three dual-lattice rays $R,U,L$ in counterclockwise order, with $\ell=2$. Each crossed-edge label represents a complete order-four hopping operator, including its finite decorations. The two transport histories differ by the exchange phase in Eq.~\eqref{eq:spin}. Barred hopping operators give the opposite phase.}
\label{fig:exchange}
\end{figure}

\paragraph{Exchange with order-four hopping operators.}
For three outward rays ordered counterclockwise around a common plaquette, the hopping-operator exchange test is
\begin{equation}\label{eq:exchangeidentity}
    W_1W_2^\dagger W_3=\theta\,W_3W_2^\dagger W_1,
\end{equation}
as in the double-semion analysis of Ref.~\cite{Ellison2022}. With common plaquette $p(\mathbf0)$, take
\begin{equation}\label{eq:rays}
    W_R=\prod_{k=1}^{\ell}a_{v_{(k,0)}}^\dagger,\qquad
    W_U=\prod_{k=1}^{\ell}a_{h_{(0,k)}},\qquad
    W_L=\prod_{k=0}^{\ell-1}a_{v_{(-k,0)}}.
\end{equation}
The factors are written in increasing $k$. Their geometry is shown in Fig.~\ref{fig:exchange}. Direct qudit Pauli commutation gives the full operator identities
\begin{equation}\label{eq:spin}
 \begin{aligned}
 W_RW_U^\dagger W_L&=iW_LW_U^\dagger W_R,\\
 \overline W_R\overline W_U^\dagger\overline W_L
 &=-i\overline W_L\overline W_U^\dagger\overline W_R.
 \end{aligned}
\end{equation}
Conjugation by $\mathcal W$ and reduction preserve these products because every factor preserves the constraints.

\paragraph{Exchange with the Hermitian hopping operators.}
For one elementary hopping operator on each ray, let
\begin{equation}
 \mathsf a_R=\mathsf a_{v_{(1,0)}},\qquad
 \mathsf a_U=\mathsf a_{h_{(0,1)}},\qquad
 \mathsf a_L=\mathsf a_{v_{(0,0)}}.
\end{equation}
Specify the two-particle input sectors by
\begin{equation}\label{eq:transportProjector}
 \Pi_\eta=
 \frac{\id+\eta g_{\mathbf0}}2
 \frac{\id+\eta g_{\yhat}}2
 \frac{\id-\eta g_{-\xhat}}2
 \frac{\id-\eta g_{\xhat}}2,
 \qquad\eta=\pm1.
\end{equation}
Taking the $g=-1$ eigenvalue to denote an occupied endpoint, $\eta=+1$ occupies the left and right endpoints, while $\eta=-1$ occupies the common and upper endpoints. In either case both histories move two particles, and each elementary step moves a particle to an empty endpoint. The exchange relations are
\begin{equation}\label{eq:involutiveExchange}
 \begin{aligned}
 (\mathsf a_R\mathsf a_U\mathsf a_L
      -i\mathsf a_L\mathsf a_U\mathsf a_R)\Pi_\eta&=0,\\
 (\overline{\mathsf a}_R\overline{\mathsf a}_U
       \overline{\mathsf a}_L
      +i\overline{\mathsf a}_L\overline{\mathsf a}_U
       \overline{\mathsf a}_R)\overline\Pi_\eta&=0,
 \end{aligned}
\end{equation}
where $\overline\Pi_\eta$ replaces the loops by their barred counterparts. On a transporting step, $g_pg_q=-1$, so $\mathsf a_e=-ia_e$ and $a_e^\dagger=-a_e$ on the relevant input sector. Relative to the identity for the order-four hopping operators, the right, up, and left factors acquire phases $i,i,-i$ in both histories. These phases cancel in the comparison and prove Eq.~\eqref{eq:involutiveExchange}. The barred calculation reverses the exchange sign. The projectors specify transport inputs and do not impose additional condensation constraints. Unlike Eq.~\eqref{eq:spin}, this exchange identity for the Hermitian hopping operators is restricted to the two-particle transport sectors. The equality $\mathsf a_e^2=\id$ describes reversing an elementary hopping step.

For the orientation of Fig.~\ref{fig:exchange}, the resulting data are
\begin{equation}\label{eq:anyondata}
    s^2=\bar s^2=1,\qquad\theta_s=i,\qquad \theta_{\bar s}=-i.
\end{equation}
Their mutual braiding is trivial because the complete hopping algebras commute. Reversing the spatial orientation reverses both exchange phases.

The complete anyon count follows from the parent toric-code condensation. An anyon $(a,b)=e^am^b$ has spin $i^{ab}$. Condensation of $(2,2)$ retains $a+b$ even and identifies $(a,b)$ with $(a+2,b+2)$. Representatives $(1,1)$ and $(1,3)$ have spins $i$ and $-i$, square to the condensate, and braid trivially with one another. The quotient has precisely the four classes $1,s,\bar s,s\bar s$. The finite-depth reduction of Sec.~\ref{sec:disentangler} preserves this anyon content, with $s$ and $\bar s$ transported by the respective chiral hopping families. The local factorization above supplies the separate full-algebra statement.

The exchange phase also forces the Hermitian hopping operators to be non-Pauli. Reversing a product of three qubit Paulis changes it by a scalar $\pm1$, which cannot give the phase $i$ in Eq.~\eqref{eq:involutiveExchange} on a nonzero transport sector. Thus, the physical qubits support non-Pauli local hopping operators.

%%%%%%%%%%%%%%%%%%%%%
\section{The semion pump and its QCA}\label{sec:microscopicQCA}

We now stack the two-dimensional system along the $z$-direction and translate one chiral factor while fixing the other. Uniformly local factorization makes this operation a QCA, and the Pauli decomposition gives its action. We then use the four-copy chirality exchange to construct its order-8 cancellation circuit.

\subsection{Pumping one chiral factor}

This pumping picture is a general construction of a QCA in one higher dimension from an ISA and its commutant~\cite{Haah2023}, also used for the $\mathbb Z_p$ ISAs in our previous work~\cite{Sun2026}. It gives an operator-algebra formulation of the connection between chiral surface order and bulk locality-preserving dynamics in earlier translation-invariant Clifford QCAs on prime-dimensional qudits~\cite{haah_QCA_23, Haah2021CliffordQCA}.

Label the layers by $n=z\in\mathbb Z$. For a local operator $O^{(n)}=\sum_\lambda u_\lambda^{(n)}v_\lambda^{(n)}$, define
\begin{equation}
    \beta_\uparrow(O^{(n)})=\sum_\lambda u_\lambda^{(n+1)}v_\lambda^{(n)}.
    \label{eq:pump}
\end{equation}
The complementary-factor property makes this independent of the chosen sum. The map is extended by complex linearity and multiplication to operators in several layers. In particular,
\begin{equation}\label{eq:normalizedPump}
 \beta_\uparrow(\mathsf a_{e,n})=\mathsf a_{e,n+1},
 \qquad \beta_\uparrow(\overline{\mathsf a}_{e,n})=
 \overline{\mathsf a}_{e,n}.
\end{equation}
The same assignments hold for the order-four hopping operators, because they are locally reconstructible polynomials in the Hermitian hopping generators. All loop functions are transported by these polynomial rules.

\begin{figure}[t]
\centering
\begin{tikzpicture}[x=1cm,y=1cm,>=Stealth,
 semion/.style={draw=blue,fill=blue!3,text=blue,inner sep=3pt},
 antisemion/.style={draw=orange!80!black,fill=orange!4,
   text=orange!80!black,inner sep=3pt}]
 \foreach \y/\n in {0/{n-1},1.65/n,3.3/{n+1}} {
  \fill[gray!3] (-.45,\y-.42) -- (6.35,\y-.42)--(7.0,\y+.42)--(.20,\y+.42)--cycle;
  \draw[gray!55] (-.45,\y-.42) -- (6.35,\y-.42)--(7.0,\y+.42)--(.20,\y+.42)--cycle;
  \node[font=\small,anchor=east] at (-.65,\y) {layer $\n$};
  \node[semion] at (1.7,\y) {$\mathfrak A_s^{(\n)}$};
  \node[antisemion] at (4.9,\y) {$\mathfrak A_{\bar s}^{(\n)}$};
 }
 \draw[->,line width=.65pt,blue] (1.7,.38)--(1.7,1.27);
 \draw[->,line width=.65pt,blue] (1.7,2.03)--(1.7,2.92);
 \node[font=\small,text=blue,anchor=west] at (1.93,.82) {shift};
 \node[font=\small,text=orange!80!black] at (4.9,.82) {fixed};
 \node[font=\small,text=orange!80!black] at (4.9,2.47) {fixed};
\end{tikzpicture}
\caption{The ISA pump translates the blue semion algebra by one layer and holds the orange anti-semion algebra fixed. The boxes denote complementary operator factors, whose locality does not require an on-site division of each physical qubit.}
\label{fig:pump}
\end{figure}
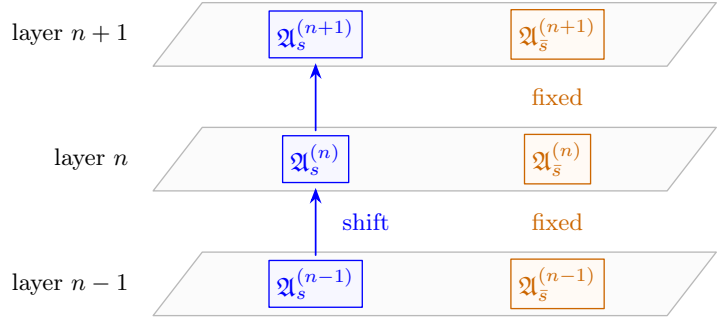

Its inverse is the downward pump,
\begin{equation}\label{eq:pumpinverse}
 \beta_\downarrow(O^{(n)}):=\beta_\uparrow^{-1}(O^{(n)})
 =\sum_\lambda u_\lambda^{(n-1)}v_\lambda^{(n)}.
\end{equation}
Both maps have uniformly bounded in-plane spread from the local factorization in Eq.~\eqref{eq:localfactorization} and move operators by at most one layer. They therefore define an exact three-dimensional qubit QCA, by the ISA construction of Ref.~\cite{Haah2023}. The downward convention will enter the TQFT comparison in Sec.~\ref{sec:TQFTcomparison}.

\subsection{Microscopic Pauli images of the pump}
\label{sec:explicitPauliQCA}

We apply the local Pauli decomposition to compute the pump's image of every single-qubit $X$ and $Z$.

The pump $\beta_\uparrow$ defined in Eq.~\eqref{eq:pump} sends every unbarred operator from layer $n$ to $n+1$ and leaves barred operators at $n$. In particular, complex linearity and multiplicativity give
\begin{equation}
     \begin{aligned}
     \beta_\uparrow\!\left(S(g_{p,n}g_{q,n})\right)
     &=\frac{1+i}{2}\id+\frac{1-i}{2}\beta_\uparrow(g_{p,n})\beta_\uparrow(g_{q,n})\\
     &=\frac{1+i}{2}\id+\frac{1-i}{2}\mathcal g_{p,n+1}\mathcal g_{q,n+1}
     =S(g_{p,n+1}g_{q,n+1}).
     \end{aligned}
\label{eq:pumpSpectralPolynomial}
\end{equation}
The same calculation fixes $S(\overline g_{p,n}\overline g_{q,n})$. Here, $\mathcal G_{v,n+1}$ is obtained by shifting every hopping operator in Eq.~\eqref{eq:loopPolynomialExpanded}. Mixed phases are likewise shifted term by term using Eq.~\eqref{eq:CZ}. These substitutions follow from the algebra automorphism in Eq.~\eqref{eq:pump} and therefore give the same image for every polynomial representing a given operator. Write $\mathsf Z_{e,n}^{\uparrow}:=\beta_\uparrow(Z_{e,n})$ and $\mathsf X_{e,n}^{\uparrow}:=\beta_\uparrow(X_{e,n})$ for the physical Pauli images. For $\mathcal V(e)=\{u,v\}$ and $j=j(e)$,
\begin{equation}
    \mathsf Z_{e,n}^{\uparrow} =\mathsf a_{j,n+1}\overline{\mathsf a}_{j,n}(g_{u,n+1}g_{v,n+1}) \CZ(g_{u,n+1},g_{v,n+1})\CZ(\overline g_{u,n},\overline g_{v,n})\CZ(g_{u,n+1},\overline g_{u,n})\CZ(g_{v,n+1},\overline g_{v,n}).
\label{eq:explicitPauliZimage}
\end{equation}
Equivalently, the shorter expression using the order-four hopping operators is
\begin{equation}
    \mathsf Z_{e,n}^{\uparrow}=a_{j(e),n+1}^{\varepsilon_e}\overline a_{j(e),n}^{-\varepsilon_e} \prod_{v\in\mathcal V(e)}S(g_{v,n+1}\overline g_{v,n})^{\varepsilon_e}.
\label{eq:auxZimageNew}
\end{equation}
For the two plaquettes $p,q\supset e$, the corresponding physical $X$ image is
\begin{equation}
    \mathsf X_{e,n}^{\uparrow} =\ww^{-1}\overline{\mathsf a}_{e,n} S(\overline g_{p,n}\overline g_{q,n})\,g_{p_+(e),n+1}\CZ(g_{p,n+1},\overline g_{p,n})\CZ(g_{q,n+1},\overline g_{q,n})\mathscr D_e[\{\mathsf Z_{e',n}^{\uparrow}\}]^\dagger.
    \label{eq:explicitPauliXimage}
\end{equation}
The functional in Eq.~\eqref{eq:explicitPauliXimage} is exactly the two explicit gate lists Eqs.~\eqref{eq:diagH} and~\eqref{eq:diagV}, now evaluated at the already specified $\mathsf Z_{e',n}^{\uparrow}$. No physical $X$ image appears on the right-hand side. Every factor acts only in layers $n,n+1$ and within a uniformly bounded in-plane neighborhood.

These identities are obtained by applying the defining ISA homomorphism to Eq.~\eqref{eq:pauliZprimary} and Eq.~\eqref{eq:pauliXprimary}. Thus, they satisfy, on the whole physical Hilbert space,
\begin{equation}
    (\mathsf X_{e,n}^{\uparrow})^2=(\mathsf Z_{e,n}^{\uparrow})^2=\id, \quad
    \mathsf X_{e,n}^{\uparrow}\mathsf Z_{e',m}^{\uparrow}=(-1)^{\delta_{ee'}\delta_{nm}}\mathsf Z_{e',m}^{\uparrow}\mathsf X_{e,n}^{\uparrow}, \quad
    [\mathsf X_{e,n}^{\uparrow},\mathsf X_{e',m}^{\uparrow}] =[\mathsf Z_{e,n}^{\uparrow},\mathsf Z_{e',m}^{\uparrow}]=0.
    \label{eq:pumpedPauliAlgebra}
\end{equation}
The inverse is obtained by replacing every unbarred $n+1$ by $n-1$, keeping all barred layers fixed. For a Pauli product, apply these assignments multiplicatively in its original order. Hence the formulas specify the entire QCA, not only its effect on the excitation labels or a selected state.

The TQFT comparison in Sec.~\ref{sec:TQFTcomparison} uses $\beta_\downarrow=\beta_\uparrow^{-1}$ and an additional source shift $\tau$, together with a finite-depth intertwiner. The formulas above are the exact action of the ISA-pumping representative. The action of that TQFT representative is obtained by applying the stated shift, downward formulas, and circuit conjugation in their specified order. Finite-depth equivalence is not equality of the raw images.

\subsection{Order-8 cancellation modulo translations}
\label{sec:eightfoldPump}

Eight applications of the semion pump are stably equivalent to four ordinary translations. The following result constructs the cancelling circuit directly from the microscopic ISA.

\begin{proposition}[Stable order-8 cancellation]\label{prop:eightfoldPump}
    Let $T_z$ translate every physical qubit upward by one layer. There is a geometrically local circuit $C$ on eight physical copies, with depth and gate diameter independent of system size, such that
    \begin{equation}\label{eq:eightfoldCircuit}
        \operatorname{Ad}_C =\bigl(\beta_\uparrow^8T_z^{-4}\bigr)\otimes\mathrm{id}^{\otimes7}.
    \end{equation}
\end{proposition}

\begin{proof}
    Appendix~\ref{app:fourCopyExchange} constructs a bounded-range QCA $\gamma_4$ exchanging the complete four-copy semion and anti-semion ISAs. Apply the same exchange in every layer to obtain $\Gamma$. Let $B=\beta_\uparrow^{\otimes4}$ pump the four semion factors upward, and let $\overline B$ instead pump the four anti-semion factors upward. Exchanging the two factors before and after pumping gives $\Gamma B\Gamma^{-1}=\overline B$.
    
    Applying both pumps moves both complementary factors upward by one layer. Every physical local operator is a finite sum of products from these factors, so their joint action moves every local observable: $B\overline B=T_z^{\otimes4}$. Physically, the complementary chiral pumps together produce ordinary translation.
    
    Write $[\alpha]$ for the stable QCA class modulo finite-depth circuits, allowing additional idle copies. The standard SWAP construction uses on-site swaps and their bounded-support image gates to implement $\alpha\otimes\alpha^{-1}$ by a finite-depth circuit~\cite[Sec.~2.1 and Appendix~A]{FreedmanHaahHastings2022}, as recalled in Appendix~\ref{app:pumpCancellationCircuit}. It then allows that an action can be transferred between registers modulo circuits. Parallel stacking and sequential composition give the same class operation. Exchanging the registers is also a local circuit, so these classes form an abelian group. Thus $[B]=[\beta_\uparrow]^4$, and conjugation by $\Gamma$ leaves $[B]$ unchanged, giving $[\overline B]=[B]$. Therefore,
    \begin{equation}\label{eq:eightfoldClassRelation}
        [\beta_\uparrow]^8=[B]^2=[B][\overline B] =[T_z^{\otimes4}]=[T_z]^4.
    \end{equation}
    Appendix~\ref{app:pumpCancellationCircuit} realizes this relation by an explicit circuit on eight copies, proving Eq.~\eqref{eq:eightfoldCircuit} with identity action on the other seven copies.
\end{proof}

The translation compensates the ordinary information flow of the one-factor pump; Appendix~\ref{app:pumpInformationFlow} proves that the uncompensated eighth power cannot be a finite-depth circuit. Thus, Eq.~\eqref{eq:eightfoldCircuit} gives order-8 cancellation modulo stable circuits and translations.

\section{Equivalence to the TQFT semion QCA}\label{sec:TQFTcomparison}\label{sec:TQFTframe}

Ref.~\cite{Shirley2022} constructed a three-dimensional semion QCA by boson condensation in a Walker-Wang model. Next, Ref.~\cite{SunNonClifford2026} gives a cubic-lattice construction from an invertible TQFT and identifies it with that QCA in Ref.~\cite{Shirley2022}.
% , using the preparation convention inverse to the disentangling convention of Ref.~\cite{Shirley2022}.
We now relate this TQFT construction to the ISA pump in Sec.~\ref{sec:microscopicQCA}.

The proof begins with separating the $xy$-face auxiliary qubits from the previously constructed QCA, then matching the remaining separators with those of the z-direction ISA pump. A final diagonal circuit corrects the remaining $X$ images. The general results that justify auxiliary qubit separation and the final correction are proved in Appendix~\ref{app:TQFTauxiliary}.

Let $\alpha_{\mathrm T}$ denote the completed TQFT QCA. The $xz$- and $yz$-face qubits at $(\rvec,n)$ correspond to the horizontal and vertical edges of the square lattice in layer $n$. We prove
\begin{equation}\label{eq:sectionNineGoal}
 \boxed{\alpha_{\mathrm T}
 =\operatorname{Ad}_{C}\circ
 \big[(\beta_\downarrow\circ\tau)\otimes\mathrm{id}_{xy}\big],}
\end{equation}
where $C$ is a finite-depth circuit. Here $\beta_\downarrow$ moves the semion factor from layer $n$ to $n-1$ while fixing the anti-semion factor. The source shift $\tau$ translates the $xz$ species by $-\xhat$ and the $yz$ species by $-\yhat$; $\mathrm{id}_{xy}$ fixes the entire auxiliary register. The equality holds on the full local operator algebra.

First, let us explain the two layer disentangling conventions. In Sec.~\ref{sec:disentangler}, the circuit $\mathcal W=VE\mathcal U$ converts the condensation constraints $D_e=1$ into $Z_e^b=1$. For consistency with the QCA construction in ~\cite{SunNonClifford2026}, we define an alternative disentangler as
\begin{equation}\label{eq:starDisentangler}
 \mathcal W_\star=\overline VE\overline{\mathcal U},
 \qquad
 \overline{\mathcal U}=\prod_v\CZ(\overline{\mathsf{G}}_v,\mathsf{W}_v).
\end{equation}
The circuit $\overline{\mathcal U}$ defined here sends $D_e$ to $\overline A_e^{\,2}$; the encoding $E$ and circuit $\overline V$ of Sec.~\ref{sec:disentangler} then send $\overline A_e^{\,2}$ to $Z_e^b$. Thus, both circuits disentangle the same condensed qubit:
\begin{equation}\label{eq:layerDisentanglerConstraints}
 \mathcal W D_e\mathcal W^\dagger
 =\mathcal W_\star D_e\mathcal W_\star^\dagger=Z_e^b.
\end{equation}
The starred choice is adapted to the cubic TQFT construction: the $xz,yz$ block of its matrix $\kappa$ is $M^\dagger$ [Eq.~\eqref{eq:kappaBlock}], which is precisely the CNOT pairing used by $\overline V$.

The explicit calculation in Appendix~\ref{app:TQFTframe} gives
\begin{equation}\label{eq:frameExact}
 \boxed{\mathcal W_\star\mathcal W^\dagger=F_\Sigma\otimes I_b.}
\end{equation}
Thus, the two descriptions after disentangling differ by a finite-depth circuit on the data alone, independently of the encoded ancilla state. Its phase is given in Eq.~\eqref{eq:frameCarry}. In particular, the semion and anti-semion loops in the starred basis are
\begin{equation}\label{eq:comparisonLoops}
 g_v^\star=F_\Sigma g_vF_\Sigma^\dagger,
 \qquad
 \overline g_v^\star=F_\Sigma\overline g_vF_\Sigma^\dagger.
\end{equation}
They remain mutually commuting local Hermitian involutions. This mapping is used twice below. In Sec.~\ref{sec:TQFTspectators}, it identifies the loop dressing of an $xy$-face separator with $g^\star$, allowing us to remove that dressing by a local controlled gate. In Sec.~\ref{sec:TQFTdata}, it expresses the remaining separators in the same layer basis as the ISA pump.

\subsection{Separating the \texorpdfstring{$xy$}{xy}-face auxiliary qubits}
\label{sec:TQFTspectators}

The TQFT construction starts with a $\mathbb Z_4$ qudit on each cubic face. The on-site encoding $E$ and the CNOT circuit $V_\kappa$ turn the condensation constraints into $Z^b=1$, leaving one physical qubit per face after the encoded $b$ qubits are removed. Here this $V_\kappa$ corresponds to the CNOT disentangler for the full cubic lattice, which will become layers of $\overline{V}$ if we restrict to the xy-direction. Appendix~\ref{app:TQFTdisentangling} recalls the parent operators and their reduction.

We use the completed local Pauli images $\mathsf Z_f^{\mathrm T}=\alpha_{\mathrm T}(Z_f)$ and $\mathsf X_f^{\mathrm T}=\alpha_{\mathrm T}(X_f)$ of Ref.~\cite{SunNonClifford2026}, Eqs.~(58) and (60). The first are the commuting separators; the second are their flippers. Our first task is to turn the $xy$ separators into ordinary on-site $Z$'s.

On each layer $\Sigma$, let $a_n$ be the integer $0$--$1$ edge cochain recording the $xz$- and $yz$-face bits, and let $w_n$ be the integer $0$--$1$ face cochain recording the $xy$ auxiliary bits. Using the signed integer coboundary $\delta$ and the cup products of Appendix~\ref{sec:cochains}, define the diagonal circuit
\begin{equation}\label{eq:sixCSphase}
 \begin{gathered}
 \mathcal F|a,w\rangle=i^{\Phi(a,w)}|a,w\rangle,\\
 \Phi(a,w)=\sum_n\int_\Sigma
 \bigl[-a_n\cup_1w_n+\delta a_n\cup_2w_{n+1}\bigr].
 \end{gathered}
\end{equation}
The first term couples each $xy$ qubit to the top and left edges of its plaquette; the second couples the auxiliary bit in the next layer to the signed plaquette boundary. The local rules in Eqs.~\eqref{eq:cup12rule}--\eqref{eq:cuptoprule} give six patterns of controlled-$S$ or controlled-$S^\dagger$ gates. These commuting gates assemble into a bounded-depth circuit.

Since $\overline{\mathcal U}$ commutes with the parent loops, Eqs.~\eqref{eq:starDisentangler} and~\eqref{eq:frameExact} imply that the semion loop $g^\star$ in qubit setting is $G$ reduced directly using $\overline VE$ in the joint $\overline A_e^{\,2}=1$ sector:
\[
 g^\star_{\rvec,n}
 ={}_b\langle0|\,\overline VE\,G_{\rvec,n}
 E^\dagger\overline V^\dagger\,|0\rangle_b,
\]
with the analogous identity for $\overline g^\star$ and $\overline G$. Thus the loop in the TQFT reduction is equivalent to ISA semion loop up to the layer circuit $F_\Sigma$.

The encoded Pauli calculation in Appendix~\ref{app:TQFTdisentangling} gives
\begin{equation}\label{eq:horizontalSep}
 \boxed{\mathcal F^\dagger\mathsf Z^{\mathrm T}_{xy,\rvec,n}\mathcal F
       =Z_{xy,\rvec,n}\,g^\star_{\rvec,n-1}.}
\end{equation}
Thus, $\mathcal F$ reduces the $xy$ separator to an on-site auxiliary $Z$ multiplied by the semion loop in the preceding layer. The semion loop decoration can be removed by the local controlled circuit
\begin{equation}\label{eq:spectatorJ}
 \mathcal J=\prod_{\rvec,n}
 \CZ\!\left(g^\star_{\rvec,n},X_{xy,\rvec,n+1}\right).
\end{equation}
Here $\CZ$ is the spectral controlled phase of Eq.~\eqref{eq:CZ}. Its conjugation sends $Z_{xy}$ to $gZ_{xy}$ and hence sends $gZ_{xy}$ back to $Z_{xy}$. Applying this identity to Eq.~\eqref{eq:horizontalSep} yields the adjusted QCA
\begin{equation}\label{eq:rhoDefinition}
 \rho:=\operatorname{Ad}_{\mathcal J\mathcal F^\dagger}\circ\alpha_{\mathrm T},
 \qquad \rho(Z_{xy,\rvec,n})=Z_{xy,\rvec,n}.
\end{equation}

The auxiliary-bit sectors are now invariant. Let $\rho_0$ be the induced QCA on the $xz$- and $yz$-face qubits in the sector where every auxiliary has $Z_{xy}=+1$. Both $\rho$ and $\rho^{-1}$ preserve this sector and retain their locality bounds. Fixing the auxiliary $Z$'s does not yet fix their $X$'s or remove the dependence of data images on the auxiliary bits. Lemma~\ref{prop:spectator}, proved in Appendix~\ref{app:TQFTauxiliary}, supplies a finite-depth circuit $\mathcal V$ such that
\begin{equation}\label{eq:spectatorFactorization}
 \boxed{\rho=\operatorname{Ad}_{\mathcal V}
                    \circ(\rho_0\otimes\mathrm{id}_{xy}).}
\end{equation}
After removing the finite-depth circuit, the QCA acts trivially on the $xy$ register. Consequently,
\begin{equation}\label{eq:TQFTdataReduction}
 \boxed{\alpha_{\mathrm T}
 =\operatorname{Ad}_{\mathcal F\mathcal J\mathcal V}
                    \circ(\rho_0\otimes\mathrm{id}_{xy}).}
\end{equation}
This is the required separation of the $xy$-face auxiliary qubits.

\subsection{Identifying the QCA with the ISA pump}
\label{sec:TQFTdata}

After separating the $xy$ auxiliaries, $\rho_0$ acts on a stack of copies of the square lattice $\Sigma$. Identify the $xz$- and $yz$-face qubits with its horizontal and vertical edges:
$Z_{h_\rvec,n}=Z_{xz,\rvec,n}$ and $Z_{v_\rvec,n}=Z_{yz,\rvec,n}$.
All cochains, coboundaries, and cup products below are two-dimensional, with the integer conventions of Sec.~\ref{sec:transport} and Appendix~\ref{sec:cochains}. The integral $\int_\Sigma$ is within one layer; the separate label $n$ records transport between layers.

We use the starred basis of Eq.~\eqref{eq:frameExact}. Let $F_{\Sigma,n}$ be the basis-change circuit on layer $n$, and continue to write $F_\Sigma=\prod_nF_{\Sigma,n}$ for its layerwise product. In addition to the starred loops, define
\begin{equation}
 a_{e,n}^\star=F_{\Sigma,n}a_{e,n}F_{\Sigma,n}^\dagger,
 \qquad
 \overline a_{e,n}^\star=F_{\Sigma,n}\overline a_{e,n}F_{\Sigma,n}^\dagger.
\end{equation}
These are the order-four hopping operators of Eq.~\eqref{eq:reducedhops} in the common layer basis. By Lemma~\ref{prop:hopOnly}, they generate the same local algebras as the Hermitian hopping operators. The downward pump in this basis is
\begin{equation}\label{eq:pumpFrame}
 \begin{gathered}
 \beta_\downarrow^\star
 =\operatorname{Ad}_{F_\Sigma}\circ\beta_\downarrow
                  \circ\operatorname{Ad}_{F_\Sigma^\dagger},\\
 \beta_\downarrow^\star(Z_{e,n})
 =F_\Sigma\beta_\downarrow(Z_{e,n})F_\Sigma^\dagger.
 \end{gathered}
\end{equation}
The second equality uses the diagonality of $F_\Sigma$. Thus semion operators still move from $n$ to $n-1$, while anti-semion operators stay at $n$.

The ordinary cup pairing identifies each vertex $v$ with the face $f(v)$ having lower-left vertex $v$: $\int_\Sigma\mathbf v\cup\mathbf f=\delta_{f,f(v)}$. The product of the two chiral loops is the corresponding plaquette parity,
\begin{equation}\label{eq:comparisonPlaquetteParity}
 g^\star_{v,n}\overline g^\star_{v,n}
 =\prod_{e'}Z_{e',n}^{\int_\Sigma\mathbf v\cup\delta\mathbf e'}
 =\prod_{e'\in\partial f(v)}Z_{e',n}.
\end{equation}
Its signed integer exponents can be reduced modulo two because the physical $Z$'s are involutions. The basis change fixes this product.

We further introduce here the selector of Eq.~\eqref{eq:cupselector},
\begin{equation}\label{eq:comparisonCupSelectors}
 \chi_{e'}(v)=\int_\Sigma\mathbf v\cup
       \bigl(\delta\mathbf e'\cup_2\delta\mathbf e'\bigr).
\end{equation}
It equals one at the lower-left vertices of the two faces incident to $e'$ and zero elsewhere. For the edge selected by the ordinary cup pairing, these vertices are precisely the endpoints of $e$.

Writing $I_{e,n}=\beta_\downarrow^\star(Z_{e,n})$, the Pauli factorization in Eq.~\eqref{eq:ZreducedFactorization} gives
\begin{equation}\label{eq:comparisonCupPump}
 \begin{aligned}
 I_{e,n}
 &=\prod_{e'}\left (a^\star_{e',n-1}
       \overline a^{\star \dagger}_{e',n}\prod_{v}
 S\left(g^\star_{v,n-1}\overline g^\star_{v,n}\right)^{
       \chi_{e'}(v)}\right )^ {\int_\Sigma\mathbf e\cup\mathbf e'}.
 \end{aligned}
\end{equation}
Only one edge $e'$ contributes. Thus the hopping product gives the oriented hopping factor, and the product of spectral phases supplies the two endpoint phases with the same orientation sign. This proves the cup-product form directly from Eq.~\eqref{eq:ZreducedFactorization}. The operator order and the integer signs are retained; in particular, the exponent of $S$ is not reduced modulo two. The layer labels display the transported semion factor and the stationary anti-semion factor.

Writing $e+\xhat+\yhat$ for the translate of $e$, the calculation in Appendix~\ref{app:TQFTdisentangling} combines the two data separators into one identity:
\begin{equation}\label{eq:reducedTwoSeparators}
 \boxed{\begin{aligned}
 \rho_0(Z_{e+\xhat+\yhat,n})
 &=I_{e,n}\prod_{e',v}\Bigl[
   (\overline g^\star_{v,n-1}g^\star_{v,n})^{\chi_{e'}(v)}
   (g^\star_{v,n-1}\overline g^\star_{v,n})^{
       \int_\Sigma\mathbf{f(v)}\cup_1\mathbf e'}
   \Bigr]^{\int_\Sigma\mathbf e\cup\mathbf e'}.
 \end{aligned}}
\end{equation}
The cup-one pairing $\int_\Sigma\mathbf{f(v)}\cup_1\mathbf e'$ selects the positively incident face of $e'$. Combined with the ordinary cup pairing, it gives $v=\rvec$ for $e=h_\rvec$ and $v=\rvec+\yhat$ for $e=v_\rvec$. All loop products in Eq.~\eqref{eq:reducedTwoSeparators} are commuting involutions, so the outer orientation sign does not change their values. The endpoint factors $\overline g^\star_{v,n-1}g^\star_{v,n}$ will be supplied by an output circuit; the remaining factor $g^\star_{v,n-1}\overline g^\star_{v,n}$ is the pumped image of an input plaquette parity.

First, use the layerwise circuit
\begin{equation}\label{eq:Kloop}
 \mathcal K=\prod_{v,n}\CZ(g^\star_{v,n},\overline g^\star_{v,n})
\end{equation}
to supply the endpoint factors. The operator $I_{e,n}$ flips $g^\star_{v,n-1}$ and $\overline g^\star_{v,n}$ at the two endpoints of $e$, and commutes with the other chiral loops. Consequently,
\begin{equation}
 \begin{aligned}
 \mathcal K I_{e,n}\mathcal K^\dagger
   &=I_{e,n}\prod_{e',v}
      (\overline g^\star_{v,n-1}g^\star_{v,n})^{
        \chi_{e'}(v)\int_\Sigma\mathbf e\cup\mathbf e'},\\
 \beta_\downarrow^\star\!\left(g^\star_{v,n}\overline g^\star_{v,n}\right)
   &=g^\star_{v,n-1}\overline g^\star_{v,n}.
 \end{aligned}
\end{equation}
Using the plaquette parity in Eq.~\eqref{eq:comparisonPlaquetteParity}, define the source change by
\begin{equation}\label{eq:sourceCupMap}
 \delta_{\rm source}(Z_{e+\xhat+\yhat,n})
   =Z_{e,n}\prod_{e',e''}Z_{e'',n}^{
       \left(\int_\Sigma\mathbf e\cup\mathbf e'\right)
       \left(\int_\Sigma\delta\mathbf e''\cup_1\mathbf e'\right)}.
\end{equation}
The ordinary cup pairing selects the transverse edge, and the cup-one pairing selects the boundary of its positively incident face. This proves that the extra factor is precisely the input plaquette parity required by Eq.~\eqref{eq:reducedTwoSeparators}; all Pauli exponents may be reduced modulo two. The map stays within layer $n$. Since $\mathcal K$ commutes with every $g^\star_{v,n-1}\overline g^\star_{v,n}$, the separator comparison becomes
\begin{equation}\label{eq:rho0match}
 \rho_0(Z_{e,n})
 =\mathcal K\,\beta_\downarrow^\star\!\bigl(\delta_{\rm source}(Z_{e,n})\bigr)\mathcal K^\dagger.
\end{equation}

To extend the source change to the full Pauli algebra, let $e_h$ and $e_v$ range over horizontal and vertical edges and define
\begin{equation}\label{eq:sourceCupCircuits}
 \begin{aligned}
 U_{\rm cn}
   &=\prod_{e_h,e_v,n}
     \CNOT_{(e_h,n)\to(e_v,n)}^{
       \int_\Sigma\mathbf e_v\cup_1\delta\mathbf e_h},\\
 L_{\rm cn}
   &=\prod_{e_h,e_v,n}
     \CNOT_{(e_v,n)\to(e_h,n)}^{
       \int_\Sigma\mathbf e_h\cup_1\delta\mathbf e_v}.
 \end{aligned}
\end{equation}
The integer exponents are reduced modulo two. The local cup-one rule gives the two control patterns $h_\rvec,h_{\rvec+\yhat}\to v_\rvec$ for $U_{\rm cn}$ and $v_{\rvec-\yhat},v_{\rvec+\xhat-\yhat}\to h_\rvec$ for $L_{\rm cn}$, all at the same $n$. Gates within each product commute, and each product has a two-layer schedule. Their composition with the species shift is
\begin{equation}\label{eq:sourceCircuit}
 \delta_{\rm source}=\operatorname{Ad}_{L_{\rm cn}}\circ\tau
                         \circ\operatorname{Ad}_{U_{\rm cn}},
\end{equation}
where $\tau(Z_{h_\rvec,n})=Z_{h_{\rvec-\xhat},n}$ and $\tau(Z_{v_\rvec,n})=Z_{v_{\rvec-\yhat},n}$, with the same shifts on $X$ and the identity on the $xy$ auxiliaries. Reading the composition from right to left reproduces Eq.~\eqref{eq:sourceCupMap}, as also checked in Appendix~\ref{app:TQFTdisentangling}. Thus the two-dimensional cup-product formulas identify all data $Z$ images using reversible local operations; the next subsection completes the comparison on the full operator algebra.

\subsection{Completing the equivalence on the full operator algebra}
\label{sec:TQFTfull}

We now combine the auxiliary separation with the data comparison. The source CNOT circuits can be transferred to the output side of the pump, giving
\begin{equation}
 C_{\rm sep}=\mathcal F\,\mathcal J\,\mathcal V\,\mathcal K\,
 \beta_\downarrow^\star(L_{\rm cn})(\beta_\downarrow^\star\circ\tau)(U_{\rm cn})\,F_\Sigma.
\label{eq:explicitMatchingCircuit}
\end{equation}
The product is read from right to left. The factor $F_\Sigma$ changes the ISA layer basis; the two image CNOT circuits implement the source change; $\mathcal K$ adds the loop factors. Finally, $\mathcal F\mathcal J\mathcal V$ restores the coupling to the $xy$ auxiliaries, as in Eq.~\eqref{eq:TQFTdataReduction}.

Applying a QCA to a circuit gate means taking its local unitary image. For example, $\CNOT_{c\to t}=(1+Z_c+X_t-Z_cX_t)/2$, so its image is obtained by substituting the known Pauli images. The ISA mapping in Sec.~\ref{sec:completeness} provides these images. Gates in a disjoint input layer have commuting images with uniformly bounded supports. A finite coloring schedules them in bounded depth. All data-only operations are extended by the identity on the auxiliary register.

Set
\begin{equation}
 \eta=(\beta_\downarrow\circ\tau)\otimes\mathrm{id}_{xy}.
\end{equation}
Eqs.~\eqref{eq:TQFTdataReduction} and~\eqref{eq:rho0match} imply
\begin{equation}\label{eq:missingBridge}
    C_{\rm sep}\,\eta(Z_j)\,C_{\rm sep}^\dagger =\mathsf Z_j^{\mathrm T}\qquad\text{for every data or auxiliary qubit }j.
\end{equation}
For the source factors, the required identity is
\begin{equation}
 \beta_\downarrow^\star\circ\delta_{\rm source}
 =\operatorname{Ad}_{\,\beta_\downarrow^\star(L_{\rm cn})
                  (\beta_\downarrow^\star\circ\tau)(U_{\rm cn})}
 \circ\beta_\downarrow^\star\circ\tau.
\end{equation}
Eq.~\eqref{eq:pumpFrame}, together with the fact that $F_\Sigma$ fixes every input $Z$, then converts the starred pump to $\eta$. This verifies the order of all factors in $C_{\rm sep}$.

The two QCAs now agree on every $Z$ image, including both data species. To isolate the remaining difference in their $X$ images, define
\begin{equation}
 \gamma=\eta^{-1}\circ\operatorname{Ad}_{C_{\rm sep}^\dagger}\circ\alpha_{\mathrm T}.
\end{equation}
Eq.~\eqref{eq:missingBridge} gives $\gamma(Z_j)=Z_j$ for every microscopic qubit. Corollary~\ref{prop:fixZ} in Appendix~\ref{app:TQFTauxiliary} therefore gives a finite-depth diagonal circuit $G$ with $\gamma=\operatorname{Ad}_G$. Hence
\begin{equation}\label{eq:bridgeConclusion}
 \boxed{\alpha_{\mathrm T}
   =\operatorname{Ad}_{\,C_{\rm sep}\eta(G)}\circ\eta.}
\end{equation}
As for the image CNOT circuits, locality of $\eta$ gives a bounded-depth schedule for $\eta(G)$. Thus $C=C_{\rm sep}\eta(G)$ is a finite-depth circuit, and Eq.~\eqref{eq:bridgeConclusion} identifies the action on both Pauli families and therefore on every local operator. One added $xy$ auxiliary per ISA cell, together with the stated source shifts, gives the claimed stable equivalence.

\begin{theorem}[Explicit semion factorization and pumping]\label{thm:main}
The construction has the following properties.
\begin{enumerate}[leftmargin=*,itemsep=4pt,topsep=3pt]
 \item On the infinite square lattice with one qubit per edge, the local Hermitian involutions $\mathsf a_e$ and $\overline{\mathsf a}_e$ in Eq.~\eqref{eq:involutivehop} generate complementary local $*$-algebras $\Asalg$ and $\Abalg$. They are mutual commutants within the physical local operator algebra. Every local operator is a finite sum of products from the two factors with a uniform bound on support enlargement. Their string excitations have order-two fusion and semion and anti-semion exchange phases $i$ and $-i$.
 \item Translating $\Asalg$ by one layer while fixing $\Abalg$ defines a three-dimensional qubit QCA with microscopic Pauli images given by Eqs.~\eqref{eq:explicitPauliZimage} and~\eqref{eq:explicitPauliXimage}. Its eighth power differs from four whole-layer translations by the stable finite-depth circuit constructed in Eq.~\eqref{eq:eightfoldCircuit}.
 \item After adding one $xy$ auxiliary qubit per cubic unit cell, the inverse pump is related to the completed TQFT semion QCA by the finite-depth circuit and source reparametrizations in Eqs.~\eqref{eq:missingBridge} and~\eqref{eq:bridgeConclusion}.
\end{enumerate}
\end{theorem}

\section{Discussion}\label{sec:discussion}

In this work, we constructed complementary semion and anti-semion algebras on unrestricted physical qubits, with explicit local generators and uniformly local reconstruction of every operator. Their Pauli decomposition makes the three-dimensional non-Clifford layer pump explicit. We established its eightfold cancellation modulo stable circuits and lattice translations, and related its inverse to the TQFT semion QCA on the full local operator algebra~\cite{SunNonClifford2026}. The construction also extends to spatial dimensions $d=4k{-}2$ and to a $U(1)_4$ ISA on physical $\mathbb Z_4$ qudits. Together with earlier odd-prime ISAs, finite stacking represents every pointed Witt class of bosonic Abelian anyon theories.

A central follow-up is to determine which relations between Abelian anyon theories lift to stable equivalences of ISAs or their layer pumps. The explicit local factorizations provide examples in which these relations can be tested microscopically, at the lattice-level. For translation-invariant Pauli stabilizer systems, boundary operator modules and generalized Witt groups give a related algebraic description of bulk-boundary correspondence~\cite{RubaYang2025WittBoundary}. Comparing these descriptions with the spectrum framework for general QCAs~\cite{JiYang2026Spectrum} could help formulate the corresponding questions for non-Pauli ISAs. Constructing boundary algebras from the Pauli decomposition could also clarify how chiral anomalies constrain finite-depth realizations of the resulting QCAs. The relation between chirality pumping and the three-fermion QCA~\cite{FidkowskiHastings2024PumpingChirality} motivates a comparison with dynamical pumps of chiral phases.

In higher dimensions, natural next steps are to derive the extended-excitation statistics, compare the resulting algebras with the higher-form semion QCA boundaries of Ref.~\cite{SunNonClifford2026}, and determine stable cancellation relations for the pumps. Berry-phase processes and their Pauli formulations provide concrete tools for this analysis~\cite{Kobayashi2026GeneralizedStatistics,Feng2026PauliTQFT}. In particular, the semionic-membrane processes in Ref.~\cite{Feng2026PauliTQFT} offer a natural comparison for the six-dimensional construction. Related Pontryagin statistics of membranes~\cite{Feng2026AnyonicMembranes} raise the question of which additional higher-form ISAs can be constructed. Further cyclic parent models may also lead to a systematic treatment of condensation and local disentangling beyond the two examples considered here.

More conjecturally, the existence of ISAs for non-Abelian chiral theories remains to be explored. Operator-algebraic descriptions of Levin--Wen and Walker-Wang boundaries~\cite{JonesNaaijkensPenneys2025Holography}, together with disentangling constructions for modular Walker-Wang models that allow auxiliary fermions~\cite{Bauer2023ModularWalkerWang}, provide concrete settings in which to investigate this question.

\section*{Acknowledgments}

Z.W. thanks Jeongwan Haah for many helpful discussions while Z.W. was a Ph.D. student at the California Institute of Technology. Z.W. and Y.-A.C. are supported by the National Natural Science Foundation of China (Grant No.~12474491) and the Central University Fundamental Research Funds (Peking University).

%%%%%%%%%%%%%%%%%%%%%%%%
\appendix

\section{Local completeness and microscopic Pauli identities}
\label{app:pauliDerivations}

This appendix proves uniformly local generation of the parent constraint commutant and the resulting physical commutants, and then derives the microscopic Pauli mapping.

\subsection{Parent hopping patterns}\label{app:parentPatterns}

The finite-support notation in this appendix encodes the explicit strings in Eq.~\eqref{eq:parentHops}. It is used only to prove the support bound. The construction is specified by those strings.

A Laurent polynomial is a finite sum $\sum_{m,n}c_{mn}x^my^n$, allowing both positive and negative powers. Here, $x^my^n$ places the coefficient at the displacement $m\xhat+n\yhat$: $x,y$ translate forward, while their inverses translate backward. Finite sums therefore describe finite support. The coefficients are Pauli exponents in $\mathbb Z_4$, since $X^4=Z^4=\id$. We use integer lifts of these coefficients for the divisibility arguments below, reducing modulo four afterward.

An edge pattern is a two-component column, ordered horizontal then vertical. Matrix multiplication combines translated patterns. In $M$, each column gives the $Z$ decorations of a hopping operator of that orientation, and the rows specify the orientations of the decorated edges. For example, the entry $M_{hh}=y-1$ places $Z$ on the horizontal edge one step above the hopping edge and $Z^{-1}$ on the hopping edge. The Laurent adjoint $\dagger$ transposes a matrix and replaces $x,y$ by $x^{-1},y^{-1}$. The $Z$ decorations in the unbarred and barred families are the columns of $M$ and $M^\dagger$, respectively:
\begin{equation}\label{eq:M}
 M=\begin{pmatrix}
 y-1 & x^{-1}y\\
 1-x-y^{-1} & x^{-1}-1
 \end{pmatrix}.
\end{equation}
Thus, a product with exponent pattern $a$ has Pauli exponents $(a,Ma)$ in the first family, while a barred product with pattern $b$ has exponents $(b,M^\dagger b)$: $A(a)\propto X^aZ^{Ma}$ and $\overline A(b)\propto X^bZ^{M^\dagger b}$.

\subsection{Uniformly local factorization of constraint-preserving Paulis}\label{app:localPauliFactorization}

This subsection proves the parent factorization used in Sec.~\ref{sec:physicalFactorization}. We use the translation-based representation of the hopping patterns in Appendix~\ref{app:parentPatterns}. For a finitely supported exponent pattern $u$, $A(u)$ denotes an ordered product of the $A_e$ with those exponents. The order affects only the scalar phase. The notation $\overline A(w)$ is defined in the same way.

Over the integer Laurent-polynomial ring, introduce
\begin{equation}\label{eq:dpc}
 d=\begin{pmatrix}x^{-1}-1\\y^{-1}-1\end{pmatrix},\qquad
 c=\begin{pmatrix}1-y\\x-1\end{pmatrix},\qquad
 p=\begin{pmatrix}1+y^{-1}&1+x^{-1}\end{pmatrix}.
\end{equation}
The columns $d$ and $c$ describe the star and plaquette patterns. Set
\begin{equation}\label{eq:KLJ}
 K=M-M^\dagger,\qquad L=2I+dp,\qquad
 J=\begin{pmatrix}0&x^{-1}\\-y^{-1}&0\end{pmatrix}.
\end{equation}
The Pauli exponent columns of $C_e$ are $\binom{L}{ML}$. Direct integer matrix multiplication gives
\begin{equation}\label{eq:integeridentities}
 \begin{gathered}
 Md=c,\qquad M^\dagger d=-c,\qquad KL=4J,
 \qquad L^\dagger K=-4J^\dagger,\\
 \det K=4,\qquad\det L=4x^{-1}y^{-1}.
 \end{gathered}
\end{equation}
These identities precede reduction modulo four.

\begin{lemma}[Local Pauli factorization]\label{prop:factor}
Every finitely supported qudit Pauli commuting with all $C_e$ is, up to scalar phase, an ordered product of $A_e$ operators followed by an ordered product of $\overline A_e$ operators. Both products are supported in a fixed enlargement of the original support.
\end{lemma}
\begin{proof}
Write the Pauli as $P(u,v)=X^uZ^v$. Its commutation condition is
\begin{equation}\label{eq:centralizerOriginal}
 L^\dagger(v-M^\dagger u)=0\pmod4.
\end{equation}
Using $L^\dagger(M-M^\dagger)=-4J^\dagger$, this is equivalent to
\begin{equation}\label{eq:centralizer}
 L^\dagger(v-Mu)=0\pmod4.
\end{equation}
Choose finite-support integer lifts of $u,v$ and define
\begin{equation}\label{eq:tandbeta}
 t=v-Mu,\qquad\beta=\frac{L^\dagger t}{4},\qquad
 w=(J^\dagger)^{-1}\beta.
\end{equation}
The divisibility in Eq.~\eqref{eq:centralizer} makes $\beta$ an integer pattern. Eq.~\eqref{eq:integeridentities} implies
\begin{equation}
 L^\dagger(t+Kw)=L^\dagger t-4J^\dagger w=0.
\end{equation}
The integer Laurent ring is torsion-free and $\det L=4x^{-1}y^{-1}$, so $L^\dagger$ is injective and $t=-Kw$. Thus, \begin{equation}\label{eq:splitexponents}
 \binom{u}{v}
 =\binom{u-w}{M(u-w)}+\binom{w}{M^\dagger w}\pmod4,
\end{equation}
which gives
\begin{equation}\label{eq:factorformula}
 P(u,v)\propto A(u-w)\,\overline A(w).
\end{equation}
Every operation in Eq.~\eqref{eq:tandbeta} is multiplication by a fixed finite Laurent matrix, coefficientwise division by four, or a signed translation. Their support enlargement is bounded independently of the input support.
\end{proof}

This proves $\mathfrak C^\perp=\mathfrak H$ in Eq.~\eqref{eq:parentCommutants} before selecting a constraint sector. Indeed, conjugation by the Pauli generators $C_e$ is diagonal in the qudit Pauli basis, so every operator in $\mathfrak C^\perp$ is a finite linear combination of individually constraint-preserving Paulis. Lemma~\ref{prop:factor} factors each of them into the hopping families.

The reverse commutant also has a uniformly local proof. A Pauli $P(u,v)$ commuting with both hopping families obeys
\begin{equation}\label{eq:jointParentCentralizer}
 v=Mu=M^\dagger u\pmod4,\qquad Ku=0\pmod4.
\end{equation}
Choose finite-support integer lifts and set
\begin{equation}\label{eq:parentCenterLift}
 q=\frac{Ku}{4},\qquad t=J^{-1}q.
\end{equation}
Using $KL=4J$, one gets $K(u-Lt)=0$ over the integers. Since $\det K=4$, the map $K$ is injective on finite-support integer patterns, so $u=Lt$. The columns of $\binom{L}{ML}$ are the exponent patterns of $C_e$. Consequently, $P(u,v)\propto C(t)$, is an ordered product of condensates. The same diagonal-Pauli-basis argument extends this conclusion to arbitrary operators in $\mathfrak H^\perp$. This proves $\mathfrak H^\perp=\mathfrak C$. All operations have finite range, so the support enlargement is uniform. On a periodic lattice the integer pattern group remains torsion-free, and multiplication by $\det K=4$ and $\det L=4x^{-1}y^{-1}$ remains injective. Both commutant equalities therefore hold there as well on the full parent Hilbert space, before any constraint sector is selected. Finally, $\mathfrak C\subseteq\mathfrak H$ is commutative, giving $\operatorname{Center}(\mathfrak H)=\mathfrak C$.

\subsection{Periodic factorization and uniformly local commutants}\label{app:localCommutants}

We now justify the finite- and infinite-lattice conclusions following Eq.~\eqref{eq:localfactorization}.

The factorization also identifies the commutants, with the same locality bound. First consider a sufficiently large periodic lattice. The integer lifting argument remains valid in its periodic group ring: multiplication by $\det L=4x^{-1}y^{-1}$ is injective on the free abelian group of integer patterns, even though the ring has other zero divisors. Consequently Eq.~\eqref{eq:localfactorization} holds for all operators in this finite system.

Let $\mathfrak A_{s,N}$ and $\mathfrak A_{\bar s,N}$ be the two periodic algebras. They commute and jointly generate the full physical matrix algebra. A central element of either one commutes with both and hence with the full matrix algebra, so it is scalar. The finite-dimensional tensor-factor decomposition therefore gives
\begin{equation}\label{eq:finiteISA}
 \mathfrak A_{s,N}^\perp=\mathfrak A_{\bar s,N},\qquad\mathfrak A_{\bar s,N}^\perp=\mathfrak A_{s,N}.
\end{equation}

To retain locality on the infinite plane, let $O$ be a local operator commuting with $\Asalg$, and use Eq.~\eqref{eq:localfactorization} with every $u_\lambda,v_\lambda$ in the uniformly enlarged neighborhood of $\supp O$. Embed this neighborhood in a torus large enough that no relevant support wraps. Then $O$ commutes with $\mathfrak A_{s,N}$: the generators overlapping it agree with their infinite-lattice representatives, and all others have disjoint support. Apply the finite tracial conditional expectation onto $\mathfrak A_{\bar s,N}$ to the local factorization. Since the algebras are complementary matrix factors, it sends each separated product to
\begin{equation}
 u_\lambda v_\lambda
 \longmapsto\tau_N(u_\lambda)v_\lambda,
\end{equation}
where $\tau_N$ is the normalized physical trace. It fixes $O$, giving
\begin{equation}\label{eq:localCommutantExpectation}
 O=\sum_\lambda\tau_N(u_\lambda)v_\lambda.
\end{equation}
The right-hand side retains the original uniform support bound. It is therefore an identity in the infinite local algebra and shows $O\in\Abalg$. Interchanging the two families proves the other inclusion. Finite local linear relations can likewise be tested in a sufficiently large torus, so multiplication identifies the algebraic tensor product of the two factors with the full local algebra.

\subsection{\texorpdfstring{The Pauli-$Z$ lift and normalization}{The Pauli-Z lift and normalization}}

We derive the physical Pauli-$Z$'s decomposition by lifting $Z$ through the disentangler and reducing its parent factorization.

The encoded $Z_e^a$ commutes with $V$, and $E^\dagger Z_e^aE=Z_e^2$. Its conjugation by $\mathcal U$ is
\begin{equation}\label{eq:pauliZlift}
 \mathcal W^\dagger(Z_e^a\otimes\id_b)\mathcal W
 =\mathcal U Z_e^2\mathcal U
 =Z_e^2\prod_{v\in\mathcal V(e)}\mathsf{W}_v.
\end{equation}
At precisely these vertices $Z_e^2$ changes the sign of $\mathsf{G}_v$. The controlled-loop gate attaches $\mathsf{W}_v$. Multiplying the strings in Eqs.~\eqref{eq:parentHops} and~\eqref{eq:toricStarPlaquette} gives
\begin{equation}
 Z_e^2=A_{j(e)}^{\varepsilon_e}\overline A_{j(e)}^{-\varepsilon_e}
 \prod_{v\in\mathcal V(e)}W_v^{\varepsilon_e}.
\end{equation}
For an order-four $W$, $W\mathsf{W}=S(W^2)$ and $W^{-1}\mathsf{W}=S(W^2)^\dagger$. Thus, the full-qudit lift is
\begin{equation}
 A_{j(e)}^{\varepsilon_e}\overline A_{j(e)}^{-\varepsilon_e}
 \prod_{v\in\mathcal V(e)}S(W_v^2)^{\varepsilon_e}.
\end{equation}
Reduction uses $\mathcal R(W_v^2)=g_v\overline g_v$ and proves Eq.~\eqref{eq:pauliZaux}.

Substituting Eq.~\eqref{eq:recoverAuxiliary} into that equation proves Eq.~\eqref{eq:pauliZprimary}. The scalar phase identity needed here is, for commuting involutions $b_1,b_2,c_1,c_2$,
\begin{align}
 &S(b_1b_2)^\varepsilon S(c_1c_2)^{-\varepsilon}
 S(b_1c_1)^\varepsilon S(b_2c_2)^\varepsilon\nonumber\\
 &\hspace{7mm}=b_1b_2\CZ(b_1,b_2)\CZ(c_1,c_2)
 \CZ(b_1,c_1)\CZ(b_2,c_2),\qquad\varepsilon=\pm1.
 \label{eq:pauliZphase}
\end{align}
To see this algebraically, write $b_i=(-1)^{x_i}$ and $c_i=(-1)^{y_i}$ on their joint eigenspaces. The exponent of $i$ on the left is
\begin{equation}\label{eq:pauliZbinaryPhase}
 \begin{split}
 \varepsilon\big[(x_1\oplus x_2)-(y_1\oplus y_2)
             +(x_1\oplus y_1)+(x_2\oplus y_2)\big]
 \equiv 2\big(x_1+x_2+x_1x_2+y_1y_2+x_1y_1+x_2y_2\big)\pmod4.
 \end{split}
\end{equation}
The right-hand side is precisely the phase of the loop product and four controlled phases in Eq.~\eqref{eq:pauliZphase}. The identity holds for either sign of $\varepsilon$ and therefore, by the joint spectral decomposition, for the commuting operators themselves. No loop is fixed to $+1$.

\subsection{\texorpdfstring{The Pauli-$X$ action and ancillary sector}{The Pauli-X action and ancillary sector}}

We derive the physical Pauli-$X$ identity while retaining the ancillary sector until the final reduction.

The first useful identity is valid on the whole encoded qubit space:
\begin{equation}\label{eq:pauliDisentanglerIdentity}
 VE\overline A_eE^\dagger V^\dagger
 =\ww\,X_e^a\CZ(Z_e^a,Z_e^b)\mathscr D_e[Z^a].
\end{equation}
Indeed, $EX_eE^\dagger=X_e^a\CZ_{a_eb_e}$ and $EZ_{e'}E^\dagger=X_{e'}^bS_{a_{e'}}$. Conjugating $X_e^a$ by $V$ attaches
\begin{equation*}
 \prod_{e'}(X_{e'}^b)^{\int(\mathbf e\cup\mathbf e'-\delta\mathbf e\cup_1\mathbf e')},
\end{equation*}
which cancels the corresponding string from the clocks in $\overline A_e$. Moving this string through the on-site $\CZ$ gives one $Z_e^a$. It is canceled by the other controlled phase, since $\int(\mathbf e\cup\mathbf e-\delta\mathbf e\cup_1\mathbf e)=-1$ for either edge orientation. The remaining terms are precisely Eq.~\eqref{eq:diagFunctional} and the retained $\CZ(Z_e^a,Z_e^b)$.

The second identity concerns $\mathcal U$. Put $s_{ve}=\delta\mathbf e(p(v))\in\{\pm1\}$ for an incident plaquette. Since $\overline A_e$ commutes with $\mathsf{G}_v$ and obeys $W_v\overline A_e=i^{s_{ve}}\overline A_eW_v$, one obtains
\begin{equation}\label{eq:pauliUbar}
 \mathcal U\overline A_e\mathcal U
 =\overline A_e\prod_{v:\,p(v)\supset e}
 \CZ\!\left(\mathsf{G}_v,(-1)^{(1-s_{ve})/2}W_v^2\right).
\end{equation}
This is a full-qudit identity. The high-bit rules behind it are
\begin{equation}
 \mathsf{iW}=W^2\mathsf{W},\qquad
 \mathsf{-iW}=-W^2\mathsf{W},
\end{equation}
and multiplication of two controlled phases with the same first argument combines their second arguments.

After imposing the already specified $D=1$ constraints, $\mathsf{G}_v$ reduces to $g_v$ and $W_v^2$ to $g_v\overline g_v$. Consequently,
\begin{equation}\label{eq:pauliReducedG}
 \mathcal R(\mathcal U\overline A_e\mathcal U)
 =\overline a_e\,g_{p_+(e)}\CZ(g_p,\overline g_p)\CZ(g_q,\overline g_q),
 \qquad p,q\supset e.
\end{equation}
On the other hand, $\mathcal W=VE\mathcal U$ cancels both $\mathcal U$'s in this reduced operator. Setting only the ancillary $Z^b=1$ in Eq.~\eqref{eq:pauliDisentanglerIdentity} gives
\begin{equation}
 \mathcal R(\mathcal U\overline A_e\mathcal U)
 =\ww X_e\mathscr D_e[Z].
\end{equation}
Combining the last two equations and multiplying by $\mathscr D_e[Z]^\dagger$ on the right proves Eq.~\eqref{eq:pauliXaux}. The ancillary factor in Eq.~\eqref{eq:pauliDisentanglerIdentity} must not be dropped before this reduction.

Every step is an identity in the local operator algebra: the on-site encoding gives Eq.~\eqref{eq:pauliDisentanglerIdentity}, the spectral normalization gives Eq.~\eqref{eq:pauliUbar}, and their common reduced operator gives Eq.~\eqref{eq:pauliXaux}. Since the reduction is a $*$-homomorphism on constraint-preserving operators, this argument retains all scalar phases and holds on the whole physical-qubit Hilbert space.

\FloatBarrier

\section{Local qubit gate products}
\label{app:gates}

This appendix gives deterministic, ancilla-free gate products for the order-four hopping operators $a_e,\overline a_e$ and the Hermitian hopping operators $\mathsf a_e,\overline{\mathsf a}_e$ defined in Eqs.~\eqref{eq:reducedhops} and~\eqref{eq:involutivehop}. All gates act on the surviving physical qubits. We first give the order-four hopping products, then compile the spectral normalization into ordinary gates on the same support.

\subsection{Gate products for the order-four hopping operators}
\label{sec:auxiliaryGates}

We compile the reduced order-four hopping operators into explicit gate products on the physical qubits.

Only two spectral gates contribute to a hopping operator. Directly from Eq.~\eqref{eq:parentHops},
\begin{equation}\label{eq:twoU}
 [u_{\rvec},A_e]=[u_{\rvec},\overline A_e]=0
 \qquad\text{if }e\notin\partial p(\rvec).
\end{equation}
Writing $p,q\supset e$ for the two incident plaquettes, the reduction can therefore be evaluated from ${}_b\langle0|VE\,u_pu_q A_eu_qu_p\,E^\dagger V^\dagger|0\rangle_b$ and its barred counterpart. The result is the following ancilla-free gate product
\begin{equation}\label{eq:explicitqubitA}
 a_{t_{\rvec}}
 =\zeta_t\,\mathcal N_t^\dagger\mathcal C_t\mathcal D_t^+\mathcal N_t,
 \qquad t=h,v,
\end{equation}
\begin{equation}\label{eq:explicitqubitAbar}
 \overline a_{t_{\rvec}}
 =\zeta_t\,\mathcal N_t^\dagger\mathcal C_t\mathcal D_t^-\mathcal N_t,
 \qquad t=h,v.
\end{equation}
Here, $\zeta_h=e^{3\pi i/4}$ and $\zeta_v=e^{-\pi i/4}$. There are three circuit blocks: $\mathcal N_t$ computes plaquette parities, $\mathcal D_t^\pm$ applies diagonal phases, and $\mathcal C_t$ permutes the computational basis using CNOTs and bit flips. Their complete ordinary-gate products are given below. Products act from right to left: apply $\mathcal N_t$, then $\mathcal D_t^\pm$, then $\mathcal C_t$, and finally undo $\mathcal N_t$. The two chiralities differ only in the single-qubit phases within $\mathcal D_t^\pm$.

\begin{figure}[t]
\centering
\begin{subfigure}[b]{.42\textwidth}\centering
\begin{tikzpicture}[x=1.20cm,y=1.20cm,
  qnum/.style={circle,draw=black!50,line width=.5pt,text=blue,fill=white,inner sep=.35pt,minimum size=4.9mm,font=\scriptsize},
  tick/.style={font=\scriptsize,text=black!35}]
\draw[gridline] (-1,-2) -- (-1,1);
\node[tick,below] at (-1,-2.15) {$ -1 $};
\draw[gridline] (0,-2) -- (0,1);
\node[tick,below] at (0,-2.15) {$ 0 $};
\draw[gridline] (1,-2) -- (1,1);
\node[tick,below] at (1,-2.15) {$ 1 $};
\draw[gridline] (-1,-2) -- (1,-2);
\node[tick,left] at (-1.12,-2) {$ -2 $};
\draw[gridline] (-1,-1) -- (1,-1);
\node[tick,left] at (-1.12,-1) {$ -1 $};
\draw[gridline] (-1,0) -- (1,0);
\node[tick,left] at (-1.12,0) {$ 0 $};
\draw[gridline] (-1,1) -- (1,1);
\node[tick,left] at (-1.12,1) {$ 1 $};
\draw[edgefocus,line width=.75pt] (0,0) -- (1,0);
\node[qnum] at (-0.5,-1) {1};
\node[qnum] at (-0.5,0) {2};
\node[qnum] at (-0.5,1) {3};
\node[qnum] at (0.5,-1) {4};
\node[qnum] at (0.5,0) {5};
\node[qnum] at (0.5,1) {6};
\node[qnum] at (-1,-1.5) {7};
\node[qnum] at (-1,-0.5) {8};
\node[qnum] at (-1,0.5) {9};
\node[qnum] at (0,-1.5) {10};
\node[qnum] at (0,-0.5) {11};
\node[qnum] at (0,0.5) {12};
\node[qnum] at (1,-0.5) {13};
\node[qnum] at (1,0.5) {14};
\fill[black] (0,0) circle (1.4pt);
\node[font=\scriptsize,text=black,below left,inner sep=.35pt] at (0,0) {$\rvec$};
\node[tick,below] at (1.25,-2.15) {$x$};
\node[tick,left] at (-1.12,1.28) {$y$};
\end{tikzpicture}
\caption{Horizontal hopping operators: $a_{h_\rvec},\overline a_{h_\rvec}$.}
\end{subfigure}\hfill
\begin{subfigure}[b]{.54\textwidth}\centering
\begin{tikzpicture}[x=1.15cm,y=1.15cm,
  qnum/.style={circle,draw=black!50,line width=.5pt,text=blue,fill=white,inner sep=.35pt,minimum size=4.9mm,font=\scriptsize},
  tick/.style={font=\scriptsize,text=black!35}]
\draw[gridline] (-2,-1) -- (-2,1);
\node[tick,below] at (-2,-1.15) {$ -2 $};
\draw[gridline] (-1,-1) -- (-1,1);
\node[tick,below] at (-1,-1.15) {$ -1 $};
\draw[gridline] (0,-1) -- (0,1);
\node[tick,below] at (0,-1.15) {$ 0 $};
\draw[gridline] (1,-1) -- (1,1);
\node[tick,below] at (1,-1.15) {$ 1 $};
\draw[gridline] (-2,-1) -- (1,-1);
\node[tick,left] at (-2.12,-1) {$ -1 $};
\draw[gridline] (-2,0) -- (1,0);
\node[tick,left] at (-2.12,0) {$ 0 $};
\draw[gridline] (-2,1) -- (1,1);
\node[tick,left] at (-2.12,1) {$ 1 $};
\draw[edgefocus,line width=.75pt] (0,0) -- (0,1);
\node[qnum] at (-1.5,0) {1};
\node[qnum] at (-1.5,1) {2};
\node[qnum] at (-0.5,0) {3};
\node[qnum] at (-0.5,1) {4};
\node[qnum] at (0.5,0) {5};
\node[qnum] at (0.5,1) {6};
\node[qnum] at (-2,-0.5) {7};
\node[qnum] at (-2,0.5) {8};
\node[qnum] at (-1,-0.5) {9};
\node[qnum] at (-1,0.5) {10};
\node[qnum] at (0,-0.5) {11};
\node[qnum] at (0,0.5) {12};
\node[qnum] at (1,0.5) {13};
\fill[black] (0,0) circle (1.4pt);
\node[font=\scriptsize,text=black,below left,inner sep=.35pt] at (0,0) {$\rvec$};
\node[tick,below] at (1.25,-1.15) {$x$};
\node[tick,left] at (-2.12,1.28) {$y$};
\end{tikzpicture}
\caption{Vertical hopping operators: $a_{v_\rvec},\overline a_{v_\rvec}$.}
\end{subfigure}
\caption{Numbered physical-qubit supports for the gate products in Eqs.~\eqref{eq:explicitqubitA} and~\eqref{eq:explicitqubitAbar}. The marked vertex is the origin $\rvec$; the emphasized edge is the hopping edge. Labels in circles identify the remaining $a$-qubits, not ancillary $b$-qubits. The two chiralities use the same support and numbering. All gates act on the numbered qubits of the appropriate panel; unlabeled edges carry the identity.}
\label{fig:qubitSupport}
\end{figure}
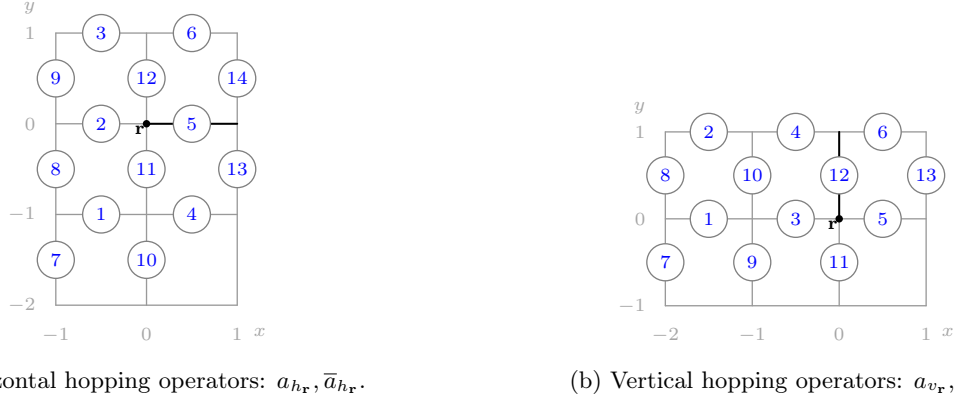

\paragraph{Physical-qubit labels.}
Use the numbering in Fig.~\ref{fig:qubitSupport}, separately for each orientation. All coordinates are relative to the hopping anchor $\rvec$. Thus, $h_{(u,v)}$ in that figure means the surviving qubit on $h_{\rvec+u\xhat+v\yhat}$, and similarly for a vertical edge. In the formulas below $X_j$, $S_j$, and the gate labels refer to these numbered physical qubits. The hopping edge is qubit $5$ in the horizontal panel and qubit $12$ in the vertical panel.

\paragraph{CNOT and bit-flip factors.}
The first factor computes the two incident plaquette parities into existing physical qubits:
\begin{align}
 \mathcal N_h
 &=\left(\prod_{i\in\{4,5,11\}}\CNOT_{i\to13}\right)
   \left(\prod_{i\in\{5,6,12\}}\CNOT_{i\to14}\right),\label{eq:Nh}\\
 \mathcal N_v
 &=\left(\prod_{i\in\{3,10,12\}}\CNOT_{i\to4}\right)
   \left(\prod_{i\in\{5,6,12\}}\CNOT_{i\to13}\right).\label{eq:Nv}
\end{align}
The six CNOTs within each $\mathcal N_t$ commute, so $\mathcal N_t^\dagger=\mathcal N_t$. If $z_i$ are the incoming physical bits and $y=\mathcal N_t z$, then
\[
 \begin{array}{c|cc}
 t&\text{first plaquette parity}&\text{second plaquette parity}\\\hline
 h&y_{13}=z_4\oplus z_5\oplus z_{11}\oplus z_{13}
  &y_{14}=z_5\oplus z_6\oplus z_{12}\oplus z_{14}\\
 v&y_4=z_3\oplus z_4\oplus z_{10}\oplus z_{12}
  &y_{13}=z_5\oplus z_6\oplus z_{12}\oplus z_{13}.
 \end{array}
\]
All other bits are unchanged. The intermediate use of a physical qubit for a parity is a reversible change of basis, not a constraint or an initialization.

The computational-basis permutations are
\begin{align}
 \mathcal C_h
 &=X_1X_4X_5X_{10}X_{11}X_{13}X_{14}
   \left(\prod_{i\in\{1,4,10,11\}}\CNOT_{13\to i}\right)
   \left(\prod_{i\in\{2,5,11,12\}}\CNOT_{14\to i}\right),\label{eq:CandXh}\\
 \mathcal C_v
 &=X_3X_4X_5X_{11}X_{13}
   \left(\prod_{i\in\{1,3,9,10\}}\CNOT_{4\to i}\right)
   \left(\prod_{i\in\{3,5,11,12\}}\CNOT_{13\to i}\right).\label{eq:CandXv}
\end{align}
\begin{samepage}
The CNOTs within each block commute. They are applied before its $X$ gates. The order between the different circuit blocks in Eqs.~\eqref{eq:explicitqubitA} and~\eqref{eq:explicitqubitAbar} must also be retained.

\paragraph{Single-qubit phases and the common diagonal gates.}
Using the qubit phase $S$ from Eq.~\eqref{eq:S}, define the full diagonal blocks by
\begin{equation}\label{eq:gateDiagonal}
 \begin{aligned}
 \mathcal D_t^\pm
 &=\left(\text{single-qubit product below}\right)\\
 &\quad{}\times\prod_{(i,j)\in\mathcal E_t}\CZ_{ij}
 \prod_{k\in\mathcal K_t}\ \prod_{(i,j)\in\mathcal T_{t,k}}
                  \mathrm{CCZ}_{ijk}.
 \end{aligned}
\end{equation}
\[
\begin{array}{c|cc}
 t&+&-\\\hline
 h&Z_1Z_2Z_8S_{11}S_{13}^\dagger S_{14}&Z_1Z_2Z_8S_{11}^\dagger Z_{14}\\[2pt]
 v&S_3^\dagger S_4S_{13}&S_3Z_4Z_{13}
\end{array}
\]
\end{samepage}
Here, $\mathcal K_h=\{13,14\}$ and $\mathcal K_v=\{4,13\}$. The $\CZ$ and $\mathrm{CCZ}$ factors are common to both chiralities. The $\CZ_{ij}$ gates are given by Eq.~\eqref{eq:CZ} with arguments $Z_i,Z_j$, and $\mathrm{CCZ}_{ijk}|y\rangle=(-1)^{y_iy_jy_k}|y\rangle$. The following lists give \emph{every gate} in Eq.~\eqref{eq:gateDiagonal}.

\begin{equation}\label{eq:Ehgates}
\begin{aligned}
 \mathcal E_h=\{&(1,2),(1,7),(1,8),(1,10),(4,10),(7,10),\\
                &(1,11),(2,11),(5,11),(8,11),(1,13),(2,13),(8,13),\\
                &(3,14),(6,14),(8,14),(9,14),(11,14),(13,14)\},
\end{aligned}
\end{equation}
\begin{equation}\label{eq:Evgates}
\begin{aligned}
 \mathcal E_v=\{&(1,4),(2,4),(4,8),(3,9),(4,9),(4,10),(3,11),\\
                &(5,11),(9,11),(4,12),(3,13),(4,13),(6,13)\}.
\end{aligned}
\end{equation}
For the horizontal hopping operator, the two lists of controlled-$\CZ$ pairs are
\begin{align}
 \mathcal T_{h,13}=\{&(1,2),(4,5),(1,7),(1,8),(1,10),(4,10),\notag\\
                    &(7,10),(1,11),(2,11),(4,11),(5,11),(8,11)\},\label{eq:Th13}\\
 \mathcal T_{h,14}=\{&(2,3),(5,6),(2,8),(2,9),(2,11),(5,11),(8,11),\notag\\
                    &(2,12),(3,12),(5,12),(6,12),(9,12),(2,13),(5,13),(8,13)\}.
                    \label{eq:Th14}
\end{align}
For the vertical hopping operator, they are
\begin{align}
 \mathcal T_{v,4}=\{&(1,2),(1,7),(1,8),(1,9),(3,9),(7,9),\notag\\
                   &(1,10),(2,10),(3,10),(8,10),(3,12),(10,12)\},\label{eq:Tv4}\\
 \mathcal T_{v,13}=\{&(5,6),(3,9),(4,9),(3,11),(5,11),(9,11),\notag\\
                    &(3,12),(5,12),(6,12)\}.\label{eq:Tv13}
\end{align}
For example, $(1,2)\in\mathcal T_{h,13}$ means the single gate $\mathrm{CCZ}_{1,2,13}$. Within $\mathcal D_t^\pm$ all gates commute, including gates from different $k$ lists that overlap. This gives $19$ $\CZ$ and $27$ $\mathrm{CCZ}$ gates for $h$, and $13$ $\CZ$ and $21$ $\mathrm{CCZ}$ gates for $v$. Each hopping operator uses $20$ CNOTs in total, including $\mathcal N_t$ and $\mathcal N_t^\dagger$, on $14$ or $13$ physical qubits. No ancilla qubit or spectral-controlled gate is required. These are explicit local products, not claims of optimal gate count or nearest-neighbor depth.

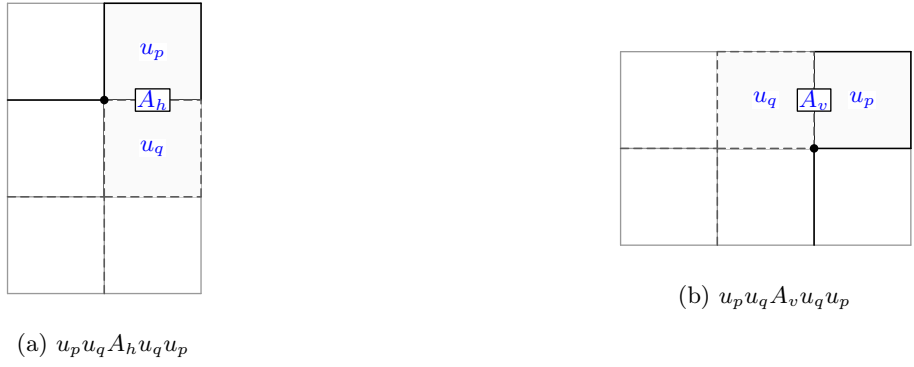
\begin{figure}[t]
\centering
\begin{subfigure}[c]{.47\linewidth}\centering
\begin{tikzpicture}[x=1.28cm,y=1.28cm,line cap=round]
\draw[gridline] (-1,-2)--(-1,1);
\draw[gridline] (0,-2)--(0,1);
\draw[gridline] (1,-2)--(1,1);
\draw[gridline] (-1,-2)--(1,-2);
\draw[gridline] (-1,-1)--(1,-1);
\draw[gridline] (-1,0)--(1,0);
\draw[gridline] (-1,1)--(1,1);
\fill[black!2] (0,0) rectangle (1,1);
\draw[black,line width=.58pt] (-1,0)--(0,0);
\draw[black,line width=.58pt] (0,0)--(1,0);
\draw[black,line width=.58pt] (0,-1)--(0,0);
\draw[black,line width=.58pt] (0,0)--(0,1);
\draw[black,line width=.58pt] (0,1)--(1,1);
\draw[black,line width=.58pt] (1,0)--(1,1);
\node[font=\small,fill=white,inner sep=.35pt,text=blue] at (0.5,0.5) {$u_p$};
\fill[black!2] (0,-1) rectangle (1,0);
\draw[black!65,line width=.58pt,densely dashed] (-1,-1)--(0,-1);
\draw[black!65,line width=.58pt,densely dashed] (0,-1)--(1,-1);
\draw[black!65,line width=.58pt,densely dashed] (0,-2)--(0,-1);
\draw[black!65,line width=.58pt,densely dashed] (0,-1)--(0,0);
\draw[black!65,line width=.58pt,densely dashed] (0,0)--(1,0);
\draw[black!65,line width=.58pt,densely dashed] (1,-1)--(1,0);
\node[font=\small,fill=white,inner sep=.35pt,text=blue] at (0.5,-0.5) {$u_q$};
\node[text=blue,draw=black,fill=white,inner sep=.6pt,font=\small] at (0.5,0) {$A_h$};
\fill[black] (0,0) circle (1.6pt);
\path[use as bounding box] (-1.22,-2.22) rectangle (1.22,1.22);
\end{tikzpicture}
\caption{$u_pu_q A_hu_qu_p$}
\end{subfigure}
\hfill
\begin{subfigure}[c]{.47\linewidth}\centering
\begin{tikzpicture}[x=1.28cm,y=1.28cm,line cap=round]
\draw[gridline] (-2,-1)--(-2,1);
\draw[gridline] (-1,-1)--(-1,1);
\draw[gridline] (0,-1)--(0,1);
\draw[gridline] (1,-1)--(1,1);
\draw[gridline] (-2,-1)--(1,-1);
\draw[gridline] (-2,0)--(1,0);
\draw[gridline] (-2,1)--(1,1);
\fill[black!2] (0,0) rectangle (1,1);
\draw[black,line width=.58pt] (-1,0)--(0,0);
\draw[black,line width=.58pt] (0,0)--(1,0);
\draw[black,line width=.58pt] (0,-1)--(0,0);
\draw[black,line width=.58pt] (0,0)--(0,1);
\draw[black,line width=.58pt] (0,1)--(1,1);
\draw[black,line width=.58pt] (1,0)--(1,1);
\node[font=\small,fill=white,inner sep=.35pt,text=blue] at (0.5,0.5) {$u_p$};
\fill[black!2] (-1,0) rectangle (0,1);
\draw[black!65,line width=.58pt,densely dashed] (-2,0)--(-1,0);
\draw[black!65,line width=.58pt,densely dashed] (-1,0)--(0,0);
\draw[black!65,line width=.58pt,densely dashed] (-1,-1)--(-1,0);
\draw[black!65,line width=.58pt,densely dashed] (-1,0)--(-1,1);
\draw[black!65,line width=.58pt,densely dashed] (-1,1)--(0,1);
\draw[black!65,line width=.58pt,densely dashed] (0,0)--(0,1);
\node[font=\small,fill=white,inner sep=.35pt,text=blue] at (-0.5,0.5) {$u_q$};
\node[text=blue,draw=black,fill=white,inner sep=.6pt,font=\small] at (0,0.5) {$A_v$};
\fill[black] (0,0) circle (1.6pt);
\path[use as bounding box] (-2.22,-1.22) rectangle (1.22,1.22);
\end{tikzpicture}
\caption{$u_pu_q A_vu_qu_p$}
\end{subfigure}
\caption{Origin of the explicit physical-qubit gates. Only the two local spectral gates $u_p,u_q$ enter before the encoding and CNOT reduction. Solid and dashed edges distinguish their complete six-edge supports. The box denotes the entire qudit hopping operator from Eq.~\eqref{eq:parentHops}, not a single-edge $X$. Performing this conjugation and reduction gives the ordinary qubit-gate products in Eqs.~\eqref{eq:explicitqubitA} and~\eqref{eq:explicitqubitAbar}, with the complete gate lists above.}
\label{fig:dressing}
\end{figure}

\paragraph{Algebraic derivation.}
The two-loop-gate geometry is shown in Fig.~\ref{fig:dressing}. We first derive the reduced action of a general constraint-preserving Pauli, and then apply it to the dressed hopping operator. On a finite periodic lattice with $N_e$ edges, the inverse encoding restricted to $A_e^2=1$ is the isometry
\begin{equation}\label{eq:disentanglingIsometry}
 \iota_M|z\rangle
 :=E^\dagger V^\dagger(|z\rangle_a\otimes|0\rangle_b)
 =2^{-N_e/2}\sum_{t\in\mathbb F_2^{N_e}}
      (-1)^{t\cdot r_M(z)}|z+2t\rangle,
 \qquad r_M(z):=[M^Tz]_2.
\end{equation}
Here $[\cdot]_2$ denotes componentwise reduction to the canonical $0$--$1$ representative. The same computation applies locally on the infinite lattice: factors outside the Pauli support cancel. The matrix $M$ is the site-indexed incidence matrix corresponding to Eq.~\eqref{eq:M}; its transpose reverses translations in the Laurent presentation.

Let $P=X^uZ^v$ and write $u=u^{(0)}+2u^{(1)}$ using binary vectors. Commutation with every $A_e^2$ is equivalent to $v=M^Tu\pmod2$. For an input $|z+2t\rangle$, the low output bit is $z'=z\oplus u^{(0)}$, the high bit is $t'=t\oplus s(z)$ with $s(z)=[u^{(1)}+u^{(0)}z]_2$, and the clock phase is $i^{v\cdot z}(-1)^{v\cdot t}$. Products of binary vectors in this expression are componentwise. Reindexing the sum in Eq.~\eqref{eq:disentanglingIsometry}, and using $r_M(z')=[r_M(z)+v]_2$, yields
\begin{equation}\label{eq:reducedPauliIntertwining}
 X^uZ^v\iota_M|z\rangle
 =i^{v\cdot z}(-1)^{s(z)\cdot r_M(z')}\iota_M|z'\rangle.
\end{equation}
Therefore the exact reduced action is
\begin{equation}\label{eq:hopGateDisentangle}
 \begin{split}
 {}_b\langle0|VE X^uZ^vE^\dagger V^\dagger|0\rangle_b\,|z\rangle
 &=i^{p_{u,v}(z)}|z\oplus u^{(0)}\rangle,\\
 p_{u,v}(z)
 &=\sum_i v_i z_i
 +2\sum_i\big(u_i^{(1)}+u_i^{(0)}z_i\big)
                       \big((M^Tz)_i+v_i\big)\pmod4.
 \end{split}
\end{equation}
Any scalar multiplying the canonical Pauli is retained separately. This derivation fixes the phase for every binary input; it uses no condition on the surviving loop eigenvalues.

For an incident face labeled by $v$, let $s_{ve}=\pm1$ be the coefficient of $e$ in its oriented boundary. The local Pauli commutation relations are
\[
 G_vA_e=-A_eG_v,\qquad [G_v,\overline A_e]=0,\qquad
 W_vA_e=i^{s_{ve}}A_eW_v,\qquad
 W_v\overline A_e=i^{s_{ve}}\overline A_eW_v.
\]
Together with $\mathsf{i^{s_{ve}}W_v}=s_{ve}W_v^2\mathsf{W}_v$, they give the one-face dressing identities, with the displayed order retained
\begin{align*}
 u_vA_eu_v
 &=A_e\mathsf{W}_v\left[\frac{1-\mathsf{G}_v}{2}
       +s_{ve}W_v^2\frac{1+\mathsf{G}_v}{2}\right],\\
 u_v\overline A_eu_v
 &=\overline A_e\left[\frac{1+\mathsf{G}_v}{2}
       +s_{ve}W_v^2\frac{1-\mathsf{G}_v}{2}\right].
\end{align*}
Apply these identities at the two incident faces and expand $u_pu_qA_eu_qu_p$ using $u_v=(1+\mathsf{G}_v+\mathsf{W}_v-\mathsf{G}_v\mathsf{W}_v)/2$ and $\mathsf{W}=(1-i)W/2+(1+i)W^{-1}/2$. Collecting canonical qudit Paulis gives $36$ terms. Since the complete dressed hopping operator commutes with every $A_{e'}^2$, and conjugation by $A_{e'}^2$ is diagonal in the canonical Pauli basis, each nonzero collected Pauli string separately commutes with every $A_{e'}^2$. It can therefore be reduced by Eq.~\eqref{eq:hopGateDisentangle}. Their coefficients are collected using
\begin{equation}\label{eq:binaryReductionRules}
    x\oplus y=x+y-2xy,\qquad x^2=x,\qquad i^{2x}=(-1)^x,\qquad x,y\in\{0,1\}.
\end{equation}
For clarity, the resulting phase polynomial is given explicitly below. Write $a^+_e=a_e$, $a^-_e=\overline a_e$ and $y=\mathcal N_tz$. Let $C_t(y)$ be the affine bit permutation implemented by $\mathcal C_t$ in Eqs.~\eqref{eq:CandXh} and~\eqref{eq:CandXv}, including its bit flips. The reduced action is
\begin{equation}\label{eq:auxiliarySymbolicAction}
    a^\pm_{t_\rvec}|z\rangle =\zeta_t i^{\Theta_t^\pm(y)}\big|\mathcal N_t^{-1}C_t(y)\big\rangle,\qquad y=\mathcal N_tz,
\end{equation}
where all the coefficients of the phase are specified by
\begin{equation}\label{eq:auxiliarySymbolicPhase}
 \Theta_t^\pm(y)
 =\ell_t^\pm(y)
 +2\sum_{(i,j)\in\mathcal E_t}y_i y_j
 +2\sum_{k\in\mathcal K_t}\sum_{(i,j)\in\mathcal T_{t,k}}y_i y_j y_k
 \pmod4,
\end{equation}
\begin{equation}\label{eq:auxiliaryLinearPhases}
 \begin{aligned}
 \ell_h^+(y)&=2y_1+2y_2+2y_8+y_{11}-y_{13}+y_{14},&
 \ell_h^-(y)&=2y_1+2y_2+2y_8-y_{11}+2y_{14},\\
 \ell_v^+(y)&=-y_3+y_4+y_{13},&
 \ell_v^-(y)&=y_3+2y_4+2y_{13}.
 \end{aligned}
\end{equation}
Here, $\mathcal E_t$, $\mathcal K_t$, and $\mathcal T_{t,k}$ are the finite sets in Eqs.~\eqref{eq:gateDiagonal}--\eqref{eq:Tv13}. The reduction by Eq.~\eqref{eq:binaryReductionRules} leaves the output mask in Eq.~\eqref{eq:auxiliarySymbolicAction}. The constant phase, including the original $\ww$, is $\zeta_h=e^{3\pi i/4}$ or $\zeta_v=e^{-\pi i/4}$. The linear, quadratic, and cubic terms are respectively the $S$, $\CZ$, and $\mathrm{CCZ}$ factors of the displayed circuit. Its permutation block $\mathcal C_t$ gives exactly the same output mask. Thus, Eqs.~\eqref{eq:explicitqubitA} and~\eqref{eq:explicitqubitAbar} agree with the reduced operators on an arbitrary basis vector, and hence are operator identities on the entire physical Hilbert space.

The exact algebraic properties survive reduction as
\begin{equation}
    [a_e,\overline a_{e'}]=0,\qquad a_e^4=\overline a_e^4=\id.
\label{eq:physicalcross}
\end{equation}
Their loop reductions use only $VE$, because $\mathcal U$ commutes with every $G_p$ and $\overline G_p$. The Hermitian hopping operators $\mathsf a_e,\overline{\mathsf a}_e$ require the normalization in Appendix~\ref{sec:normalizerGates}.

\subsection{Ancilla-free gate products for the Hermitian hopping operators}
\label{sec:normalizerGates}

The normalization factor $S(a_e^2)^\dagger$ can be compiled into ordinary gates on the same $14$ or $13$ physical qubits. In this subsection write $a_e^+=a_e$, $a_e^-=\overline a_e$ and $\mathsf a_e^+=\mathsf a_e$, $\mathsf a_e^-=\overline{\mathsf a}_e$. All qubit numbers refer to Fig.~\ref{fig:qubitSupport}; the superscripts $+$ and $-$ label the chirality. Define five-CNOT basis changes
\begin{equation}\label{eq:normalizerG}
 \mathcal A_h=\prod_{j\in\{2,4,5,10,12\}}\CNOT_{1\to j},\qquad
 \mathcal A_v=\prod_{j\in\{5,9,10,11,12\}}\CNOT_{1\to j}.
\end{equation}
Each $\mathcal A_t$ is self-inverse. Let $\mathrm{CS}_{ij}=\operatorname{diag}(1,1,1,i)$ on qubits $i,j$ and define diagonal gates
\begin{align}
 \mathcal L_h^+
 &=\mathrm{CS}_{1,13}^{\dagger}\mathrm{CS}_{1,14}^{\dagger}
   \prod_{j\in\mathcal Z_h}\CZ_{1j}
   \prod_{(i,j)\in\mathcal P_h}\mathrm{CCZ}_{1ij},\label{eq:Lhplus}\\
 \mathcal L_h^-
 &=\mathrm{CS}_{1,13}\mathrm{CS}_{1,14}
   \prod_{j\in\mathcal Z_h}\CZ_{1j}
   \prod_{(i,j)\in\mathcal P_h}\mathrm{CCZ}_{1ij},\label{eq:Lhminus}\\
 \mathcal L_v^+
 &=\mathrm{CS}_{1,4}\mathrm{CS}_{1,13}^{\dagger}
   \prod_{j\in\mathcal Z_v}\CZ_{1j}
   \prod_{(i,j)\in\mathcal P_v}\mathrm{CCZ}_{1ij},\label{eq:Lvplus}\\
 \mathcal L_v^-
 &=\mathrm{CS}_{1,4}^{\dagger}\mathrm{CS}_{1,13}
   \prod_{j\in\mathcal Z_v}\CZ_{1j}
   \prod_{(i,j)\in\mathcal P_v}\mathrm{CCZ}_{1ij},\label{eq:Lvminus}
\end{align}
where the complete gate lists are
\begin{equation}\label{eq:normalizerZlists}
 \mathcal Z_h=\{2,3,5,9,11,12\},\qquad
 \mathcal Z_v=\{2,3,8,9\},
\end{equation}
\begin{equation}\label{eq:normalizerHlist}
\begin{aligned}
 \mathcal P_h=\{&(2,3),(4,5),(5,6),(2,8),(2,9),(4,10),(7,10),\, (4,11),\\
                &(2,12),(3,12),(5,12),(6,12),(9,12),\, (5,13),\, (11,13),\\
                &(6,14),(12,14)\},
\end{aligned}
\end{equation}
\begin{equation}\label{eq:normalizerVlist}
\begin{aligned}
 \mathcal P_v=\{&(3,4),(5,6),(7,9),(2,10),(3,10),(4,10),(8,10),\\
                &(3,11),(5,11),(9,11),(5,12),(6,12),(10,12),\\
                &(6,13),(12,13)\}.
\end{aligned}
\end{equation}
Every gate within a given $\mathcal L_t^\pm$ is diagonal and hence commutes with the others.

Using $H_1$ for the Hadamard on physical qubit $1$, the exact normalization factor is
\begin{equation}\label{eq:explicitNormalizer}
 \begin{aligned}
 \Lambda_t^\pm&:=S\bigl((a_{t_\rvec}^\pm)^2\bigr)^\dagger\\
 &=\mathcal N_t^\dagger\mathcal A_t\mathcal L_t^\pm
 H_1S_1^\dagger H_1
 (\mathcal L_t^\pm)^\dagger\mathcal A_t^\dagger\mathcal N_t.
 \end{aligned}
\end{equation}
The Hermitian hopping generators have the explicit gate products
\begin{equation}\label{eq:explicitInvolutiveGates}
 \begin{aligned}
 \mathsf a_{t_\rvec}^\pm
 &=a_{t_\rvec}^\pm\Lambda_t^\pm\\
 &=\zeta_t\mathcal N_t^\dagger\mathcal C_t\mathcal D_t^\pm\mathcal A_t\mathcal L_t^\pm
       H_1S_1^\dagger H_1
       (\mathcal L_t^\pm)^\dagger\mathcal A_t^\dagger\mathcal N_t.
 \end{aligned}
\end{equation}
In the second line the adjacent $\mathcal N_t\mathcal N_t^\dagger$ from the order-four hopping operator and the normalization have canceled. The products act from right to left. All gates act on the same numbered support as the order-four hopping operator. The normalized chiralities differ in the diagonal blocks $\mathcal D_t^\pm$ and $\mathcal L_t^\pm$.

The controlled-$S$ gates are also ordinary qubit gates and may be eliminated using
\begin{equation}\label{eq:CSdecomposition}
 \mathrm{CS}_{ij}^{\,\epsilon}
 =T_i^{\epsilon}T_j^{\epsilon}\CNOT_{i\to j}
  T_j^{-\epsilon}\CNOT_{i\to j},
 \qquad \epsilon=\pm1,\qquad T=\operatorname{diag}(1,e^{i\pi/4}).
\end{equation}
Indeed, the phase exponent is
$\epsilon\pi(z_i+z_j-(z_i\oplus z_j))/4 =\epsilon\pi z_iz_j/2$. Thus, Eq.~\eqref{eq:explicitInvolutiveGates} contains only $X,H,S,T$, CNOT, $\CZ$, and $\mathrm{CCZ}$, with inverses allowed.

\paragraph{Algebraic derivation of the normalization.}
The square of an order-four hopping operator is a Hermitian involution. In the basis obtained from $\mathcal N_t$ and $\mathcal A_t$, its action is particularly simple. Write $\mathcal L_t^\pm|u\rangle=i^{u_1p_t^\pm(u)}|u\rangle$, where $p_t^\pm$ is independent of $u_1$. The coefficients of $p_t^\pm$ are read directly from the gates above: each $\mathrm{CS}_{1j}^{\epsilon}$ contributes $\epsilon u_j$, each $\CZ_{1j}$ contributes $2u_j$, and each $\mathrm{CCZ}_{1ij}$ contributes $2u_iu_j$. Direct squaring of Eqs.~\eqref{eq:explicitqubitA} and~\eqref{eq:explicitqubitAbar} gives
\begin{equation}\label{eq:normalizerPhaseProof}
 \mathcal A_t^\dagger\mathcal N_t(a_{t_\rvec}^\pm)^2
 \mathcal N_t^\dagger\mathcal A_t|u\rangle
 =i^{(1-2u_1)p_t^\pm(u)}|u\oplus e_1\rangle
 =\mathcal L_t^\pm X_1(\mathcal L_t^\pm)^\dagger|u\rangle.
\end{equation}
To finish the derivation, set
$R_t^\pm=\mathcal N_t^\dagger\mathcal A_t\mathcal L_t^\pm$.
Eq.~\eqref{eq:normalizerPhaseProof} says $(a^\pm_{t_\rvec})^2=R_t^\pm X_1(R_t^\pm)^\dagger$. Functional calculus for this involution, together with $S(X_1)^\dagger=H_1S_1^\dagger H_1$, gives
\begin{equation}\label{eq:normalizerFunctionalProof}
 S\big((a^\pm_{t_\rvec})^2\big)^\dagger
 =R_t^\pm H_1S_1^\dagger H_1(R_t^\pm)^\dagger,
\end{equation}
which is Eq.~\eqref{eq:explicitNormalizer}, including its scalar phase. Multiplying by the order-four hopping product proves Eq.~\eqref{eq:explicitInvolutiveGates}. Since $a^\pm_e$ commutes with every function of its square and $S((a^\pm_e)^2)^{\dagger\,2}=(a^\pm_e)^2$, the resulting product has $(\mathsf a^\pm_e)^2=\id$. This proves the normalization on all physical states without a selected loop sector. Eqs.~\eqref{eq:explicitqubitA} and~\eqref{eq:explicitqubitAbar} implement the order-four hopping operators, whereas Eq.~\eqref{eq:explicitInvolutiveGates} implements the Hermitian hopping operators.

\section{Four-copy chirality reversal and pump cancellation}
\label{app:eightfoldParent}

We construct the local chirality reversal used in the eightfold circuit identity, Eq.~\eqref{eq:eightfoldCircuit}. The construction acts on the complete physical operator algebra and preserves all surviving loop sectors.

\subsection{Exchange of the chiral factors at the lattice-level}
\label{app:fourCopyExchange}

\begin{lemma}[Four-copy chirality reversal]\label{prop:fourCopyChirality}
There is a bounded-range QCA $\gamma_4$ on four copies of the physical square lattice, with bounded-range inverse, such that
\begin{equation}\label{eq:fourCopyReducedHops}
 \gamma_4\bigl(a_e^{(i)}\bigr)
 =\prod_{j=1}^4\bigl(\overline a_e^{(j)}\bigr)^{U_{ji}},
 \qquad
 \gamma_4\bigl(\overline a_e^{(i)}\bigr)
 =\prod_{j=1}^4\bigl(a_e^{(j)}\bigr)^{U_{ji}},
\end{equation}
where
\begin{equation}\label{eq:fourCopyU}
 U=\begin{pmatrix}
 0&1&1&1\\
 -1&0&1&-1\\
 -1&-1&0&1\\
 -1&1&-1&0
 \end{pmatrix},\qquad U^{\mathsf T}U=3I.
\end{equation}
The products in Eq.~\eqref{eq:fourCopyReducedHops} run over different copies, so their factors commute. Consequently, $\gamma_4$ exchanges $\Asalg^{\otimes4}$ and $\Abalg^{\otimes4}$ by uniformly local changes of generators.
\end{lemma}

\begin{proof}
Write the parent hopping operators in canonical $X$-then-$Z$ order as
\begin{equation}\label{eq:fourCopyParentHops}
 \begin{aligned}
 A_e&=\ww X_e\prod_{e'}Z_{e'}^{\int(\mathbf e'\cup\mathbf e-\delta\mathbf e'\cup_1\mathbf e)},\\
 \overline A_e&=\ww X_e\prod_{e'}Z_{e'}^{\int(\mathbf e\cup\mathbf e'-\delta\mathbf e\cup_1\mathbf e')}.
 \end{aligned}
\end{equation}
Let $M$ be the finite-range integer matrix recording these unbarred clock decorations, as in Eq.~\eqref{eq:M}. Its rows label clock edges and its columns label hopping edges; each entry is the corresponding signed cup pairing above. Transposition includes the edge positions and exchanges the two input cochains. The diagonal pairing is $-1$ for both edge orientations. Thus, $H:=M+M^{\mathsf T}$ is a symmetric finite-range integer matrix with $H_{ee}=-2$. In the translation notation of Appendix~\ref{app:parentPatterns}, $M^{\mathsf T}$ is $M^\dagger$.

First mix the four species on each edge by the on-site basis permutation
\begin{equation}\label{eq:fourCopyBasisPermutation}
 \mathcal P_U|\mathbf j\rangle=|U\mathbf j\bmod4\rangle,
 \qquad \mathbf j\in(\mathbb Z/4\mathbb Z)^4.
\end{equation}
Since $U^{-\mathsf T}=3U\pmod4$, this sends Pauli exponents $(u,v)$ to $(Uu,3Uv)$ without a scalar phase. Define the negative family by complex conjugation in the clock basis,
\begin{equation}
 \begin{aligned}
 A_e^-&=\ww^{-1}X_e\prod_{e'}Z_{e'}^{-\int(\mathbf e'\cup\mathbf e-\delta\mathbf e'\cup_1\mathbf e)},\\
 \overline A_e^-&=\ww^{-1}X_e\prod_{e'}Z_{e'}^{-\int(\mathbf e\cup\mathbf e'-\delta\mathbf e\cup_1\mathbf e')}.
 \end{aligned}
\end{equation}
For the raw negative hopping operator, the Pauli commutation relation gives
\begin{align*}
 h_e^-&=X_e\prod_{e'}Z_{e'}^{-\int(\mathbf e'\cup\mathbf e-\delta\mathbf e'\cup_1\mathbf e)},\\
 (h_e^-)^{-1}&=iX_e^{-1}\prod_{e'}Z_{e'}^{\int(\mathbf e'\cup\mathbf e-\delta\mathbf e'\cup_1\mathbf e)}.
\end{align*}
Every column of $U$ has three nonzero entries and an odd number $n_-$ of negative entries. The scalar in the corresponding product of normalized negative hopping operators is therefore $\ww^{-3+4n_-}=\ww$. This checks the normalization as well as the Pauli exponents, and yields the exact identity
\begin{equation}\label{eq:fourCopyNegativeHops}
 \mathcal P_U A_e^{(i)}\mathcal P_U^\dagger
 =\prod_j\bigl(A_e^{-,(j)}\bigr)^{U_{ji}}.
\end{equation}
The same argument applies to the barred family, whose diagonal cup pairing is also $-1$.

Next apply, separately in each species, the diagonal Clifford circuit for $\mathbb Z_4$ qudits
\begin{equation}\label{eq:fourCopyDiagonalCircuit}
 \mathcal D_H|\{j_e\}\rangle=i^{q(\mathbf j)}|\{j_e\}\rangle,
 \qquad
 q(\mathbf j)=-\sum_ej_e^2+2\sum_ej_e
       +\sum_{e<e'}H_{e'e}j_{e'}j_e\pmod4.
\end{equation}
The ordering $e<e'$ counts each pair once. The phase is independent of the integer representatives of the $j_e$. Its on-site terms and bounded-range controlled-clock terms have a finite-depth schedule obtained by coloring their supports. The same local gates define the circuit on the infinite plane. Finite differencing gives
\begin{equation}\label{eq:fourCopyDiagonalConjugation}
 \mathcal D_H X_e\mathcal D_H^\dagger
 =iX_e\prod_{e'}Z_{e'}^{H_{e'e}},\qquad
 \mathcal D_H Z_e\mathcal D_H^\dagger=Z_e.
\end{equation}
Since $H-M=M^{\mathsf T}$ and $i\ww^{-1}=\ww$, the circuit converts the negative family exactly into the barred family, and the negative barred family into the unbarred one. Thus, the bounded-depth parent circuit $\mathcal V:=\mathcal D_H^{\otimes4}\mathcal P_U$ obeys
\begin{equation}\label{eq:fourCopyParentImages}
 \mathcal V A_e^{(i)}\mathcal V^\dagger
 =\prod_j\bigl(\overline A_e^{(j)}\bigr)^{U_{ji}},
 \qquad
 \mathcal V\overline A_e^{(i)}\mathcal V^\dagger
 =\prod_j\bigl(A_e^{(j)}\bigr)^{U_{ji}}.
\end{equation}

Let us verify that this circuit preserves the condensation sector, including its scalar phases. Let $d_\rvec,c_\rvec$ be the integer star and plaquette patterns, so $T_\rvec=X^{d_\rvec}$ and $W_\rvec=Z^{c_\rvec}$. The explicit patterns give
\begin{equation}\label{eq:fourCopyStarPatterns}
 Md_\rvec=c_\rvec,\qquad
 M^{\mathsf T}d_\rvec=-c_\rvec,\qquad
 d_\rvec^{\mathsf T}c_\rvec=0,\qquad
 \sum_e(d_\rvec)_e=0.
\end{equation}
The first two are also recorded in Eq.~\eqref{eq:integeridentities}. Hence, $Hd_\rvec=0$ and $q(d_\rvec)=\tfrac12d_\rvec^{\mathsf T}Hd_\rvec+2\sum_e(d_\rvec)_e=0$. There is no inverse scalar in $G_\rvec^-=X^{d_\rvec}Z^{-c_\rvec}$, because $d_\rvec^{\mathsf T}c_\rvec=0$. The species permutation therefore sends $G_\rvec^{(i)}$ to $\prod_j(G_\rvec^{-,(j)})^{U_{ji}}$ exactly, and Eq.~\eqref{eq:fourCopyDiagonalCircuit} sends $G_\rvec^-$ to $\overline G_\rvec$. Interchanging the two signs gives the converse. We obtain
\begin{equation}\label{eq:fourCopyLoopImages}
 \mathcal V G_\rvec^{(i)}\mathcal V^\dagger
 =\prod_j\bigl(\overline G_\rvec^{(j)}\bigr)^{U_{ji}},
 \qquad
 \mathcal V\overline G_\rvec^{(i)}\mathcal V^\dagger
 =\prod_j\bigl(G_\rvec^{(j)}\bigr)^{U_{ji}}.
\end{equation}
Squaring Eq.~\eqref{eq:fourCopyParentImages} gives the corresponding exact images of $A_e^2$ and $\overline A_e^{\,2}$. Since these squares commute with the loops, the common condensate identity in Eq.~\eqref{eq:Cold} now yields
\begin{equation}\label{eq:fourCopyCondensateImages}
    \mathcal V C_e^{(i)}\mathcal V^\dagger=\prod_j\bigl(C_e^{(j)}\bigr)^{U_{ji}}.
\end{equation}
The invertibility of $U$ modulo four implies that $\mathcal V$ maps the entire commuting constraint algebra onto itself and preserves its simultaneous $C_e=1$ space. By Lemma~\ref{prop:samespace}, this is exactly the $D_e=1$ space.

The disentangler $\mathcal W$ maps this space onto all the physical data qubits tensored with $|0\rangle_b$, as in Eq.~\eqref{eq:reducedH}. On a sufficiently large periodic lattice define
\begin{equation}\label{eq:fourCopyPhysicalUnitary}
 \psi={}_b\langle0|\,\mathcal W^{\otimes4}\mathcal V
            (\mathcal W^{\otimes4})^\dagger\,|0\rangle_b,
 \qquad \gamma_4=\operatorname{Ad}_\psi.
\end{equation}
Because $\mathcal V$ preserves that whole subspace, $\psi$ is a unitary on the unrestricted four-copy data Hilbert space. Its adjoint action has uniformly bounded range: lift a local data operator through $\mathcal W^\dagger$, conjugate through $\mathcal V$ and $\mathcal W$, and take the ancillary matrix element. The circuits enlarge support by a fixed amount, and the matrix element cannot enlarge it. The same argument with $\mathcal V^\dagger$ proves locality of the inverse. The resulting local formulas define $\gamma_4$ on the infinite plane.

Reducing Eq.~\eqref{eq:fourCopyParentImages} proves Eq.~\eqref{eq:fourCopyReducedHops}. The order-four hopping operators generate the complete chiral algebras by Lemma~\ref{prop:hopOnly}. $U^{-1}$ modulo four makes both images surjective. No physical loop eigenvalue has been fixed anywhere in this reduction.
\end{proof}

The restriction in Eq.~\eqref{eq:fourCopyPhysicalUnitary} proves that $\gamma_4$ is a QCA. It does not by itself prove that $\gamma_4$ tensored with ancillas is a circuit on their unrestricted operator algebras. The doubled-SWAP construction in Appendix~\ref{app:pumpCancellationCircuit} uses only the locality of $\gamma_4$ and its inverse, and supplies that stronger circuit conclusion for the compensated eightfold pump.

As an algebraic consequence, the chirality reversal on the second group of four factors, followed by the complementarity in Eq.~\eqref{eq:ISA} and the uniformly local factorization of Proposition~\ref{prop:physicalFactorization}, gives a uniformly local $*$-isomorphism
\begin{equation}\label{eq:eightCopyAlgebraCancellation}
    \Asalg^{\otimes8}\ \cong\ \Asalg^{\otimes4}\otimes\Abalg^{\otimes4}\ \cong\ \mathcal M_{\mathrm{qubits}}^{\otimes4}.
\end{equation}

\subsection{Explicit circuit for the compensated eightfold pump}
\label{app:pumpCancellationCircuit}

We retain a constructive realization of the stable-group relation in Eq.~\eqref{eq:eightfoldClassRelation}. The standard doubled-SWAP identity implements a QCA and its inverse on two registers. Applying it to the four-copy exchange gives the eight-copy circuit stated in Proposition~\ref{prop:eightfoldPump}.

\begin{proof}
\smallskip\noindent\emph{Exchanging four chiral factors.}\quad
Let $\gamma_4$ be the four-copy chirality-exchange QCA constructed in Appendix~\ref{app:fourCopyExchange}.

Stack this map identically in every layer and call the resulting QCA $\Gamma$. Set $B=\beta_\uparrow^{\otimes4}$, and let $\overline B$ pump the four anti-semion factors upward while fixing the semion factors. Since the two factors commute and generate the full physical algebra,
\begin{equation}\label{eq:fourCopyPumpCommutator}
 \Gamma B\Gamma^{-1}=\overline B,
 \qquad B\overline B=T_z^{\otimes4},
 \qquad [B,\Gamma]=B^2\bigl(T_z^{\otimes4}\bigr)^{-1}.
\end{equation}
Here, $[B,\Gamma]=B\Gamma B^{-1}\Gamma^{-1}$ is the automorphism commutator. The same chirality exchange is used in every layer, so these identities hold for all local operators, not only for transport generators.

\smallskip\noindent\emph{Implementing the commutator.}\quad
We use the doubled-SWAP construction~\cite{FreedmanHaahHastings2022}, whose required identities can be checked directly. For any bounded-range QCA $\alpha$, introduce a second identical register. The label $q$ specifies a physical qubit, including its lattice position and any copy index, and $q'$ denotes its partner in the added register. Let $\SWAP_q$ exchange these two qubits and write $\SWAP=\prod_q\SWAP_q$ for the on-site register swap. Define
\begin{equation}\label{eq:inversePairCircuit}
 E_\alpha=(\alpha\otimes\mathrm{id})(\SWAP)\SWAP,
 \qquad \operatorname{Ad}_{E_\alpha}
 =\alpha\otimes\alpha^{-1}.
\end{equation}
\begin{figure}[t]
\centering
\begin{tikzpicture}[x=1cm,y=1cm,
 register/.style={draw=black!65,line width=.55pt},
 circuitlayer/.style={draw=black,fill=white,rounded corners=1pt,
   minimum width=1.8cm,minimum height=1.4cm,align=center,font=\small},
 formalmap/.style={draw=blue,dashed,fill=white,rounded corners=1pt,
   minimum width=1.1cm,minimum height=.6cm,text=blue,font=\normalsize},
 reglabel/.style={anchor=east,align=right,font=\small}]
\node[anchor=west,font=\small] at (-2.35,1.8)
 {(a) Implemented circuit $E_\alpha$};
\node[reglabel] at (-.2,0) {original register};
\node[reglabel] at (-.2,-.9) {added register};
\draw[register] (0,0)--(11.2,0);
\draw[register] (0,-.9)--(11.2,-.9);
\node[circuitlayer,minimum width=1.65cm] at (1.2,-.45)
 {on-site\\$\SWAP$ gates};
\node[circuitlayer,fill=blue!3,minimum width=7.3cm] at (7.15,-.45)
 {image $\SWAP$ gates\\scheduled in bounded depth};
\node[font=\small] at (7.15,1.02)
 {individual image gate: $(\alpha\otimes\mathrm{id})(\SWAP_q)$};
\draw[-{Stealth[length=1.8mm]},black!65] (0,-1.5)--(11.2,-1.5)
 node[midway,below=2pt,font=\small] {execution order};

\node[anchor=west,font=\small] at (-2.35,-2.25)
 {(b) Algebraic identity for the net action};
\node[reglabel] at (-.2,-3) {original register};
\node[reglabel] at (-.2,-3.9) {added register};
\draw[register] (0,-3)--(6.85,-3);
\draw[register] (0,-3.9)--(6.85,-3.9);
\foreach \xx in {1,4.2} {
 \draw[black] (\xx,-3)--(\xx,-3.9);
 \foreach \yy in {-3,-3.9} {
  \draw[black] (\xx-.09,\yy-.09)--(\xx+.09,\yy+.09);
  \draw[black] (\xx-.09,\yy+.09)--(\xx+.09,\yy-.09);
 }
 \node[font=\small] at (\xx,-2.65) {$\SWAP$};
}
\node[formalmap] at (2.65,-3) {$\alpha^{-1}$};
\node[formalmap] at (5.8,-3) {$\alpha$};
\node at (7.5,-3.45) {$=$};
\draw[register] (8.1,-3)--(11.2,-3);
\draw[register] (8.1,-3.9)--(11.2,-3.9);
\node[formalmap] at (9.65,-3) {$\alpha$};
\node[formalmap] at (9.65,-3.9) {$\alpha^{-1}$};
\node[font=\small,align=center] at (5.1,-4.55)
 {Dashed boxes denote formal QCA actions on operators.};
\end{tikzpicture}
\caption{The doubled-SWAP construction in Eq.~\eqref{eq:inversePairCircuit}. Each horizontal line represents a full register, and execution runs left to right. (a) Apply all on-site swaps $\SWAP_q$ first, then the local image gates $(\alpha\otimes\mathrm{id})(\SWAP_q)$. The single image-gate block represents a schedule obtained by a finite coloring of the overlapping supports: gates of each color have disjoint supports, and the number of colors is independent of system size. (b) At the level of operator maps, the image-SWAP stage is $\alpha^{-1}$ on the original register, followed by a register swap and then $\alpha$. Together with the first swap, this places $\alpha$ on the original register and $\alpha^{-1}$ on the added register. The dashed boxes are formal QCA actions. The implemented gates are those in (a).}
\label{fig:inversePairCircuit}
\end{figure}
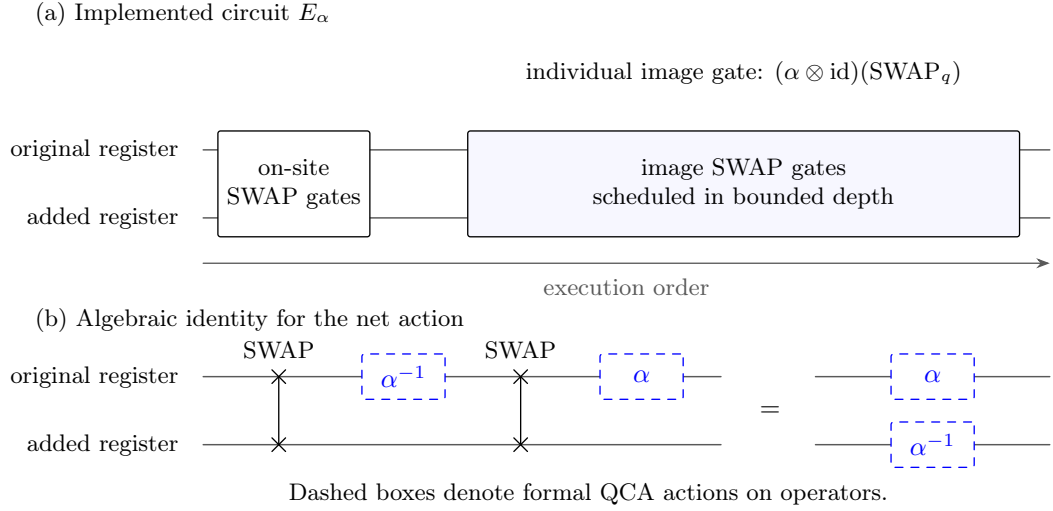

Fig.~\ref{fig:inversePairCircuit}(a) shows the circuit in execution order: first apply the on-site swaps, then apply the image gates $(\alpha\otimes\mathrm{id})(\SWAP_q)$. The notation $(\alpha\otimes\mathrm{id})(\SWAP)$ denotes their product. Since $\alpha\otimes\mathrm{id}$ preserves unitarity, each image gate is a unitary supported near $q$ in the original register and on $q'$ in the added register. Their supports have uniformly bounded diameter, and the gates commute because the original swaps commute. A finite coloring of these supports therefore schedules the image gates in bounded depth. On the infinite lattice, both products denote locally finite circuits acting by conjugation on local operators. The formal sequence in Fig.~\ref{fig:inversePairCircuit}(b) proves the second identity in Eq.~\eqref{eq:inversePairCircuit}.

For another QCA $\eta$, construct $(\eta\otimes\mathrm{id})(E_\alpha^\dagger)$ from the gates of $E_\alpha^\dagger$ one layer at a time. Within each layer, the image gates still commute and have uniformly bounded support, so a finite coloring schedules them in bounded depth. Retaining the order of the original layers gives a finite-depth circuit for the entire image. Direct composition yields
\begin{equation}\label{eq:commutatorCircuit}
 C_{\alpha,\eta}
 :=E_\alpha(\eta\otimes\mathrm{id})(E_\alpha^\dagger),
 \qquad \operatorname{Ad}_{C_{\alpha,\eta}}
 =[\alpha,\eta]\otimes\mathrm{id}.
\end{equation}
In particular, $C_K:=C_{B,\Gamma}$ acts on eight physical copies and implements $[B,\Gamma]$ on the first four, with net identity action on the other four. This is a circuit on their unrestricted operator algebras. No ancillary state is selected.

\begin{figure}[t]
\centering
\begin{tikzpicture}[x=1.35cm,y=1.02cm,
 action/.style={font=\normalsize,inner sep=2pt},
 result/.style={action,draw=blue,fill=blue!4,rounded corners=1pt,
                minimum width=1.20cm,minimum height=.48cm},
 stage/.style={font=\small,fill=white,inner sep=2pt}]
\node[action,anchor=east] at (-.55,0) {copy $1$};
\node[action,anchor=east] at (-.55,-.7) {copy $2$};
\node[action,anchor=east] at (-.55,-1.4) {copy $3$};
\node[action,anchor=east] at (-.55,-2.1) {copy $4$};
\node[action,anchor=east] at (-.55,-3) {copies $5$--$8$};
\node[stage] at (0,1.12) {start};
\node[stage] at (8,1.12) {$\operatorname{Ad}_C$};
\foreach \left/\right/\circuit in
 {0/2/{E_{14}^\dagger},2/4/{E_{13}^\dagger},4/6/{E_{12}^\dagger},6/8/{C_K}} {
 \draw[-{Stealth[length=1.8mm]},black!65] (\left+.3,.63)--(\right-.3,.63)
  node[midway,above=2pt,stage] {$\circuit$};
}
\foreach \col/\one/\two/\three/\four in
 {0/{\mathrm{id}}/{\mathrm{id}}/{\mathrm{id}}/{\mathrm{id}},
  2/{\delta}/{\mathrm{id}}/{\mathrm{id}}/{\delta^{-1}},
  4/{\delta^2}/{\mathrm{id}}/{\delta^{-1}}/{\delta^{-1}},
  6/{\delta^3}/{\delta^{-1}}/{\delta^{-1}}/{\delta^{-1}}} {
 \node[action] at (\col,0) {$\one$};
 \node[action] at (\col,-.7) {$\two$};
 \node[action] at (\col,-1.4) {$\three$};
 \node[action] at (\col,-2.1) {$\four$};
 \node[action] at (\col,-3) {$\mathrm{id}$};
}
\node[result,text=blue] at (8,0) {$\delta^4$};
\foreach \yy in {-.7,-1.4,-2.1,-3}
 \node[result] at (8,\yy) {$\mathrm{id}$};
\draw[black!25] (-.35,-2.55)--(8.55,-2.55);
\node[font=\small,text=blue] at (4,-3.65)
 {$\delta=\beta_\uparrow^2T_z^{-1},\qquad
   \delta^4=\beta_\uparrow^8T_z^{-4}$};
\end{tikzpicture}
\caption{Combining the four parallel pumps onto copy $1$. Each column records the cumulative automorphism after the circuit above its incoming arrow has been completed. Execution runs left to right, as in the rightmost-first product $C=C_KE_{12}^\dagger E_{13}^\dagger E_{14}^\dagger$. Each $E_{1j}^\dagger$ adds $\delta$ on copy $1$ and $\delta^{-1}$ on copy $j$. The final $C_K$ adds $\delta$ on copies $1$--$4$. Its circuit uses all eight copies, with net identity action on copies $5$--$8$. The result is $\delta^4$ on copy $1$ and identity on the other seven.}
\label{fig:pumpMerge}
\end{figure}
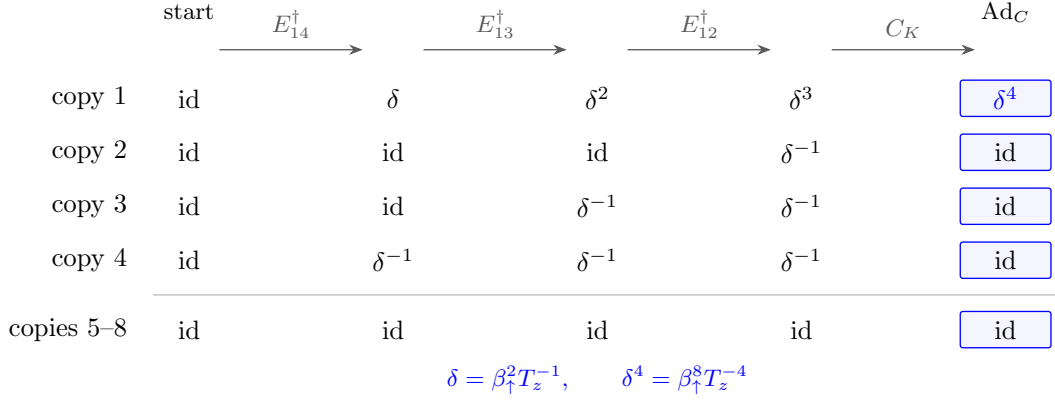

\smallskip\noindent\emph{Combining parallel pumps.}\quad
Write $\delta=\beta_\uparrow^2T_z^{-1}$ and let $\delta_j$ act on copy $j$. Eqs.~\eqref{eq:fourCopyPumpCommutator} and~\eqref{eq:commutatorCircuit} give $\operatorname{Ad}_{C_K}=\delta_1\delta_2\delta_3\delta_4$, with identity action on copies $5,\ldots,8$. For $j=2,3,4$, let $E_{1j}$ be the circuit of Eq.~\eqref{eq:inversePairCircuit} with $\alpha=\delta^{-1}$ on the ordered pair of copies $(1,j)$. It implements $\Delta_{1j}:=\delta_1^{-1}\delta_j$. The automorphisms obey
\begin{equation}\label{eq:parallelPumpMerge}
 \delta_1\delta_2\delta_3\delta_4
 =\delta_1^4\Delta_{14}\Delta_{13}\Delta_{12}.
\end{equation}
Hence, $C=C_KE_{12}^\dagger E_{13}^\dagger E_{14}^\dagger$ implements $\delta^4$ on copy $1$ and the identity on all seven other copies. Fig.~\ref{fig:pumpMerge} tracks the accumulated automorphism in the execution order $E_{14}^\dagger$, $E_{13}^\dagger$, $E_{12}^\dagger$, $C_K$. The pump commutes with whole-layer translation, so $\delta^4=\beta_\uparrow^8T_z^{-4}$. This proves Eq.~\eqref{eq:eightfoldCircuit}.
\end{proof}

\subsection{Ordinary information flow of the uncompensated pump}
\label{app:pumpInformationFlow}

The residual translation is required by the one-dimensional information-flow index on a transverse torus. This also excludes a circuit for the uncompensated eighth power without assuming spatial periodicity.

Compactify each transverse plane to an $L\times L$ torus larger than all local supports. The two chiral algebras are complementary matrix factors by Eq.~\eqref{eq:ISA}; write them as $\Mat(d_s)$ and $\Mat(d_{\bar s})$. Lemma~\ref{prop:fourCopyChirality} also holds on these tori, so $d_s^4=d_{\bar s}^4$. There are $2L^2$ physical edge qubits, hence $d_sd_{\bar s}=2^{2L^2}$ and therefore
\begin{equation}
    d_s=d_{\bar s}=2^{L^2}.
\label{eq:eightfoldFactorDimensions}
\end{equation}
Regarding a plane as one site of a one-dimensional chain, $\beta_\uparrow$ shifts the first matrix factor upward and fixes the second. In the support-algebra convention of Ref.~\cite{Gross2012GNVWindex}, an operator shift from $n$ to $n+1$ of a factor $\Mat(d)$ has index $d^{-1}$. Consequently,
\begin{equation}\label{eq:eightfoldFlux}
 \operatorname{ind}(\beta_\uparrow^8)=2^{-8L^2}
 =\operatorname{ind}(T_z^4),\qquad
 \operatorname{ind}(\beta_\uparrow^8T_z^{-4})=1.
\end{equation}
Every one-dimensional finite-depth circuit has index one.

This also rules out a nonperiodic three-dimensional finite-depth circuit for the unmodified $\beta_\uparrow^8$, even with a uniformly bounded number of idle ancillary qubits per unit cell. Suppose such a circuit existed, with fixed depth and gate diameter. Retain its gates inside a transverse fundamental square, deleting gates that cross its boundary, and identify the opposite sides. This defines a circuit on the cylinder that agrees with $\beta_\uparrow^8$ outside a seam of fixed width. Their quotient $\eta_{\rm seam}$ fixes every bulk operator and hence acts only on the seam algebra, with $N_{\rm seam}=O(L)$ qubits per plane. If its range along $z$ is $r$, the one-dimensional support-algebra bound gives $|\log_2\operatorname{ind}\eta_{\rm seam}|\leq c\,rN_{\rm seam}=O(L)$, where $c$ and $r$ are independent of $L$. Multiplicativity and the circuit index, however, require $|\log_2\operatorname{ind}\eta_{\rm seam}|=8L^2$, a contradiction for large $L$. The uniformly bounded ancillary register adds only $O(L)$ seam qubits. Thus, the unmodified eighth power has ordinary transport flow, while the compensated power in Eq.~\eqref{eq:eightfoldCircuit} admits the stated circuit.

\section{Local identities for the TQFT comparison}
\label{app:TQFTlocal}

This appendix supplies the local calculations and the two general QCA results used in Sec.~\ref{sec:TQFTcomparison}. We derive the auxiliary and data separator identities, prove the relation between the layer condensation maps, and establish the auxiliary-register lemma and diagonal-QCA corollary. All operator products retain their displayed order, and integer cup products are reduced modulo four only after their evaluation.

\subsection{Parent operators and reduced separators}
\label{app:TQFTdisentangling}

The TQFT construction of Ref.~\cite{SunNonClifford2026} places a qudit on every cubic-lattice face. Its condensation operators and unencoded separator are
\begin{align}
 C_f^{\mathrm T}&=X_f^2\prod_gZ_g^{2\kappa_{gf}},
 &\kappa_{gf}&=\int(g\cup_1 f+f\cup_2\delta g),\label{eq:TQFTconstraint}\\
 G_\epsilon^{\mathrm T}&=X_{\delta\epsilon}
                 \prod_gZ_g^{\int\delta\epsilon\cup_1g},
 &\widetilde{\mathsf Z}_f^{\mathrm T}
 &=Z_f^2\prod_\epsilon(G_\epsilon^{\mathrm T})^{\int\epsilon\cup f}.
 \label{eq:TQFTseparator}
\end{align}
The qubit reduction is the on-site encoding $E$, followed by $V_\kappa=\prod_{f,g}\CNOT_{a_g\to b_f}^{\kappa_{gf}}$ and removal of the $b$ qubits. The completed local flippers are those in Eq.~(58) of Ref.~\cite{SunNonClifford2026}. Together with the separators they define the QCA in its Eq.~(60). We denote these flippers by $\mathsf X_f^{\mathrm T}$ and use their explicit local matrices in the circuit below.

For a binary edge cochain $a$, use its canonical integer $0$--$1$ lift and define the carry cochain
\begin{equation}
    K_2(a):=[a\cup a+a\cup_1\delta a]_2,\qquad \delta a\equiv[\delta a]_2+2K_2(a)\pmod4.
\label{eq:Kcarry}
\end{equation}
Here, $[\cdot]_2$ denotes reduction modulo two with the same $0$--$1$ lift. The second identity follows from the signed square rules in Appendix~\ref{sec:cochains}: $K_2(a)$ records the carry in the integer plaquette flux.

The carry identity relates the TQFT response to its boundary descendant
\begin{equation}\label{eq:sharedDescent}
 \frac14\int_{M_4}(B\cup B+B\cup_1\delta B)
 \equiv\int_{\partial M_4}
 \left[\frac14a\cup\delta a+\frac12\delta a\cup_1K_2(a)\right]\pmod1,
 \qquad B=[\delta a]_2.
\end{equation}
This is the transgression in Eqs.~(B.4) and (B.20) of Ref.~\cite{SunNonClifford2026}. The bare square-lattice kernels also obey
\begin{equation}\label{eq:bareBoundaryDictionary}
 \begin{aligned}
 H_e^\pm&=X_e\prod_{e'}Z_{e'}^{\pm\int\mathbf e'\cup\mathbf e},\\
 A_e&=\ww H_e^+\prod_fW_f^{-\int\mathbf f\cup_1\mathbf e},
 &\overline A_e&=\ww H_e^-\prod_fW_f^{-\int\mathbf e\cup_1\mathbf f}.
 \end{aligned}
\end{equation}
These are useful exact dictionaries, but neither is being used as a substitute for a full-space equivalence proof.

We recall the physical Pauli-$Z$ factorization of the ISA. For an edge $e$, define
\begin{equation}\label{eq:perpEdge}
 \begin{array}{c|c|c|c}
 e&j(e)&\varepsilon_e&\mathcal V(e)\\\hline
 h_\rvec&v_{\rvec+\xhat}&+1&\{\rvec,\rvec+\xhat\}\\
 v_\rvec&h_{\rvec+\yhat}&-1&\{\rvec,\rvec+\yhat\}.
 \end{array}
\end{equation}
Conjugating the physical $Z_e$ back through the ancilla disentangler gives the full-qudit identity
\begin{equation}\label{eq:ZliftFactorization}
 \mathcal W^\dagger Z_e^a\mathcal W
 =Z_e^2\prod_{v\in\mathcal V(e)}\mathsf{W}_v
 =A_{j(e)}^{\varepsilon_e}\overline A_{j(e)}^{-\varepsilon_e}
           \prod_{v\in\mathcal V(e)}S(W_v^2)^{\varepsilon_e}.
\end{equation}
Since $W_v^2=G_v\overline G_v^{-1}$, the physical ISA identity is
\begin{equation}\label{eq:ZreducedFactorization}
 Z_e=a_{j(e)}^{\varepsilon_e}\overline a_{j(e)}^{-\varepsilon_e}
           \prod_{v\in\mathcal V(e)}S(g_v\overline g_v)^{\varepsilon_e}.
\end{equation}
The upward pump therefore has the physical separator
\begin{equation}\label{eq:ISAseparatorExplicit}
 \mathsf Z^{\uparrow}_{e,n}
 =a_{j(e),n+1}^{\varepsilon_e}\overline a_{j(e),n}^{-\varepsilon_e}
   \prod_{v\in\mathcal V(e)}S(g_{v,n+1}\overline g_{v,n})^{\varepsilon_e}.
\end{equation}
The downward pump replaces $n+1$ by $n-1$. All formulas retain the indicated operator order.

The Laurent variable $z$ translates by one layer in the positive cubic direction. The adjoint transposes a matrix and inverts all three variables $x,y,z$. For reproducibility, the cubic cup products of Sun et al.~\cite[Appendix B]{Sun2026} give the following integer Laurent matrices, with face order $(xz,yz,xy)$ and edge order $(x,y,z)$:
\begin{align}
 D_0&=\begin{pmatrix}
 1-z^{-1}&0&x^{-1}-1\\
 0&1-z^{-1}&y^{-1}-1\\
 1-y^{-1}&x^{-1}-1&0
 \end{pmatrix},
 &H&=\begin{pmatrix}0&y&-z^{-1}\\-x&0&-xz^{-1}\\y^{-1}&1&0\end{pmatrix},\\
 J_0&=\begin{pmatrix}0&x^{-1}&0\\-y^{-1}&0&0\\0&0&z^{-1}\end{pmatrix},
 &c_3&=\begin{pmatrix}1-y\\x-1\\z-1\end{pmatrix}.
\end{align}
Here $D_0$ is coboundary, $H$ sends the first input of $\int f\cup_1g$ to the second, and $J_0$ is the signed edge--face cup pairing. Therefore
\begin{equation}\label{eq:cubicMatrices}
 \kappa=H^\dagger+c_3(y^{-1},1,1),
 \qquad N=D_0J_0,
 \qquad P=2I+HN.
\end{equation}
The columns of $(N,P)$ are the canonical $X$ and $Z$ exponents of Eq.~\eqref{eq:TQFTseparator}. In particular,
\begin{equation}\label{eq:kappaBlock}
 \kappa=\begin{pmatrix}
 M^\dagger&\binom{1}{x}\\[1mm]
 (-z+zy^{-1}-y^{-1},\;z-1-zx^{-1})&z-1
 \end{pmatrix}.
\end{equation}
The scalar phases are $+1$: the relevant $G_\epsilon$ self-pairings vanish, including the edge with negative cup-pairing sign.

To derive Eq.~\eqref{eq:horizontalSep}, reduce each qudit Pauli directly. Write $u_i=u_i^{(0)}+2u_i^{(1)}$ with $u_i^{(0)},u_i^{(1)}\in\{0,1\}$. For a Pauli $X^uZ^v$ preserving the even constraints, $v=\kappa^\dagger u\pmod2$, and the reduced action is
\begin{equation}\label{eq:reducePhase}
 |a\rangle\longmapsto
 i^{p_{u,v}(a)}|a\oplus u^{(0)}\rangle,
 \quad
 p_{u,v}(a)=\sum_i v_i a_i+
 2\sum_i\big(u_i^{(1)}+u_i^{(0)}a_i\big)
 \big((\kappa^\dagger a)_i+v_i\big)\pmod4.
\end{equation}
Any separate canonical Pauli scalar is retained. Insert the $xy$ column of Eq.~\eqref{eq:cubicMatrices} and subtract $\Phi(a\oplus u^{(0)},w)-\Phi(a,w)$. The data part is exactly $G_\rvec$ reduced using $M^\dagger$, while the remaining phase is $(-1)^{w_{\rvec,n}}$. This gives Eq.~\eqref{eq:horizontalSep}. The derivation of Eq.~\eqref{eq:reducePhase} is the isometry calculation in Eqs.~\eqref{eq:disentanglingIsometry}--\eqref{eq:hopGateDisentangle}, with $M$ replaced by $\kappa$. Thus, the phase comparison holds for arbitrary data and auxiliary bits, before any auxiliary sector is selected.

For the comparison calculation below, retain the starred loops. Their joint product is the plaquette Pauli-$Z$ product displayed in Sec.~\ref{sec:TQFTdata}. For $e=h_\rvec$ or $v_\rvec$, let $p=\rvec$ and let $q$ be the other endpoint of $e$.

Let $A_j^\circ$ denote the order-four parent hopping operator $A_j$ reduced directly in the joint $\overline A_e^{\,2}=1$ sector with $\overline VE$, without $\overline{\mathcal U}$. This hopping operator preserves that sector since it belongs to the opposite commuting algebra. Let $f_+(j)$ and $f_-(j)$ be its incident faces with oriented incidence $+1$ and $-1$. The comparison-frame semion hopping operator is
\begin{equation}\label{eq:bareComparisonHop}
 a_j^\star=A_j^\circ\,\overline g^\star_{f_+(j)}
       \prod_{f\supset j}\CZ(g^\star_f,\overline g^\star_f).
\end{equation}
To derive this, $A_j$ commutes with $\overline G_v$ and shifts $W_v$ by $i^{(\delta\mathbf j)(p(v))}$. The high bit of a fourth root changes by its low bit for a $+1$ step, and by one plus its low bit for a $-1$ step. Conjugating the two gates of $\overline{\mathcal U}$ that meet $j$ therefore yields Eq.~\eqref{eq:bareComparisonHop}.

There is also a compact formula for the entire $\mathcal F$-conjugated vertical-face separator, before restricting the auxiliary register. For either edge orientation define
\begin{equation}
    H_e=A^\circ_{j(e),n-1}(A^\circ_{j(e),n})^\dagger, \quad m=f_-(j(e)),\quad O_{e,0}=Z_{e,n}\,g^\star_{m,n-1}\overline g^\star_{m,n-1}\,H_e.
\end{equation}
Both orientations use this definition of $H_e$: their different oriented Pauli exponents are accounted for by the carry phases after disentangling and the ordered loop factor in $O_{e,0}$.
The ordered expression is
\begin{equation}\label{eq:allSectorVertical}
 \mathcal F^\dagger\mathsf Z^{\mathrm T}_{\nu,\rvec+\xhat+\yhat,n}\mathcal F
  =X_{xy,p,n}X_{xy,q,n}\,O_{e,0}
  \prod_{u\in\{p,q\}}\biggl[
      i^{w_{u,n}-w_{u,n+1}}(-1)^{w_{u,n}w_{u,n+1}}
      \left(\prod_{\ell=n-1}^{n}g^\star_{u,\ell}\overline g^\star_{u,\ell}\right)^{w_{u,n}}\biggr],
\end{equation}
where $e=h_\rvec$ for $\nu=xz$ and $e=v_\rvec$ for $\nu=yz$. The bits in the phase factors are spectral number operators of the physical auxiliary $Z$'s. The $X_{xy}$ factors are written on the left. Substitution in Eq.~\eqref{eq:reducePhase} proves this identity, including its scalar, for both face orientations.

The isometry $\mathcal J|0\rangle_{xy}$ records each $g^\star_{v,n-1}$ eigenvalue into $w_{v,n}$. Consequently Eq.~\eqref{eq:allSectorVertical} gives
\begin{equation}
 \begin{gathered}
 \rho_0(Z_{\nu,\rvec+\xhat+\yhat,n})
   =O_{e,0}\prod_{u\in\{p,q\}}\Xi_u,\\
 \Xi_u=S(g^\star_{u,-})S(g^\star_{u,+})^\dagger
       \CZ(g^\star_{u,-},g^\star_{u,+})
       \CZ\!\left(g^\star_{u,-},
        g^\star_{u,-}\overline g^\star_{u,-}g^\star_{u,+}\overline g^\star_{u,+}\right).
 \end{gathered}
\end{equation}
Here, $-$ and $+$ mean $n-1$ and $n$. All substituted loop observables commute, and their functions are kept to the right of $O_{e,0}$. Using $S(PQ)=S(P)S(Q)\CZ(P,Q)$ for commuting involutions simplifies this to
\begin{equation}
 \Xi_u=S(g^\star_{u,-})^\dagger S(g^\star_{u,+})^\dagger
       \CZ(g^\star_{u,-},\overline g^\star_{u,-})\CZ(g^\star_{u,-},\overline g^\star_{u,+}).
\end{equation}
Conjugating Eq.~\eqref{eq:ZreducedFactorization} by $F_\Sigma$ replaces its hopping operators and loops by their starred versions and leaves its Pauli $Z_e$ unchanged. We combine this comparison-frame identity with Eq.~\eqref{eq:bareComparisonHop} and the preceding expression, keeping the loop functions in their stated order.
Define the loop products
\[
 Q_{v,n}:=\overline g^\star_{v,n-1}g^\star_{v,n},\qquad
 P_{v,n}:=g^\star_{v,n-1}\overline g^\star_{v,n}.
\]
For a face $f(v)$ with lower-left vertex $v$, set $P_{f(v),n}:=P_{v,n}$.
With these definitions, the common multiplier is
\[
 \rho_0(Z_{\nu,\rvec+\xhat+\yhat,n})I_{e,n}^{-1}
 =Q_{p,n}Q_{q,n}P_{f_+(j(e)),n}.
\]
The positively incident face $f_+(j(e))$ has lower-left vertex $p$ for a horizontal edge and $q$ for a vertical edge. With loop subscripts written using these vertex labels, the two orientations are
\begin{equation}
 \begin{array}{c|c|c}
 e&m&\rho_0(Z_{\nu,\rvec+\xhat+\yhat,n})I_{e,n}^{-1}\\\hline
 h_\rvec&p(\rvec+\xhat)&Q_{p,n}Q_{q,n}P_{p,n}\\
 v_\rvec&p(\rvec)&Q_{p,n}Q_{q,n}P_{q,n}.
 \end{array}
\end{equation}
This calculation uses only the local Pauli commutation relations and the preceding spectral identity. Applying these identities with the displayed operator order yields the two multipliers in the table and proves Eq.~\eqref{eq:reducedTwoSeparators}. Both eigenvalues of every physical loop remain present.

The source automorphism in Sec.~\ref{sec:TQFTcomparison} has the following binary Laurent representation of its $Z$ images:
\begin{equation}\label{eq:sourceFactorization}
 [\delta_{\rm source}]_Z
 =\begin{pmatrix}1&0\\(1+x)y^{-1}&1\end{pmatrix}
  \begin{pmatrix}x^{-1}&0\\0&y^{-1}\end{pmatrix}
  \begin{pmatrix}1&1+y\\0&1\end{pmatrix}
 \quad\text{over }\mathbb F_2[x^{\pm1},y^{\pm1}].
\end{equation}

\subsection{Relation between the layer condensation maps}\label{app:TQFTframe}

For a binary edge configuration $a$ in a layer, set
\begin{equation}\label{eq:frameCarry}
 F_\Sigma|a\rangle
 =(-1)^{\int\delta a\cup_2K_2(a)}|a\rangle.
\end{equation}
Products over layers will also be denoted $F_\Sigma$.

For an individual square with bottom, right, top and left bits $(a_b,a_r,a_t,a_l)$, its exponent in Eq.~\eqref{eq:frameCarry} is
\begin{equation}
 \sum_{i<j<k}a_i a_j a_k+a_t+a_l
 +(a_b+a_r)(a_t+a_l)\pmod2.
\end{equation}
Thus, $F_\Sigma$ is a product of explicit four-edge diagonal gates, decomposable into $Z$, $\CZ$ and $\mathrm{CCZ}$ gates. It fixes every physical $Z_e$.

To prove Eq.~\eqref{eq:frameExact}, write the parent qudit configuration before the encoding $E$ as $q=a+2b$, with binary edge cochains $a,b$. Eq.~\eqref{eq:Kcarry} gives $\delta q\equiv[\delta a]_2+2(\delta b+K_2(a))\pmod4$. Since $W_f$ has eigenvalue $i^{(\delta q)(f)}$ and the normalization in Eq.~\eqref{eq:hatQ} sends $i^{j_0+2j_1}$ to $(-1)^{j_1}$,
\begin{equation}\label{eq:RhatCarry}
 \mathsf{W}_f|a,b\rangle
 =(-1)^{(\delta b+K_2(a))(f)}|a,b\rangle.
\end{equation}
This holds on the full parent Hilbert space, without a flatness constraint. Also, $\overline G_v=G_vW_v^2$ and $\overline G_v^2=G_v^2$, so their common spectral correction gives $\overline{\mathsf{G}}_v=\mathsf{G}_vW_v^2$. Since all loop observables commute,
\begin{equation}
 \overline{\mathcal U}\mathcal U
 =\prod_v\CZ(W_v^2,\mathsf{W}_v).
\end{equation}
A gate in this product therefore has phase $(-1)^{(\delta a)(f)(\delta b+K_2(a))(f)}$. The part linear in $b$ shifts the Fourier-encoded ancilla by $\delta^\dagger[\delta a]_2$. Modulo two, $M-M^\dagger=\delta^\dagger\delta$, so this shift is precisely canceled by changing $V$ to $\overline V$. The remaining phase is Eq.~\eqref{eq:frameCarry}, independently of the incoming ancilla bit. This proves Eq.~\eqref{eq:frameExact} without postselection.

\subsection{Auxiliary-register disentangling}
\label{app:TQFTauxiliary}

The following result upgrades invariance of the auxiliary-bit sectors to an identity action on the full auxiliary register, up to a finite-depth circuit.

\begin{lemma}[Local disentangling of auxiliary registers]\label{prop:spectator}
Let $\rho$ be a bounded-range QCA on data qubits and auxiliary qubits. Denote the auxiliary Pauli operators by $X_{xy,j},Z_{xy,j}$. Suppose $\rho(Z_{xy,j})=Z_{xy,j}$ for every auxiliary qubit, and let $\rho_0$ be its induced QCA in the all-zero auxiliary sector. There is a finite-depth circuit $\mathcal V$, diagonal in the auxiliary bits but not necessarily in the data bits, such that
\[
 \rho=\operatorname{Ad}_{\mathcal V}
             \circ(\rho_0\otimes\mathrm{id}_{xy}).
\]
The circuit can be chosen to act as the identity in the all-zero auxiliary sector and can be constructed from the local operators $\rho(X_{xy,j})$.
\end{lemma}

\begin{proof}
The gate formula requires only the known local operators
\begin{equation}\label{eq:linkOperator}
 L_j:=\rho(X_{xy,j})X_{xy,j}
  =\mathcal J\mathcal F^\dagger\mathsf X_{xy,j}^{\mathrm T}
       \mathcal F\mathcal J\,X_{xy,j}
\end{equation}
for the present application. Each $L_j$ is unitary, supported in a uniformly bounded neighborhood, and commutes with every auxiliary $Z_{xy,k}$.

Color the auxiliary sites with finitely many colors so that the support neighborhoods for two sites of the same color are disjoint. For example, if every $L_j$ is supported in an $\ell_\infty$ ball of radius $R$, a periodic coloring of period $2R+1$ in every direction suffices. Let $\lambda_j$ be the following \emph{spectral block} of $L_j$: set its own auxiliary bit to one, set auxiliary bits of later colors to zero, and retain the earlier-color bits as operator-valued arguments. No other site of its own color occurs in its support. Set
\begin{equation}\label{eq:untwistingGates}
 v_j=\frac{1+Z_{xy,j}}{2}
       +\frac{1-Z_{xy,j}}{2}\,\lambda_j,
 \qquad
 \mathcal V=\mathcal V_m\cdots\mathcal V_1,
 \quad \mathcal V_c=\prod_{j:\,\mathrm{color}(j)=c}v_j.
\end{equation}
Taking a spectral block of $L_j$ produces a unitary on the remaining local degrees of freedom. Thus, $v_j$ is a local unitary. Each color is a disjoint-support gate layer, and the number of colors is independent of system size. Eqs.~\eqref{eq:linkOperator}--\eqref{eq:untwistingGates} determine every gate from a fixed-size local matrix.

It remains to prove the circuit identity in the local operator algebra. For an auxiliary configuration $w$, let $L_j(w)$ denote the spectral block of $L_j$. The relations $\rho(X_{xy,j})^2=1$ and $[\rho(X_{xy,j}),\rho(X_{xy,k})]=0$ imply
\begin{equation}\label{eq:spectatorCocycle}
 \begin{aligned}
 L_j(w\oplus e_j)&=L_j(w)^\dagger,\\
 L_j(w)L_k(w\oplus e_j)&=L_k(w)L_j(w\oplus e_k).
 \end{aligned}
\end{equation}
Each block depends on finitely many auxiliary bits. For a configuration with finitely many occupied bits, introduce them starting from zero. The second identity makes the ordered product of links independent of the order of introduction. Introducing the bits color by color gives precisely the block $\mathcal V(w)$ of Eq.~\eqref{eq:untwistingGates}: later-color bits vanish, and no other same-color site enters the support of the current link. Introducing site $j$ last, or using the first identity if $w_j=0$, gives
\begin{equation}\label{eq:spectatorIntegratedLinks}
 L_j(w)=\mathcal V(w)\mathcal V(w\oplus e_j)^\dagger.
\end{equation}
Because $\mathcal V$ has bounded depth and gate range, the relative block $\mathcal V(w)\mathcal V(w\oplus e_j)^\dagger$ is local: it is the diagonal coefficient of $\mathcal V X_{xy,j}\mathcal V^\dagger X_{xy,j}$. Thus, both sides depend on only finitely many bits, and validity on finite-support configurations proves the identity for every auxiliary configuration. Therefore $\operatorname{Ad}_{\mathcal V^\dagger}\circ\rho$ fixes both $X_{xy,j}$ and $Z_{xy,j}$ for every auxiliary. Its data images commute with all auxiliary Paulis and consequently act only on the data register. Moreover, $\mathcal V(0)=1$, so their restriction to the zero sector is $\rho_0$. This proves Eq.~\eqref{eq:spectatorFactorization} directly on the infinite lattice.
\end{proof}

\begin{corollary}[A QCA fixing all $Z$'s is a diagonal circuit]\label{prop:fixZ}
If a bounded-range QCA $\gamma$ fixes $Z_j$ for every microscopic qubit, then $\gamma=\operatorname{Ad}_G$ for a finite-depth diagonal circuit $G$.
\end{corollary}
\begin{proof}
Apply Lemma~\ref{prop:spectator} with no data register. The all-zero sector is then one-dimensional, so its operator algebra has only the identity automorphism. The resulting circuit is diagonal in every computational-basis bit.
\end{proof}

Every circuit factor is specified by local operator formulas. The small geometric pieces are controlled-$S$, loop-controlled, carry-phase, and CNOT gates. The potentially larger gates of $\mathcal V$ are the fixed-size spectral blocks in Eqs.~\eqref{eq:linkOperator} and~\eqref{eq:untwistingGates}, evaluated using the stated TQFT flippers. They are ordinary local unitaries because each block belongs to a unitary commuting with all auxiliary $Z$'s.

These results justify the order used in Sec.~\ref{sec:TQFTcomparison}: Eq.~\eqref{eq:frameExact} first supplies the common layer basis. Eq.~\eqref{eq:horizontalSep} and Lemma~\ref{prop:spectator} then separate the $xy$ auxiliary register. And, Eq.~\eqref{eq:reducedTwoSeparators} identifies the remaining data separators with the ISA pump. Corollary~\ref{prop:fixZ} supplies the final diagonal correction.

\FloatBarrier

\section{Cup-product formulation of the full construction}\label{sec:cochains}

The hopping operators, original and corrected condensates, and both parts of the disentangling circuit admit a uniform higher-cup-product description. The condensation correction is a local spectral phase in commuting operator-valued cochains. The resulting condensate remains non-Pauli.

\subsection{Integer cubical conventions and the incident-face selector}

As in the main text, $v,e,f$ label cells, whereas $\mathbf v,\mathbf e,\mathbf f$ are their elementary integer cochains, equal to $+1$ on the indicated cell and zero elsewhere. In the two-dimensional formulas, $\int$ pairs a degree-two cochain with the oriented spatial surface $\Sigma$, whose label is suppressed. We use the corrected integer boundary and cubical cup-product conventions of Sun \textit{et al.}~\cite[Appendix B]{Sun2026} throughout. On a square with bottom, top, left, and right edges denoted by $h_b,h_t,v_l,v_r$, the needed local rules are
\begin{align}
 (a_1\cup b_1)(f)&=a_1(h_b)b_1(v_r)-a_1(v_l)b_1(h_t),\label{eq:cup11rule}\\
 (a_2\cup_1 b_1)(f)&=a_2(f)\bigl[b_1(h_b)+b_1(v_r)\bigr],\\
 (a_1\cup_1 b_2)(f)&=-\bigl[a_1(v_l)+a_1(h_t)\bigr]b_2(f),\label{eq:cup12rule}\\
 (a_2\cup_2 b_2)(f)&=a_2(f)b_2(f).
 \label{eq:cuptoprule}
\end{align}
Also,
$\delta a_1(f)=a_1(h_b)-a_1(h_t)+a_1(v_r)-a_1(v_l)$ and
$\delta a_0(e_{ij})=a_0(j)-a_0(i)$.

For an edge basis cochain $\mathbf e$, define
\begin{equation}\label{eq:cupselector}
 \phi_e:=\delta\mathbf e\cup_2\delta\mathbf e,
 \qquad
 \chi_e(v):=\int\mathbf v\cup\phi_e.
\end{equation}
Because $\phi_e(f)=[(\delta\mathbf e)(f)]^2$, it is the \emph{unsigned} sum of the two incident faces. Thus, $\chi_e(v)$ is one at their lower-left vertices and zero elsewhere. More explicitly,
\begin{equation}
 \phi_{h_{\rvec}}=\mathbf f_{\rvec}+\mathbf f_{\rvec-\yhat},\qquad
 \phi_{v_{\rvec}}=\mathbf f_{\rvec}+\mathbf f_{\rvec-\xhat}.
\end{equation}
Consequently, a product over faces $f\supset e$ can be written as a product over vertices with exponent $\chi_e(v)$, using the indexing convention $G_{p(v)}=G_v$ and its analogues. Cochains and cup products are evaluated over integers first. Pauli exponents may subsequently be reduced modulo four, and CNOT exponents modulo two. The phase correction below requires an integer lift and must not first be reduced modulo two.

\subsection{Original hopping operators and dyon loops}

The order-four hopping operators are
\begin{align}
 A_e&=\ww X_e\prod_{e'}Z_{e'}^{\int(\mathbf e'\cup\mathbf e-\delta\mathbf e'\cup_1\mathbf e)},\label{eq:rawcupA}\\
 \overline A_e&=\ww X_e\prod_{e'}Z_{e'}^{\int(\mathbf e\cup\mathbf e'-\delta\mathbf e\cup_1\mathbf e')}.
 \label{eq:rawcupAb}
\end{align}
The scalar $\ww$ fixes the operator phases as in Eq.~\eqref{eq:parentHops}. The signed clock decorations agree with the Laurent matrices $M$ and $M^\dagger$ of Eq.~\eqref{eq:M}, respectively, using the conventions in Eqs.~\eqref{eq:cup11rule}--\eqref{eq:cuptoprule}.

For compact supports, integration of the cup-one identity gives
\begin{equation}
 \int\bigl(\delta a\cup_1 b-a\cup_1\delta b\bigr)
 =\int(a\cup b+b\cup a),\qquad a,b\in C^1(\Sigma,\mathbb Z).
\end{equation}
In particular,
\begin{equation}\label{eq:cupTranspose}
 \int\bigl(-\mathbf e'\cup\mathbf e-\mathbf e\cup_1\delta\mathbf e'\bigr)
 =\int\bigl(\mathbf e\cup\mathbf e'-\delta\mathbf e\cup_1\mathbf e'\bigr).
\end{equation}
This exchanged pairing proves $[A_e,\overline A_{e'}]=0$ directly in cochain notation. For either edge orientation, $\int(\mathbf e\cup\mathbf e-\delta\mathbf e\cup_1\mathbf e)=-1$. Together with the scalar in Eq.~\eqref{eq:parentHops}, this gives
\begin{equation}\label{eq:cupHopSquares}
 \begin{aligned}
 A_e^2&=X_e^2\prod_{e'}Z_{e'}^{2\int(\mathbf e'\cup\mathbf e-\delta\mathbf e'\cup_1\mathbf e)},\\
 \overline A_e^{\,2}&=X_e^2\prod_{e'}Z_{e'}^{2\int(\mathbf e\cup\mathbf e'-\delta\mathbf e\cup_1\mathbf e')}.
 \end{aligned}
\end{equation}

For a face $f$, let $W_f=Z_{\partial f}=\prod_{e'}Z_{e'}^{(\delta\mathbf e')(f)}$. The pairing $\int\mathbf v\cup\mathbf f$ selects the unique square whose lower-left vertex is $v$. Thus, \begin{align}
 T_v&=X_{\delta\mathbf v},\qquad
 W_v:=\prod_f W_f^{\int\mathbf v\cup\mathbf f}
      =\prod_{e'}Z_{e'}^{\int\mathbf v\cup\delta\mathbf e'},\label{eq:cupR}\\
 G_v&=X_{\delta\mathbf v}\prod_{e'}Z_{e'}^{\int\mathbf v\cup\delta\mathbf e'},\qquad
 \overline G_v=X_{\delta\mathbf v}\prod_{e'}Z_{e'}^{-\int\mathbf v\cup\delta\mathbf e'}.
 \label{eq:cupB}
\end{align}
These are the six-edge loops in Eqs.~\eqref{eq:toricStarPlaquette} and~\eqref{eq:B}, with $v$ now a vertex. The sign and displacement agree with $Md=c$ and $M^\dagger d=-c$. In particular,
\begin{equation}\label{eq:cupPhi}
 \Phi_v:=G_v^2=\overline G_v^2
 =X_{2\delta\mathbf v}\prod_{e'}Z_{e'}^{2\int\mathbf v\cup\delta\mathbf e'}.
\end{equation}
Every $\Phi_v$ is a Hermitian involution commuting with both hopping families. For example, $A_e G_v=(-1)^{\int\mathbf v\cup\delta\mathbf e}G_v A_e$, so the square of $G_v$ commutes with $A_e$. The cross-family relations give the remaining commutators.

\subsection{The order-four and order-two condensates}

The exact factorization of the original condensate becomes
\begin{equation}
    C_e=A_e^2\prod_v G_v^{\chi_e(v)} =\overline A_e^{\,2} \prod_v \overline{G}_v^{\chi_e(v)}.
\label{eq:cupColdFactor}
\end{equation}
The corresponding ordered Pauli expression is
\begin{equation}
    C_e=X_e^2\prod_{e'}Z_{e'}^{\int(2 \mathbf{e}'\cup\mathbf{e}-\delta\mathbf{e}\cup_2\delta\mathbf{e}')}\prod_vX_{\delta\mathbf{v}}^{\int\mathbf{v}\cup(\delta\mathbf{e}\cup_2\delta\mathbf{e})}.
\label{eq:rawcupC}
\end{equation}
The written order is part of this formula. It agrees, including its scalar phase, with Eq.~\eqref{eq:cupColdFactor}. Its square is
\begin{equation}
    C_e^2=\prod_v\Phi_v^{\chi_e(v)}.
\label{eq:cupCsq}
\end{equation}

Use the spectral function $S$ from Eq.~\eqref{eq:S}, and set
$\mathsf{G}_v=G_vS(\Phi_v)^\dagger$ and
$\overline{\mathsf{G}}_v=\overline G_vS(\Phi_v)^\dagger$. The corrected condensation operators are
\begin{equation}
    D_e =A_e^2\prod_v\mathsf{G}_v^{\int\mathbf{v}\cup(\delta\mathbf{e}\cup_2\delta\mathbf e)} = \overline A_e^{\,2}\prod_v\overline{\mathsf{G}}_v^{\int\mathbf{v}\cup(\delta\mathbf e\cup_2\delta\mathbf{e})}.
\label{eq:cupDfact}
\end{equation}
This is a single formula for both edge orientations. Every factor in the unbarred expression is a commuting involution, so $D_e^2=1$. The four Pauli strings in the expansion of $D_e$ are displayed in Figs.~\ref{fig:Dh} and~\ref{fig:Dv}, with every single-edge operator and scalar phase explicit.

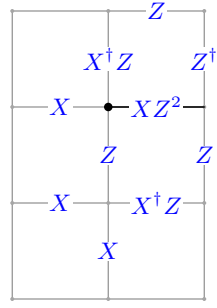
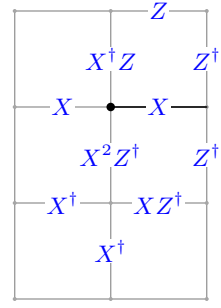
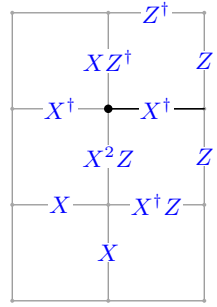
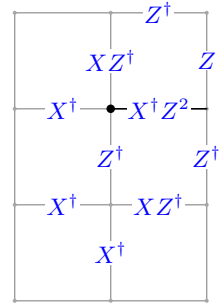
\begin{figure}[t]
\centering
\begin{subfigure}[c]{.47\linewidth}\centering
\begin{tikzpicture}[x=1.27cm,y=1.27cm,line cap=round,line join=round]
\draw[gridline] (-1,-2) -- (-1,1);
\draw[gridline] (0,-2) -- (0,1);
\draw[gridline] (1,-2) -- (1,1);
\draw[gridline] (-1,-2) -- (1,-2);
\draw[gridline] (-1,-1) -- (1,-1);
\draw[gridline] (-1,0) -- (1,0);
\draw[gridline] (-1,1) -- (1,1);
\fill[black!35] (-1,-2) circle (0.75pt);
\fill[black!35] (-1,-1) circle (0.75pt);
\fill[black!35] (-1,0) circle (0.75pt);
\fill[black!35] (-1,1) circle (0.75pt);
\fill[black!35] (0,-2) circle (0.75pt);
\fill[black!35] (0,-1) circle (0.75pt);
\fill[black!35] (0,0) circle (0.75pt);
\fill[black!35] (0,1) circle (0.75pt);
\fill[black!35] (1,-2) circle (0.75pt);
\fill[black!35] (1,-1) circle (0.75pt);
\fill[black!35] (1,0) circle (0.75pt);
\fill[black!35] (1,1) circle (0.75pt);
\draw[edgefocus] (0,0) -- (1,0);
\node[fill=white,inner sep=.35pt,text=blue,oplabel] at (-0.5,-1) {${X}$};
\node[fill=white,inner sep=.35pt,text=blue,oplabel] at (-0.5,0) {${X}$};
\node[fill=white,inner sep=.35pt,text=blue,oplabel] at (0.5,-1) {${X^{\dagger}}{Z}$};
\node[fill=white,inner sep=.35pt,text=blue,oplabel] at (0.5,0) {${X}{Z^{2}}$};
\node[fill=white,inner sep=.35pt,text=blue,oplabel] at (0.5,1) {${Z}$};
\node[fill=white,inner sep=.35pt,text=blue,oplabel] at (0,-1.5) {${X}$};
\node[fill=white,inner sep=.35pt,text=blue,oplabel] at (0,-0.5) {${Z}$};
\node[fill=white,inner sep=.35pt,text=blue,oplabel] at (0,0.5) {${X^{\dagger}}{Z}$};
\node[fill=white,inner sep=.35pt,text=blue,oplabel] at (1,-0.5) {${Z}$};
\node[fill=white,inner sep=.35pt,text=blue,oplabel] at (1,0.5) {${Z^{\dagger}}$};
\fill[black] (0,0) circle (1.6pt);
\path[use as bounding box] (-1.33,-2.25) rectangle (1.33,1.25);
\end{tikzpicture}
\caption{$P_h^{++}=A_h^2G_pG_q$}
\end{subfigure}
\hfill
\begin{subfigure}[c]{.47\linewidth}\centering
\begin{tikzpicture}[x=1.27cm,y=1.27cm,line cap=round,line join=round]
\draw[gridline] (-1,-2) -- (-1,1);
\draw[gridline] (0,-2) -- (0,1);
\draw[gridline] (1,-2) -- (1,1);
\draw[gridline] (-1,-2) -- (1,-2);
\draw[gridline] (-1,-1) -- (1,-1);
\draw[gridline] (-1,0) -- (1,0);
\draw[gridline] (-1,1) -- (1,1);
\fill[black!35] (-1,-2) circle (0.75pt);
\fill[black!35] (-1,-1) circle (0.75pt);
\fill[black!35] (-1,0) circle (0.75pt);
\fill[black!35] (-1,1) circle (0.75pt);
\fill[black!35] (0,-2) circle (0.75pt);
\fill[black!35] (0,-1) circle (0.75pt);
\fill[black!35] (0,0) circle (0.75pt);
\fill[black!35] (0,1) circle (0.75pt);
\fill[black!35] (1,-2) circle (0.75pt);
\fill[black!35] (1,-1) circle (0.75pt);
\fill[black!35] (1,0) circle (0.75pt);
\fill[black!35] (1,1) circle (0.75pt);
\draw[edgefocus] (0,0) -- (1,0);
\node[fill=white,inner sep=.35pt,text=blue,oplabel] at (-0.5,-1) {${X^{\dagger}}$};
\node[fill=white,inner sep=.35pt,text=blue,oplabel] at (-0.5,0) {${X}$};
\node[fill=white,inner sep=.35pt,text=blue,oplabel] at (0.5,-1) {${X}{Z^{\dagger}}$};
\node[fill=white,inner sep=.35pt,text=blue,oplabel] at (0.5,0) {${X}$};
\node[fill=white,inner sep=.35pt,text=blue,oplabel] at (0.5,1) {${Z}$};
\node[fill=white,inner sep=.35pt,text=blue,oplabel] at (0,-1.5) {${X^{\dagger}}$};
\node[fill=white,inner sep=.35pt,text=blue,oplabel] at (0,-0.5) {${X^{2}}{Z^{\dagger}}$};
\node[fill=white,inner sep=.35pt,text=blue,oplabel] at (0,0.5) {${X^{\dagger}}{Z}$};
\node[fill=white,inner sep=.35pt,text=blue,oplabel] at (1,-0.5) {${Z^{\dagger}}$};
\node[fill=white,inner sep=.35pt,text=blue,oplabel] at (1,0.5) {${Z^{\dagger}}$};
\fill[black] (0,0) circle (1.6pt);
\path[use as bounding box] (-1.33,-2.25) rectangle (1.33,1.25);
\end{tikzpicture}
\caption{$P_h^{+-}=A_h^2G_pG_q^{-1}$}
\end{subfigure}
\par\vspace{5mm}
\begin{subfigure}[c]{.47\linewidth}\centering
\begin{tikzpicture}[x=1.27cm,y=1.27cm,line cap=round,line join=round]
\draw[gridline] (-1,-2) -- (-1,1);
\draw[gridline] (0,-2) -- (0,1);
\draw[gridline] (1,-2) -- (1,1);
\draw[gridline] (-1,-2) -- (1,-2);
\draw[gridline] (-1,-1) -- (1,-1);
\draw[gridline] (-1,0) -- (1,0);
\draw[gridline] (-1,1) -- (1,1);
\fill[black!35] (-1,-2) circle (0.75pt);
\fill[black!35] (-1,-1) circle (0.75pt);
\fill[black!35] (-1,0) circle (0.75pt);
\fill[black!35] (-1,1) circle (0.75pt);
\fill[black!35] (0,-2) circle (0.75pt);
\fill[black!35] (0,-1) circle (0.75pt);
\fill[black!35] (0,0) circle (0.75pt);
\fill[black!35] (0,1) circle (0.75pt);
\fill[black!35] (1,-2) circle (0.75pt);
\fill[black!35] (1,-1) circle (0.75pt);
\fill[black!35] (1,0) circle (0.75pt);
\fill[black!35] (1,1) circle (0.75pt);
\draw[edgefocus] (0,0) -- (1,0);
\node[fill=white,inner sep=.35pt,text=blue,oplabel] at (-0.5,-1) {${X}$};
\node[fill=white,inner sep=.35pt,text=blue,oplabel] at (-0.5,0) {${X^{\dagger}}$};
\node[fill=white,inner sep=.35pt,text=blue,oplabel] at (0.5,-1) {${X^{\dagger}}{Z}$};
\node[fill=white,inner sep=.35pt,text=blue,oplabel] at (0.5,0) {${X^{\dagger}}$};
\node[fill=white,inner sep=.35pt,text=blue,oplabel] at (0.5,1) {${Z^{\dagger}}$};
\node[fill=white,inner sep=.35pt,text=blue,oplabel] at (0,-1.5) {${X}$};
\node[fill=white,inner sep=.35pt,text=blue,oplabel] at (0,-0.5) {${X^{2}}{Z}$};
\node[fill=white,inner sep=.35pt,text=blue,oplabel] at (0,0.5) {${X}{Z^{\dagger}}$};
\node[fill=white,inner sep=.35pt,text=blue,oplabel] at (1,-0.5) {${Z}$};
\node[fill=white,inner sep=.35pt,text=blue,oplabel] at (1,0.5) {${Z}$};
\fill[black] (0,0) circle (1.6pt);
\path[use as bounding box] (-1.33,-2.25) rectangle (1.33,1.25);
\end{tikzpicture}
\caption{$P_h^{-+}=A_h^2G_p^{-1}G_q$}
\end{subfigure}
\hfill
\begin{subfigure}[c]{.47\linewidth}\centering
\begin{tikzpicture}[x=1.27cm,y=1.27cm,line cap=round,line join=round]
\draw[gridline] (-1,-2) -- (-1,1);
\draw[gridline] (0,-2) -- (0,1);
\draw[gridline] (1,-2) -- (1,1);
\draw[gridline] (-1,-2) -- (1,-2);
\draw[gridline] (-1,-1) -- (1,-1);
\draw[gridline] (-1,0) -- (1,0);
\draw[gridline] (-1,1) -- (1,1);
\fill[black!35] (-1,-2) circle (0.75pt);
\fill[black!35] (-1,-1) circle (0.75pt);
\fill[black!35] (-1,0) circle (0.75pt);
\fill[black!35] (-1,1) circle (0.75pt);
\fill[black!35] (0,-2) circle (0.75pt);
\fill[black!35] (0,-1) circle (0.75pt);
\fill[black!35] (0,0) circle (0.75pt);
\fill[black!35] (0,1) circle (0.75pt);
\fill[black!35] (1,-2) circle (0.75pt);
\fill[black!35] (1,-1) circle (0.75pt);
\fill[black!35] (1,0) circle (0.75pt);
\fill[black!35] (1,1) circle (0.75pt);
\draw[edgefocus] (0,0) -- (1,0);
\node[fill=white,inner sep=.35pt,text=blue,oplabel] at (-0.5,-1) {${X^{\dagger}}$};
\node[fill=white,inner sep=.35pt,text=blue,oplabel] at (-0.5,0) {${X^{\dagger}}$};
\node[fill=white,inner sep=.35pt,text=blue,oplabel] at (0.5,-1) {${X}{Z^{\dagger}}$};
\node[fill=white,inner sep=.35pt,text=blue,oplabel] at (0.5,0) {${X^{\dagger}}{Z^{2}}$};
\node[fill=white,inner sep=.35pt,text=blue,oplabel] at (0.5,1) {${Z^{\dagger}}$};
\node[fill=white,inner sep=.35pt,text=blue,oplabel] at (0,-1.5) {${X^{\dagger}}$};
\node[fill=white,inner sep=.35pt,text=blue,oplabel] at (0,-0.5) {${Z^{\dagger}}$};
\node[fill=white,inner sep=.35pt,text=blue,oplabel] at (0,0.5) {${X}{Z^{\dagger}}$};
\node[fill=white,inner sep=.35pt,text=blue,oplabel] at (1,-0.5) {${Z^{\dagger}}$};
\node[fill=white,inner sep=.35pt,text=blue,oplabel] at (1,0.5) {${Z}$};
\fill[black] (0,0) circle (1.6pt);
\path[use as bounding box] (-1.33,-2.25) rectangle (1.33,1.25);
\end{tikzpicture}
\caption{$P_h^{--}=A_h^2G_p^{-1}G_q^{-1}$}
\end{subfigure}
\caption{Full Pauli-string expansion of the corrected \emph{horizontal} condensate. Here $e=h_{\rvec}$, $p=\rvec$, and $q=\rvec-\yhat$. The operator is $D_h=(-iP_h^{++}+P_h^{+-}+P_h^{-+}+iP_h^{--})/2$. Panel (a) is $C_h$. All labels are single-edge qudit operators, with $X$ written before $Z$. These four diagrams specify the complete non-Pauli operator, not only its support.}
 \label{fig:Dh}
\end{figure}
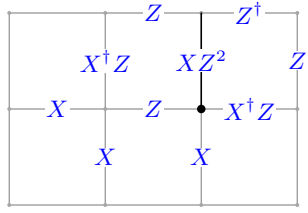
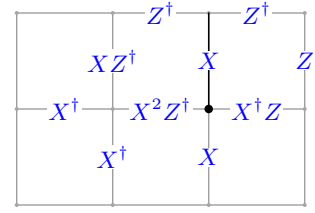
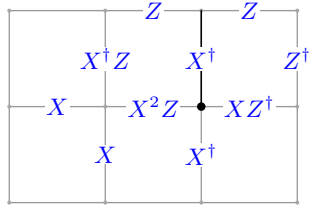
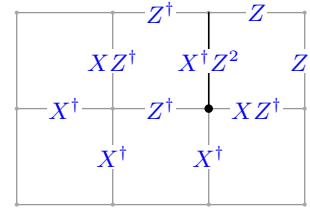
\begin{figure}[t]
\centering
\begin{subfigure}[c]{.47\linewidth}\centering
\begin{tikzpicture}[x=1.27cm,y=1.27cm,line cap=round,line join=round]
\draw[gridline] (-2,-1) -- (-2,1);
\draw[gridline] (-1,-1) -- (-1,1);
\draw[gridline] (0,-1) -- (0,1);
\draw[gridline] (1,-1) -- (1,1);
\draw[gridline] (-2,-1) -- (1,-1);
\draw[gridline] (-2,0) -- (1,0);
\draw[gridline] (-2,1) -- (1,1);
\fill[black!35] (-2,-1) circle (0.75pt);
\fill[black!35] (-2,0) circle (0.75pt);
\fill[black!35] (-2,1) circle (0.75pt);
\fill[black!35] (-1,-1) circle (0.75pt);
\fill[black!35] (-1,0) circle (0.75pt);
\fill[black!35] (-1,1) circle (0.75pt);
\fill[black!35] (0,-1) circle (0.75pt);
\fill[black!35] (0,0) circle (0.75pt);
\fill[black!35] (0,1) circle (0.75pt);
\fill[black!35] (1,-1) circle (0.75pt);
\fill[black!35] (1,0) circle (0.75pt);
\fill[black!35] (1,1) circle (0.75pt);
\draw[edgefocus] (0,0) -- (0,1);
\node[fill=white,inner sep=.35pt,text=blue,oplabel] at (-1.5,0) {${X}$};
\node[fill=white,inner sep=.35pt,text=blue,oplabel] at (-0.5,0) {${Z}$};
\node[fill=white,inner sep=.35pt,text=blue,oplabel] at (-0.5,1) {${Z}$};
\node[fill=white,inner sep=.35pt,text=blue,oplabel] at (0.5,0) {${X^{\dagger}}{Z}$};
\node[fill=white,inner sep=.35pt,text=blue,oplabel] at (0.5,1) {${Z^{\dagger}}$};
\node[fill=white,inner sep=.35pt,text=blue,oplabel] at (-1,-0.5) {${X}$};
\node[fill=white,inner sep=.35pt,text=blue,oplabel] at (-1,0.5) {${X^{\dagger}}{Z}$};
\node[fill=white,inner sep=.35pt,text=blue,oplabel] at (0,-0.5) {${X}$};
\node[fill=white,inner sep=.35pt,text=blue,oplabel] at (0,0.5) {${X}{Z^{2}}$};
\node[fill=white,inner sep=.35pt,text=blue,oplabel] at (1,0.5) {${Z}$};
\fill[black] (0,0) circle (1.6pt);
\path[use as bounding box] (-2.33,-1.25) rectangle (1.33,1.25);
\end{tikzpicture}
\caption{$P_v^{++}=A_v^2G_pG_q$}
\end{subfigure}
\hfill
\begin{subfigure}[c]{.47\linewidth}\centering
\begin{tikzpicture}[x=1.27cm,y=1.27cm,line cap=round,line join=round]
\draw[gridline] (-2,-1) -- (-2,1);
\draw[gridline] (-1,-1) -- (-1,1);
\draw[gridline] (0,-1) -- (0,1);
\draw[gridline] (1,-1) -- (1,1);
\draw[gridline] (-2,-1) -- (1,-1);
\draw[gridline] (-2,0) -- (1,0);
\draw[gridline] (-2,1) -- (1,1);
\fill[black!35] (-2,-1) circle (0.75pt);
\fill[black!35] (-2,0) circle (0.75pt);
\fill[black!35] (-2,1) circle (0.75pt);
\fill[black!35] (-1,-1) circle (0.75pt);
\fill[black!35] (-1,0) circle (0.75pt);
\fill[black!35] (-1,1) circle (0.75pt);
\fill[black!35] (0,-1) circle (0.75pt);
\fill[black!35] (0,0) circle (0.75pt);
\fill[black!35] (0,1) circle (0.75pt);
\fill[black!35] (1,-1) circle (0.75pt);
\fill[black!35] (1,0) circle (0.75pt);
\fill[black!35] (1,1) circle (0.75pt);
\draw[edgefocus] (0,0) -- (0,1);
\node[fill=white,inner sep=.35pt,text=blue,oplabel] at (-1.5,0) {${X^{\dagger}}$};
\node[fill=white,inner sep=.35pt,text=blue,oplabel] at (-0.5,0) {${X^{2}}{Z^{\dagger}}$};
\node[fill=white,inner sep=.35pt,text=blue,oplabel] at (-0.5,1) {${Z^{\dagger}}$};
\node[fill=white,inner sep=.35pt,text=blue,oplabel] at (0.5,0) {${X^{\dagger}}{Z}$};
\node[fill=white,inner sep=.35pt,text=blue,oplabel] at (0.5,1) {${Z^{\dagger}}$};
\node[fill=white,inner sep=.35pt,text=blue,oplabel] at (-1,-0.5) {${X^{\dagger}}$};
\node[fill=white,inner sep=.35pt,text=blue,oplabel] at (-1,0.5) {${X}{Z^{\dagger}}$};
\node[fill=white,inner sep=.35pt,text=blue,oplabel] at (0,-0.5) {${X}$};
\node[fill=white,inner sep=.35pt,text=blue,oplabel] at (0,0.5) {${X}$};
\node[fill=white,inner sep=.35pt,text=blue,oplabel] at (1,0.5) {${Z}$};
\fill[black] (0,0) circle (1.6pt);
\path[use as bounding box] (-2.33,-1.25) rectangle (1.33,1.25);
\end{tikzpicture}
\caption{$P_v^{+-}=A_v^2G_pG_q^{-1}$}
\end{subfigure}
\par\vspace{5mm}
\begin{subfigure}[c]{.47\linewidth}\centering
\begin{tikzpicture}[x=1.27cm,y=1.27cm,line cap=round,line join=round]
\draw[gridline] (-2,-1) -- (-2,1);
\draw[gridline] (-1,-1) -- (-1,1);
\draw[gridline] (0,-1) -- (0,1);
\draw[gridline] (1,-1) -- (1,1);
\draw[gridline] (-2,-1) -- (1,-1);
\draw[gridline] (-2,0) -- (1,0);
\draw[gridline] (-2,1) -- (1,1);
\fill[black!35] (-2,-1) circle (0.75pt);
\fill[black!35] (-2,0) circle (0.75pt);
\fill[black!35] (-2,1) circle (0.75pt);
\fill[black!35] (-1,-1) circle (0.75pt);
\fill[black!35] (-1,0) circle (0.75pt);
\fill[black!35] (-1,1) circle (0.75pt);
\fill[black!35] (0,-1) circle (0.75pt);
\fill[black!35] (0,0) circle (0.75pt);
\fill[black!35] (0,1) circle (0.75pt);
\fill[black!35] (1,-1) circle (0.75pt);
\fill[black!35] (1,0) circle (0.75pt);
\fill[black!35] (1,1) circle (0.75pt);
\draw[edgefocus] (0,0) -- (0,1);
\node[fill=white,inner sep=.35pt,text=blue,oplabel] at (-1.5,0) {${X}$};
\node[fill=white,inner sep=.35pt,text=blue,oplabel] at (-0.5,0) {${X^{2}}{Z}$};
\node[fill=white,inner sep=.35pt,text=blue,oplabel] at (-0.5,1) {${Z}$};
\node[fill=white,inner sep=.35pt,text=blue,oplabel] at (0.5,0) {${X}{Z^{\dagger}}$};
\node[fill=white,inner sep=.35pt,text=blue,oplabel] at (0.5,1) {${Z}$};
\node[fill=white,inner sep=.35pt,text=blue,oplabel] at (-1,-0.5) {${X}$};
\node[fill=white,inner sep=.35pt,text=blue,oplabel] at (-1,0.5) {${X^{\dagger}}{Z}$};
\node[fill=white,inner sep=.35pt,text=blue,oplabel] at (0,-0.5) {${X^{\dagger}}$};
\node[fill=white,inner sep=.35pt,text=blue,oplabel] at (0,0.5) {${X^{\dagger}}$};
\node[fill=white,inner sep=.35pt,text=blue,oplabel] at (1,0.5) {${Z^{\dagger}}$};
\fill[black] (0,0) circle (1.6pt);
\path[use as bounding box] (-2.33,-1.25) rectangle (1.33,1.25);
\end{tikzpicture}
\caption{$P_v^{-+}=A_v^2G_p^{-1}G_q$}
\end{subfigure}
\hfill
\begin{subfigure}[c]{.47\linewidth}\centering
\begin{tikzpicture}[x=1.27cm,y=1.27cm,line cap=round,line join=round]
\draw[gridline] (-2,-1) -- (-2,1);
\draw[gridline] (-1,-1) -- (-1,1);
\draw[gridline] (0,-1) -- (0,1);
\draw[gridline] (1,-1) -- (1,1);
\draw[gridline] (-2,-1) -- (1,-1);
\draw[gridline] (-2,0) -- (1,0);
\draw[gridline] (-2,1) -- (1,1);
\fill[black!35] (-2,-1) circle (0.75pt);
\fill[black!35] (-2,0) circle (0.75pt);
\fill[black!35] (-2,1) circle (0.75pt);
\fill[black!35] (-1,-1) circle (0.75pt);
\fill[black!35] (-1,0) circle (0.75pt);
\fill[black!35] (-1,1) circle (0.75pt);
\fill[black!35] (0,-1) circle (0.75pt);
\fill[black!35] (0,0) circle (0.75pt);
\fill[black!35] (0,1) circle (0.75pt);
\fill[black!35] (1,-1) circle (0.75pt);
\fill[black!35] (1,0) circle (0.75pt);
\fill[black!35] (1,1) circle (0.75pt);
\draw[edgefocus] (0,0) -- (0,1);
\node[fill=white,inner sep=.35pt,text=blue,oplabel] at (-1.5,0) {${X^{\dagger}}$};
\node[fill=white,inner sep=.35pt,text=blue,oplabel] at (-0.5,0) {${Z^{\dagger}}$};
\node[fill=white,inner sep=.35pt,text=blue,oplabel] at (-0.5,1) {${Z^{\dagger}}$};
\node[fill=white,inner sep=.35pt,text=blue,oplabel] at (0.5,0) {${X}{Z^{\dagger}}$};
\node[fill=white,inner sep=.35pt,text=blue,oplabel] at (0.5,1) {${Z}$};
\node[fill=white,inner sep=.35pt,text=blue,oplabel] at (-1,-0.5) {${X^{\dagger}}$};
\node[fill=white,inner sep=.35pt,text=blue,oplabel] at (-1,0.5) {${X}{Z^{\dagger}}$};
\node[fill=white,inner sep=.35pt,text=blue,oplabel] at (0,-0.5) {${X^{\dagger}}$};
\node[fill=white,inner sep=.35pt,text=blue,oplabel] at (0,0.5) {${X^{\dagger}}{Z^{2}}$};
\node[fill=white,inner sep=.35pt,text=blue,oplabel] at (1,0.5) {${Z^{\dagger}}$};
\fill[black] (0,0) circle (1.6pt);
\path[use as bounding box] (-2.33,-1.25) rectangle (1.33,1.25);
\end{tikzpicture}
\caption{$P_v^{--}=A_v^2G_p^{-1}G_q^{-1}$}
\end{subfigure}
\caption{Full Pauli-string expansion of the corrected \emph{vertical} condensate. Here $e=v_{\rvec}$, $p=\rvec$, and $q=\rvec-\xhat$. The operator is $D_v=(-iP_v^{++}+P_v^{+-}+P_v^{-+}+iP_v^{--})/2$. Panel (a) is $C_v$. Together, Figs.~\ref{fig:Dh} and~\ref{fig:Dv} give a translation-invariant prescription for every corrected condensation term on the square lattice.}
\label{fig:Dv}
\end{figure}

A compact form of the non-Pauli correction uses the commuting operator-valued zero-cochain
\begin{equation}\label{eq:operatorn}
    \mathsf{N}_0:=\sum_v n_v\mathbf{v},\qquad n_v:=\frac{1-\Phi_v}{2}.
\end{equation}
The coefficients $n_v$ are local commuting projectors with eigenvalues zero and one; they are not additional degrees of freedom or imposed constraints. Cup products with these coefficients are well defined by using operator multiplication in their commuting algebra. Then
\begin{equation}
    D_e=C_e \exp \left[-\frac{i\pi}{2}\int \mathsf{N}_0\cup(\delta\mathbf{e}\cup_2\delta\mathbf{e}) \right].
\label{eq:cupDcompact}
\end{equation}
Indeed, the integral is exactly the sum of the two $n_v$'s at the incident faces, so the exponential is the product of the two $S(\Phi_v)^\dagger$ factors. Writing
$N_e:=\int\mathsf N_0\cup\phi_e$, one has
\begin{equation}
    C_e^2=e^{i\pi N_e},\qquad D_e^2=C_e^2e^{-i\pi N_e}=1.
\end{equation}
The correction commutes with both hopping families, giving the full-space identities
$[D_e,A_{e'}]=[D_e,\overline A_{e'}]=[D_e,D_{e'}]=0$.

\paragraph{The integer lift matters.}
The integral in Eq.~\eqref{eq:cupDcompact} is an integer-valued operator, not its reduction modulo two. When both incident $\Phi$'s equal $-1$, it equals two and the correction is $-1$. For two incident loops $p,q$, the exact relation is
\begin{equation}
    S(\Phi_p)^\dagger S(\Phi_q)^\dagger =(-1)^{n_pn_q}S(\Phi_p\Phi_q)^\dagger.
\label{eq:rootproductwarning}
\end{equation}
Thus, replacing the correction by $S(C_e^2)^\dagger$ alone would omit a controlled sign. The sitewise lift in Eq.~\eqref{eq:operatorn} selects the correction used in the finite-depth disentangler. The central Pauli string and its spectral correction are displayed in Fig.~\ref{fig:correctioncontrol}.

\begin{figure}[t]
\centering
\begin{minipage}[c]{.44\linewidth}\centering
\begin{tikzpicture}[x=1.65cm,y=1.65cm,line cap=round,line join=round]
\fill[black!2] (0,0) rectangle (1,1);
\draw[gridline] (-1,-1) -- (-1,1);
\draw[gridline] (0,-1) -- (0,1);
\draw[gridline] (1,-1) -- (1,1);
\draw[gridline] (-1,-1) -- (1,-1);
\draw[gridline] (-1,0) -- (1,0);
\draw[gridline] (-1,1) -- (1,1);
\fill[black!35] (-1,-1) circle (0.75pt);
\fill[black!35] (-1,0) circle (0.75pt);
\fill[black!35] (-1,1) circle (0.75pt);
\fill[black!35] (0,-1) circle (0.75pt);
\fill[black!35] (0,0) circle (0.75pt);
\fill[black!35] (0,1) circle (0.75pt);
\fill[black!35] (1,-1) circle (0.75pt);
\fill[black!35] (1,0) circle (0.75pt);
\fill[black!35] (1,1) circle (0.75pt);
\node[fill=white,inner sep=.35pt,text=blue,oplabel] at (-0.5,0) {${X^{2}}$};
\node[fill=white,inner sep=.35pt,text=blue,oplabel] at (0.5,0) {${X^{2}}{Z^{2}}$};
\node[fill=white,inner sep=.35pt,text=blue,oplabel] at (0.5,1) {${Z^{2}}$};
\node[fill=white,inner sep=.35pt,text=blue,oplabel] at (0,-0.5) {${X^{2}}$};
\node[fill=white,inner sep=.35pt,text=blue,oplabel] at (0,0.5) {${X^{2}}{Z^{2}}$};
\node[fill=white,inner sep=.35pt,text=blue,oplabel] at (1,0.5) {${Z^{2}}$};
\fill[black] (0,0) circle (1.6pt);
\path[use as bounding box] (-1.33,-1.25) rectangle (1.33,1.25);
\end{tikzpicture}
\par\vspace{-1mm}
 {\small $G_{\rvec}^2=T_{\rvec}^2W_{\rvec}^2$}
 \end{minipage}\hfill
 \begin{minipage}[c]{.53\linewidth}\centering
 \renewcommand{\arraystretch}{1.25}
 \begin{tabular}{cc|cc}
 $G_p^2$ & $G_q^2$ & $C_e^2$ & $D_e/C_e$\\\hline
 $+1$ & $+1$ & $+1$ & $+1$\\
 $-1$ & $+1$ & $-1$ & $-i$\\
 $+1$ & $-1$ & $-1$ & $-i$\\
 $-1$ & $-1$ & $+1$ & $-1$
 \end{tabular}
 \end{minipage}
 \caption{The local central control for the correction. Left: the complete Pauli string $G_{\rvec}^2$. Right: its contribution at the two plaquettes $p,q\supset e$. Here $D_e/C_e$ denotes the commuting phase factor $S(G_p^2)^\dagger S(G_q^2)^\dagger$, with $D_e^2=1$ in every row.}
 \label{fig:correctioncontrol}
\end{figure}

The proof that the constrained subspace is unchanged also has a useful cochain form:
\begin{equation}
     \prod_e D_e^{(\delta\mathbf v)(e)}=\Phi_v,\qquad
     \prod_e C_e^{(\delta\mathbf v)(e)}=\Phi_{v-\xhat-\yhat}.
\end{equation}
The translation in the second identity is fixed by the lower-left cup pairing. These are exactly Eqs.~\eqref{eq:Dstar} and~\eqref{eq:Cstar}, since signs do not matter for the order-two $D_e$.

\subsection{Cup-product formulas for the disentangling circuits}

We express the two disentangling circuits in cup-product notation. The spectral loop-controlled circuit is a cup-product phase. Define
\begin{equation}\label{eq:operatorqr}
 \mathsf g_0:=\sum_v\frac{1-\mathsf{G}_v}{2}\,\mathbf v,
 \qquad
 \mathsf w_2:=\sum_f\frac{1-\mathsf{W}_f}{2}\,\mathbf f,
 \qquad
 \mathsf{W}_f:=W_fS(W_f^2)^\dagger.
\end{equation}
All coefficients commute because all toric-code stars and plaquette loops commute. The same vertex--face pairing then gives
\begin{equation}\label{eq:cupU}
    \mathcal U=\exp\!\left[i\pi\int\mathsf g_0\cup\mathsf w_2\right] =\prod_v\CZ(\mathsf{G}_v,\mathsf{W}_v).
\end{equation}
This is the circuit of Eq.~\eqref{eq:U}. The conjugation identities needed to evaluate this circuit are
\begin{equation}
 A_e^2\mathsf{W}_vA_e^2=(-1)^{\chi_e(v)}\mathsf{W}_v,
 \qquad A_e^2\mathsf{G}_vA_e^2=\mathsf{G}_v.
\end{equation}
They give $\mathcal UA_e^2\mathcal U^\dagger=D_e$ and
$\mathcal UD_e\mathcal U^\dagger=A_e^2$.

After the on-site encoding $E$ in Eq.~\eqref{eq:encoding}, the CNOT circuit is
\begin{equation}\label{eq:cupVproduct}
    V=\prod_{e,e'}\CNOT_{a_{e'}\to b_e}^{\int(\mathbf e'\cup\mathbf e-\delta\mathbf e'\cup_1\mathbf e)}.
\end{equation}
The alternate disentangling circuit $\overline V$ exchanges the two edge labels in each gate, and hence
\begin{equation}\label{eq:cupVbar}
 \overline V=\prod_{e,e'}\CNOT_{a_{e'}\to b_e}^{\int
 (\mathbf e\cup\mathbf e'-\delta\mathbf e\cup_1\mathbf e')}.
\end{equation}
The circuit $V$ also has a single exponential form. On the two-qubit Hilbert space, introduce
\begin{equation}
 \mathsf z_1^a:=\sum_e\frac{1-Z_e^a}{2}\,\mathbf e,
 \qquad
 \mathsf x_1^b:=\sum_e\frac{1-X_e^b}{2}\,\mathbf e.
\end{equation}
Their coefficients commute, and
$\CNOT_{a\to b}=\exp[i\pi(1-Z_a)(1-X_b)/4]$. Consequently,
\begin{equation}\label{eq:cupVexponential}
    V=\exp \left[ i\pi\int \left(\mathsf z_1^a\cup\mathsf x_1^b - \delta\mathsf z_1^a\cup_1\mathsf x_1^b \right) \right].
\end{equation}
The operator-valued cochains in Eq.~\eqref{eq:cupVexponential} act on the encoded qubits, whereas those in Eq.~\eqref{eq:cupU} act on the original qudits. Combining them with $E$ in the stated order gives
\begin{equation}
 \mathcal W=V E\mathcal U,\qquad
 \mathcal W D_e\mathcal W^\dagger=Z_e^b.
\end{equation}
Thus, the cup-product formulation retains the explicit removal of exactly one ancilla qubit per edge.

\subsection{Reduced hopping operators and scope of the formulas}

We give the reduced hopping formulas and specify which ancillary sector enters the reduction. The order-four hopping operators have the cochain implementation
\begin{equation}
 a_e={}_b\langle0|\mathcal W A_e\mathcal W^\dagger|0\rangle_b,
 \qquad
 \overline a_e={}_b\langle0|\mathcal W\overline A_e\mathcal W^\dagger|0\rangle_b.
\end{equation}
These are unitary operators on the unrestricted remaining qubits because the unreduced hopping operators preserve every $D_e$ sector. A strictly local version replaces $\mathcal U$ in the conjugation of a single hopping operator by
\begin{equation}
 \mathcal U_e:=\prod_v u_v^{\chi_e(v)}
 =\exp\!\left[i\pi\int
 \mathsf g_0\cup(\mathsf w_2\cup_2\phi_e)\right],
\end{equation}
which contains exactly the two gates in Fig.~\ref{fig:dressing}. Letting $g_v=\mathcal R(G_v)$ and $\overline g_v=\mathcal R(\overline G_v)$, the fusion and cross-commutation relations become
\begin{equation}
 [a_e,\overline a_{e'}]=0,\qquad
 a_e^2=\prod_v g_v^{\int\mathbf v\cup\phi_e},\qquad
 \overline a_e^2=\prod_v\overline g_v^{\int\mathbf v\cup\phi_e}.
\end{equation}
The Hermitian hopping generators have the equally local cup-product prescription
\begin{equation}
    \begin{aligned}
        \mathsf a_e&=a_e S\!\left(\prod_v g_v^{\int\mathbf v\cup\phi_e}\right)^\dagger,\\
        \overline{\mathsf a}_e&=\overline a_e S\!\left(\prod_v\overline g_v^{\int\mathbf v\cup\phi_e}\right)^\dagger.
    \end{aligned}
\label{eq:cupInvolutive}
\end{equation}
Here, $\phi_e=\delta\mathbf e\cup_2\delta\mathbf e$. The spectral function is applied to the \emph{product} of the two endpoint loops, not separately to each loop. These are distinct operations and generally differ by a controlled sign. Equivalently, normalize $A_e$ by $S(A_e^2)^\dagger$ before applying the same reduction. The normalized hopping operators alone generate the same algebras by Lemma~\ref{prop:hopOnly}.

The formulas in this appendix are exact on the specified square lattice. Their use of the one-to-one pairing $\int\mathbf v\cup\mathbf f$ and the explicit signed cubical identities is substantive. Appendix~\ref{app:hdsem} performs the loop, constraint, disentangler, and completeness checks for the extension to spatial dimensions $4k-2$, using a bilinear flux cup-two pairing. Other cellulations still require their own local identities and completeness proof.

\subsection{Coordinate dictionary and integer identities}

We connect the cup-product formulas to the coordinate expressions and verify the integer identities used in the local factorization. The exponent matrices of the order-four condensates are
\begin{equation}\label{eq:Lexpanded}
 L=\begin{pmatrix}
 1-y^{-1}+x^{-1}+x^{-1}y^{-1}&x^{-2}-1\\
 y^{-2}-1&1-x^{-1}+y^{-1}+x^{-1}y^{-1}
 \end{pmatrix},
\end{equation}
\begin{equation}\label{eq:MLexpanded}
 ML=\begin{pmatrix}
 y-2+y^{-1}&-y+1+x^{-1}y+x^{-1}\\
 -x+xy^{-1}+1-3y^{-1}&x-2+x^{-1}
 \end{pmatrix}.
\end{equation}
Both columns give scalar phase $+1$ in canonical $X$-then-$Z$ order. The useful additional identities are
\begin{equation}\label{eq:additionalLaurent}
 K+cp=2J,\qquad pd=2x^{-1}y^{-1}-2,\qquad\det L=4x^{-1}y^{-1}.
\end{equation}
Substituting the local cubical rules in Eqs.~\eqref{eq:cup11rule}--\eqref{eq:cuptoprule} reproduces the condensate diagrams in Eq.~\eqref{eq:Cold} and the Laurent identities in Eq.~\eqref{eq:integeridentities}. These integer identities retain the signs before reduction modulo four. The scalar phases follow from the written $X$-then-$Z$ order and the on-site Pauli commutation relation $ZX=iXZ$.

\section{Higher-dimensional semion invertible subalgebras}
\label{app:hdsem}

The $\mathbb Z_4$ qudit condensation construction extends to spatial dimensions $d=4k{-}2$, with $\mathbb Z_4$ qudits on $(2k{-}1)$-cells. We prove the parent mutual-commutant relations, decorate the condensates into order-two constraints, and give a finite-depth circuit that removes one ancillary qubit per cell. The remaining qubits carry complementary invertible subalgebras. $k=2$ gives the six-dimensional construction.

Generalized semion non-Clifford QCAs in $4k-1$ spatial dimensions, including the $(7+1)$-dimensional spacetime case, were constructed in Ref.~\cite{SunNonClifford2026}. The ISAs below live one spatial dimension lower and define QCA pumps by layer transport, as in Sec.~\ref{sec:microscopicQCA}. They provide candidate local algebras for the higher-form semion boundary order of those QCAs. Their microscopic identification requires a separate comparison.

The disentangler uses the bilinear cup-two pairing of flux cochains. In two dimensions these are top-degree cochains, so their cup-two product is pointwise and the circuit reduces to one gate between each vertex loop and its paired plaquette flux. In higher dimensions, the flux cochains have degree $2k$, and cup two is required to make their pairing have degree $d=4k{-}2$. Its off-diagonal cell pairings must then be retained. The signed higher-cup identities below relate this disentangler to the original condensates on the full parent Hilbert space.

\subsection{Cochains and parent operators}
\label{app:hdsem:conventions}

Set $\ell=2k-1$, so $d=2\ell$. Work on the infinite oriented hypercubic lattice $\Sigma^d$, or on a periodic lattice larger than the supports of the local operators. Write $c_\ell,c_{\ell-1},c_{\ell+1}$ for cells of the indicated degrees, and $\mathbf c_\ell,\mathbf c_{\ell-1}, \mathbf c_{\ell+1}$ for their indicator cochains. All cochains used in integer identities have finite support, or live on a finite torus.

The construction builds on the higher-cup-product formalism and its lattice Hamiltonian applications developed in Refs.~\cite{ChenTata2023,chen2023exactly}. We use the corrected integer boundary and cup-product sign conventions of Ref.~\cite{Sun2026} throughout. For $\alpha\in C^m$ and $\beta\in C^n$, the signed recursive identity reads
\begin{align}
 \delta(\alpha\cup_i\beta)
 &=\delta\alpha\cup_i\beta+(-1)^m\alpha\cup_i\delta\beta
   +(-1)^{m+n-i}\alpha\cup_{i-1}\beta
   +(-1)^{mn+m+n}\beta\cup_{i-1}\alpha .
 \label{eq:hdsem:cupidentity}
\end{align}
Here, $\cup_0=\cup$ and $\cup_{-1}=0$. Integrals pair degree-$d$ cochains with the fundamental chain.

The ordinary cubical cup pairing between complementary degrees is a signed pairing of cells with bounded translations. For each $q$-cell $c_q$, let $p(c_q)$ be its displaced complementary $(d-q)$-cell, oriented so that its indicator cochain $\mathbf p(c_q)$ satisfies
\begin{equation}
 \int\mathbf c'_q\cup\mathbf p(c_q)
       =\begin{cases}1,&c'_q=c_q,\\0,&c'_q\ne c_q.\end{cases}
 \label{eq:hdsem:cupdual}
\end{equation}
In two dimensions, $p(c_0)$ is the plaquette $p(\rvec)$ with lower-left vertex $c_0=\rvec$, as in Sec.~\ref{sec:construction}. This is a geometric complementary-cell pairing with bounded displacement. In particular, $\mathbf p(c_{\ell-1})$ has degree $\ell+1$.

Place a $\mathbb Z_4$ qudit on each $c_\ell$, with $ZX=iXZ$ and $X^4=Z^4=1$. For an $\ell$-cochain $\alpha$, write $X_\alpha=\prod_{c_\ell}X_{c_\ell}^{\alpha(c_\ell)}$, and always place shifts before clocks. Generalizing Eq.~\eqref{eq:parentCupHops}, define the two hopping families by
\begin{align}
 A_{c_\ell}
 &=\epsilon X_{c_\ell}
   \prod_{c'_\ell}Z_{c'_\ell}^{\int(
     \mathbf c'_\ell\cup\mathbf c_\ell
       -\delta\mathbf c'_\ell\cup_1\mathbf c_\ell)},\nonumber\\
 \overline A_{c_\ell}
 &=\epsilon X_{c_\ell}
   \prod_{c'_\ell}Z_{c'_\ell}^{\int(
     \mathbf c_\ell\cup\mathbf c'_\ell
       -\delta\mathbf c_\ell\cup_1\mathbf c'_\ell)},\nonumber\\
 \epsilon
 &=\begin{cases}
     \ww,&d=2,\\
     1,&d=6,10,14,\ldots .
   \end{cases}
 \label{eq:hdsem:hops}
\end{align}
The prefactor compensates the ordering phase between the shift $X_{c_\ell}$ and its own clock factor. That clock power is $\int(\mathbf c_\ell\cup\mathbf c_\ell-\delta\mathbf c_\ell\cup_1\mathbf c_\ell)$. The ordinary cup term vanishes: it requires disjoint free coordinate directions, whereas both inputs have the same directions. The cup-one term pairs $c_\ell$ with its cofaces in $\delta\mathbf c_\ell$. A cubical cup-one term requires exactly one shared free direction, but these cells share all $\ell$ directions. For $\ell=1$, an edge and its adjacent face share one direction, and the conventions of Ref.~\cite{Sun2026} give $\int\delta\mathbf c_1\cup_1\mathbf c_1=1$. For $\ell=3$, a three-cell and its four-cell coface share three directions, so the same term vanishes. It likewise vanishes for all $\ell\ge3$.

Thus, the on-site clock power is $-1$ in two dimensions and $0$ in six dimensions and above, independently of the cell's orientation. Using $ZX=iXZ$, the square of the bare hopping word has scalar $-i$ in the first case and $1$ in the second. The factor $\ww^2=i$ cancels the former. These choices give the analogue of Eq.~\eqref{eq:parentHopSquares}:
\begin{equation}
 A_{c_\ell}^2=X_{c_\ell}^2
   \prod_{c'_\ell}Z_{c'_\ell}^{2\int(
     \mathbf c'_\ell\cup\mathbf c_\ell
       -\delta\mathbf c'_\ell\cup_1\mathbf c_\ell)},
 \qquad A_{c_\ell}^4=\overline A_{c_\ell}^{\,4}=1.
 \label{eq:hdsem:squares}
\end{equation}
The barred square has the reversed hopping pairing. The two clock decorations are transposes, so $[A_{c_\ell},\overline A_{c'_\ell}]=0$.

For an $(\ell-1)$-cochain $\eta$, generalize the dyon loops of Eq.~\eqref{eq:Gcup} by
\begin{equation}
 G(\eta)=X_{\delta\eta}
    \prod_{c_\ell}Z_{c_\ell}^{\int\eta\cup\delta\mathbf c_\ell},
 \qquad
 \overline G(\eta)=X_{\delta\eta}
    \prod_{c_\ell}Z_{c_\ell}^{-\int\eta\cup\delta\mathbf c_\ell}.
 \label{eq:hdsem:loops}
\end{equation}
Write $G_{c_{\ell-1}}=G(\mathbf c_{\ell-1})$, and similarly for $\overline G$. The ordinary flux generalizes Eq.~\eqref{eq:toricStarPlaquette}, and the common loop square obeys the same relation as in Eq.~\eqref{eq:Bprops}
\begin{equation}
 W_{c_{\ell+1}}
   =\prod_{c_\ell}Z_{c_\ell}^{(\delta\mathbf c_\ell)(c_{\ell+1})},
 \qquad
 G_{c_{\ell-1}}^2=\overline G_{c_{\ell-1}}^{\,2}.
 \label{eq:hdsem:flux}
\end{equation}
Stokes' identity and $\delta^2=0$ give the exact, phase-sensitive relations
\begin{equation}
 G(\eta+\eta')=G(\eta)G(\eta'),\qquad
 G(\delta\xi)=1,\qquad [G(\eta),W_{c_{\ell+1}}]=0.
 \label{eq:hdsem:additive}
\end{equation}
The barred loops obey the same relations. Integrated cup one also gives
\begin{align}
 \int(\beta\cup\delta\eta-\delta\beta\cup_1\delta\eta)
       &=\int\eta\cup\delta\beta,\nonumber\\
 \int(\delta\eta\cup\beta-\delta^2\eta\cup_1\beta)
       &=-\int\eta\cup\delta\beta .
 \label{eq:hdsem:loopidentities}
\end{align}
Thus, each loop belongs to its corresponding hopping algebra. The loops commute internally, their squares are central Hermitian involutions in the joint hopping algebra, and every hopping square
commutes with all loops.

\subsection{Condensates and parent mutual commutants}
\label{app:hdsem:parent}

Generalizing Eq.~\eqref{eq:Cold}, define the original condensates
by the mixed cup-one expression
\begin{align}
 C_{c_\ell}
 &=A_{c_\ell}^2\prod_{c_{\ell-1}}
   G_{c_{\ell-1}}^{\int(
     \mathbf p(c_{\ell-1})\cup_1\mathbf c_\ell
        -\mathbf c_\ell\cup_1\mathbf p(c_{\ell-1}))}
 \nonumber\\
 &=\overline A_{c_\ell}^{\,2}\prod_{c_{\ell-1}}
   \overline G_{c_{\ell-1}}^{\int(
     \mathbf p(c_{\ell-1})\cup_1\mathbf c_\ell
        -\mathbf c_\ell\cup_1\mathbf p(c_{\ell-1}))}.
 \label{eq:hdsem:C}
\end{align}
The equality includes the scalar phase in the written $X$-then-$Z$ order. We compute the commutators directly, keeping the written integer exponents before reducing modulo four. For an integer $\ell$-cochain $\beta$, the squared hopping operator in $C_{c_\ell}$ contributes
\[
 2\int(\beta\cup\mathbf c_\ell
       -\delta\beta\cup_1\mathbf c_\ell
       -\mathbf c_\ell\cup\beta
       +\delta\mathbf c_\ell\cup_1\beta)
\]
to the unreduced integer commutator exponent with the hopping product $\prod_{c'_\ell}A_{c'_\ell}^{\beta(c'_\ell)}$. The loop dressing contributes $2\int(\delta\beta\cup_1\mathbf c_\ell-\mathbf c_\ell\cup_1\delta\beta)$, by ordinary cup duality and Eq.~\eqref{eq:hdsem:loopidentities}. Their sum is therefore
\begin{equation}
 2\int(\beta\cup\mathbf c_\ell-\mathbf c_\ell\cup\beta
       +\delta\mathbf c_\ell\cup_1\beta
       -\mathbf c_\ell\cup_1\delta\beta)
       =4\int\beta\cup\mathbf c_\ell .
 \label{eq:hdsem:commutatorexponent}
\end{equation}
The equality follows from integrated cup one, for odd-degree
$\alpha,\beta$
\begin{equation}
 \int(\delta\alpha\cup_1\beta-\alpha\cup_1\delta\beta)
       =\int(\alpha\cup\beta+\beta\cup\alpha).
 \label{eq:hdsem:cupone}
\end{equation}
In particular, taking $\beta=\mathbf c'_\ell$ gives
\[
 C_{c_\ell}A_{c'_\ell}
   =i^{4\int\mathbf c'_\ell\cup\mathbf c_\ell}
                       A_{c'_\ell}C_{c_\ell}
   =A_{c'_\ell}C_{c_\ell}.
\]
The barred expression in Eq.~\eqref{eq:hdsem:C} gives the barred commutator in the same way. Thus, every $C_{c_\ell}$ commutes with both hopping families.

Completeness rests on two facts: the integer hopping pairing can distinguish all shift patterns, and condensate products realize every fourfold integer shift locally. We keep the written Pauli exponents as integers, before reducing modulo four. Eq.~\eqref{eq:hdsem:loopidentities} ensures that a condensate product has the unbarred hopping clock decoration for its accumulated
shifts, up to scalar.

The antisymmetric pairing is the unbarred hopping clock decoration minus the barred one in Eq.~\eqref{eq:hdsem:hops}. If a cochain $u$ pairs to zero with every test, it also pairs to zero with the integer shifts of each condensate. Reversing the arguments in Eq.~\eqref{eq:hdsem:commutatorexponent} gives $-4\int u\cup\mathbf c_\ell=0$ for every cell. Ordinary cup duality forces $u=0$. Thus, equal responses to all hopping tests imply equal integer shift patterns.

To realize $4\beta$, pair each hopping test against $\beta$. Ordinary cup duality expresses this as an ordinary cup pairing, with the test on the left and the dual cochain on the right. Take the components of this dual cochain as the powers of the $C$ generators. By Eq.~\eqref{eq:hdsem:commutatorexponent}, the resulting product has four times the original pairing with every test, exactly as $4\beta$ does. Nondegeneracy then identifies its integer shifts with $4\beta$.

It follows that an integer linear functional vanishing on every condensate shift vanishes on every $4\beta$, and hence on every $\beta$. This is the completeness fact used below. The construction uses only local cups, coboundaries, and bounded complementary-cell displacements, so it enlarges support by a uniformly bounded distance and also applies on a finite torus.

\begin{theorem}[Parent mutual commutants]
\label{thm:hdsem:parent}
In the full parent local operator algebra,
\begin{equation}
 \mathfrak H:=\operatorname{Alg}\{A_{c_\ell},\overline A_{c_\ell}\},
 \qquad \mathfrak C:=\operatorname{Alg}\{C_{c_\ell}\}
 \label{eq:hdsem:algebras}
\end{equation}
satisfy
\begin{equation}
 \mathfrak C^\perp=\mathfrak H,\qquad
 \mathfrak H^\perp=\mathfrak C,\qquad
 \operatorname{Center}(\mathfrak H)=\mathfrak C .
 \label{eq:hdsem:parentcommutants}
\end{equation}
Every reconstruction enlarges support by a uniformly bounded distance. The statements also hold on finite periodic lattices.
\end{theorem}

\begin{proof}
We first remove a Pauli's shifts and reconstruct its remaining clock factors. We then identify the Paulis commuting with both
hopping families.

Let a local Pauli $P\in\mathfrak C^\perp$ have integer shift cochain $u$. Multiply it by the inverse of an unbarred hopping product with the same shifts. The remaining operator is a pure clock Pauli and still commutes with every condensate. Its clock powers evaluated on the written integer shifts of each $C_{c_\ell}$ are therefore divisible by four. Divide these integers by four, and use ordinary cup duality to choose $b$ with $\int b\cup\mathbf c_\ell$ equal to the corresponding quotient.

An unbarred hopping product with shifts $-b$ followed by a barred product with shifts $b$ is also pure clock. Its clock decoration is the negative antisymmetric hopping pairing with $b$. Eq.~\eqref{eq:hdsem:commutatorexponent} therefore gives the same integer clock pairings with every condensate as the remaining clock Pauli. Their difference vanishes on all condensate shifts, and hence identically by the fourfold spanning proved above. Thus, $P$ factors, up to scalar, into an unbarred hopping product with shifts $u-b$ and a barred product with shifts $b$. Since the hopping families already commute with $\mathfrak C$, this proves $\mathfrak C^\perp=\mathfrak H$.

Conversely, if $P\in\mathfrak H^\perp$, its clock powers agree modulo four with both hopping decorations for its shift $u$. Their difference is divisible by four, so ordinary cup duality determines a local integer cochain $a$ by
\[
 \int\beta\cup a
 =\frac14\int(\beta\cup u-\delta\beta\cup_1u
                              -u\cup\beta+\delta u\cup_1\beta)
 \quad\hbox{for every }\beta .
\]
By Eq.~\eqref{eq:hdsem:commutatorexponent}, $\prod_{c_\ell}C_{c_\ell}^{a(c_\ell)}$ has the same integer hopping pairing as $u$. Nondegeneracy identifies their shifts. Its clock powers have the unbarred hopping decoration, which agrees with those of $P$ modulo four. Thus $P$ is this condensate product up to scalar, proving $\mathfrak H^\perp=\mathfrak C$.

Pauli conjugation is diagonal in the Pauli basis, so both conclusions extend to finite linear combinations. All reconstruction steps use local commutators, division by four, and ordinary cup duality, and therefore have a uniform support bound on the infinite lattice and finite tori. Finally, $\mathfrak C\subseteq\mathfrak H$ gives the
center statement.
\end{proof}

\subsection{Equivalent order-two constraints}
\label{app:hdsem:D}

Take the spectral function $S$ of Eq.~\eqref{eq:S}. For any order-four unitary $O$, write $\mathsf{O}=OS(O^2)^{-1}$, as in Eq.~\eqref{eq:hatQ}. It is a Hermitian involution. In particular, normalize the loops $G_{c_{\ell-1}},\overline G_{c_{\ell-1}}$ and fluxes $W_{c_{\ell+1}}$.

\begin{lemma}[Local spectral correction]
\label{lem:hdsem:decoration}
Generalizing Eqs.~\eqref{eq:D} and~\eqref{eq:Dbar}, the
cup-product decoration
\begin{align}
 D_{c_\ell}
 &:=A_{c_\ell}^2\prod_{c_{\ell-1}}
     \mathsf{G}_{c_{\ell-1}}^{\int\mathbf p(c_{\ell-1})
                                      \cup_2\delta\mathbf c_\ell}
 \nonumber\\
 &=\overline A_{c_\ell}^{\,2}\prod_{c_{\ell-1}}
     \overline{\mathsf{G}}_{c_{\ell-1}}^{\int\mathbf p(c_{\ell-1})
                                      \cup_2\delta\mathbf c_\ell}
 \nonumber\\
 &=C_{c_\ell}\prod_{c_{\ell-1}}
   \bigl[S(G_{c_{\ell-1}}^2)^\dagger\bigr]^{
      \int(\mathbf p(c_{\ell-1})\cup_2\delta\mathbf c_\ell
                    -2\mathbf c_\ell\cup_1\mathbf p(c_{\ell-1}))}
 \label{eq:hdsem:D}
\end{align}
is a central Hermitian involution. Moreover,
\begin{equation}
 \operatorname{Alg}\{D_{c_\ell}\}=\mathfrak C
 \label{eq:hdsem:equalconstraints}
\end{equation}
with a uniformly local change of generators, and the simultaneous
$+1$ sectors of the two constraint families coincide.
\end{lemma}

\begin{proof}
Integrated cup two, for degrees $\ell+1$ and $\ell$, gives the exact integer identity
\begin{equation}
 \int(\mathbf p(c_{\ell-1})\cup_1\mathbf c_\ell
                   -\mathbf c_\ell\cup_1\mathbf p(c_{\ell-1}))
 =\int(\mathbf p(c_{\ell-1})\cup_2\delta\mathbf c_\ell
       -2\mathbf c_\ell\cup_1\mathbf p(c_{\ell-1})
       +\delta\mathbf p(c_{\ell-1})\cup_2\mathbf c_\ell).
 \label{eq:hdsem:integercuptwo}
\end{equation}
For $\ell\ge3$, Stokes' identity and complementary cup duality yield
\[
 \delta\mathbf p(c_{\ell-1})
 =\sum_{c_{\ell-2}}(\delta\mathbf c_{\ell-2})(c_{\ell-1})
                                  \mathbf p(c_{\ell-2}) .
\]
Consequently, the last coefficients in
Eq.~\eqref{eq:hdsem:integercuptwo} form the exact $(\ell-1)$-cochain
\[
 \delta\!\left[\sum_{c_{\ell-2}}\mathbf c_{\ell-2}
                   \int\mathbf p(c_{\ell-2})\cup_2\mathbf c_\ell\right].
\]
Its loop product is exactly one by Eq.~\eqref{eq:hdsem:additive}. For $\ell=1$ this term is absent. Substituting the other two terms into Eq.~\eqref{eq:hdsem:C} and using $S(G^2)^2=G^2$ proves Eq.~\eqref{eq:hdsem:D}, including all phases. The barred derivation is identical. The correction is a local function of central loop squares. The final expressions are products of commuting involutions. The exponents of $S(G^2)$ must be evaluated over the integers: their even parts contribute factors of $G^2$.

For $\ell=1$, $\mathbf p(c_0)=\mathbf f$ for the face $f=p(c_0)$ paired with the vertex $c_0$, and $c_1=e$. The normalized-loop exponent is the signed incidence $(\delta\mathbf e)(f)$; its sign is irrelevant because $\mathsf{G}_f^{-1}=\mathsf{G}_f$. Meanwhile, $\int(\mathbf f\cup_2\delta\mathbf e-2\mathbf e\cup_1\mathbf f)$ equals $1$ on the two incident faces and $0$ elsewhere. With $G_{c_0}=G_f$, Eq.~\eqref{eq:hdsem:D} therefore reduces exactly to Eq.~\eqref{eq:D}.

Both constraint families locally recover the loop squares:
\begin{equation}
 \prod_{c_\ell}D_{c_\ell}^{(\delta\mathbf c_{\ell-1})(c_\ell)}
       =G_{c_{\ell-1}}^2,\qquad
 \prod_{c_\ell}C_{c_\ell}^{(\delta\mathbf c_{\ell-1})(c_\ell)}
       =G_{c_{\ell-1}^{\leftarrow}}^2 .
 \label{eq:hdsem:stars}
\end{equation}
The cell $c_{\ell-1}^{\leftarrow}$ is translated one step backwards in every coordinate, as selected by the reversed ordinary cup pairing. For the first identity, $\delta^2=0$ cancels all normalized loops, while $\prod A_{c_\ell}^{2(\delta\mathbf c_{\ell-1})(c_\ell)}
=G_{c_{\ell-1}}^2$ exactly.

For the second identity, multiply the $C$ constraints with the same signed incidence weights around $c_{\ell-1}$. The squared hopping operators again produce $G_{c_{\ell-1}}^2$. The cochain recording the signed powers of the elementary loops in their remaining dressing satisfies
\[
 \sum_{c'_{\ell-1}}\mathbf c'_{\ell-1}
   \int(\mathbf p(c'_{\ell-1})\cup_1\delta\mathbf c_{\ell-1}
               -\delta\mathbf c_{\ell-1}\cup_1
                                      \mathbf p(c'_{\ell-1}))
 \equiv 2\mathbf c_{\ell-1}^{\leftarrow}-2\mathbf c_{\ell-1}
                  \pmod{\operatorname{im}\delta} .
\]
This follows from two applications of integrated cup one. The congruence is over integer cochains. The difference is a local coboundary, whose loop operator is exactly one by Eq.~\eqref{eq:hdsem:additive}. Thus, the accumulated dressing consists of two copies of the translated loop and two inverse copies of the original loop
\[
 \prod_{c_\ell}
   \bigl(C_{c_\ell}A_{c_\ell}^{-2}\bigr)^{
                      (\delta\mathbf c_{\ell-1})(c_\ell)}
   =G_{c_{\ell-1}^{\leftarrow}}^2G_{c_{\ell-1}}^{-2}.
\]
Its original-loop factor cancels the $G_{c_{\ell-1}}^2$ supplied by the squares of the hopping operators, leaving precisely the translated square in Eq.~\eqref{eq:hdsem:stars}. All these equalities include the scalar phases, since these squares commute with every loop and the loop representation is exactly additive.

The backward translation has a local inverse. Thus each loop square is a local polynomial in either constraint family. Eq.~\eqref{eq:hdsem:D} and its inverse prove Eq.~\eqref{eq:hdsem:equalconstraints}. In either simultaneous $+1$ sector the star identities force every $G_{c_{\ell-1}}^2=1$, at which point the decoration is the identity and $C_{c_\ell}=D_{c_\ell}$. Individual dyon loops remain unrestricted.
\end{proof}

\subsection{A cup-product disentangler and ancillary-qubit removal}
\label{app:hdsem:disentangler}

Generalize the loop--flux circuit of Eqs.~\eqref{eq:ugate} and~\eqref{eq:U}, using the original $(\ell-1)$-cell loops and the controlled phase of Eq.~\eqref{eq:CZ}:
\begin{equation}
 \boxed{\displaystyle
 \mathcal U=\prod_{c_{\ell-1},c'_{\ell+1}}
   \CZ(\mathsf{G}_{c_{\ell-1}},\mathsf{W}_{c'_{\ell+1}})^{
          \int\mathbf p(c_{\ell-1})\cup_2\mathbf c'_{\ell+1}} .}
 \label{eq:hdsem:U}
\end{equation}
Here $\mathbf p(c_{\ell-1})$ is the complementary cochain paired with the loop label $c_{\ell-1}$, and $c'_{\ell+1}$ independently labels the flux. Cup two specifies which loop--flux pairs enter the circuit. In two dimensions it is pointwise, reducing this expression to $\prod_f\CZ(\mathsf{G}_f,\mathsf{W}_f)$ as in Eq.~\eqref{eq:U}. In higher dimensions the off-diagonal pairings remain. All loops and fluxes commute, so the gates commute. Cup products and complementary-cell displacements have bounded range; coloring the overlap graph gives a finite-depth schedule for fixed $d$. Each gate is an ordinary local unitary on its bounded support. Also $\mathcal U^\dagger=\mathcal U$ and $\mathcal U^2=1$.

The hopping square fixes all normalized dyon loops and flips the normalized flux by
\begin{equation}
 A_{c_\ell}^2\mathsf{W}_{c_{\ell+1}}A_{c_\ell}^2
   =(-1)^{(\delta\mathbf c_\ell)(c_{\ell+1})}
                                      \mathsf{W}_{c_{\ell+1}} .
 \label{eq:hdsem:fluxflip}
\end{equation}
Spectral normalization respects this flip because $\mathsf{(-W)}=-\mathsf{W}$. Conjugation by each controlled phase therefore attaches its loop whenever its flux is flipped. Summing the cup-two exponents gives
\begin{align}
 \mathcal U A_{c_\ell}^2\mathcal U^\dagger
   &=A_{c_\ell}^2\prod_{c_{\ell-1}}
      \mathsf{G}_{c_{\ell-1}}^{
           \int\mathbf p(c_{\ell-1})\cup_2\delta\mathbf c_\ell}
     =D_{c_\ell},\nonumber\\
 \mathcal U D_{c_\ell}\mathcal U^\dagger&=A_{c_\ell}^2 .
 \label{eq:hdsem:Uaction}
\end{align}

Encode each $\mathbb Z_4$ qudit into two qubits using the same on-site map as in Eq.~\eqref{eq:encoding}:
\begin{equation}
 E:\ |a+2b\rangle\longmapsto |a\rangle_a\otimes H|b\rangle_b,
 \qquad a,b\in\{0,1\},
 \label{eq:hdsem:encoding}
\end{equation}
where $H$ is the one-qubit Hadamard. Then
\begin{align}
 EX_{c_\ell}^2E^\dagger&=Z_{c_\ell}^b,\qquad
 EZ_{c_\ell}^2E^\dagger=Z_{c_\ell}^a,\nonumber\\
 EA_{c_\ell}^2E^\dagger
 &=Z_{c_\ell}^b\prod_{c'_\ell}(Z_{c'_\ell}^a)^{
      \int(\mathbf c'_\ell\cup\mathbf c_\ell
                  -\delta\mathbf c'_\ell\cup_1\mathbf c_\ell)} .
 \label{eq:hdsem:encoded}
\end{align}
The commuting CNOT circuit generalizing Eq.~\eqref{eq:VM}
\begin{equation}
 V=\prod_{c_\ell,c'_\ell}
   \operatorname{CNOT}_{a_{c'_\ell}\to b_{c_\ell}}^{
      \int(\mathbf c'_\ell\cup\mathbf c_\ell
                  -\delta\mathbf c'_\ell\cup_1\mathbf c_\ell)}
 \label{eq:hdsem:V}
\end{equation}
removes the clock-square decorations. Its controls are all $a$ qubits and its targets all $b$ qubits, so it again admits a bounded-depth schedule. As in Eq.~\eqref{eq:W}, the complete disentangler is
\begin{equation}
 \mathcal W:=VE\mathcal U,\qquad
 \boxed{\mathcal W D_{c_\ell}\mathcal W^\dagger=Z_{c_\ell}^b .}
 \label{eq:hdsem:disentangler}
\end{equation}
Thus, the constraints become independent ancillary Pauli operators. On a finite torus with $N$ $\mathbb Z_4$ qudits, the common $+1$ sector has dimension $4^N/2^N=2^N$. Removing the fixed $b$ qubits leaves one unrestricted physical qubit per $\ell$-cell.

\subsection{Local factorization on the physical qubits}
\label{app:hdsem:physical}

For an operator commuting with the constraints, use the reduction of Eq.~\eqref{eq:reduction}:
\begin{equation}
 \mathcal R(O):={}_b\langle0|\mathcal W O\mathcal W^\dagger|0\rangle_b .
 \label{eq:hdsem:reduction}
\end{equation}
This is a unital $*$-homomorphism on $\mathfrak C^\perp$, because the conjugated inputs preserve every joint $Z^b$ sector. It maps onto the full local physical operator algebra. A local physical $O_a$ has the local preimage $\mathcal W^\dagger(O_a\otimes1_b)\mathcal W$. Define the reduced hopping operators and their algebras as in Eqs.~\eqref{eq:reducedhops} and~\eqref{eq:algebras}:
\begin{equation}
 a_{c_\ell}=\mathcal R(A_{c_\ell}),\quad
 \overline a_{c_\ell}=\mathcal R(\overline A_{c_\ell}),\quad
 \mathfrak A_+=\operatorname{Alg}\{a_{c_\ell}\},\quad
 \mathfrak A_-=\operatorname{Alg}\{\overline a_{c_\ell}\}.
 \label{eq:hdsem:physicalalgebras}
\end{equation}

\begin{theorem}[Higher-dimensional ISA]
\label{thm:hdsem:ISA}
The two physical algebras are complementary mutual commutants:
\begin{equation}
 [\mathfrak A_+,\mathfrak A_-]=0,\qquad
 \mathfrak A_+\vee\mathfrak A_-=\mathcal M_a,\qquad
 \mathfrak A_+^\perp=\mathfrak A_-,\quad
 \mathfrak A_-^\perp=\mathfrak A_+ .
 \label{eq:hdsem:ISA}
\end{equation}
Every local $O_a$ has a finite decomposition
\begin{equation}
 O_a=\sum_\lambda u_\lambda v_\lambda,\qquad
 u_\lambda\in\mathfrak A_+,\quad v_\lambda\in\mathfrak A_-,\qquad
 \operatorname{supp}(u_\lambda),\operatorname{supp}(v_\lambda)
     \subseteq N_R(\operatorname{supp}O_a),
 \label{eq:hdsem:localISA}
\end{equation}
with $R$ depending only on $d$. Multiplication identifies $\mathfrak A_+\otimes\mathfrak A_-$ with $\mathcal M_a$. Thus, $\mathfrak A_+$ is an ISA~\cite{Haah2023}.
\end{theorem}

\begin{proof}
The parent algebras commute, so their reductions commute. Lift $O_a$ by the finite-depth disentangler and apply the uniformly local Pauli reconstruction of Theorem~\ref{thm:hdsem:parent}. Reduction gives Eq.~\eqref{eq:hdsem:localISA}, retaining both eigenvalues of
each surviving physical loop.

On a finite torus, the commuting reduced algebras jointly generate the full physical matrix algebra. A central element of either commutes with both, hence is scalar. Finite-dimensional matrix-factor decomposition makes them mutual commutants and gives injectivity of tensor-product multiplication.

On the infinite lattice, take a local operator commuting with $\mathfrak A_+$ and its uniformly local separated decomposition. Embed its neighborhood in a torus with a sufficient buffer. Overlapping generators agree with those on the infinite lattice. All other generators commute by disjoint support. The tracial conditional expectation onto the torus minus factor sends $u_\lambda v_\lambda$ to $\operatorname{tr}_+(u_\lambda)v_\lambda$ and fixes the original operator. This retains the same support bound and puts it in $\mathfrak A_-$. Exchanging the factors proves the other commutant identity. Finite linear relations and tensor-product injectivity likewise transfer from a sufficiently large torus.
\end{proof}

\section{Extension to a \texorpdfstring{$U(1)_4$}{U(1)4} invertible subalgebra}
\label{app:u14}
\addtocontents{toc}{\protect\enlargethispage{12pt}}

The semion construction condenses $e^2m^2$ in the $\mathbb Z_4$ toric code and removes one of the two qubits in each parent edge register. Here we instead condense $e^4m^4$ in the $\mathbb Z_8$ toric code. Removing one ancillary qubit from each eight-state register leaves one unrestricted physical $\mathbb Z_4$ qudit per edge. The dyons $em$ and $em^{-1}$ become order-four charges with spins $e^{\pm\pi i/4}$, giving $U(1)_4$ and its orientation reverse. We first give the square-lattice construction and its local factorization, then extend the same operator-algebra construction to the dimensions considered in Appendix~\ref{app:hdsem}.

We use the notation of Secs.~\ref{sec:construction}--\ref{sec:completeness}: $A_e,\overline A_e$ are parent hopping operators, $G_v,\overline G_v$ are dyon loops, and $C_e,D_e$ are the original and involutive condensates. Their physical reductions are $a_e,\overline a_e$ and $g_v,\overline g_v$. Within this appendix, the parent Paulis are $X_8,Z_8$, whereas $X_4,Z_4$ act on the surviving qudits. Vertices may also be written as $v=\rvec$, and $p(v)$ is the plaquette with lower-left vertex $v$. Bold cell symbols denote the integer indicator cochains of Appendix~\ref{sec:cochains}. Products act rightmost first; Pauli words are written with shifts before clocks. We prove
\begin{equation}
 \mathfrak A_{4,+}^{\perp}=\mathfrak A_{4,-},\qquad
 \mathfrak A_{4,-}^{\perp}=\mathfrak A_{4,+},\qquad
 \mathfrak A_{4,+}\otimes\mathfrak A_{4,-}\cong\mathcal M_4.
 \label{app:u14:isa}
\end{equation}
Here, $\mathcal M_4$ is the full local operator algebra of the physical $\mathbb Z_4$ qudits, and the tensor-product map is multiplication, with a uniform bound on support enlargement.

\subsection{Eight-state parent and independent constraints}
\label{app:u14:parent}

Set $\ww=e^{\pi i/4}$ and $\nu_4=e^{3\pi i/8}$, and take $X_8|j\rangle=|j+1\bmod8\rangle$ and $Z_8|j\rangle=\ww^j|j\rangle$. Thus, $Z_8X_8=\ww X_8Z_8$ and $X_8^8=Z_8^8=1$. For an integer edge cochain $u$, write $X_{8,u}=\prod_eX_{8,e}^{u(e)}$, and similarly for physical shifts. The hopping decorations are exactly those of Eq.~\eqref{eq:parentCupHops}:
\begin{equation}
 m(e',e):=\int\bigl(\mathbf e'\cup\mathbf e
                  -\delta\mathbf e'\cup_1\mathbf e\bigr),
 \qquad m(e',e)=M_{e'e}.
 \label{app:u14:cup-kernel}
\end{equation}
Define
\begin{align}
 A_e&=\nu_4X_{8,e}\prod_{e'}Z_{8,e'}^{m(e',e)},\nonumber\\
 \overline A_e&=\nu_4X_{8,e}\prod_{e'}Z_{8,e'}^{m(e,e')}.
 \label{app:u14:parent-hops}
\end{align}
The exchanged cup pairing gives $[A_e,\overline A_{e'}]=0$. Since $m(e,e)=-1$ and $\nu_4^4\ww^{-6}=1$, their fourth powers have no ordering scalar:
\begin{equation}
 A_e^4=X_{8,e}^4\prod_{e'}Z_{8,e'}^{4m(e',e)},\qquad
 \overline A_e^{\,4}=X_{8,e}^4\prod_{e'}Z_{8,e'}^{4m(e,e')}.
 \label{app:u14:hopFourthPowers}
\end{equation}
They are Hermitian involutions, and
$A_e^8=\overline A_e^{\,8}=1$.

The analogue of Eqs.~\eqref{eq:toricStarPlaquette} and~\eqref{eq:Gcup} is
\begin{gather}
 T_v=X_{8,\delta\mathbf v},\qquad
 W_f=\prod_eZ_{8,e}^{(\delta\mathbf e)(f)},\qquad
 W_v:=W_{p(v)}
     =\prod_eZ_{8,e}^{\int\mathbf v\cup\delta\mathbf e},\nonumber\\
 G_v=T_vW_v,\qquad \overline G_v=T_vW_v^{-1},\qquad
 G_v^4=\overline G_v^{\,4}.
 \label{app:u14:loops-parent}
\end{gather}
All star shifts and fluxes commute. The fourth powers $G_v^4$ are central Hermitian involutions in the joint hopping algebra. The raw loops themselves have eighth-root spectra.

Use the incident-face selector $\chi_e(v)$ of Eq.~\eqref{eq:cupselector}. Equivalently, the signed cup rules give
\[
 \chi_e(v)=\int\bigl(\mathbf p(v)\cup_1\mathbf e
                         -\mathbf e\cup_1\mathbf p(v)\bigr).
\]
The original condensate, analogous to Eq.~\eqref{eq:Cold}, is
\begin{equation}
 C_e=A_e^4\prod_vG_v^{2\chi_e(v)}
    =\overline A_e^{\,4}\prod_v\overline G_v^{\,2\chi_e(v)}.
 \label{app:u14:C}
\end{equation}
Both expressions include their scalar phases. They put $C_e$ in both commuting hopping algebras, so it is central. Its power and signed-star relations are
\begin{gather}
 C_e^2=\prod_v(G_v^4)^{\chi_e(v)},\qquad C_e^4=1,\nonumber\\
 \prod_e(A_e^4)^{(\delta\mathbf v)(e)}=G_v^4,\qquad
 \prod_eC_e^{(\delta\mathbf v)(e)}=G_{v^\leftarrow}^4.
 \label{app:u14:Cstars}
\end{gather}
Here, $v^\leftarrow=v-\xhat-\yhat$.

Normalizing the order-four \emph{squared} loops using the same function $S$ as in Eqs.~\eqref{eq:S} and~\eqref{eq:hatQ}, we find
\[
 \mathsf{G_v^2}=G_v^2S(G_v^4)^\dagger,\qquad
 \overline{\mathsf{G_v^{\,2}}}
       =\overline G_v^{\,2}S(G_v^4)^\dagger .
\]
These are commuting Hermitian involutions. The counterpart of Eq.~\eqref{eq:D} has the cup-product form
\begin{align}
 D_e
 &=A_e^4\prod_v\mathsf{G_v^2}^{\,
                 \int\mathbf p(v)\cup_2\delta\mathbf e}
  =\overline A_e^{\,4}\prod_v
       \overline{\mathsf{G_v^{\,2}}}^{\,
                 \int\mathbf p(v)\cup_2\delta\mathbf e},
 \nonumber\\
 &=C_e\prod_v[S(G_v^4)^\dagger]^{
          \int(\mathbf p(v)\cup_2\delta\mathbf e
                       -2\mathbf e\cup_1\mathbf p(v))}.
 \label{app:u14:D}
\end{align}
The normalized-loop exponent is signed incidence. Its sign is irrelevant for an involution. The last exponent is precisely $\chi_e(v)$ over the integers, so on the square lattice  $D_e=C_e\prod_vS(G_v^4)^{-\chi_e(v)}$. Thus, $D_e$ is a central Hermitian involution. Moreover,
\begin{equation}
 \prod_eD_e^{(\delta\mathbf v)(e)}=G_v^4,\qquad
 \mathfrak C_4:=\Alg\{C_e\}=\Alg\{D_e\}.
 \label{app:u14:constraint-algebra}
\end{equation}
The star identities recover every $G_v^4$ locally from either family, and Eq.~\eqref{app:u14:D} then changes between the generators locally. Their simultaneous $+1$ sectors coincide. Both impose $G_v^4=1$, after which $C_e=D_e$. Individual dyon loops remain unrestricted.

For comparison with the coordinate formulas in Appendix~\ref{app:localPauliFactorization}, the shift matrix is twice the semion one:
\begin{equation}
 L_4=4I+2dp=2L,\qquad C_e=X_8^{L_4e}Z_8^{ML_4e}.
 \label{app:u14:L}
\end{equation}
With $K=M-M^\dagger$ and the integer matrices of Eq.~\eqref{eq:M}, Eq.~\eqref{eq:dpc}, and Eq.~\eqref{eq:KLJ}, one has
\begin{gather}
 Md=c,\qquad M^\dagger d=-c,\qquad d^\dagger c=0,\qquad
 pd=-2+2x^{-1}y^{-1},\nonumber\\
 KL_4=8J,\qquad L_4^\dagger K=-8J^\dagger,\qquad
 \det K=4,\qquad \det L_4=16x^{-1}y^{-1}.
 \label{app:u14:identities}
\end{gather}
In these matrix expressions, $e$ denotes its unit edge pattern.

\subsection{Finite-depth disentangler}
\label{app:u14:disentangler}

The loop--flux circuit of Eq.~\eqref{eq:U} now needs a branch observable for an eighth-root flux:
\begin{equation}
 F_4(\ww^j)=(-1)^{\lfloor[j]_8/4\rfloor},
 \qquad [j]_8\in\{0,\ldots,7\}.
 \label{app:u14:branch}
\end{equation}
Functional calculus gives a Hermitian involution $F_4(W_f)$, with $F_4(-W_f)=-F_4(W_f)$. Define
\begin{equation}
 \mathcal U
 =\prod_{v,f}\CZ\!\left(\mathsf{G_v^2},F_4(W_f)\right)^{
                         \int\mathbf p(v)\cup_2\mathbf f}
 =\prod_v\CZ\!\left(\mathsf{G_v^2},F_4(W_v)\right).
 \label{app:u14:U}
\end{equation}
Cup two is pointwise for these two-dimensional flux cochains. The gates commute and have bounded support, so a finite coloring gives a finite-depth circuit. Each $A_e^4$ fixes the normalized loops and sends $W_f$ to $(-1)^{(\delta\mathbf e)(f)}W_f$. So, $\mathcal U A_e^4\mathcal U^\dagger=D_e$ and $\mathcal U D_e\mathcal U^\dagger=A_e^4$. The branch $F_4(W)$ is essential: $W^2$ is unchanged by $W\mapsto-W$ and cannot supply this conjugation.

Decompose each parent register into a physical $\mathbb Z_4$ qudit and an ancillary qubit, generalizing Eq.~\eqref{eq:encoding}
\begin{equation}
 E:|a+4b\rangle\longmapsto |a\rangle_a\otimes H|b\rangle_b,
 \qquad a\in\{0,1,2,3\},\quad b\in\{0,1\}.
 \label{app:u14:E}
\end{equation}
Write $a=a_0+2a_1$ and $P_{0,e}|a\rangle=(-1)^a|a\rangle$. Then, $EX_{8,e}^4E^\dagger=Z_e^b$, $EZ_{8,e}^4E^\dagger=P_{0,e}$, and
\[
 EA_e^4E^\dagger
 =Z_e^b\prod_{e'}P_{0,e'}^{m(e',e)}.
\]
The counterpart of Eq.~\eqref{eq:VM} is the commuting CNOT circuit
\begin{equation}
 V=\prod_{e,e'}\CNOT_{a_{0,e'}\to b_e}^{m(e',e)},\qquad
 \mathcal W=VE\mathcal U,\qquad
 \mathcal W D_e\mathcal W^\dagger=Z_e^b.
 \label{app:u14:W}
\end{equation}
CNOT exponents are reduced modulo two. All controls are physical low bits and all targets are ancillary bits, giving bounded depth. The constraints are independent, and removing the fixed ancillary qubits leaves one unrestricted physical $\mathbb Z_4$ qudit per edge.

Using precisely the reduction of Eq.~\eqref{eq:reduction}, we have
\begin{equation}
 \mathcal R(O):={}_b\langle0|\mathcal W O\mathcal W^\dagger
                                     |0\rangle_b,
 \qquad O\in\mathfrak C_4^\perp.
 \label{app:u14:R}
\end{equation}
It is a local unital $*$-homomorphism on the constraint commutant. As in Eqs.~\eqref{eq:reducedhops} and~\eqref{eq:algebras}, set
\begin{gather}
 a_e=\mathcal R(A_e),\quad
 \overline a_e=\mathcal R(\overline A_e),\quad
 g_v=\mathcal R(G_v),\quad
 \overline g_v=\mathcal R(\overline G_v),\nonumber\\
 \mathfrak A_{4,+}=\Alg\{a_e\},\qquad
 \mathfrak A_{4,-}=\Alg\{\overline a_e\}.
 \label{app:u14:algebras}
\end{gather}
Since $\mathcal R(G_v^4)=1$, $\mathcal R(\mathsf{G_v^2})=g_v^2$. This identifies the normalized squared loop used below.

\subsection{Explicit physical generators}
\label{app:u14:physical}

\begin{figure}[t]
\centering
\begin{subfigure}[c]{.47\linewidth}\centering
\begin{tikzpicture}[x=1.15cm,y=1.15cm]
\foreach \xx in {0,1}{\draw[gridline] (\xx,-1)--(\xx,1);}
\foreach \yy in {-1,0,1}{\draw[gridline] (0,\yy)--(1,\yy);}
\foreach \xx in {0,1}{\foreach \yy in {-1,0,1}{\fill[black!35] (\xx,\yy) circle (.75pt);}}
\draw[edgefocus] (0,0)--(1,0);
\node[op] at (.5,0) {$X_{4}R_4^{-1}$};
\node[op] at (.5,1) {$R_4$};
\node[op] at (0,.5) {$R_4$};
\node[op] at (1,.5) {$R_4^{-1}$};
\node[op] at (0,-.5) {$R_4^{-1}$};
\fill[black] (0,0) circle (1.6pt);
\path[use as bounding box] (-.4,-1.25) rectangle (1.4,1.25);
\end{tikzpicture}
\caption{$\nu_4^{-1}\mathscr B_{h_{\rvec}}$}
\end{subfigure}
\hfill
\begin{subfigure}[c]{.47\linewidth}\centering
\begin{tikzpicture}[x=1.15cm,y=1.15cm]
\foreach \xx in {-1,0,1}{\draw[gridline] (\xx,0)--(\xx,1);}
\foreach \yy in {0,1}{\draw[gridline] (-1,\yy)--(1,\yy);}
\foreach \xx in {-1,0,1}{\foreach \yy in {0,1}{\fill[black!35] (\xx,\yy) circle (.75pt);}}
\draw[edgefocus] (0,0)--(0,1);
\node[op] at (0,.5) {$X_{4}R_4^{-1}$};
\node[op] at (-.5,1) {$R_4$};
\node[op] at (-1,.5) {$R_4$};
\fill[black] (0,0) circle (1.6pt);
\path[use as bounding box] (-1.35,-.75) rectangle (1.35,1.75);
\end{tikzpicture}
\caption{$\nu_4^{-1}\mathscr B_{v_{\rvec}}$}
\end{subfigure}
\medskip %%%%%%
\begin{subfigure}[c]{.47\linewidth}\centering
\begin{tikzpicture}[x=1.15cm,y=1.15cm]
\foreach \xx in {0,1}{\draw[gridline] (\xx,-1)--(\xx,1);}
\foreach \yy in {-1,0,1}{\draw[gridline] (0,\yy)--(1,\yy);}
\foreach \xx in {0,1}{\foreach \yy in {-1,0,1}{\fill[black!35] (\xx,\yy) circle (.75pt);}}
\draw[edgefocus] (0,0)--(1,0);
\node[op] at (.5,0) {$X_{4}R_4^{-1}$};
\node[op] at (.5,-1) {$R_4$};
\node[op] at (1,-.5) {$R_4$};
\fill[black] (0,0) circle (1.6pt);
\path[use as bounding box] (-.4,-1.25) rectangle (1.4,1.25);
\end{tikzpicture}
\caption{$\nu_4^{-1}\overline{\mathscr B}_{h_{\rvec}}$}
\end{subfigure}
\hfill
\begin{subfigure}[c]{.47\linewidth}\centering
\begin{tikzpicture}[x=1.15cm,y=1.15cm]
\foreach \xx in {-1,0,1}{\draw[gridline] (\xx,0)--(\xx,1);}
\foreach \yy in {0,1}{\draw[gridline] (-1,\yy)--(1,\yy);}
\foreach \xx in {-1,0,1}{\foreach \yy in {0,1}{\fill[black!35] (\xx,\yy) circle (.75pt);}}
\draw[edgefocus] (0,0)--(0,1);
\node[op] at (0,.5) {$X_{4}R_4^{-1}$};
\node[op] at (.5,0) {$R_4$};
\node[op] at (-.5,0) {$R_4^{-1}$};
\node[op] at (.5,1) {$R_4^{-1}$};
\node[op] at (1,.5) {$R_4$};
\fill[black] (0,0) circle (1.6pt);
\path[use as bounding box] (-1.35,-.75) rectangle (1.35,1.75);
\end{tikzpicture}
\caption{$\nu_4^{-1}\overline{\mathscr B}_{v_{\rvec}}$}
\end{subfigure}
\caption{The four shift/phase factors for the $U(1)_4$ construction, with the common scalar $\nu_4=e^{3\pi i/8}$ removed. Each black dot marks $\rvec$ and each emphasized edge carries $X_4R_4^{-1}$, with $R_4^{-1}$ acting first. The full generators are $a_e=\mathscr B_e\Gamma_eF_{e,\rvec_1}F_{e,\rvec_2}K_{e,\rvec_1}K_{e,\rvec_2}$ and $\overline a_e=\overline{\mathscr B}_e\Gamma_eK_{e,\rvec_1}K_{e,\rvec_2}$, where $\rvec_1,\rvec_2$ label the two incident faces. The dressing factors enlarge the support beyond the edges labelled here.}
\label{app:u14:fig-stencils}
\end{figure}
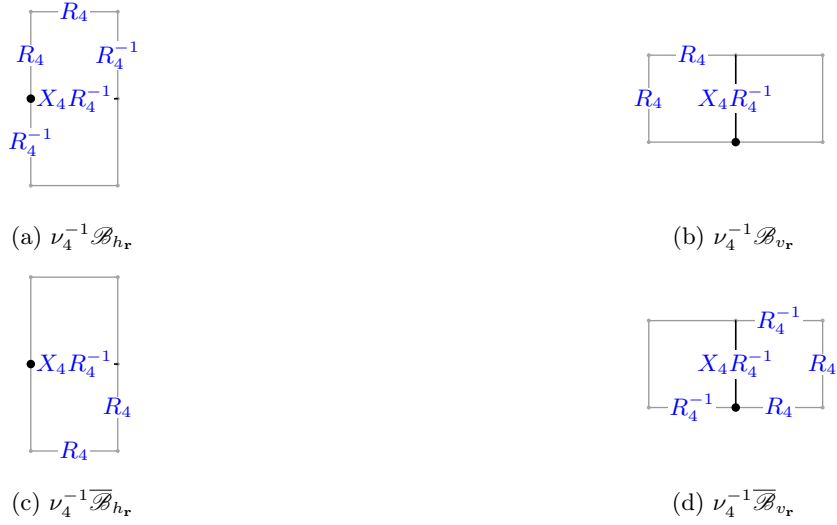

We specify the reduced hopping operators in three steps: the shift--phase factors, the carry correction, and the remaining factors from the disentangling circuit. They are assembled into the full operators in Eq.~\eqref{app:u14:full-generators}. All factors act on the unrestricted physical $\mathbb Z_4$ qudits. Set $X_4|a\rangle=|a+1\bmod4\rangle$ and $Z_4|a\rangle=i^a|a\rangle$, so $Z_4X_4=iX_4Z_4$. The diagonal gates are
\begin{equation}
 R_4|a\rangle=\ww^a|a\rangle,\quad
 P_0|a\rangle=(-1)^a|a\rangle,\quad
 P_1|a\rangle=(-1)^{\lfloor a/2\rfloor}|a\rangle,\quad
 P_{01}|a\rangle=(-1)^{[a=3]}|a\rangle.
 \label{app:u14:physical-gates}
\end{equation}
Write $a=a_0+2a_1$, with $a_0,a_1\in\{0,1\}$. On these two bits, $P_0=Z_{a_0}$, $P_1=Z_{a_1}$, $P_{01}=\CZ_{a_0,a_1}$, and $X_4^2=X_{a_1}$. Both bits belong to the physical $\mathbb Z_4$ qudit and are distinct from the removed ancillary qubit $b$. The shift--phase factors have the cup-product form
\begin{align}
 \mathscr B_e&=\nu_4X_{4,e}\prod_{e'}R_{4,e'}^{m(e',e)},\nonumber\\
 \overline{\mathscr B}_e
             &=\nu_4X_{4,e}\prod_{e'}R_{4,e'}^{m(e,e')}.
 \label{app:u14:physical-cup-hops}
\end{align}
Expanding the same integer cups gives the four shift--phase factors
\begin{align}
 \mathscr B_{h_\rvec}&=\nu_4 X_{4,h_\rvec}R_{4,h_{\rvec+\yhat}}R_{4,h_\rvec}^{-1}
                    R_{4,v_\rvec}R_{4,v_{\rvec+\xhat}}^{-1}R_{4,v_{\rvec-\yhat}}^{-1},
 \nonumber\\
 \mathscr B_{v_\rvec}&=\nu_4 X_{4,v_\rvec}R_{4,h_{\rvec-\xhat+\yhat}}
                    R_{4,v_{\rvec-\xhat}}R_{4,v_\rvec}^{-1},\nonumber\\
 \overline{\mathscr B}_{h_\rvec}&=\nu_4 X_{4,h_\rvec}R_{4,h_{\rvec-\yhat}}R_{4,h_\rvec}^{-1}
                    R_{4,v_{\rvec+\xhat-\yhat}},\nonumber\\
 \overline{\mathscr B}_{v_\rvec}&=\nu_4 X_{4,v_\rvec}R_{4,h_\rvec}R_{4,h_{\rvec-\xhat}}^{-1}
                    R_{4,h_{\rvec+\yhat}}^{-1}R_{4,v_{\rvec+\xhat}}R_{4,v_\rvec}^{-1}.
 \label{app:u14:S}
\end{align}
The four factors are drawn in Fig.~\ref{app:u14:fig-stencils}. The remaining corrections are specified below.

The carry correction has the cup-product expression
\begin{equation}
 \Gamma_e=\prod_{e'\ne e}
       \CZ(P_{01,e},P_{0,e'})^{m(e',e)},\qquad
 \eta_e=\prod_{e'}P_{0,e'}^{m(e',e)}.
 \label{app:u14:carry}
\end{equation}
Its off-site targets are $\mathcal N(h_\rvec)=\{h_{\rvec+\yhat},v_\rvec,v_{\rvec+\xhat},v_{\rvec-\yhat}\}$ and $\mathcal N(v_\rvec)=\{h_{\rvec-\xhat+\yhat},v_{\rvec-\xhat}\}$. Thus, $\eta_e=P_{0,e}\prod_{e'\in\mathcal N(e)}P_{0,e'}$. Each factor of $\Gamma_e$ is a CCZ on the two physical bits of $e$ and the low bit of $e'$. These target lists apply to both chiralities. Define
\begin{gather}
 \Omega_v=\prod_{e'}R_{4,e'}^{\int\mathbf v\cup\delta\mathbf e'},
 \qquad w_v=\Omega_v^2,\qquad
 \Pi_{v,t}=\frac14\sum_{m=0}^3i^{-tm}w_v^m,\nonumber\\
 g_v^2=X_{4,2\delta\mathbf v}\,w_v
       \prod_{e':\,(\delta\mathbf v)(e')\ne0}
        \CZ\!\left((\delta\mathbf v)(e')P_{1,e'},\eta_{e'}\right).
 \label{app:u14:H}
\end{gather}

Here $d_{e'v}=(\delta\mathbf v)(e')$ is $+1$ on incoming star edges and $-1$ on outgoing ones. This completely specifies $g_v^2$ as a nine-edge physical word. In particular,
\[
 \CZ(d_{e'\rvec}P_{1,e'},\eta_{e'})
 =\CZ(d_{e'\rvec}P_{1,e'},P_{0,e'})
   \prod_{e''\in\mathcal N(e')}\CZ(d_{e'\rvec}P_{1,e'},P_{0,e''}),
\]
so the on-site factors are included. One has $g_\rvec^4=\id$ and $[g_\rvec^2,g_s^2]=[g_\rvec^2,w_s]=0$.

The signed face incidences are
\begin{equation}
\begin{array}{c|cc}
 e&(\rvec_1,\epsilon_{e\rvec_1})&(\rvec_2,\epsilon_{e\rvec_2})\\ \hline
 h_\rvec&(\rvec,+1)&(\rvec-\yhat,-1)\\
 v_\rvec&(\rvec,-1)&(\rvec-\xhat,+1).
\end{array}
\label{app:u14:incidence}
\end{equation}
These incidences agree with the boundary-column coefficients:
$\epsilon_{ev}=(\delta\mathbf e)(p(v))=\int\mathbf v\cup\delta\mathbf e=c_{ev}$.
For each incident pair, set
\begin{equation}
 F_{e,s}=F_4(\ww^{\epsilon_{es}}\Omega_s),\qquad
 K_{e,s}=\id+(g_s^2-\id)\Pi_{s,t(\epsilon_{es})},\qquad
 t(+1)=3,\quad t(-1)=0.
 \label{app:u14:FK}
\end{equation}
The complete generators are
\begin{equation}
    \boxed{
    \begin{aligned}
        a_e&=\mathscr B_e\Gamma_e F_{e,\rvec_1}F_{e,\rvec_2}K_{e,\rvec_1}K_{e,\rvec_2},\\
        \overline a_e&=\overline{\mathscr B}_e\Gamma_e K_{e,\rvec_1}K_{e,\rvec_2}.
    \end{aligned}}
    \label{app:u14:full-generators}
\end{equation}
The rightmost conditional factors $K_{e,s}$ act first, followed by the branch phases $F_{e,s}$ for the unbarred hopping operator. Their order is essential: these factors cannot generally be interchanged. The support bounds are fourteen edge registers for a horizontal hopping operator and thirteen for a vertical hopping operator. Eqs.~\eqref{app:u14:S}--\eqref{app:u14:full-generators} specify every gate without ancillary registers or a physical subspace restriction.

\paragraph{Derivation of the carry and conditional factors.}
On a finite torus with $n_e$ edges, the $A_e^4=1$ block has isometry
\begin{equation}
 \mathcal I|a\rangle=2^{-n_e/2}\sum_{b\in\mathbb Z_2^{n_e}}
       (-1)^{b\cdot M^Ta}|a+4b\rangle,
 \qquad \mathcal R(O)=\mathcal I^\dagger \mathcal UO\mathcal U^\dagger\mathcal I.
 \label{app:u14:Jisometry}
\end{equation}
Here $M^T$ is the transpose of the finite-lattice matrix. With $a'=[a+u]_4$ and $k=(a+u-a')/4$, direct summation gives
\begin{equation}
 \mathcal I^\dagger X_8^uZ_8^v\mathcal I|a\rangle
 =\ww^{v\cdot a}(-1)^{k\cdot M^Ta'}
   \delta_{v+M^Tu\,=\,0\ (\mathrm{mod}\ 2)}|a'\rangle.
 \label{app:u14:compression}
\end{equation}
For a hopping step, the carry occurs when the physical digit on $e$ is three, giving precisely $\Gamma_e$. The barred hopping operator preserves this block. For the unbarred hopping operator, the product with the two shifted branch functions preserves it because $(M+M^T)e+\sum_{s\in\mathcal{P}(e)}c_s=0\pmod2$ where $\mathcal{P}(e) := \{v : \chi_e(v) = 1\}$. One must compress that entire product. Taking $(u,v)=(2d_\rvec,2c_\rvec)$ gives Eq.~\eqref{app:u14:H}, since the carry parity on a star edge is $(1-d_{e'\rvec}P_{1,e'})/2$.

For the remaining factors from the disentangling circuit, put $\ell=F_4(W_s)$, $\ell'=F_4(\ww^{\epsilon}W_s)$ and $B=(\id-\ell'\ell)/2$. The elementary identities
\begin{align}
 \CZ(-\mathsf{G_s^2},\ell')\CZ(\mathsf{G_s^2},\ell)
     &=\ell'[\id+(\mathsf{G_s^2}-\id)B],\nonumber\\
 \CZ(\mathsf{G_s^2},\ell')\CZ(\mathsf{G_s^2},\ell)
     &=\id+(\mathsf{G_s^2}-\id)B
 \label{app:u14:spectral}
\end{align}
apply to the unbarred and barred hopping operators, respectively. On the joint $A_e^4=1$ block, $\mathsf{G_s^2}$ becomes $g_s^2$ and $B$ becomes $\Pi_{s,3}$ for $\epsilon=+1$, or $\Pi_{s,0}$ for $\epsilon=-1$. This proves the displayed order in Eq.~\eqref{app:u14:full-generators}.

\subsection{Local completeness and the ISA property}
\label{app:u14:proof}

We give the eight-state version of the local argument in Appendices~\ref{app:localPauliFactorization} and~\ref{app:localCommutants}. Let $\mathfrak H_4=\Alg\{A_e,\overline A_e\}$. A parent Pauli $X_8^uZ_8^v$ commutes with $\mathfrak C_4$ exactly when $L_4^\dagger(v-Mu)=0\pmod8$. For finite integer representatives set
\begin{equation}
 t=v-Mu,\qquad \beta=L_4^\dagger t/8,\qquad
 z=(J^\dagger)^{-1}\beta.
 \label{app:u14:factor}
\end{equation}
These are finite integer patterns and $L_4^\dagger(t+Kz)=0$. The determinant in Eq.~\eqref{app:u14:identities} implies injectivity on integer patterns, so $t=-Kz$ and
\begin{equation}
    X_8^uZ_8^v\ \propto\ A(u-z)\overline A(z).
\label{app:u14:factor-word}
\end{equation}
Ordered words are understood on the right. Every operation in Eq.~\eqref{app:u14:factor} has fixed finite range. Division by eight is coefficientwise exact and introduces no support. Pauli conjugation is diagonal in the Pauli basis, so this proves $\mathfrak C_4^\perp=\mathfrak H_4$ for arbitrary local operators. Conversely, commuting with both hopping families gives $v=Mu=M^\dagger u\pmod8$. With $q=Ku/8$ and $t_0=J^{-1}q$, one has $K(u-L_4t_0)=0$. Since $\det K=4$, the Pauli is a word in $C_e$, proving $\mathfrak H_4^\perp=\mathfrak C_4$.

Every physical $O$ has a local lift $\widetilde O=\mathcal W^\dagger(O\otimes \id_b)\mathcal W\in\mathfrak C_4^\perp$. Expand it in Paulis, apply Eq.~\eqref{app:u14:factor-word}, and restrict. For a uniform constant $R$ this gives
\begin{equation}
    O=\sum_\lambda u_\lambda v_\lambda,\qquad u_\lambda\in\mathfrak A_{4,+}, \quad v_\lambda\in\mathfrak A_{4,-}, \qquad \supp(u_\lambda),\supp(v_\lambda)\subseteq N_R(\supp O).
\label{app:u14:local-reconstruction}
\end{equation}
This proves local surjectivity of multiplication.

For injectivity, the same proof applies on sufficiently large tori: integer periodic patterns are torsion free, so the two determinant arguments remain valid. The two commuting finite-dimensional algebras generate the full physical matrix algebra. Each has scalar center and is therefore a matrix factor. Multiplication from their tensor product is an isomorphism. Any proposed plane kernel element involves only finitely many generator words. Embed those words and their supports, with a buffer, into a large torus. This embedding is faithful on their finite-dimensional spans, so torus injectivity implies plane injectivity.

Finally, suppose that a local operator $O$ commutes with the plus algebra. Embed its reconstruction Eq.~\eqref{app:u14:local-reconstruction} into a torus large enough that every generator overlapping the reconstructed support has the same local action as on the plane. Those generators commute with $O$ by assumption, and all others commute by disjoint support. Thus $O$ lies in the torus commutant of the plus factor. Conditional expectation onto the minus factor gives $O=\sum_\lambda\tau_+(u_\lambda)v_\lambda$, where $\tau_+$ is the normalized trace on the plus factor. The right-hand side retains the same bounded support, so the identity transfers back to the plane. Exchanging signs proves both commutant equalities and completes Eq.~\eqref{app:u14:isa}.

\subsection{Loops, charges, and statistics}
\label{app:u14:statistics}

The ordered physical loops are the counterparts of Eqs.~\eqref{eq:Bloop} and~\eqref{eq:barBloop}. They are
\begin{align}
 g_\rvec&=\ww^{-1}a_{h_{\rvec-\xhat}}a_{h_\rvec}^{-1}
                         a_{v_{\rvec-\yhat}}a_{v_\rvec}^{-1},\nonumber\\
 \overline g_\rvec
        &=\ww\,\overline a_{h_{\rvec-\xhat}}\overline a_{h_\rvec}^{-1}
                         \overline a_{v_{\rvec-\yhat}}\overline a_{v_\rvec}^{-1}.
 \label{app:u14:physical-loops}
\end{align}
They agree with $\mathcal R(G_\rvec)$ and $\mathcal R(\overline G_\rvec)$. All loops commute, and restriction gives
\begin{gather}
 \overline g_v^{\,2}=g_v^2w_v^{-2},\qquad
 g_v^4=\overline g_v^{\,4}=1,\nonumber\\
 a_e^4=\prod_v(g_v^2)^{\chi_e(v)},\qquad
 \overline a_e^{\,4}
       =\prod_v(\overline g_v^{\,2})^{\chi_e(v)},\qquad
 a_e^8=\overline a_e^{\,8}=1,\nonumber\\
 a_eg_v=i^{-c_{ev}}g_va_e,\qquad
 \overline a_e\,\overline g_v=i^{c_{ev}}\overline g_v\,\overline a_e,
 \qquad c_{ev}=\int\mathbf v\cup\delta\mathbf e .
 \label{app:u14:endpoints}
\end{gather}
Opposite chiralities commute. A charge $a$ has loop eigenvalue $i^a$ in the unbarred sector and $i^{-a}$ in the barred sector. A hopping operator transfers one unit between neighboring faces. The loops have exact order four, whereas the displayed hopping operators have exact order eight and fourth powers equal to local squared loops. The charge fusion is therefore of order four. These hopping operators are not the involutive semion generators of Sec.~\ref{sec:primary}.

For completeness, the commuting loop projectors are $P_v^+=\frac14\sum_{j=0}^3g_v^j$ and $P_v^-=\frac14\sum_{j=0}^3\overline g_v^{\,j}$. On a sufficiently large periodic square lattice with $n$ vertices, $\prod_vg_v=\prod_v\overline g_v=1$. Independent unbarred and barred hopping operators act transitively on the $4^{2n-2}$ pairs of zero-total loop characters. At least one simultaneous eigenspace exists, so all these eigenspaces occur with equal multiplicity. The physical dimension is $4^{2n}$, and the common $+1$ eigenspace has dimension sixteen. Thus, the joint loop model is frustration free.

On the plane, the syndrome records the changes of loop eigenvalues produced by an ordered product of hopping operators. A finite pattern of hopping exponents has zero syndrome precisely when $c^\dagger u=0\pmod4$. Filling a finite dual-lattice cycle over $\mathbb Z_4$ gives $u=dv+4w$ with finite integer patterns $v,w$. The ordered star relation and the fourth-power relation in Eq.~\eqref{app:u14:endpoints} reduce every such word to a loop word, up to scalar. Averaging loop conjugations projects arbitrary local expressions onto their syndrome components. The zero-syndrome algebras within $\mathfrak A_{4,+}$ and $\mathfrak A_{4,-}$ are $\Alg\{g_v\}$ and $\Alg\{\overline g_v\}$, respectively; in the full physical algebra, the joint zero-syndrome algebra is $\Alg\{g_\rvec,\overline g_\rvec\}$.

Torus vacuum limits exist, and their local expectations are uniquely fixed: nonzero-syndrome words have expectation zero, while loop words are evaluated at $g_v=\overline g_v=1$. Words with the same syndrome differ only by loop words on this vacuum.

Finite ray-prefix conjugations stabilize on each local observable, because remote hopping operators commute with it and all commutators of hopping operators are scalar. Thus finite strings and their ray limits have charge labels $\mathbb Z_4$ in each chirality, with one-dimensional fusion channels. A large boundary loop measures their total charge. These are the explicit string-charge sectors of this loop model.

To fix the exchange phase, take the three outward parent hopping operators from the face at the origin,
\begin{equation}
 W_R=A_{v_{(1,0)}}^{-1},\qquad W_U=A_{h_{(0,1)}},\qquad
 W_L=A_{v_{(0,0)}}.
\end{equation}
Writing $O_1O_2=\ww^{\chi(O_1,O_2)}O_2O_1$, their exponents are
\[
 \chi(W_R,W_U^\dagger)=1,\qquad \chi(W_R,W_L)=-1,\qquad
 \chi(W_U^\dagger,W_L)=1.
\]
Hence
\begin{equation}
    W_RW_U^\dagger W_L=\ww W_LW_U^\dagger W_R.
\label{app:u14:exchange}
\end{equation}
Restriction preserves this identity. With the orientation convention of Sec.~\ref{sec:statistics} and Eq.~\eqref{eq:exchangeidentity}, the spin is $\ww$. The barred family conjugates the phase, and taking $a$th powers multiplies each commutator exponent by $a^2$. Therefore
\begin{equation}
 \theta_\pm(a)=e^{\pm\pi i a^2/4},\qquad
 B_\pm(a,b)=e^{\pm\pi iab/2},\qquad a,b\in\mathbb Z_4.
 \label{app:u14:quadratic}
\end{equation}
The braiding is nondegenerate, and the cross braiding is trivial. These are $U(1)_4$ and its orientation reverse. Their Gauss sums are $\frac12\sum_a\theta_\pm(a)=e^{\pm\pi i/4}$, giving $c_-=\pm1\pmod8$.

As an independent check, condensing the order-two boson $(4,4)$ in the parent $\mathbb Z_8$ toric theory keeps labels with $e+m$ even and identifies $(e,m)\sim(e+4,m+4)$. The isomorphism
\begin{equation}
 (a,b)\in\mathbb Z_4^2\longmapsto[(a+b,a-b)]
 \in\{(e,m)\in\mathbb Z_8^2:e+m\text{ even}\}/\langle(4,4)\rangle
\end{equation}
pulls the spin $\ww^{em}$ back to $\ww^{a^2-b^2}$. This agrees with the lattice-level calculation. The ISA proof above also establishes the required unconstrained physical operator algebra.

\subsection{Completion of the pointed Witt generators}
\label{app:u14:witt}

We now state precisely the completeness supplied by this extension. Bosonic Abelian anyon theories are described by nondegenerate finite quadratic forms $(A,q)$. Their Witt group is the pointed subgroup $W_{\mathrm{pt}}$ of the Witt group of nondegenerate braided fusion categories. Orthogonal sum is stacking and $(A,q^{-1})$ gives the inverse. Isotropic reduction replaces $(A,q)$ by $(H^\perp/H,\bar q)$ for an isotropic subgroup $H\subset A$, meaning $q(h)=1$ for every $h\in H$. Here $H^\perp=\{a\in A:B(a,h)=1\text{ for all }h\in H\}$. This reduction preserves the Witt class. The classification is~\cite[Sec.~5.3]{Davydov2013Witt}
\begin{equation}
 W_{\mathrm{pt}}\cong
 (\mathbb Z_8\oplus\mathbb Z_2)
 \oplus\!\bigoplus_{\substack{p\text{ prime}\\p\equiv1\ (\mathrm{mod}\ 4)}}
          (\mathbb Z_2\oplus\mathbb Z_2)
 \oplus\!\bigoplus_{\substack{p\text{ prime}\\p\equiv3\ (\mathrm{mod}\ 4)}}
          \mathbb Z_4.
 \label{app:u14:Wpt}
\end{equation}
Every element has only finitely many nonzero primary components.

The earlier odd-prime ISAs~\cite[Sec.~V.B]{Sun2026} realize the cyclic quadratic data
\begin{equation}
 \mathsf C_p^{(k)}=
 \left(\mathbb Z_p,\ q_k(a)=e^{2\pi i k a^2/p}\right),
 \qquad k\in\mathbb F_p^\times.
 \label{app:u14:odd-prime}
\end{equation}
A relabeling $a\mapsto ua$ multiplies $k$ by a square. For $p\equiv1\pmod4$, the square and nonsquare classes give the two $\mathbb Z_2$ generators. Both are included in the earlier family. For $p\equiv3\pmod4$, either class generates $\mathbb Z_4$ and the other is its inverse. Thus the odd-prime primary components are covered.

The remaining component is $W_{\mathrm{pt}}(2)\cong\mathbb Z_8\oplus \mathbb Z_2$. Write
\begin{equation}
 \mathsf S=(\mathbb Z_2,q(1)=i),\qquad
 \mathsf C_4=(\mathbb Z_4,q(a)=e^{\pi i a^2/4}).
\end{equation}
The semion class $[\mathsf S]$ generates the $\mathbb Z_8$ factor. The present construction realizes the missing class: in coordinates with $[\mathsf S]=(1,0)$, one may take
\begin{equation}
 [\mathsf C_4]=(1,1),\qquad
 [\mathsf C_4]-[\mathsf S]=(0,1),\qquad
 2[\mathsf C_4]=2[\mathsf S].
 \label{app:u14:witt-generators}
\end{equation}
The last relation also has a direct condensation proof. In $\mathsf C_4\boxtimes\mathsf C_4$, the subgroup $H=\langle(2,2)\rangle$ is isotropic. Its orthogonal complement has $a+b$ even, and $H^\perp/H$ is generated by the classes of $(1,1)$ and $(1,3)$. Both have spin $i$ and their mutual braiding is one, so the reduced metric group is $\mathsf S\boxtimes\mathsf S$.

To see that the extra class is nonzero, use two Witt invariants of a $2$-primary metric group $(A,q)$
\begin{equation}
 \delta_2(A)=v_2(|A|)\pmod2,\qquad
 \xi(A,q)=|A|^{-1/2}\sum_{a\in A}q(a).
\end{equation}
Isotropic reduction changes $|A|$ by the square $|H|^2$ and preserves $\xi$. The semion has $(\delta_2,\xi)=(1,\ww)$, whereas $\mathsf C_4$ has $(0,\ww)$. If $[\mathsf C_4]=n[\mathsf S]$, the first invariant requires $n$ even, while the second requires $n=1\pmod8$, a contradiction. Primary decomposition prevents odd-prime classes from removing this obstruction. Hence
\begin{equation}
 W_{\mathrm{pt}}\big/\left\langle[\mathsf S],
       [\mathsf C_p^{(k)}]:p\text{ odd}\right\rangle\cong\mathbb Z_2,
\end{equation}
and $[\mathsf C_4]$ supplies its generator.

Finally, finite stacking preserves the ISA property. That is, tensor the local factorizations and take the maximum of their reconstruction radii. Orientation reversal supplies inverse Witt classes. It follows that \emph{the semion ISA, the present $U(1)_4$ ISA, and the earlier odd-prime ISAs give, by finite stacking, explicit local ISAs whose string-charge theories represent every class in the pointed Witt group of bosonic Abelian anyon theories}.

\subsection{Higher-dimensional construction}
\label{app:u14:higherdimensions}

The eight-state construction also extends to spatial dimensions $d=4k{-}2$. Set $\ell=2k-1$, so $d=2\ell$, and use the oriented hypercubic lattice, signed integer cup products, and complementary cells $p(c_q)$ of Appendix~\ref{app:hdsem}. In particular, $\mathbf p(c_{\ell-1})$ has degree $\ell+1$ and obeys Eq.~\eqref{eq:hdsem:cupdual}. Work on the infinite lattice or a periodic lattice larger than all local supports. Each $\ell$-cell carries a parent $\mathbb Z_8$ qudit. After condensation and local disentangling, it leaves one unrestricted physical $\mathbb Z_4$ qudit. The construction below establishes the parent commutants, independent constraints, and uniformly local physical factorization.

\paragraph{Parent hopping operators and loops.} Let $Z_8X_8=\ww X_8Z_8$, where $\ww=e^{\pi i/4}$, and write $X_{8,\alpha}=\prod_{c_\ell}X_{8,c_\ell}^{\alpha(c_\ell)}$ for an integer $\ell$-cochain $\alpha$. Shifts precede clocks. The parent hopping operators use the same cup kernels as in Eq.~\eqref{eq:hdsem:hops}:
\begin{align}
 A_{c_\ell}
 &=\nu(d)X_{8,c_\ell}
   \prod_{c'_\ell}Z_{8,c'_\ell}^{\int(
      \mathbf c'_\ell\cup\mathbf c_\ell
       -\delta\mathbf c'_\ell\cup_1\mathbf c_\ell)},\nonumber\\
 \overline A_{c_\ell}
 &=\nu(d)X_{8,c_\ell}
   \prod_{c'_\ell}Z_{8,c'_\ell}^{\int(
      \mathbf c_\ell\cup\mathbf c'_\ell
       -\delta\mathbf c_\ell\cup_1\mathbf c'_\ell)},\nonumber\\
 \nu(d)&=
 \begin{cases}
 e^{3\pi i/8},&d=2,\\
 1,&d=6,10,14,\ldots .
 \end{cases}
 \label{app:u14:hd:hops}
\end{align}
The decorations are transposes, so the two hopping families commute with one another. Their on-site clock power is $-1$ for $\ell=1$ and $0$ for $\ell\ge3$, by the cubical support argument following Eq.~\eqref{eq:hdsem:hops}. Raising a bare hopping operator to the fourth power produces the scalar $\ww^{6m}$ when this clock power is $m$. The stated prefactor cancels it. Thus,
\begin{equation}
 A_{c_\ell}^4=X_{8,c_\ell}^4
   \prod_{c'_\ell}Z_{8,c'_\ell}^{4\int(
      \mathbf c'_\ell\cup\mathbf c_\ell
       -\delta\mathbf c'_\ell\cup_1\mathbf c_\ell)},
 \qquad A_{c_\ell}^8=\overline A_{c_\ell}^{\,8}=1.
 \label{app:u14:hd:fourthpowers}
\end{equation}
The barred fourth power has the transposed decoration. All these fourth powers are commuting Hermitian involutions.

For an $(\ell-1)$-cochain $\eta$, define the raw dyon loops and ordinary fluxes by
\begin{align}
 G(\eta)&=X_{8,\delta\eta}
       \prod_{c_\ell}Z_{8,c_\ell}^{\int\eta\cup\delta\mathbf c_\ell},
 &\overline G(\eta)&=X_{8,\delta\eta}
       \prod_{c_\ell}Z_{8,c_\ell}^{-\int\eta\cup\delta\mathbf c_\ell},
 \nonumber\\
 W_{c_{\ell+1}}&=\prod_{c_\ell}
       Z_{8,c_\ell}^{(\delta\mathbf c_\ell)(c_{\ell+1})}.
 \label{app:u14:hd:loops}
\end{align}
Write $G_{c_{\ell-1}}=G(\mathbf c_{\ell-1})$, and similarly for $\overline G$. Stokes' identity gives zero integer clock--shift pairings between any two loops and between a loop and a flux. Consequently, the relations
\begin{equation}
 G(\eta+\eta')=G(\eta)G(\eta'),\qquad
 G(\delta\xi)=1,\qquad
 G_{c_{\ell-1}}^4=\overline G_{c_{\ell-1}}^{\,4}
 \label{app:u14:hd:additive}
\end{equation}
are exact, including their scalar phases. The barred loops are also additive, and all raw loops and fluxes commute. The integrated cup-one identities in Eq.~\eqref{eq:hdsem:loopidentities} place each loop in its corresponding hopping algebra. A hopping operator has loop commutator $i^{-\int\eta\cup\delta\mathbf c_\ell}$ in the unbarred family and its conjugate in the barred family. Hence every loop fourth power is central in the joint hopping algebra, and every hopping fourth power commutes with all raw loops.

\paragraph{Condensates and parent completeness.}
Double the mixed cup-one dressing of Eq.~\eqref{eq:hdsem:C}:
\begin{align}
 C_{c_\ell}
 &=A_{c_\ell}^4\prod_{c_{\ell-1}}G_{c_{\ell-1}}^{
       2\int(\mathbf p(c_{\ell-1})\cup_1\mathbf c_\ell
                     -\mathbf c_\ell\cup_1\mathbf p(c_{\ell-1}))}
 \nonumber\\
 &=\overline A_{c_\ell}^{\,4}
       \prod_{c_{\ell-1}}\overline G_{c_{\ell-1}}^{
       2\int(\mathbf p(c_{\ell-1})\cup_1\mathbf c_\ell
                     -\mathbf c_\ell\cup_1\mathbf p(c_{\ell-1}))}.
 \label{app:u14:hd:C}
\end{align}
Both expressions have the same shifts. Testing the difference of their clock decorations against $\beta$ gives $8\int\beta\cup\mathbf c_\ell$, so the clocks agree modulo eight. The crossed powers in $A^4G^{2\eta}$ are multiples of eight. Thus, the equality also includes the scalar phase. To check centrality, retain integer exponents before reduction modulo eight. The commutator of $C_{c_\ell}$ with a hopping product having integer shifts $\beta$ is
\begin{align}
 &4\int(\beta\cup\mathbf c_\ell
       -\mathbf c_\ell\cup\beta
       +\delta\mathbf c_\ell\cup_1\beta
       -\mathbf c_\ell\cup_1\delta\beta)
   =8\int\beta\cup\mathbf c_\ell .
 \label{app:u14:hd:commutatorexponent}
\end{align}
This uses Eq.~\eqref{eq:hdsem:cupone}. The resulting phase is one. The barred expression proves centrality in the other family. Moreover,
\begin{equation}
 C_{c_\ell}^{\,2}
 =\prod_{c_{\ell-1}}(G_{c_{\ell-1}}^4)^{
       \int(\mathbf p(c_{\ell-1})\cup_1\mathbf c_\ell
                     -\mathbf c_\ell\cup_1\mathbf p(c_{\ell-1}))},
 \qquad C_{c_\ell}^{\,4}=1.
 \label{app:u14:hd:Csquares}
\end{equation}

The integer completeness argument of Thm~\ref{thm:hdsem:parent} applies with eight in place of four. We spell out the reconstruction. The antisymmetric hopping pairing is
\begin{equation}
 \int(\beta\cup\alpha-\delta\beta\cup_1\alpha
                       -\alpha\cup\beta+\delta\alpha\cup_1\beta).
 \label{app:u14:hd:pairing}
\end{equation}
A condensate word has the unbarred hopping clock decoration for its accumulated integer shifts, up to scalar. If an integer shift pattern has zero antisymmetric hopping pairing with every test, it pairs to zero with every condensate shift. By Eq.~\eqref{app:u14:hd:commutatorexponent}, ordinary cup duality then forces the pattern to vanish. Conversely, for any integer pattern $\beta$, ordinary cup duality represents its hopping pairing as an ordinary cup pairing, with the hopping test on the left and a locally determined cup-dual cochain on the right. The condensate word labelled by that cochain has the same hopping responses as $8\beta$. Nondegeneracy identifies its integer shifts with $8\beta$. Thus, condensate shifts span every eightfold integer pattern locally.

Let a parent Pauli commuting with all $C$ have shifts $u$. Remove an unbarred hopping word with the same shifts. The remaining pure clock Pauli evaluates to a multiple of eight on each condensate shift. Divide these evaluations by eight, and use ordinary cup duality to choose $b$ whose pairing $\int b\cup\mathbf c_\ell$ equals the corresponding quotient. An unbarred word with shifts $-b$ followed by a barred word with shifts $b$ has precisely the same clock responses on the condensate shifts. The difference vanishes on all eightfold patterns and hence on every integer pattern. The original Pauli therefore factors, up to scalar, into an unbarred word with shifts $u-b$ and a barred word with shifts $b$.

Conversely, a Pauli commuting with both hopping families has clocks agreeing modulo eight with both hopping decorations for its shifts $u$. Their difference is divisible by eight. Choose a local cochain $a$ by ordinary cup duality so that, for every $\beta$,
\begin{equation}
 \int\beta\cup a=\frac18\int(
        \beta\cup u-\delta\beta\cup_1u
                         -u\cup\beta+\delta u\cup_1\beta).
 \label{app:u14:hd:centralreconstruction}
\end{equation}
The condensate word with exponents $a(c_\ell)$ has the same integer hopping responses as $u$, and therefore the same shifts. Its clocks agree with those of the Pauli modulo eight. This identifies the Pauli as a condensate word up to scalar.

Pauli conjugation is diagonal in the Pauli basis, so these reconstructions extend to all local operators. Writing $\mathfrak H_4=\Alg\{A_{c_\ell},\overline A_{c_\ell}\}$ and $\mathfrak C_4=\Alg\{C_{c_\ell}\}$, they prove
\begin{equation}
 \mathfrak C_4^\perp=\mathfrak H_4,\qquad
 \mathfrak H_4^\perp=\mathfrak C_4,
 \qquad\operatorname{Center}(\mathfrak H_4)=\mathfrak C_4.
 \label{app:u14:hd:parentcommutants}
\end{equation}
Every operation uses local cups, coboundaries, complementary-cell displacements, or exact coefficientwise division by eight. Hence support enlargement is uniformly bounded. Ordinary cup duality and the integer arguments remain valid on sufficiently large finite tori.

\paragraph{Equivalent involutive constraints.}
Normalize the order-four loop squares using the spectral function
$S$ of Eq.~\eqref{eq:S}:
\begin{equation}
 \mathsf{G_{c_{\ell-1}}^2}
   :=G_{c_{\ell-1}}^2S(G_{c_{\ell-1}}^4)^\dagger,
 \qquad
 \overline{\mathsf{G_{c_{\ell-1}}^{\,2}}}
   :=\overline G_{c_{\ell-1}}^{\,2}
                  S(G_{c_{\ell-1}}^4)^\dagger.
 \label{app:u14:hd:normalizedloops}
\end{equation}
These are commuting Hermitian involutions. The cup-two correction is
\begin{align}
 D_{c_\ell}
 &:=A_{c_\ell}^4\prod_{c_{\ell-1}}
       \mathsf{G_{c_{\ell-1}}^2}^{
              \int\mathbf p(c_{\ell-1})\cup_2\delta\mathbf c_\ell}
 \nonumber\\
 &=\overline A_{c_\ell}^{\,4}\prod_{c_{\ell-1}}
       \overline{\mathsf{G_{c_{\ell-1}}^{\,2}}}^{
              \int\mathbf p(c_{\ell-1})\cup_2\delta\mathbf c_\ell}
 \nonumber\\
 &=C_{c_\ell}\prod_{c_{\ell-1}}
       [S(G_{c_{\ell-1}}^4)^\dagger]^{
          \int(\mathbf p(c_{\ell-1})\cup_2\delta\mathbf c_\ell
                      -2\mathbf c_\ell\cup_1\mathbf p(c_{\ell-1}))}.
 \label{app:u14:hd:D}
\end{align}
To prove the last equality, use the exact integer cup-two identity in Eq.~\eqref{eq:hdsem:integercuptwo}. Its final term, involving $\delta\mathbf p(c_{\ell-1})$, gives an exact lower-degree cochain in the loop exponents. Its raw loop product is one by Eq.~\eqref{app:u14:hd:additive}. The remaining mixed cup-one dressing is therefore the cup-two dressing in raw loop squares, times fourth powers with exponent $-\int\mathbf c_\ell\cup_1\mathbf p(c_{\ell-1})$. Using $S(G^4)^2=G^4$ gives the stated correction, including its scalar phases. Applying the same argument to the barred loops proves the second equality. The last line is a local function of central loop fourth powers multiplying $C_{c_\ell}$, so $D_{c_\ell}$ is central. The first line is a product of commuting involutions, so $D_{c_\ell}^2=1$. The spectral-correction exponents in the last line are integers; their even parts contribute loop fourth powers and must be retained.

Both constraint families locally recover the central loop fourth powers
\begin{equation}
 \prod_{c_\ell}D_{c_\ell}^{(\delta\mathbf c_{\ell-1})(c_\ell)}
       =G_{c_{\ell-1}}^4,
 \qquad
 \prod_{c_\ell}C_{c_\ell}^{(\delta\mathbf c_{\ell-1})(c_\ell)}
       =G_{c_{\ell-1}^{\leftarrow}}^4.
 \label{app:u14:hd:stars}
\end{equation}
Here, $c_{\ell-1}^{\leftarrow}$ is the all-coordinate backward translate selected by the reversed ordinary cup pairing, as in Eq.~\eqref{eq:hdsem:stars}. For the first identity, $\delta^2=0$ cancels the normalized loop dressing. The hopping fourth powers give $G_{c_{\ell-1}}^4$ exactly. For the second, the integer congruence underlying Eq.~\eqref{eq:hdsem:stars}, multiplied by two, makes the raw loop dressing $G_{c_{\ell-1}^{\leftarrow}}^4G_{c_{\ell-1}}^{-4}$; it cancels the original loop fourth power. All crossed phases are one. Hopping fourth powers crossing one another give multiples of sixteen in the Pauli exponent, while their clocks crossing squared-loop shifts give multiples of eight.

The backward translation has a local inverse. Hence the star relations and Eq.~\eqref{app:u14:hd:D} prove
\begin{equation}
 \Alg\{C_{c_\ell}\}=\Alg\{D_{c_\ell}\}=\mathfrak C_4
 \label{app:u14:hd:equalconstraints}
\end{equation}
with uniformly local changes of generators. In either simultaneous $+1$ sector, Eq.~\eqref{app:u14:hd:stars} first forces every $G_{c_{\ell-1}}^4=1$. The correction in Eq.~\eqref{app:u14:hd:D} is then the identity, so the two sectors coincide. Individual loops and their squares remain unrestricted. For $\ell=1$, the normalized-loop exponent is the signed face incidence, whose sign is immaterial for an involution. The last exponent in Eq.~\eqref{app:u14:hd:D} is one on each incident face. Thus, this constraint reduces to the two-dimensional constraint above.

\paragraph{A cup-product disentangler.}
Retain the eighth-root branch observable used in the two-dimensional circuit:
\begin{equation}
 F_4(\ww^j)=(-1)^{\lfloor[j]_8/4\rfloor},\qquad
 [j]_8\in\{0,1,\ldots,7\}.
 \label{app:u14:hd:branch}
\end{equation}
It is a Hermitian involution and obeys $F_4(-W)=-F_4(W)$.
Using the controlled phase of Eq.~\eqref{eq:CZ}, set
\begin{equation}
 \mathcal U=\prod_{c_{\ell-1},c'_{\ell+1}}
   \CZ\!\left(\mathsf{G_{c_{\ell-1}}^2},
                         F_4(W_{c'_{\ell+1}})\right)^{
          \int\mathbf p(c_{\ell-1})\cup_2\mathbf c'_{\ell+1}}.
 \label{app:u14:hd:U}
\end{equation}
The loop and flux labels are independent. All gates commute, have bounded support, and square to one. A finite coloring of their overlap graph schedules them in bounded depth for fixed $d$. In two dimensions the cup-two pairing of top-degree cochains is pointwise, giving the loop--flux circuit above. In higher dimensions its off-diagonal pairings remain part of the circuit.

A hopping fourth power fixes the normalized loops and acts on a flux by
\begin{equation}
 A_{c_\ell}^4W_{c_{\ell+1}}A_{c_\ell}^4
     =(-1)^{(\delta\mathbf c_\ell)(c_{\ell+1})}
                                           W_{c_{\ell+1}}.
 \label{app:u14:hd:fluxflip}
\end{equation}
The same sign flips the branch observable. Conjugating the controlled phases therefore attaches precisely the normalized loops selected by the cup-two pairing:
\begin{equation}
 \mathcal U A_{c_\ell}^4\mathcal U^\dagger=D_{c_\ell},
 \qquad
 \mathcal U D_{c_\ell}\mathcal U^\dagger=A_{c_\ell}^4.
 \label{app:u14:hd:Uaction}
\end{equation}

Decompose each parent register into a physical $\mathbb Z_4$ qudit and an ancillary qubit as
\begin{equation}
 E:|a+4b\rangle\longmapsto|a\rangle_a\otimes H|b\rangle_b,
 \qquad a\in\{0,1,2,3\},\quad b\in\{0,1\}.
 \label{app:u14:hd:E}
\end{equation}
Write $a=a_0+2a_1$ for the two physical bits, and define $P_{0,c_\ell}|a\rangle=(-1)^a|a\rangle$. Thus, $P_{0,c_\ell}=Z_{4,c_\ell}^2=Z_{a_{0,c_\ell}}$, with $Z_4|a\rangle=i^a|a\rangle$ and $X_4|a\rangle=|a+1\bmod4\rangle$. The decomposition gives
\begin{align}
 EX_{8,c_\ell}^4E^\dagger=Z_{c_\ell}^b, \quad
 EZ_{8,c_\ell}^4E^\dagger=P_{0,c_\ell}, \quad
 EA_{c_\ell}^4E^\dagger
 =Z_{c_\ell}^b\prod_{c'_\ell}P_{0,c'_\ell}^{
          \int(\mathbf c'_\ell\cup\mathbf c_\ell
                      -\delta\mathbf c'_\ell\cup_1\mathbf c_\ell)}.
 \label{app:u14:hd:encoded}
\end{align}
The commuting controlled gates
\begin{equation}
 V=\prod_{c_\ell,c'_\ell}
   \CNOT_{a_{0,c'_\ell}\to b_{c_\ell}}^{
          \int(\mathbf c'_\ell\cup\mathbf c_\ell
                      -\delta\mathbf c'_\ell\cup_1\mathbf c_\ell)}
 \label{app:u14:hd:V}
\end{equation}
remove these clock decorations. Only the parity of the exponent enters each CNOT. Every control is a physical low bit and every target an ancillary qubit, so the gates commute and admit a bounded-depth schedule. The complete disentangler is
\begin{equation}
 \mathcal W=VE\mathcal U,\qquad
 \boxed{\mathcal W D_{c_\ell}\mathcal W^\dagger=Z_{c_\ell}^b.}
 \label{app:u14:hd:W}
\end{equation}
This proves independence of the constraints. On a finite torus with $N$ parent cells, their common $+1$ sector has dimension $8^N/2^N=4^N$. Removing the fixed ancillary qubits leaves one unrestricted physical $\mathbb Z_4$ qudit per $\ell$-cell.

\paragraph{Physical algebras and local factorization.}
For a parent operator commuting with the constraints, define
\begin{equation}
 \mathcal R(O):={}_b\langle0|\mathcal W O\mathcal W^\dagger|0\rangle_b.
 \label{app:u14:hd:R}
\end{equation}
Commuting with every $Z^b$ makes this restriction a unital $*$-homomorphism on $\mathfrak C_4^\perp$. It maps onto the full local physical operator algebra: a local $O_a$ has the local lift $\mathcal W^\dagger(O_a\otimes1_b)\mathcal W$. Define
\begin{align}
 a_{c_\ell}&=\mathcal R(A_{c_\ell}),
 &\overline a_{c_\ell}&=\mathcal R(\overline A_{c_\ell}),\nonumber\\
 g_{c_{\ell-1}}&=\mathcal R(G_{c_{\ell-1}}),
 &\overline g_{c_{\ell-1}}&=\mathcal R(\overline G_{c_{\ell-1}}),\nonumber\\
 \mathfrak A_{4,+}&=\Alg\{a_{c_\ell}\},
 &\mathfrak A_{4,-}&=\Alg\{\overline a_{c_\ell}\}.
 \label{app:u14:hd:physicalalgebras}
\end{align}
The loops lie in their respective algebras. They obey $g_{c_{\ell-1}}^4=\overline g_{c_{\ell-1}}^{\,4}=1$, while the reduced hopping operators obey $a_{c_\ell}^8=\overline a_{c_\ell}^{\,8}=1$. Since every gate in $\mathcal W$ has explicit bounded support, the reductions in Eq.~\eqref{app:u14:hd:physicalalgebras} specify local operators on the unrestricted four-state physical registers.

\begin{theorem}[Higher-dimensional four-state ISA]
\label{thm:u14:hdISA}
For every $d=4k{-}2$, the construction gives complementary local
$*$-algebras on one physical $\mathbb Z_4$ qudit per $(2k{-}1)$-cell.
Within the physical local operator algebra $\mathcal M_4$,
\begin{equation}
 [\mathfrak A_{4,+},\mathfrak A_{4,-}]=0,\qquad
 \mathfrak A_{4,+}^\perp=\mathfrak A_{4,-},\qquad
 \mathfrak A_{4,-}^\perp=\mathfrak A_{4,+},
 \label{app:u14:hd:commutants}
\end{equation}
and multiplication gives an isomorphism
\begin{equation}
 \mathfrak A_{4,+}\otimes_{\mathrm{alg}}\mathfrak A_{4,-}
       \xrightarrow[\ u\otimes v\mapsto uv\ ]{\ \cong\ }\mathcal M_4.
 \label{app:u14:hd:ISA}
\end{equation}
Every local physical operator has a finite decomposition
\begin{equation}
 O_a=\sum_\lambda u_\lambda v_\lambda,\qquad
 u_\lambda\in\mathfrak A_{4,+},\quad v_\lambda\in\mathfrak A_{4,-},\qquad
 \supp u_\lambda,\supp v_\lambda
       \subseteq N_R(\supp O_a),
 \label{app:u14:hd:localfactorization}
\end{equation}
where $R$ depends only on $d$. The statements hold on the
infinite lattice and sufficiently large periodic lattices.
\end{theorem}

\begin{proof}
The parent hopping families commute, so their reduced algebras commute. Lift $O_a$ using $\mathcal W^\dagger(O_a\otimes1_b)\mathcal W$. The lift belongs to $\mathfrak C_4^\perp$. Expand it in parent Paulis and apply the uniformly local reconstruction proving Eq.~\eqref{app:u14:hd:parentcommutants}. Each Pauli is an unbarred hopping word times a barred hopping word, up to scalar. Restriction by the homomorphism $\mathcal R$ gives Eq.~\eqref{app:u14:hd:localfactorization}. The finite-depth lifts, parent reconstruction, and restriction all enlarge support by bounded distances, so their composition has the stated uniform bound.

On a finite torus the two commuting reduced algebras jointly generate the full matrix algebra on the $4^N$-dimensional physical space. A central element of either commutes with both and is therefore scalar. Each algebra is a matrix factor. Finite-dimensional matrix-factor decomposition then proves their mutual-commutant relations and injectivity of tensor-product multiplication.

For the infinite lattice, let a local $O_a$ commute with $\mathfrak A_{4,+}$ and take its separated reconstruction in Eq.~\eqref{app:u14:hd:localfactorization}. Embed this neighborhood, with a buffer larger than all generator ranges, into a finite torus. Generators overlapping the support have the same action as on the infinite lattice and commute with $O_a$. All others commute by disjoint support. Thus $O_a$ belongs to the torus commutant of the plus factor. Conditional expectation onto the minus factor fixes it and sends each separated term to $\operatorname{tr}_+(u_\lambda)v_\lambda$, where $\operatorname{tr}_+$ is the normalized factor trace. This retains the support bound and expresses $O_a$ in $\mathfrak A_{4,-}$ on the infinite lattice. Interchanging the signs proves the other commutant relation.

Finally, every proposed kernel element of tensor-product multiplication involves finite-dimensional spans of finitely many local generator words. Their buffered torus embedding is faithful on these spans, since all words act identically on the same finite physical registers. Torus injectivity therefore proves injectivity on the infinite lattice as well.
\end{proof}

For $k=1$, this recovers the two-dimensional $U(1)_4$ and orientation-reversed ISAs above. For higher $k$, the theorem establishes their local algebraic generalization on unrestricted physical $\mathbb Z_4$ qudits. As for the semion construction in Appendix~\ref{app:hdsem}, layer transport defines QCA pumps in one higher spatial dimension~\cite{Haah2023}. The higher-dimensional excitation statistics and comparison with TQFT boundary algebras require their own analysis.

\bibliographystyle{utphys}
\bibliography{references}

\end{document}